\documentclass[a4paper,11pt]{article}
\pdfoutput=1 
\usepackage{graphicx}
\graphicspath{ {figures/} }
\usepackage{jinstpub} 
\usepackage{ensuremath}
\usepackage{xspace}
\usepackage{url}

\def\argon40{${}^{40}$Ar}       
\def\Ar39{$^{39}$Ar}
\def\Cl40{$^{40}$Cl}
\def\K40{$^{40}$K}
\def\B8{$^{8}$B}

\newcommand{\fdth}{feedthrough\xspace} 

\newcommand{\lsim}{{\;\raise0.3ex\hbox{$<$\kern-0.75em\raise-1.1ex\hbox{$\sim$}}\;}}
\newcommand{\gsim}{{\;\raise0.3ex\hbox{$>$\kern-0.75em\raise-1.1ex\hbox{$\sim$}}\;}}
\newcommand{\beq}{\begin{equation}}
\newcommand{\eeq}{\end{equation}}
\newcommand{\bea}{\begin{eqnarray}}
\newcommand{\eea}{\end{eqnarray}}

\mathchardef\minus="002D

\newcommand{\rrt}[1]{}

\newcommand{\pdsp}{ProtoDUNE-SP\xspace} 
\newcommand{\pddp}{ProtoDUNE-DP\xspace} 
\newcommand{\pdspalt}{NP-04\xspace}

\usepackage{caption}
\usepackage{subcaption}

\usepackage[colorinlistoftodos]{todonotes}
\usepackage{lineno}

\title{Design, construction and operation of the ProtoDUNE-SP Liquid Argon TPC}

\newcommand{\Abilene}{Abilene Christian University, Abilene, TX 79601, USA}

\newcommand{\Albanysuny}{University of Albany, SUNY, Albany, NY 12222, USA}

\newcommand{\Amsterdam}{University of Amsterdam, NL-1098 XG Amsterdam, The Netherlands}

\newcommand{\Antalya}{Antalya Bilim University, 07190 D{\"o}{\c{s}}emealt{\i}/Antalya, Turkey}

\newcommand{\Antananarivo}{University of Antananarivo, Antananarivo 101, Madagascar}

\newcommand{\AntonioNarino}{Universidad Antonio Nari{\~n}o, Bogot{\'a}, Colombia}

\newcommand{\Argonne}{Argonne National Laboratory, Argonne, IL 60439, USA}

\newcommand{\Arizona}{University of Arizona, Tucson, AZ 85721, USA}

\newcommand{\Asuncion}{Universidad Nacional de Asunci{\'o}n, San Lorenzo, Paraguay}

\newcommand{\Athens}{University of Athens, Zografou GR 157 84, Greece}

\newcommand{\Atlantico}{Universidad del Atl{\'a}ntico, Barranquilla, Atl{\'a}ntico, Colombia}

\newcommand{\Augustana}{Augustana University, Sioux Falls, SD 57197, USA}

\newcommand{\Banaras}{Banaras Hindu University, Varanasi - 221 005, India}

\newcommand{\Basel}{University of Basel, CH-4056 Basel, Switzerland}

\newcommand{\Bern}{University of Bern, CH-3012 Bern, Switzerland}

\newcommand{\Beykent}{Beykent University, Istanbul, Turkey}

\newcommand{\Birmingham}{University of Birmingham, Birmingham B15 2TT, United Kingdom}

\newcommand{\BolognaUniversity}{Universit{\`a} del Bologna, 40127 Bologna, Italy}

\newcommand{\Boston}{Boston University, Boston, MA 02215, USA}

\newcommand{\Bristol}{University of Bristol, Bristol BS8 1TL, United Kingdom}

\newcommand{\Brookhaven}{Brookhaven National Laboratory, Upton, NY 11973, USA}

\newcommand{\Bucharest}{University of Bucharest, Bucharest, Romania}

\newcommand{\CBPF}{Centro Brasileiro de Pesquisas F\'isicas, Rio de Janeiro, RJ 22290-180, Brazil}

\newcommand{\CEASaclay}{IRFU, CEA, Universit{\'e} Paris-Saclay, F-91191 Gif-sur-Yvette, France}

\newcommand{\CERN}{CERN, The European Organization for Nuclear Research, 1211 Meyrin, Switzerland}

\newcommand{\CIEMAT}{CIEMAT, Centro de Investigaciones Energ{\'e}ticas, Medioambientales y Tecnol{\'o}gicas, E-28040 Madrid, Spain}

\newcommand{\CUSB}{Central University of South Bihar, Gaya, 824236, India }

\newcommand{\CalBerkeley}{University of California Berkeley, Berkeley, CA 94720, USA}

\newcommand{\CalDavis}{University of California Davis, Davis, CA 95616, USA}

\newcommand{\CalIrvine}{University of California Irvine, Irvine, CA 92697, USA}

\newcommand{\CalLosangeles}{University of California Los Angeles, Los Angeles, CA 90095, USA}

\newcommand{\CalRiverside}{University of California Riverside, Riverside CA 92521, USA}

\newcommand{\CalSantabarbara}{University of California Santa Barbara, Santa Barbara, California 93106 USA}

\newcommand{\Caltech}{California Institute of Technology, Pasadena, CA 91125, USA}

\newcommand{\Cambridge}{University of Cambridge, Cambridge CB3 0HE, United Kingdom}

\newcommand{\Campinas}{Universidade Estadual de Campinas, Campinas - SP, 13083-970, Brazil}

\newcommand{\CataniaUniversitadi}{Universit{\`a} di Catania, 2 - 95131 Catania, Italy}

\newcommand{\Charles}{Institute of Particle and Nuclear Physics of the Faculty of Mathematics and Physics of the Charles University, 180 00 Prague 8, Czech Republic }

\newcommand{\Chicago}{University of Chicago, Chicago, IL 60637, USA}

\newcommand{\ChungAng}{Chung-Ang University, Seoul 06974, South Korea}

\newcommand{\Cincinnati}{University of Cincinnati, Cincinnati, OH 45221, USA}

\newcommand{\Cinvestav}{Centro de Investigaci{\'o}n y de Estudios Avanzados del Instituto Polit{\'e}cnico Nacional (Cinvestav), Mexico City, Mexico}

\newcommand{\Colima}{Universidad de Colima, Colima, Mexico}

\newcommand{\ColoradoBoulder}{University of Colorado Boulder, Boulder, CO 80309, USA}

\newcommand{\ColoradoState}{Colorado State University, Fort Collins, CO 80523, USA}

\newcommand{\Columbia}{Columbia University, New York, NY 10027, USA}

\newcommand{\CzechAcademyofSciences}{Institute of Physics, Czech Academy of Sciences, 182 00 Prague 8, Czech Republic}

\newcommand{\CzechTechnical}{Czech Technical University, 115 19 Prague 1, Czech Republic}

\newcommand{\DakotaState}{Dakota State University, Madison, SD 57042, USA}

\newcommand{\Dallas}{University of Dallas, Irving, TX 75062-4736, USA}

\newcommand{\DannecyleVieux}{Laboratoire d{\textquoteright}Annecy de Physique des Particules, Univ. Grenoble Alpes, Univ. Savoie Mont Blanc, CNRS, LAPP-IN2P3, 74000 Annecy, France}

\newcommand{\Daresbury}{Daresbury Laboratory, Cheshire WA4 4AD, United Kingdom}

\newcommand{\Drexel}{Drexel University, Philadelphia, PA 19104, USA}

\newcommand{\Duke}{Duke University, Durham, NC 27708, USA}

\newcommand{\Durham}{Durham University, Durham DH1 3LE, United Kingdom}

\newcommand{\EIA}{Universidad EIA, Envigado, Antioquia, Colombia}

\newcommand{\ETH}{ETH Zurich, Zurich, Switzerland}

\newcommand{\Edinburgh}{University of Edinburgh, Edinburgh EH8 9YL, United Kingdom}

\newcommand{\FCULport}{Faculdade de Ci{\^e}ncias da Universidade de Lisboa - FCUL, 1749-016 Lisboa, Portugal}

\newcommand{\FederaldeAlfenas}{Universidade Federal de Alfenas, Po{\c{c}}os de Caldas - MG, 37715-400, Brazil}

\newcommand{\FederaldeGoias}{Universidade Federal de Goias, Goiania, GO 74690-900, Brazil}

\newcommand{\FederaldeSaoCarlos}{Universidade Federal de S{\~a}o Carlos, Araras - SP, 13604-900, Brazil}

\newcommand{\FederaldoABC}{Universidade Federal do ABC, Santo Andr{\'e} - SP, 09210-580, Brazil}

\newcommand{\FederaldoRio}{Universidade Federal do Rio de Janeiro,  Rio de Janeiro - RJ, 21941-901, Brazil}

\newcommand{\Fermi}{Fermi National Accelerator Laboratory, Batavia, IL 60510, USA}

\newcommand{\Ferrarauniv}{University of Ferrara, Ferrara, Italy}

\newcommand{\Florida}{University of Florida, Gainesville, FL 32611-8440, USA}

\newcommand{\Fluminense}{Fluminense Federal University, 9 Icara{\'\i} Niter{\'o}i - RJ, 24220-900, Brazil }

\newcommand{\Genova}{Universit{\`a} degli Studi di Genova, Genova, Italy}

\newcommand{\Georgian}{Georgian Technical University, Tbilisi, Georgia}

\newcommand{\GranSasso}{Gran Sasso Science Institute, L'Aquila, Italy}

\newcommand{\GranSassoLab}{Laboratori Nazionali del Gran Sasso, L'Aquila AQ, Italy}

\newcommand{\Granada}{University of Granada {\&} CAFPE, 18002 Granada, Spain}

\newcommand{\Grenoble}{University Grenoble Alpes, CNRS, Grenoble INP, LPSC-IN2P3, 38000 Grenoble, France}

\newcommand{\Guanajuato}{Universidad de Guanajuato, Guanajuato, C.P. 37000, Mexico}

\newcommand{\Harish}{Harish-Chandra Research Institute, Jhunsi, Allahabad 211 019, India}

\newcommand{\Harvard}{Harvard University, Cambridge, MA 02138, USA}

\newcommand{\Hawaii}{University of Hawaii, Honolulu, HI 96822, USA}

\newcommand{\Houston}{University of Houston, Houston, TX 77204, USA}

\newcommand{\Hyderabad}{University of  Hyderabad, Gachibowli, Hyderabad - 500 046, India}

\newcommand{\IFAE}{Institut de F{\'\i}sica d{\textquoteright}Altes Energies (IFAE){\textemdash}Barcelona Institute of Science and Technology (BIST), Barcelona, Spain}

\newcommand{\IFIC}{Instituto de F{\'\i}sica Corpuscular, CSIC and Universitat de Val{\`e}ncia, 46980 Paterna, Valencia, Spain}

\newcommand{\IGFAE}{Instituto Galego de Fisica de Altas Enerxias, A Coru{\~n}a, Spain}

\newcommand{\INFNBologna}{Istituto Nazionale di Fisica Nucleare Sezione di Bologna, 40127 Bologna BO, Italy}

\newcommand{\INFNCatania}{Istituto Nazionale di Fisica Nucleare Sezione di Catania, I-95123 Catania, Italy}

\newcommand{\INFNFerrara}{Istituto Nazionale di Fisica Nucleare Sezione di Ferrara, I-44122 Ferrara, Italy}

\newcommand{\INFNGenova}{Istituto Nazionale di Fisica Nucleare Sezione di Genova, 16146 Genova GE, Italy}

\newcommand{\INFNLecce}{Istituto Nazionale di Fisica Nucleare Sezione di Lecce, 73100 - Lecce, Italy}

\newcommand{\INFNMilanBicocca}{Istituto Nazionale di Fisica Nucleare Sezione di Milano Bicocca, 3 - I-20126 Milano, Italy}

\newcommand{\INFNMilano}{Istituto Nazionale di Fisica Nucleare Sezione di Milano, 20133 Milano, Italy}

\newcommand{\INFNNapoli}{Istituto Nazionale di Fisica Nucleare Sezione di Napoli, I-80126 Napoli, Italy}

\newcommand{\INFNPadova}{Istituto Nazionale di Fisica Nucleare Sezione di Padova, 35131 Padova, Italy}

\newcommand{\INFNPavia}{Istituto Nazionale di Fisica Nucleare Sezione di Pavia,  I-27100 Pavia, Italy}

\newcommand{\INFNSud}{Istituto Nazionale di Fisica Nucleare Laboratori Nazionali del Sud, 95123 Catania, Italy}

\newcommand{\INR}{Institute for Nuclear Research of the Russian Academy of Sciences, Moscow 117312, Russia}

\newcommand{\IPLyon}{Institut de Physique des 2 Infinis de Lyon, 69622 Villeurbanne, France}

\newcommand{\IPM}{Institute for Research in Fundamental Sciences, Tehran, Iran}

\newcommand{\ISTlisboa}{Instituto Superior T{\'e}cnico - IST, Universidade de Lisboa, Portugal}

\newcommand{\Idaho}{Idaho State University, Pocatello, ID 83209, USA}

\newcommand{\Illinoisinstitute}{Illinois Institute of Technology, Chicago, IL 60616, USA}

\newcommand{\Imperial}{Imperial College of Science Technology and Medicine, London SW7 2BZ, United Kingdom}

\newcommand{\IndGuwahati}{Indian Institute of Technology Guwahati, Guwahati, 781 039, India}

\newcommand{\IndHyderabad}{Indian Institute of Technology Hyderabad, Hyderabad, 502285, India}

\newcommand{\Indiana}{Indiana University, Bloomington, IN 47405, USA}

\newcommand{\Ingenieria}{Universidad Nacional de Ingenier{\'\i}a, Lima 25, Per{\'u}}

\newcommand{\Insubria }{University of Insubria, Via Ravasi, 2, 21100 Varese VA, Italy}

\newcommand{\Iowa}{University of Iowa, Iowa City, IA 52242, USA}

\newcommand{\IowaState}{Iowa State University, Ames, Iowa 50011, USA}

\newcommand{\Iwate}{Iwate University, Morioka, Iwate 020-8551, Japan}

\newcommand{\JINR}{Joint Institute for Nuclear Research, Dzhelepov Laboratory of Nuclear Problems 6 Joliot-Curie, Dubna, Moscow Region, 141980 RU }

\newcommand{\Jammu}{University of Jammu, Jammu-180006, India}

\newcommand{\Jawaharlal}{Jawaharlal Nehru University, New Delhi 110067, India}

\newcommand{\Jeonbuk}{Jeonbuk National University, Jeonrabuk-do 54896, South Korea}

\newcommand{\Jyvaskyla}{University of Jyvaskyla, FI-40014, Finland}

\newcommand{\KEK}{High Energy Accelerator Research Organization (KEK), Ibaraki, 305-0801, Japan}

\newcommand{\KISTI}{Korea Institute of Science and Technology Information, Daejeon, 34141, South Korea}

\newcommand{\KL}{K L University, Vaddeswaram, Andhra Pradesh 522502, India}

\newcommand{\Kansasstate}{Kansas State University, Manhattan, KS 66506, USA}

\newcommand{\Kavli}{Kavli Institute for the Physics and Mathematics of the Universe, Kashiwa, Chiba 277-8583, Japan}

\newcommand{\Kure}{National Institute of Technology, Kure College, Hiroshima, 737-8506, Japan}

\newcommand{\Kyiv}{Taras Shevchenko National University of Kyiv, 01601 Kyiv, Ukraine}

\newcommand{\LIP}{Laborat{\'o}rio de Instrumenta{\c{c}}{\~a}o e F{\'\i}sica Experimental de Part{\'\i}culas, 1649-003 Lisboa and 3004-516 Coimbra, Portugal}

\newcommand{\Lancaster}{Lancaster University, Lancaster LA1 4YB, United Kingdom}

\newcommand{\LawrenceBerkeley}{Lawrence Berkeley National Laboratory, Berkeley, CA 94720, USA}

\newcommand{\Liverpool}{University of Liverpool, L69 7ZE, Liverpool, United Kingdom}

\newcommand{\LosAlmos}{Los Alamos National Laboratory, Los Alamos, NM 87545, USA}

\newcommand{\Louisanastate}{Louisiana State University, Baton Rouge, LA 70803, USA}

\newcommand{\Lucknow}{University of Lucknow, Uttar Pradesh 226007, India}

\newcommand{\Madrid}{Madrid Autonoma University and IFT UAM/CSIC, 28049 Madrid, Spain}

\newcommand{\Manchester}{University of Manchester, Manchester M13 9PL, United Kingdom}

\newcommand{\Massinsttech}{Massachusetts Institute of Technology, Cambridge, MA 02139, USA}

\newcommand{\Maxplanck}{Max-Planck-Institut, Munich, 80805, Germany}

\newcommand{\Medellin}{University of Medell{\'\i}n, Medell{\'\i}n, 050026 Colombia }

\newcommand{\Michigan}{University of Michigan, Ann Arbor, MI 48109, USA}

\newcommand{\Michiganstate}{Michigan State University, East Lansing, MI 48824, USA}

\newcommand{\MilanoBicocca}{Universit{\`a} del Milano-Bicocca, 20126 Milano, Italy}

\newcommand{\MilanoUniv}{Universit{\`a} degli Studi di Milano, I-20133 Milano, Italy}

\newcommand{\Minnduluth}{University of Minnesota Duluth, Duluth, MN 55812, USA}

\newcommand{\Minntwin}{University of Minnesota Twin Cities, Minneapolis, MN 55455, USA}

\newcommand{\Mississippi}{University of Mississippi, University, MS 38677 USA}

\newcommand{\Newmexico}{University of New Mexico, Albuquerque, NM 87131, USA}

\newcommand{\Niewodniczanski}{H. Niewodnicza{\'n}ski Institute of Nuclear Physics, Polish Academy of Sciences, Cracow, Poland}

\newcommand{\Nikhef}{Nikhef National Institute of Subatomic Physics, 1098 XG Amsterdam, Netherlands}

\newcommand{\Northdakota}{University of North Dakota, Grand Forks, ND 58202-8357, USA}

\newcommand{\Northernillinois}{Northern Illinois University, DeKalb, IL 60115, USA}

\newcommand{\Northwestern}{Northwestern University, Evanston, Il 60208, USA}

\newcommand{\NotreDame}{University of Notre Dame, Notre Dame, IN 46556, USA}

\newcommand{\Occidental}{Occidental College, Los Angeles, CA  90041}

\newcommand{\Ohiostate}{Ohio State University, Columbus, OH 43210, USA}

\newcommand{\OregonState}{Oregon State University, Corvallis, OR 97331, USA}

\newcommand{\Oxford}{University of Oxford, Oxford, OX1 3RH, United Kingdom}

\newcommand{\PacificNorthwest}{Pacific Northwest National Laboratory, Richland, WA 99352, USA}

\newcommand{\Padova}{Universt{\`a} degli Studi di Padova, I-35131 Padova, Italy}

\newcommand{\Parissaclay}{Universit{\'e} Paris-Saclay, CNRS/IN2P3, IJCLab, 91405 Orsay, France}

\newcommand{\Parisuniversite}{Universit{\'e} de Paris, CNRS, Astroparticule et Cosmologie, F-75006, Paris, France}

\newcommand{\Pavia}{Universit{\`a} degli Studi di Pavia, 27100 Pavia PV, Italy}

\newcommand{\Penn}{University of Pennsylvania, Philadelphia, PA 19104, USA}

\newcommand{\PennState}{Pennsylvania State University, University Park, PA 16802, USA}

\newcommand{\PhysicalResearchLaboratory}{Physical Research Laboratory, Ahmedabad 380 009, India}

\newcommand{\Pisa}{Universit{\`a} di Pisa, I-56127 Pisa, Italy}

\newcommand{\Pitt}{University of Pittsburgh, Pittsburgh, PA 15260, USA}

\newcommand{\Pontificia}{Pontificia Universidad Cat{\'o}lica del Per{\'u}, Lima, Per{\'u}}

\newcommand{\PuertoRico}{University of Puerto Rico, Mayaguez 00681, Puerto Rico, USA}

\newcommand{\Punjab}{Punjab Agricultural University, Ludhiana 141004, India}

\newcommand{\QMUL}{Queen Mary University of London, London E1 4NS, United Kingdom }

\newcommand{\Radboud}{Radboud University, NL-6525 AJ Nijmegen, Netherlands}

\newcommand{\Rochester}{University of Rochester, Rochester, NY 14627, USA}

\newcommand{\Royalholloway}{Royal Holloway College London, TW20 0EX, United Kingdom}

\newcommand{\Rutgers}{Rutgers University, Piscataway, NJ, 08854, USA}

\newcommand{\Rutherford}{STFC Rutherford Appleton Laboratory, Didcot OX11 0QX, United Kingdom}

\newcommand{\SLAC}{SLAC National Accelerator Laboratory, Menlo Park, CA 94025, USA}

\newcommand{\SURF}{Sanford Underground Research Facility, Lead, SD, 57754, USA}

\newcommand{\Salento}{Universit{\`a} del Salento, 73100 Lecce, Italy}

\newcommand{\Sanjosestate}{San Jose State University, San Jos{\'e}, CA 95192-0106, USA}

\newcommand{\SergioArboleda}{Universidad Sergio Arboleda, 11022 Bogot{\'a}, Colombia}

\newcommand{\Sheffield}{University of Sheffield, Sheffield S3 7RH, United Kingdom}

\newcommand{\SouthDakotaSchool}{South Dakota School of Mines and Technology, Rapid City, SD 57701, USA}

\newcommand{\SouthDakotaState}{South Dakota State University, Brookings, SD 57007, USA}

\newcommand{\Southcarolina}{University of South Carolina, Columbia, SC 29208, USA}

\newcommand{\SouthernMethodist}{Southern Methodist University, Dallas, TX 75275, USA}

\newcommand{\StonyBrook}{Stony Brook University, SUNY, Stony Brook, NY 11794, USA}

\newcommand{\Sunyatsen}{Sun Yat-Sen University, Guangzhou, 510275}

\newcommand{\Sussex}{University of Sussex, Brighton, BN1 9RH, United Kingdom}

\newcommand{\Syracuse}{Syracuse University, Syracuse, NY 13244, USA}

\newcommand{\Tecnologica }{Universidade Tecnol{\'o}gica Federal do Paran{\'a}, Curitiba, Brazil}

\newcommand{\Tennknox}{University of Tennessee at Knoxville, TN, 37996, USA}

\newcommand{\TexasAMcollege}{Texas A{\&}M University, College Station, Texas 77840}

\newcommand{\TexasAMcorpuscristi}{Texas A{\&}M University - Corpus Christi, Corpus Christi, TX 78412, USA}

\newcommand{\TexasArlington}{University of Texas at Arlington, Arlington, TX 76019, USA}

\newcommand{\Texasaustin}{University of Texas at Austin, Austin, TX 78712, USA}

\newcommand{\Toronto}{University of Toronto, Toronto, Ontario M5S 1A1, Canada}

\newcommand{\Tufts}{Tufts University, Medford, MA 02155, USA}

\newcommand{\UNIST}{Ulsan National Institute of Science and Technology, Ulsan 689-798, South Korea}

\newcommand{\Unifesp}{Universidade Federal de S{\~a}o Paulo, 09913-030, S{\~a}o Paulo, Brazil}

\newcommand{\UniversityCollegeLondon}{University College London, London, WC1E 6BT, United Kingdom}

\newcommand{\ValleyCity}{Valley City State University, Valley City, ND 58072, USA}

\newcommand{\VariableEnergy}{Variable Energy Cyclotron Centre, 700 064 West Bengal, India}

\newcommand{\VirginiaTech}{Virginia Tech, Blacksburg, VA 24060, USA}

\newcommand{\Warsaw}{University of Warsaw, 02-093 Warsaw, Poland}

\newcommand{\Warwick}{University of Warwick, Coventry CV4 7AL, United Kingdom}

\newcommand{\Wellesley}{Wellesley College, Wellesley, MA 02481, USA}

\newcommand{\Wichita}{Wichita State University, Wichita, KS 67260, USA}

\newcommand{\WilliamMary}{William and Mary, Williamsburg, VA 23187, USA}

\newcommand{\Wisconsin}{University of Wisconsin Madison, Madison, WI 53706, USA}

\newcommand{\Yale}{Yale University, New Haven, CT 06520, USA}

\newcommand{\Yerevan}{Yerevan Institute for Theoretical Physics and Modeling, Yerevan 0036, Armenia}

\newcommand{\York}{York University, Toronto M3J 1P3, Canada}

\newcommand{\iAbilene}{1} \affiliation[\iAbilene]{\Abilene}
\newcommand{\iAlbanysuny}{2} \affiliation[\iAlbanysuny]{\Albanysuny}
\newcommand{\iAmsterdam}{3} \affiliation[\iAmsterdam]{\Amsterdam}
\newcommand{\iAntalya}{4} \affiliation[\iAntalya]{\Antalya}
\newcommand{\iAntananarivo}{5} \affiliation[\iAntananarivo]{\Antananarivo}
\newcommand{\iAntonioNarino}{6} \affiliation[\iAntonioNarino]{\AntonioNarino}
\newcommand{\iArgonne}{7} \affiliation[\iArgonne]{\Argonne}
\newcommand{\iArizona}{8} \affiliation[\iArizona]{\Arizona}
\newcommand{\iAsuncion}{9} \affiliation[\iAsuncion]{\Asuncion}
\newcommand{\iAthens}{10} \affiliation[\iAthens]{\Athens}
\newcommand{\iAtlantico}{11} \affiliation[\iAtlantico]{\Atlantico}
\newcommand{\iAugustana}{12} \affiliation[\iAugustana]{\Augustana}
\newcommand{\iBanaras}{13} \affiliation[\iBanaras]{\Banaras}
\newcommand{\iBasel}{14} \affiliation[\iBasel]{\Basel}
\newcommand{\iBern}{15} \affiliation[\iBern]{\Bern}
\newcommand{\iBeykent}{16} \affiliation[\iBeykent]{\Beykent}
\newcommand{\iBirmingham}{17} \affiliation[\iBirmingham]{\Birmingham}
\newcommand{\iBolognaUniversity}{18} \affiliation[\iBolognaUniversity]{\BolognaUniversity}
\newcommand{\iBoston}{19} \affiliation[\iBoston]{\Boston}
\newcommand{\iBristol}{20} \affiliation[\iBristol]{\Bristol}
\newcommand{\iBrookhaven}{21} \affiliation[\iBrookhaven]{\Brookhaven}
\newcommand{\iBucharest}{22} \affiliation[\iBucharest]{\Bucharest}
\newcommand{\iCBPF}{23} \affiliation[\iCBPF]{\CBPF}
\newcommand{\iCEASaclay}{24} \affiliation[\iCEASaclay]{\CEASaclay}
\newcommand{\iCERN}{25} \affiliation[\iCERN]{\CERN}
\newcommand{\iCIEMAT}{26} \affiliation[\iCIEMAT]{\CIEMAT}
\newcommand{\iCUSB}{27} \affiliation[\iCUSB]{\CUSB}
\newcommand{\iCalBerkeley}{28} \affiliation[\iCalBerkeley]{\CalBerkeley}
\newcommand{\iCalDavis}{29} \affiliation[\iCalDavis]{\CalDavis}
\newcommand{\iCalIrvine}{30} \affiliation[\iCalIrvine]{\CalIrvine}
\newcommand{\iCalLosangeles}{31} \affiliation[\iCalLosangeles]{\CalLosangeles}
\newcommand{\iCalRiverside}{32} \affiliation[\iCalRiverside]{\CalRiverside}
\newcommand{\iCalSantabarbara}{33} \affiliation[\iCalSantabarbara]{\CalSantabarbara}
\newcommand{\iCaltech}{34} \affiliation[\iCaltech]{\Caltech}
\newcommand{\iCambridge}{35} \affiliation[\iCambridge]{\Cambridge}
\newcommand{\iCampinas}{36} \affiliation[\iCampinas]{\Campinas}
\newcommand{\iCataniaUniversitadi}{37} \affiliation[\iCataniaUniversitadi]{\CataniaUniversitadi}
\newcommand{\iCharles}{38} \affiliation[\iCharles]{\Charles}
\newcommand{\iChicago}{39} \affiliation[\iChicago]{\Chicago}
\newcommand{\iChungAng}{40} \affiliation[\iChungAng]{\ChungAng}
\newcommand{\iCincinnati}{41} \affiliation[\iCincinnati]{\Cincinnati}
\newcommand{\iCinvestav}{42} \affiliation[\iCinvestav]{\Cinvestav}
\newcommand{\iColima}{43} \affiliation[\iColima]{\Colima}
\newcommand{\iColoradoBoulder}{44} \affiliation[\iColoradoBoulder]{\ColoradoBoulder}
\newcommand{\iColoradoState}{45} \affiliation[\iColoradoState]{\ColoradoState}
\newcommand{\iColumbia}{46} \affiliation[\iColumbia]{\Columbia}
\newcommand{\iCzechAcademyofSciences}{47} \affiliation[\iCzechAcademyofSciences]{\CzechAcademyofSciences}
\newcommand{\iCzechTechnical}{48} \affiliation[\iCzechTechnical]{\CzechTechnical}
\newcommand{\iDakotaState}{49} \affiliation[\iDakotaState]{\DakotaState}
\newcommand{\iDallas}{50} \affiliation[\iDallas]{\Dallas}
\newcommand{\iDannecyleVieux}{51} \affiliation[\iDannecyleVieux]{\DannecyleVieux}
\newcommand{\iDaresbury}{52} \affiliation[\iDaresbury]{\Daresbury}
\newcommand{\iDrexel}{53} \affiliation[\iDrexel]{\Drexel}
\newcommand{\iDuke}{54} \affiliation[\iDuke]{\Duke}
\newcommand{\iDurham}{55} \affiliation[\iDurham]{\Durham}
\newcommand{\iEIA}{56} \affiliation[\iEIA]{\EIA}
\newcommand{\iETH}{57} \affiliation[\iETH]{\ETH}
\newcommand{\iEdinburgh}{58} \affiliation[\iEdinburgh]{\Edinburgh}
\newcommand{\iFCULport}{59} \affiliation[\iFCULport]{\FCULport}
\newcommand{\iFederaldeAlfenas}{60} \affiliation[\iFederaldeAlfenas]{\FederaldeAlfenas}
\newcommand{\iFederaldeGoias}{61} \affiliation[\iFederaldeGoias]{\FederaldeGoias}
\newcommand{\iFederaldeSaoCarlos}{62} \affiliation[\iFederaldeSaoCarlos]{\FederaldeSaoCarlos}
\newcommand{\iFederaldoABC}{63} \affiliation[\iFederaldoABC]{\FederaldoABC}
\newcommand{\iFederaldoRio}{64} \affiliation[\iFederaldoRio]{\FederaldoRio}
\newcommand{\iFermi}{65} \affiliation[\iFermi]{\Fermi}
\newcommand{\iFerrarauniv}{66} \affiliation[\iFerrarauniv]{\Ferrarauniv}
\newcommand{\iFlorida}{67} \affiliation[\iFlorida]{\Florida}
\newcommand{\iFluminense}{68} \affiliation[\iFluminense]{\Fluminense}
\newcommand{\iGenova}{69} \affiliation[\iGenova]{\Genova}
\newcommand{\iGeorgian}{70} \affiliation[\iGeorgian]{\Georgian}
\newcommand{\iGranSasso}{71} \affiliation[\iGranSasso]{\GranSasso}
\newcommand{\iGranSassoLab}{72} \affiliation[\iGranSassoLab]{\GranSassoLab}
\newcommand{\iGranada}{73} \affiliation[\iGranada]{\Granada}
\newcommand{\iGrenoble}{74} \affiliation[\iGrenoble]{\Grenoble}
\newcommand{\iGuanajuato}{75} \affiliation[\iGuanajuato]{\Guanajuato}
\newcommand{\iHarish}{76} \affiliation[\iHarish]{\Harish}
\newcommand{\iHarvard}{77} \affiliation[\iHarvard]{\Harvard}
\newcommand{\iHawaii}{78} \affiliation[\iHawaii]{\Hawaii}
\newcommand{\iHouston}{79} \affiliation[\iHouston]{\Houston}
\newcommand{\iHyderabad}{80} \affiliation[\iHyderabad]{\Hyderabad}
\newcommand{\iIFAE}{81} \affiliation[\iIFAE]{\IFAE}
\newcommand{\iIFIC}{82} \affiliation[\iIFIC]{\IFIC}
\newcommand{\iIGFAE}{83} \affiliation[\iIGFAE]{\IGFAE}
\newcommand{\iINFNBologna}{84} \affiliation[\iINFNBologna]{\INFNBologna}
\newcommand{\iINFNCatania}{85} \affiliation[\iINFNCatania]{\INFNCatania}
\newcommand{\iINFNFerrara}{86} \affiliation[\iINFNFerrara]{\INFNFerrara}
\newcommand{\iINFNGenova}{87} \affiliation[\iINFNGenova]{\INFNGenova}
\newcommand{\iINFNLecce}{88} \affiliation[\iINFNLecce]{\INFNLecce}
\newcommand{\iINFNMilanBicocca}{89} \affiliation[\iINFNMilanBicocca]{\INFNMilanBicocca}
\newcommand{\iINFNMilano}{90} \affiliation[\iINFNMilano]{\INFNMilano}
\newcommand{\iINFNNapoli}{91} \affiliation[\iINFNNapoli]{\INFNNapoli}
\newcommand{\iINFNPadova}{92} \affiliation[\iINFNPadova]{\INFNPadova}
\newcommand{\iINFNPavia}{93} \affiliation[\iINFNPavia]{\INFNPavia}
\newcommand{\iINFNSud}{94} \affiliation[\iINFNSud]{\INFNSud}
\newcommand{\iINR}{95} \affiliation[\iINR]{\INR}
\newcommand{\iIPLyon}{96} \affiliation[\iIPLyon]{\IPLyon}
\newcommand{\iIPM}{97} \affiliation[\iIPM]{\IPM}
\newcommand{\iISTlisboa}{98} \affiliation[\iISTlisboa]{\ISTlisboa}
\newcommand{\iIdaho}{99} \affiliation[\iIdaho]{\Idaho}
\newcommand{\iIllinoisinstitute}{100} \affiliation[\iIllinoisinstitute]{\Illinoisinstitute}
\newcommand{\iImperial}{101} \affiliation[\iImperial]{\Imperial}
\newcommand{\iIndGuwahati}{102} \affiliation[\iIndGuwahati]{\IndGuwahati}
\newcommand{\iIndHyderabad}{103} \affiliation[\iIndHyderabad]{\IndHyderabad}
\newcommand{\iIndiana}{104} \affiliation[\iIndiana]{\Indiana}
\newcommand{\iIngenieria}{105} \affiliation[\iIngenieria]{\Ingenieria}
\newcommand{\iInsubria }{106} \affiliation[\iInsubria ]{\Insubria }
\newcommand{\iIowa}{107} \affiliation[\iIowa]{\Iowa}
\newcommand{\iIowaState}{108} \affiliation[\iIowaState]{\IowaState}
\newcommand{\iIwate}{109} \affiliation[\iIwate]{\Iwate}
\newcommand{\iJINR}{110} \affiliation[\iJINR]{\JINR}
\newcommand{\iJammu}{111} \affiliation[\iJammu]{\Jammu}
\newcommand{\iJawaharlal}{112} \affiliation[\iJawaharlal]{\Jawaharlal}
\newcommand{\iJeonbuk}{113} \affiliation[\iJeonbuk]{\Jeonbuk}
\newcommand{\iJyvaskyla}{114} \affiliation[\iJyvaskyla]{\Jyvaskyla}
\newcommand{\iKEK}{115} \affiliation[\iKEK]{\KEK}
\newcommand{\iKISTI}{116} \affiliation[\iKISTI]{\KISTI}
\newcommand{\iKL}{117} \affiliation[\iKL]{\KL}
\newcommand{\iKansasstate}{118} \affiliation[\iKansasstate]{\Kansasstate}
\newcommand{\iKavli}{119} \affiliation[\iKavli]{\Kavli}
\newcommand{\iKure}{120} \affiliation[\iKure]{\Kure}
\newcommand{\iKyiv}{121} \affiliation[\iKyiv]{\Kyiv}
\newcommand{\iLIP}{122} \affiliation[\iLIP]{\LIP}
\newcommand{\iLancaster}{123} \affiliation[\iLancaster]{\Lancaster}
\newcommand{\iLawrenceBerkeley}{124} \affiliation[\iLawrenceBerkeley]{\LawrenceBerkeley}
\newcommand{\iLiverpool}{125} \affiliation[\iLiverpool]{\Liverpool}
\newcommand{\iLosAlmos}{126} \affiliation[\iLosAlmos]{\LosAlmos}
\newcommand{\iLouisanastate}{127} \affiliation[\iLouisanastate]{\Louisanastate}
\newcommand{\iLucknow}{128} \affiliation[\iLucknow]{\Lucknow}
\newcommand{\iMadrid}{129} \affiliation[\iMadrid]{\Madrid}
\newcommand{\iManchester}{130} \affiliation[\iManchester]{\Manchester}
\newcommand{\iMassinsttech}{131} \affiliation[\iMassinsttech]{\Massinsttech}
\newcommand{\iMaxplanck}{132} \affiliation[\iMaxplanck]{\Maxplanck}
\newcommand{\iMedellin}{133} \affiliation[\iMedellin]{\Medellin}
\newcommand{\iMichigan}{134} \affiliation[\iMichigan]{\Michigan}
\newcommand{\iMichiganstate}{135} \affiliation[\iMichiganstate]{\Michiganstate}
\newcommand{\iMilanoBicocca}{136} \affiliation[\iMilanoBicocca]{\MilanoBicocca}
\newcommand{\iMilanoUniv}{137} \affiliation[\iMilanoUniv]{\MilanoUniv}
\newcommand{\iMinnduluth}{138} \affiliation[\iMinnduluth]{\Minnduluth}
\newcommand{\iMinntwin}{139} \affiliation[\iMinntwin]{\Minntwin}
\newcommand{\iMississippi}{140} \affiliation[\iMississippi]{\Mississippi}
\newcommand{\iNewmexico}{141} \affiliation[\iNewmexico]{\Newmexico}
\newcommand{\iNiewodniczanski}{142} \affiliation[\iNiewodniczanski]{\Niewodniczanski}
\newcommand{\iNikhef}{143} \affiliation[\iNikhef]{\Nikhef}
\newcommand{\iNorthdakota}{144} \affiliation[\iNorthdakota]{\Northdakota}
\newcommand{\iNorthernillinois}{145} \affiliation[\iNorthernillinois]{\Northernillinois}
\newcommand{\iNorthwestern}{146} \affiliation[\iNorthwestern]{\Northwestern}
\newcommand{\iNotreDame}{147} \affiliation[\iNotreDame]{\NotreDame}
\newcommand{\iOccidental}{148} \affiliation[\iOccidental]{\Occidental}
\newcommand{\iOhiostate}{149} \affiliation[\iOhiostate]{\Ohiostate}
\newcommand{\iOregonState}{150} \affiliation[\iOregonState]{\OregonState}
\newcommand{\iOxford}{151} \affiliation[\iOxford]{\Oxford}
\newcommand{\iPacificNorthwest}{152} \affiliation[\iPacificNorthwest]{\PacificNorthwest}
\newcommand{\iPadova}{153} \affiliation[\iPadova]{\Padova}
\newcommand{\iParissaclay}{154} \affiliation[\iParissaclay]{\Parissaclay}
\newcommand{\iParisuniversite}{155} \affiliation[\iParisuniversite]{\Parisuniversite}
\newcommand{\iPavia}{156} \affiliation[\iPavia]{\Pavia}
\newcommand{\iPenn}{157} \affiliation[\iPenn]{\Penn}
\newcommand{\iPennState}{158} \affiliation[\iPennState]{\PennState}
\newcommand{\iPhysicalResearchLaboratory}{159} \affiliation[\iPhysicalResearchLaboratory]{\PhysicalResearchLaboratory}
\newcommand{\iPisa}{160} \affiliation[\iPisa]{\Pisa}
\newcommand{\iPitt}{161} \affiliation[\iPitt]{\Pitt}
\newcommand{\iPontificia}{162} \affiliation[\iPontificia]{\Pontificia}
\newcommand{\iPuertoRico}{163} \affiliation[\iPuertoRico]{\PuertoRico}
\newcommand{\iPunjab}{164} \affiliation[\iPunjab]{\Punjab}
\newcommand{\iQMUL}{165} \affiliation[\iQMUL]{\QMUL}
\newcommand{\iRadboud}{166} \affiliation[\iRadboud]{\Radboud}
\newcommand{\iRochester}{167} \affiliation[\iRochester]{\Rochester}
\newcommand{\iRoyalholloway}{168} \affiliation[\iRoyalholloway]{\Royalholloway}
\newcommand{\iRutgers}{169} \affiliation[\iRutgers]{\Rutgers}
\newcommand{\iRutherford}{170} \affiliation[\iRutherford]{\Rutherford}
\newcommand{\iSLAC}{171} \affiliation[\iSLAC]{\SLAC}
\newcommand{\iSURF}{172} \affiliation[\iSURF]{\SURF}
\newcommand{\iSalento}{173} \affiliation[\iSalento]{\Salento}
\newcommand{\iSanjosestate}{174} \affiliation[\iSanjosestate]{\Sanjosestate}
\newcommand{\iSergioArboleda}{175} \affiliation[\iSergioArboleda]{\SergioArboleda}
\newcommand{\iSheffield}{176} \affiliation[\iSheffield]{\Sheffield}
\newcommand{\iSouthDakotaSchool}{177} \affiliation[\iSouthDakotaSchool]{\SouthDakotaSchool}
\newcommand{\iSouthDakotaState}{178} \affiliation[\iSouthDakotaState]{\SouthDakotaState}
\newcommand{\iSouthcarolina}{179} \affiliation[\iSouthcarolina]{\Southcarolina}
\newcommand{\iSouthernMethodist}{180} \affiliation[\iSouthernMethodist]{\SouthernMethodist}
\newcommand{\iStonyBrook}{181} \affiliation[\iStonyBrook]{\StonyBrook}
\newcommand{\iSunyatsen}{182} \affiliation[\iSunyatsen]{\Sunyatsen}
\newcommand{\iSussex}{183} \affiliation[\iSussex]{\Sussex}
\newcommand{\iSyracuse}{184} \affiliation[\iSyracuse]{\Syracuse}
\newcommand{\iTecnologica }{185} \affiliation[\iTecnologica ]{\Tecnologica }
\newcommand{\iTennknox}{186} \affiliation[\iTennknox]{\Tennknox}
\newcommand{\iTexasAMcollege}{187} \affiliation[\iTexasAMcollege]{\TexasAMcollege}
\newcommand{\iTexasAMcorpuscristi}{188} \affiliation[\iTexasAMcorpuscristi]{\TexasAMcorpuscristi}
\newcommand{\iTexasArlington}{189} \affiliation[\iTexasArlington]{\TexasArlington}
\newcommand{\iTexasaustin}{190} \affiliation[\iTexasaustin]{\Texasaustin}
\newcommand{\iToronto}{191} \affiliation[\iToronto]{\Toronto}
\newcommand{\iTufts}{192} \affiliation[\iTufts]{\Tufts}
\newcommand{\iUNIST}{193} \affiliation[\iUNIST]{\UNIST}
\newcommand{\iUnifesp}{194} \affiliation[\iUnifesp]{\Unifesp}
\newcommand{\iUniversityCollegeLondon}{195} \affiliation[\iUniversityCollegeLondon]{\UniversityCollegeLondon}
\newcommand{\iValleyCity}{196} \affiliation[\iValleyCity]{\ValleyCity}
\newcommand{\iVariableEnergy}{197} \affiliation[\iVariableEnergy]{\VariableEnergy}
\newcommand{\iVirginiaTech}{198} \affiliation[\iVirginiaTech]{\VirginiaTech}
\newcommand{\iWarsaw}{199} \affiliation[\iWarsaw]{\Warsaw}
\newcommand{\iWarwick}{200} \affiliation[\iWarwick]{\Warwick}
\newcommand{\iWellesley}{201} \affiliation[\iWellesley]{\Wellesley}
\newcommand{\iWichita}{202} \affiliation[\iWichita]{\Wichita}
\newcommand{\iWilliamMary}{203} \affiliation[\iWilliamMary]{\WilliamMary}
\newcommand{\iWisconsin}{204} \affiliation[\iWisconsin]{\Wisconsin}
\newcommand{\iYale}{205} \affiliation[\iYale]{\Yale}
\newcommand{\iYerevan}{206} \affiliation[\iYerevan]{\Yerevan}
\newcommand{\iYork}{207} \affiliation[\iYork]{\York}

\author[\iLiverpool,\iCERN]{A.~Abed Abud}
\author[\iOxford]{B.~Abi}
\author[\iFermi]{R.~Acciarri}
\author[\iAtlantico]{M.~A.~Acero}
\author[\iTecnologica ]{M.~R.~Adames}
\author[\iGeorgian]{G.~Adamov}
\author[\iBrookhaven]{D.~Adams}
\author[\iBristol]{M.~Adinolfi}
\author[\iHouston]{A.~Aduszkiewicz}
\author[\iLawrenceBerkeley]{J.~Aguilar}
% ...

\collaboration{The DUNE Collaboration}

\abstract{The ProtoDUNE-SP detector is a single-phase liquid argon time projection chamber (LArTPC) that was constructed and operated in the CERN North Area at the end of the H4 beamline. This detector is a prototype for the first far detector module of the Deep Underground Neutrino Experiment (DUNE), which will be constructed at the Sandford Underground Research Facility (SURF) in Lead, South Dakota, USA. The ProtoDUNE-SP detector incorporates full-size components as designed for DUNE and has an active volume of $7\times 6\times 7.2$~m$^3$.  The H4 beam delivers incident particles with well-measured momenta and high-purity particle identification. ProtoDUNE-SP's successful operation between 2018 and 2020  demonstrates the effectiveness of the single-phase far detector design.  This paper describes the design, construction, assembly and operation of the detector components.}

\keywords{Noble liquid detectors (scintillation, ionization, single-phase),
Time projection chambers, Large detector systems for particle and astroparticle physics,
Scintillators, scintillation and light emission processes (solid, gas and liquid scintillators), Photon detectors for UV, visible and IR photons (solid-state) (PIN diodes, APDs, Si-PMTs, G-APDs, CCDs, EBCCDs, EMCCDs, CMOS imagers, etc)}

\arxivnumber{1234.56789} 

\begin{document}
\maketitle
\flushbottom


\section{Introduction} 
\label{sec:intro}
      
\subsection{\pdsp{} in the Context of DUNE}

The Deep Underground Neutrino Experiment (DUNE) will be a world-class neutrino observatory and nucleon decay detector designed to answer fundamental questions about elementary particles and their role in the universe. The international DUNE experiment, hosted by the U.S. Department of Energy's Fermilab, uses a near detector located at Fermilab and a far detector, 1300 km away, located approximately 1.5\,km underground at the Sanford Underground Research Facility (SURF) in South Dakota, USA. The far detector will be a very large LArTPC with a fiducial mass of 40\,kt (total mass 68\,kt) of liquid argon (LAr), split into four modules.  DUNE has been pursuing two LArTPC technologies, single-phase (liquid only) and dual-phase (liquid and gas); this paper describes \pdsp{}, a prototype for a single-phase (SP) detector module.

Construction of \pdsp{} was proposed to the CERN Super Proton Synchrotron Committee (SPSC) in June 2015~\cite{Kutter:2022751} and, following positive recommendations by SPSC and the CERN Research Board in December 2015, was approved at CERN as experiment \pdspalt{}. \pdsp{} has since been constructed and successfully operated at the CERN Neutrino Platform (NP). The design of the \pdsp{} detector has been documented in a Technical Design Report~\cite{Abi:2017aow}.

\pdsp{} prototypes most of the components of a DUNE single-phase far detector module at 1:1 scale, with an extrapolation of about 1:20 in total LAr mass. With a total LAr mass of 0.77\,kt, it represents the largest monolithic single-phase LArTPC detector  built to date, and is a significant experiment in its own right.


The \pdsp{} detector elements, 
the time projection chamber (TPC), the cold electronics (CE), and the photon detection system (PDS), are immersed in a cryostat filled with the LAr target material. 
The TPC consists of two vertical anode planes, one vertical cathode plane, and a surrounding field cage. 
A cryogenics system maintains the LAr at a stable temperature of about 87\,K and at the required purity level through a closed-loop process that recovers the evaporated argon, recondenses, filters, and returns it to the cryostat.

\pdsp{} is located in an extension to the EHN1 hall (Experimental Hall North~1 - EHN1) in CERN's North Area, where a new, dedicated charged-particle test beamline was constructed as part of the CERN NP program. Construction and installation of \pdsp{} was completed in early  July 2018. Filling  with LAr and commissioning took place in July and August of that year. First beam was delivered to EHN1 on August 29, 2018 and the beam run was completed on November 11, 2018. The detector continued to operate through July 19, 2020, collecting data to test and validate the technologies for the future DUNE far detector modules, demonstrate operational stability, and explore operational parameters. 

The construction, installation, and operation of the  \pdsp{} detector, as described in this paper, 
has validated the design elements of the single-phase technology, the membrane cryostat technology and associated cryogenics systems, as well as instrumentation, data acquisition, and detector control. First performance results are presented in a separate paper~\cite{Abi:2020mwi}.

\subsection{Cryostat and Cryogenics}

ProtoDUNE-SP implements the first large-dimension prototype cryostat built for a particle physics detector based on the technology used for liquefied natural gas (LNG) storage and transport. Its construction and operation serves as a validation of the membrane cryostat technology and associated cryogenics.
The \pdsp{} cryostat was constructed in EHN1 without any mechanical attachment to the floor or the building side walls. Its internal dimensions 
are 8.5\,m in width and length 
and 7.9\,m in height. The cryostat and the cryogenics systems are described in detail in Section~\ref{sec:cryo}.


\subsection{Detector Components}
\begin{figure}[h]
\centering
\includegraphics[width=0.9\textwidth]{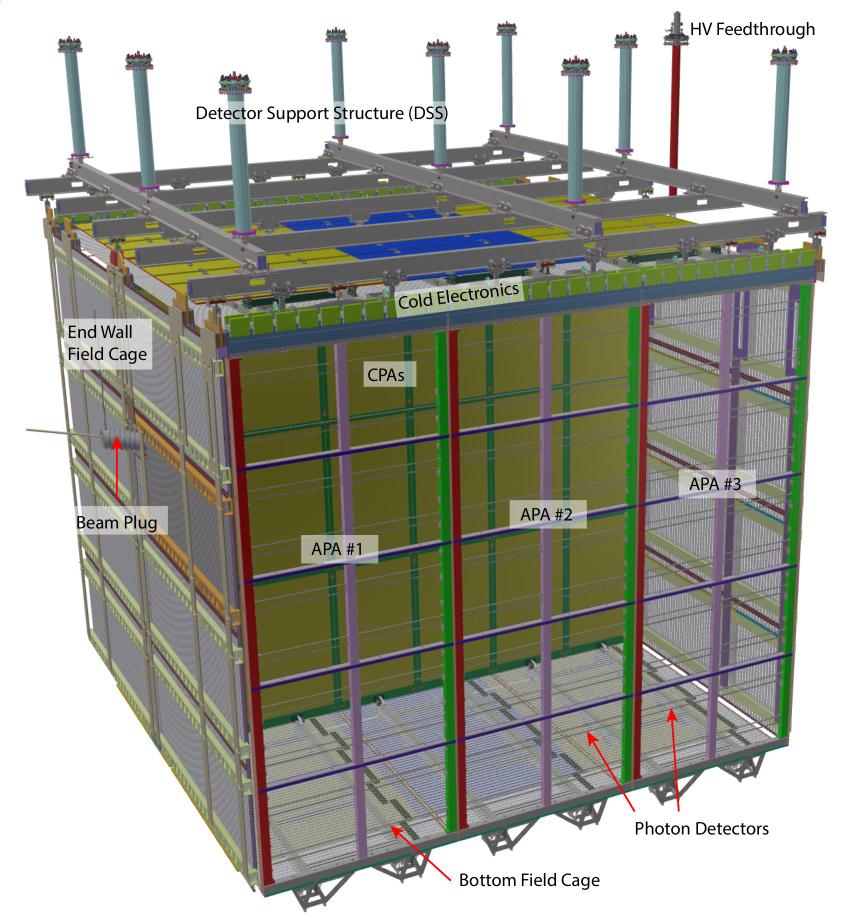}
\caption{\pdsp{} Internal Components}
\label{fig:tpc}
\end{figure}

The far detector DUNE-SP TPC components 
are designed in a modular way so that they can be transported underground. The \pdsp{} components are therefore also modular.  
Six Anode Plane Assemblies (APAs) are arranged into the two APA planes in \pdsp{}, each consisting of three side-by-side APAs. Between them,  
a central cathode plane, composed of 18 Cathode Plane Assembly (CPA) modules, splits the TPC volume into two electron-drift regions, one on each side of the cathode plane. High voltage (HV) is delivered to the cathode plane by the HV feedthrough.
A field cage (FC) completely surrounds the four
open sides of the two drift regions to ensure that the electric field within is uniform and unaffected by the presence of the cryostat walls and other nearby conductive structures. The detector dimensions are 7\,m along the beam direction ($z$ coordinate), 7.2\,m wide in the drift  direction ($x$ coordinate), and 6.1\,m high ($y$ coordinate). The detector elements are suspended from the cryostat roof by the Detector Support Structure (DSS).
The Photon Detection System (PDS) modules are embedded in the APAs. They will  collect scintillation light from ionized LAr. 
 Ten bar-shaped photon detectors with dimensions of 8.6 cm (height) 2.2~m (length) and 0.6~cm (thickness) are installed at equally spaced heights within each APA. Three different designs of photon-detector technologies are implemented in \pdsp{}, all based on light readout by silicon photomultipliers (SiPM).
Figure~\ref{fig:tpc} illustrates how these components fit together.
Cryogenic instrumentation, including a purity monitor, temperature sensors, and cameras, are located in the space between the cryostat walls and the detector elements.
The detector components are described in detail in Section~\ref{sec:detcomponents}, and  Section~\ref{sec:Assembly} presents a description of how those components were assembled together inside the cryostat.  

\subsection{Data Acquisition and Detector Control}
The Data Acquisition (DAQ) system,  a central element of \pdsp{},  interfaces with the detector's readout electronics, with external devices used for triggering (e.g., beam instrumentation), with the Detector Control System (DCS), and with the offline computing. The DCS (known historically as ``slow control'') 
is in charge of the monitoring and control of the detector and includes hardware and software elements that  give information and access to the detector subcomponents. The DAQ and DCS are described in Section~\ref{sec:detcomm} of this paper.


\section{The Neutrino Platform at CERN} 
\label{sec:NP}
\label{sec:assy:np}

Following the recommendation issued by the 2013 European Strategy for Particle Physics ~\cite{EPS:2013}
the Neutrino Platform was 
established at CERN to foster the European contribution to the next generation of long-baseline neutrino experiments. Use of this new and versatile facility is regulated through Memorandum of Understanding (MOU) agreements, making it easy to welcome contributions from new collaborators and new projects within the facility.
Since its establishment, several new projects have been included and extensions of existing ones have been approved.

The CERN NP is located at the Prévessin site, in an extension of Building 887 (EHN1) built explicitly for housing 
the two large-scale ProtoDUNEs, \pdsp{} and the dual-phase \pddp{}, where they can be exposed to charged particle beams. The Neutrino Platform 
provides 
the infrastructure needed to safely construct, install, and operate the two large LArTPCs, e.g., 
storage space for large components, dedicated overhead cranes for the handling of heavy equipment by trained personnel, logistics and transport services coherently integrated with the CERN global transport service, and mobile elevating working platform devices. 





\begin{figure}
 \centering
 \includegraphics[width=0.9\linewidth]{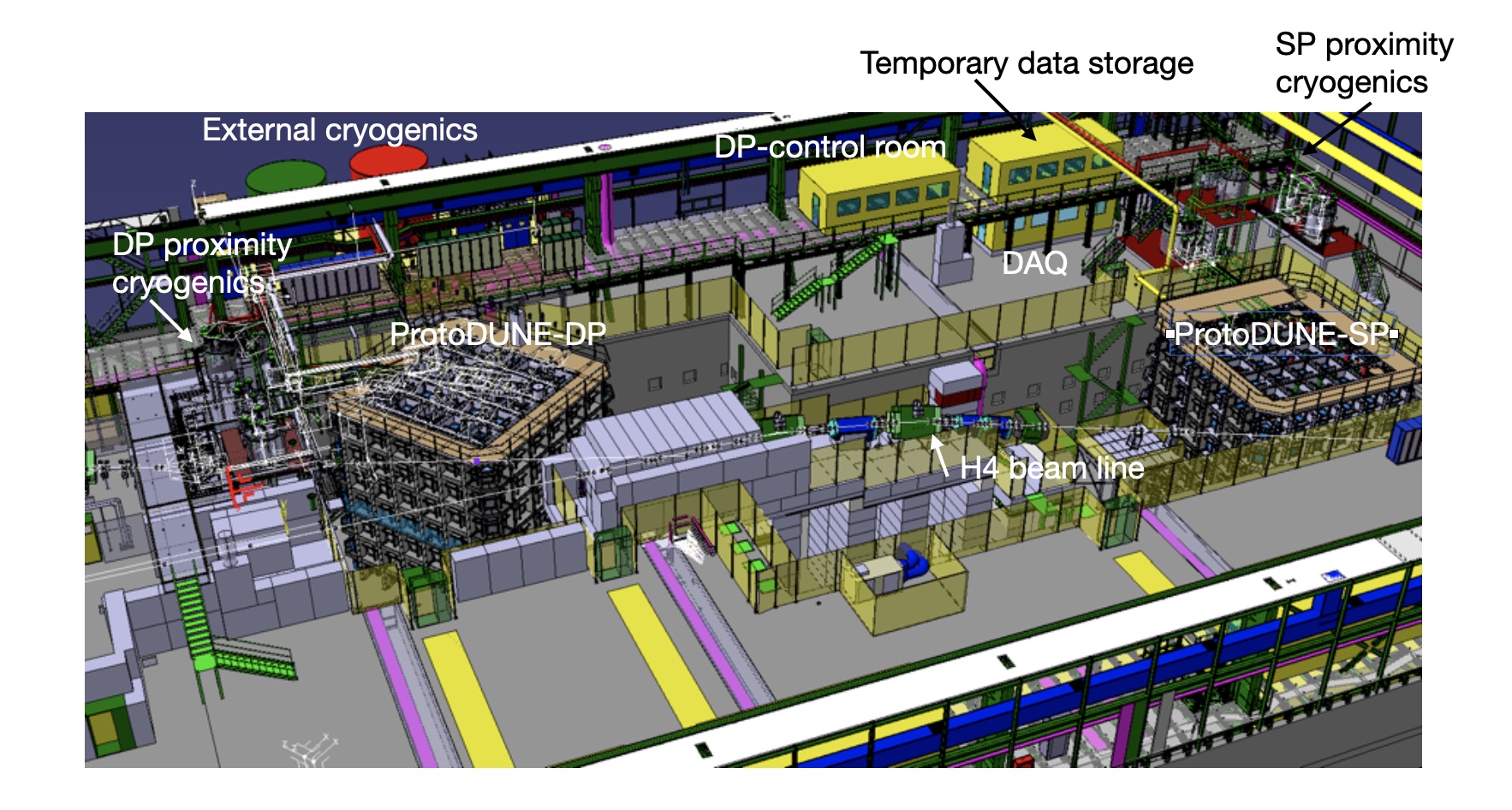}
 \caption{ \label{fig:techdrawEHN1} Layout of the EHN1 experimental area. The Dual Phase cryostat is on the left, and the Single Phase cryostat is on the right. 
The rotation of the cryostats have been defined to maximise the path of the particle beams in the drift volume given that in \pddp  the drift volume is unique, while \pdsp is divided into two parts by a central cathode.}
\end{figure}

Each of the ProtoDUNEs is installed inside a cryostat located in an open trench in EHN1  12\,m deep.
The building also houses 
prefabricated experimental control rooms and a cryogenics control room, as well as rooms dedicated to  data acquisition and computing/storage facilities. 
Figure~\ref{fig:techdrawEHN1} shows a layout of the experimental area. 


A specialised piece of infrastructure is the so-called \pdspalt{} cold box and its cryogenics system. Built for  characterisation of the APAs at  low temperature prior to their installation into the \pdsp{} cryostat, it is a tall, thin cryostat filled with cold nitrogen gas.

Two newly designed Very Low Energy (VLE) beamlines, H2 and H4, serve the two detectors \cite{PhysRevAccelBeams.20.111001}. 
Protons from the Super Proton Synchrotron (SPS) strike the T2 target about 600~m upstream of the detectors, generating secondary particles that collide with secondary targets that in turn produce particles of momenta in the region of 0.3-7~GeV/c \cite{PhysRevAccelBeams.22.061003}. The H4 beam is directed to 
\pdsp{} and the H2 beam to \pddp{}.  The beam lines and the beam instrumentation were commissioned in the summer of 2018 and the H4 beam extension was operated until the beginning of CERN's Long Shutdown~2 in November 2018 \cite{Calcutt:2021}.


Two large clean rooms next to each cryostat were used to construct, assemble, and manipulate the delicate detector components using procedures developed to ensure the required level of cleanliness of the atmosphere and equipment. 

Each cryostat has its own cryogenics system that comprises (1) the inner cryogenics and (2) the proximity cryogenics. These systems together provide and control the flow of the liquid and gases, and monitor and act to preserve both the thermodynamic conditions inside the cryostat and the liquid argon purity. The cryogenics system also has equipment external to the building that is used by both  cryostats. This external system stores and delivers the liquid argon for the cryogenics operations as described in Section~\ref{sec:cryogenics}.  


A very tight schedule and rigorous safety requirements led to very challenging  installation and  commissioning periods for the \pdsp{} cryostat and detector. A dedicated team and a set of tools and protocols were established to ensure that safety requirements were met and to coordinate parallel activities. 

During the installation phase, mechanical and structural hazards posed the most significant safety concerns, and during the operational phase, cryogenics and radiation hazards were the prominent concerns. The facility was fully equipped to ensure safety of the personnel and the equipment at all times.
%
Systems and procedures were put in place to mitigate risks associated with these hazards, including personnel training, specific risk analyses for non-standard activities, systems for oxygen deficiency hazard (ODH) monitoring and for fire detection, both linked to the hall ventilation and to the evacuation sirens and the CERN Fire Brigade. During beam operations, protocols were put in place to avoid accidental exposure to beam, e.g., access to the beam area and the trenches was controlled and interlocked with the beam. The radiation level in the hall was monitored with a series of strategically positioned radiation monitors integrated into the beam interlock system.




The CERN NP has approved Phase II of \pdsp{}, the most significant of its new projects and extensions. Starting in 2022, once the DUNE-SP detector component designs have implemented all the improvements of the past few years and are final, the components will be installed in the same cryostat used for Phase I and tested.

         \clearpage

\section{Cryostat and Cryogenics} 
\label{sec:cryo}
      
The cryostat consists of a free-standing warm steel outer structure, layers of insulation, and a cold inner membrane.  
Its design is based on the technology used for liquefied natural gas (LNG) storage and transport. The outer structure, which provides 
the mechanical support for the membrane and its insulation, consists of vertical beams that alternate with a web of metal frames. It is designed to withstand the hydrostatic pressure of the liquid argon and the pressure of the gas volumes, and to satisfy the external constraints.

\subsection{Cryostat Assembly: Design, Installation and Validation}
\label{sec:cryostat}

The LBNF/DUNE far detector membrane cryostats are required to be constructed of modular components that can be transported to the underground caverns at SURF via a 1.4~m $\times$ 3.8~m cross section shaft for assembly \emph{in situ}.
The \pdsp{} cryostat, which is intended to validate the far detector design, was constructed of modular components of the same design and size. The maximum dimensions and weight of the cryostat components were optimized to work in both the SURF and CERN locations. 

The ProtoDUNE-SP cryostat design builds on similar technologies used for liquefied natural gas (LNG) storage and transport tanker ships. A 1.2\,mm thick corrugated stainless steel membrane forms a sealed container for the liquid argon, with surrounding layers of thermal insulation and vapor barriers. Outside these layers, a steel frame forms the outer (warm) vessel. The roof of the cryostat supports most of the components and equipment within the cryostat, e.g., the time projection chamber (TPC) and photon detection system (PDS) components, electronics, and sensors. 

The ProtoDUNE-SP cryostat holds approximately 550\,m$^3$ of LAr, assuming an ullage of 5\%, at a temperature between 86.9\,K and 88.2\,K with an absolute pressure inside the volume in the range 970--1100\,mbar\footnote{the quoted pressure are referring to the pressure in the ullage, if not specified otherwise. }.
The maximum design gauge pressure is 350\,mbarg. The average heat leak is tightly controlled and kept around 8\,W/m$^2$ to avoid rapid boiling of the LAr and therefore limit the consumption of liquid nitrogen (LN$_2$) which is used during normal operations to recondense the boiled off argon via a heat exchanger. 

The \pdsp{} cryostat and associated cryogenics have successfully validated the designs for use in large-scale 
LArTPC experiments and in particular for the DUNE far detector.

\subsubsection{Cryostat Outer Structure}
\label{sec:warmstruct}

The cryostat is a free-standing, electrically isolated structure assembled in the EHN1 hall at the CERN Pr\'{e}vessin site.
The room-temperature outer structure  provides the structural integrity of the entire setup, being capable of withstanding the hydrostatic pressure of the LAr, the pressure of the argon vapor, the detector weight, and gravitational and seismic forces. 
The overall nominal outer dimensions are: $11.4~\mathrm{m} \, \times \, 11.4~\mathrm{m} \, \times \, 10.8~\mathrm{m}$ (W$\times$L$\times$H) and the nominal inner cryostat dimensions are: $8.5~\mathrm{m} \, \times \, 8.5~\mathrm{m} \, \times \, 7.9~\mathrm{m}$. 

These dimensions are dictated by several constraints: the required active volumes of LAr and ullage, the layout of the penetrations, installation needs, distances from the active volume to the cryostat inner walls, and  insulation thicknesses.

\begin{figure}
 \centering
    \includegraphics[width=0.9\textwidth]{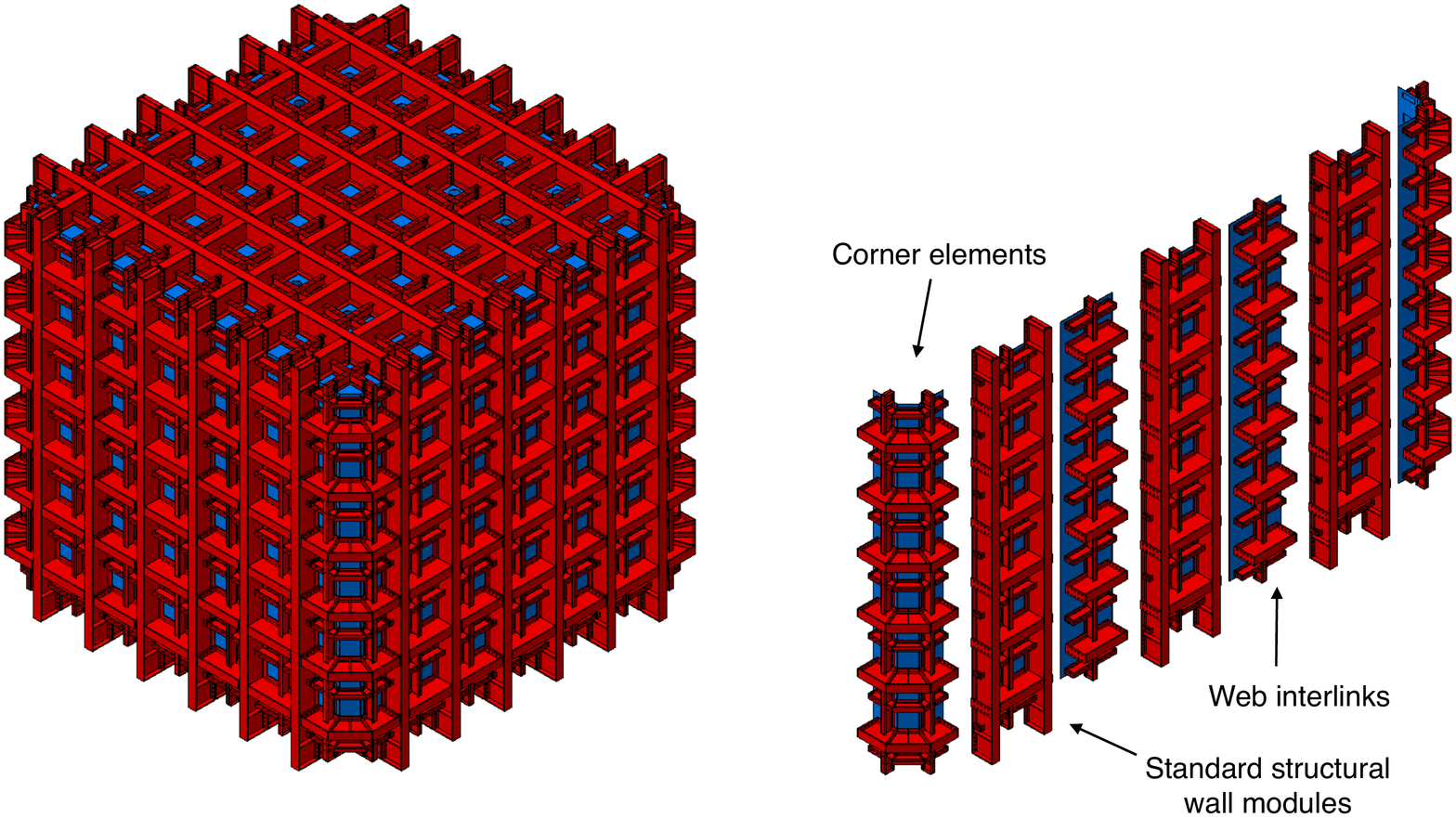}
 \caption[Warm vessel layout]{\label{fig:warmcryoout} Left: diagram of the \pdsp{} warm vessel. Right: exploded view of one cryostat face showing the corner elements at each end, with standard structural modules, and the web interlinks used to assemble each face.} 
\end{figure}

The outer structure consists of an assembly of prefabricated S460ML~\cite{EN10025-4:2004}  
steel modules of three configurations: standard structural modules used to construct the walls and floor, corner elements and a web of metallic frames to interconnect the modules. 
The complete structure and the modules used to assemble each face of the cryostat are shown in Figure~\ref{fig:warmcryoout}. 
The modules are assembled using 
standard hot-rolled profiles welded together with bolted connections. 


Three of the four cryostat outer walls are identical. The front wall, through which the beam enters, has a beam window (Section~\ref{sec:detcomp:inner:hv:bp}) and  
a separate wall section that is sealed in place over a  Temporary Construction Opening (TCO), a 7.3\,m high and 1.2\,m wide entrance used for installing detector components.

\subsubsection{Insulation and Cold Membrane}
\label{sec:coldmemb}

The inner vessel design, installed inside the warm support structure, is based on the LNG transport 
membrane technology developed by the firm GTT (Gaztransport \& Technigaz) \cite{GTT}. The overall thickness of the inner vessel is 800\,mm. The design fulfills the requirement that thermal fluxes not exceed 5-6 W/m$^2$ \footnote{This value also accounts for the aging properties of the insulation material (20~years of operation).}. The outermost layer of the inner vessel is a 10\,mm thick stainless steel membrane ~\cite{EN1993-1-4} (the tertiary membrane) that provides an effective gas enclosure and allows for control of the nitrogen atmosphere inside the insulation volume, which prevents condensation in the insulation layer.

\begin{figure}
\centering
\begin{subfigure}{.48\textwidth}
  \centering
  \includegraphics[width=.9\linewidth]{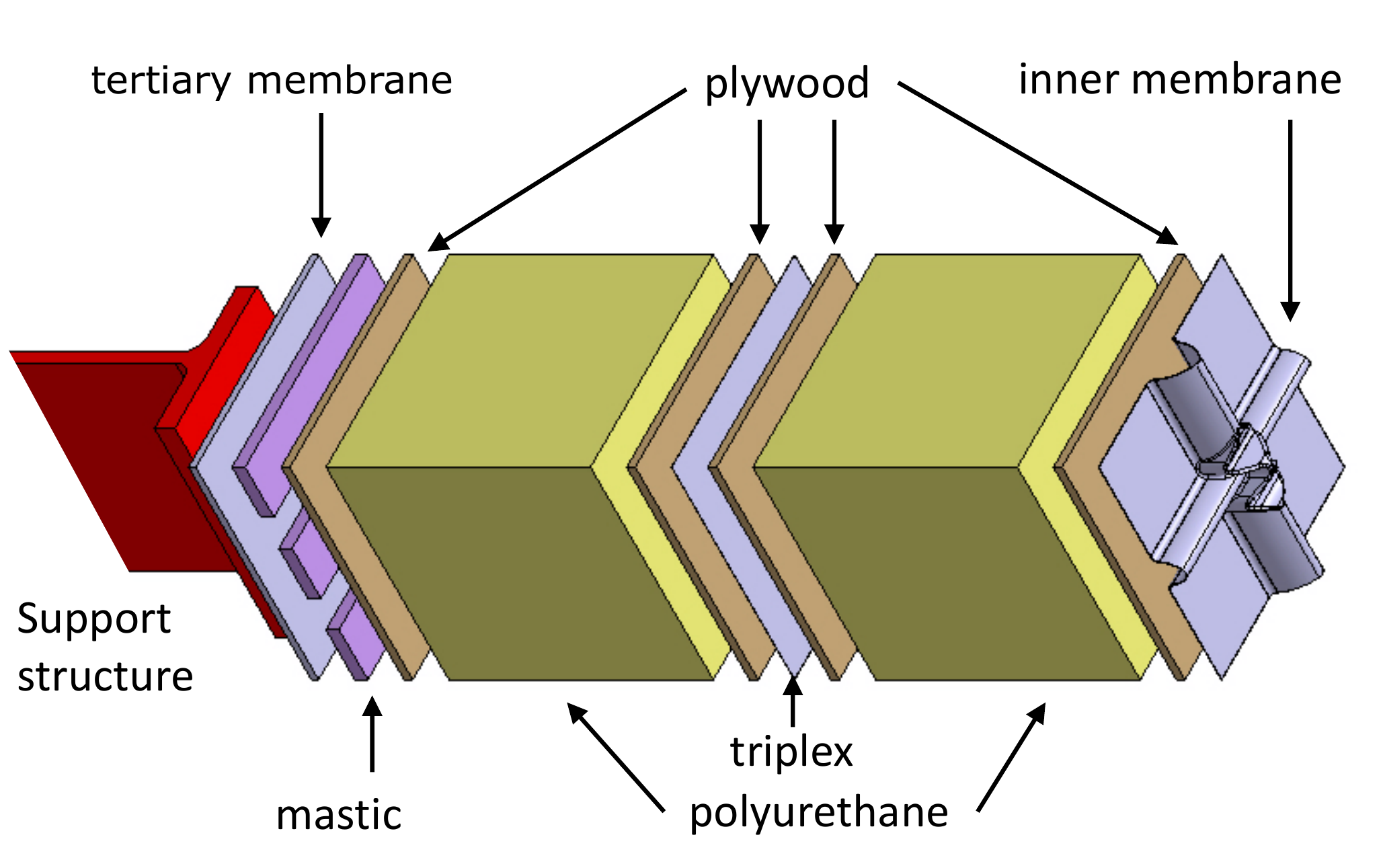}
  \caption{ \label{fig:membrane} }
\end{subfigure}%
 \vspace{.4cm}
\begin{subfigure}{.48\textwidth}
  \centering
   \includegraphics[width=.9\linewidth]{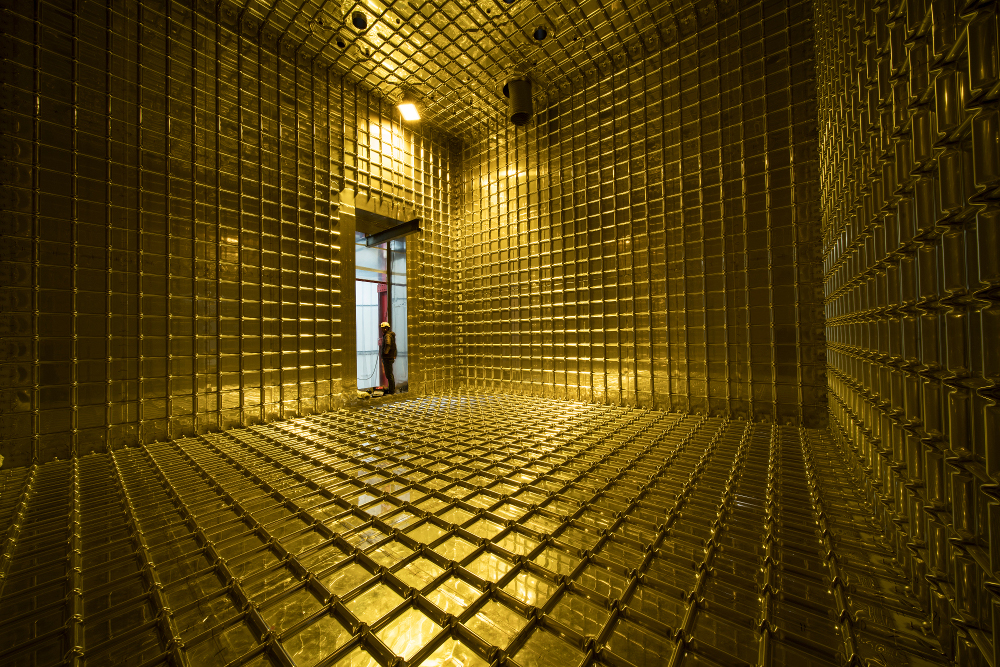}
  \caption{\label{fig:pdsp_empty} }
  \end{subfigure}
\caption[The GTT membrane layout]{\label{fig:} Left: the GTT inner vessel layers. Right: view of the inside of the cryostat \pdsp{} just after the finalisation of the primary membrane installation. The gold hue visible in the picture is an artifact of the lighting used during the installation, to protect the PDS.}
\end{figure}

Moving inward, just inside that membrane is the insulation layer that consists of two 390\,mm thick layers of a foam specially developed for this purpose,  supported by plywood plates of 9\,mm and 12\,mm thickness.  The foam material is expanded polyurethane of density 90~kg/m$^3$. The outer insulation layer is attached to the tertiary membrane by a set of rods and special mastic. Between the two insulation layers is a secondary LAr containment layer made of GTT-proprietary Triplex, a composite material consisting of a thin sheet of aluminum laminated on both sides by a layer of glass cloth and resin. In contact with the innermost side of the foam insulation is the 304L stainless steel 1.2\,mm thick primary membrane that holds the LAr. 
The primary membrane can expand or shrink in two dimensions as a function of the temperature, thanks to a special corrugation configuration that allows it to 
respond like a bi-dimensional spring. A cross sectional view of the insulation and membrane layers is shown in Figure~\ref{fig:membrane}.  Figure~\ref{fig:pdsp_empty} shows a view of the interior of the \pdsp{} cryostat before the insertion of the detector modules and the closure of the TCO. The corrugations 
of the primary membrane are visible. 


Once all the large detector components have been installed inside the cryostat, the TCO is closed.
The sealing is done in two steps since the entire length of the TCO is divided along the horizontal axis, into two parts. First the top part is bolted and sealed with the standard outer structure section. The layers of insulation are installed from inside the cryostat, following the standard sequence presented above. Once the primary membrane of the top part of the TCO is also installed, all the material needed for the bottom part is brought inside the cryostat and the sequence done for the top part is repeated. Since the sealing of the TCO involves  welding activities inside the inner volume, a temporary ``dirty room'' was built to protect the detector components during those operations.  

\begin{figure}
 \centering
    \includegraphics[width=0.6\textwidth]{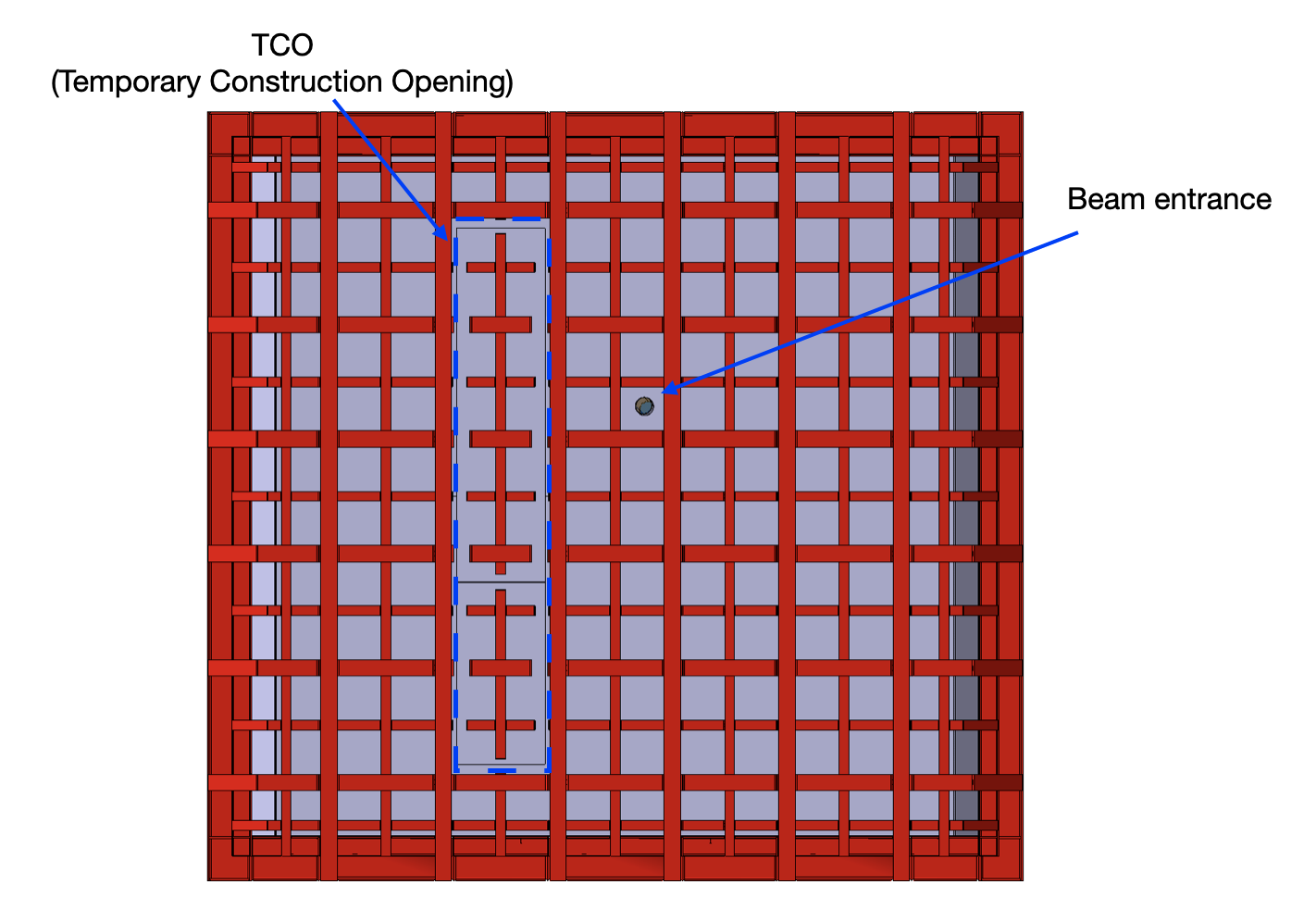}
 \caption[Warm cryostat front view]{ \label{fig:warmcryo_frontface}Front view of the \pdsp{} cryostat model. The Temporary Construction Opening and the cut-away for the beam entrance are shown.
}
\end{figure}

Figure~\ref{fig:warmcryo_frontface} shows the TCO. The circular opening for the beam entrance is also visible.


\subsubsection{Signal and Service Penetrations}
\label{sec:cryo_DSS}

The penetrations into the cryostat are installed  on the roof of the warm structure, with a couple of exceptions. 
To keep the high level of purity required, LAr is extracted from a point as low as possible in the cryostat and pushed by cryo-pumps to the external filtering system through the liquid recirculation circuit (see Section~\ref{sec:cryogenics}). A special penetration is thus required low on one of the cryostat's side walls to connect to the LAr pumps. This penetration has a dedicated system of safety valves and involves local modifications to the insulation panels and the primary membrane.

Since \pdsp{} was designed for exposure to a charged particle beam from CERN's SPS, a second modification to the cryostat was made to accommodate a beam window. The window minimises energy loss and scattering, which would be much higher if the beam were to pass through all the cryostat layers (see Section~\ref{sec:detcomp:inner:hv:bp}). 
The installation of the beam entrance window does not require any particular modification of the warm support structure. The tertiary membrane has been perforated with a hole where the beam entrance window is installed.  
The beam entrance window is composed of 175 $\mu m$-thick Mylar foil and a gate valve that opens when the beam is present and closes for safety reasons when the beam is off. 
The insulation between the secondary and the tertiary membrane is removed in the region around the beam window and the foam between the secondary and the primary membrane is replaced with a lower density foam (9 kg/m$^3$). Finally the plywood supporting the primary membrane in the vicinity of the beam window penetration is replaced with a Nomex honeycomb plate sandwiched between thin G10 layers.
Nomex is a polymer material with high thermal resistance and Nomex sandwiches are well known for their structural resistance . Thermal and stress analyses to validate the design were conducted in collaboration with GTT. The total amount of material, including the (unaffected) primary membrane, given a 0.3\,mm G10 thickness and a 0.3-mm-thick steel beam window, is equivalent to 10\% of a radiation length.

The other penetrations, all on the cryostat roof, are grouped by position, diameter, and function.
They provide feedthroughs for the TPC detector power and signals, support for the argon instrumentation devices, and feedthroughs for the instrumentation signals. Of the 55 penetrations, 43 go through to the liquid of which seven are dedicated to the cryogenics system, nine to the detector support system, six to detector charge and light readout, and one to the cathode HV feedthrough.
One is for the beamplug services, one for the diffuser fibre, 11 for the monitoring system (cameras, temperature and pressure sensors, purity monitors). There are also two manholes and five spares. The remaining 12 penetrations are limited to the insulation volume and are dedicated to either input/output of nitrogen gas or temperature and pressure sensors.

\subsubsection{Cryostat Assembly Procedure }
\label{sec:cryoIstal}

\begin{figure}
 \centering
    \includegraphics[width=0.9\textwidth]{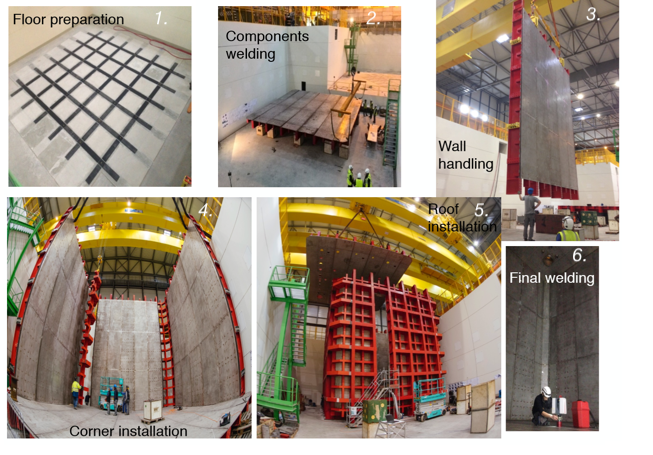}
 \caption[Cryostat installation procedure] { \label{fig:inststeps} Cryostat warm outer structure assembly sequence, left to right.  }
\end{figure}

Figure~\ref{fig:inststeps} shows steps in the assembly sequence of the outer support structure. 
Installation starts with a planarity survey of the floor and the positioning of the elastomer pads and G10 strips (for seismic protection and electrical isolation) on which the cryostat will be placed. Three pre-assembled large wall pieces are positioned horizontally on temporary  concrete supports, then three web interlink modules are positioned between them and connected together as shown in Figure~\ref{fig:warmcryoout} (right). The stainless-steel plates of the outermost (tertiary) membrane are then welded together and a helium leak test is performed to qualify the welds (see Section~\ref{sec:cryostatHelium}).
To ensure the planarity of each completed wall, a detailed survey is carried out before it is lifted and installed in its final position on the elastometer pads. Wherever needed, temporary stabilisers are added to the experimental pit concrete walls to ensure safety during this process. Once the four walls are in place, the four pre-assembled cryostat corners are installed and the stabilisers are removed. Finally, the roof modules are assembled and the roof is installed. A final helium check of the SS plate welding is performed, and a global three-dimensional scan is done to verify the internal dimensions. 

 Once the outer structure and the tertiary membrane are installed, the pre-cut inner layer sections that come pre-assembled with mastic and fasteners are installed. After each layer is completed, all voids are filled to stop any circulation paths. At the end, helium leak tests are performed once again (see Section~\ref{sec:cryostatHelium}). The installation proceeds by layers (see Figure~\ref{fig:membrane}) with the inner membrane installed  last. Installation of the \pdsp{} warm structure took about 12 weeks.

\subsection{Cryostat Structure Validation and Testing Campaigns}
\label{sec:validation}

Validation and certification of the cryostat structure has two aspects. The first, leak checking, is performed at various stages of the cryostat construction. A description of these tests is given in Section~\ref{sec:cryostatHelium}. 
The second  concerns the mechanical behaviour of the cryostat in terms of compliance with engineering safety standards and regulations. This validation is also performed multiple times throughout the course of construction, and also includes a test campaign done during and after filling with LAr.  The structural and mechanical performance validation is described in Section~\ref{sec:mechtests}.

\subsubsection{Cryostat Tightness Verification}
\label{sec:cryostatHelium}

Leak testing is performed on the external warm structure, the inner membrane, and all the penetrations. 
The tests are generally done using helium and 
a leak detector in {\it sniffing mode}. A deviation from the environmental background ($\sim2-3 \times 10^{-6}$~mbar~l/s) is considered a possible leak. 

\subsubsection*{Warm Structure}
The walls of the cryostat outer structure (see Section~\ref{sec:warmstruct}), despite their separation from the LAr volume, are required to be leak-tight for two main reasons: $(i)$ they are the last barrier between the LAr volume and the external world; and $(ii)$ any leaks on the outer surface would be sources of localised heat and humidity injection. The latter would produce water that would freeze inside the insulation volume, weakening it and making it less effective. The checks were performed during the assembly of \pdsp{} and no leaks were found.

\subsubsection*{Inner membrane}
The cryostat inner membrane (see Section~\ref{sec:coldmemb}) was tested twice for leaks. 
The first test, upon completion of the membrane installation, was the official qualification for leak-tightness by the vendor, and was
performed on all welding seams. In this test, the insulation volume, normally flushed with N$_2$ gas, undergoes a few cycles of evacuation (down to roughly 550~mbar), then is filled with helium to match the external pressure. This allows a uniform fill of the insulation with helium. 
Two test holes in the lower inner membrane are intentionally not welded at this stage, but simply closed with mastic to allow occasional checks for the continued presence of helium in the volume under test. The few small leaks found were swiftly repaired.  These test conditions also allowed for continuous monitoring of the insulation space internal pressure stability (Primary Barrier Global Test).

\begin{figure}[tb]
\centering 
\includegraphics[width=0.7\columnwidth]{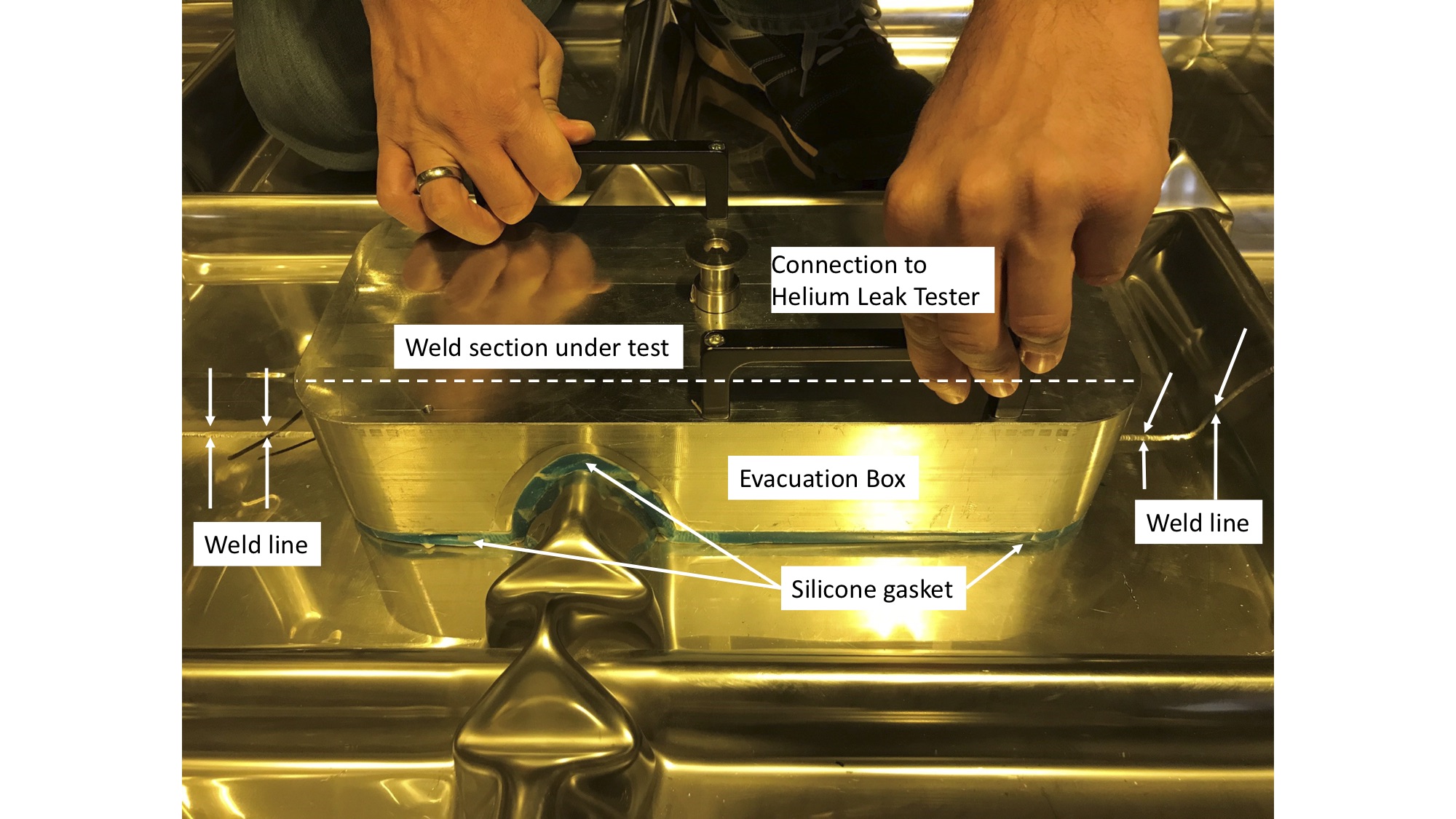}
\caption[Vacuum box]{Example of vacuum box constructed for tests on flat walls. Two classes of such boxes were produced, in SS and in Plexiglas (lighter) for upside-down use on ceiling welding lines. The silicon-based gasket was produced by the CERN polymer lab.}
\label{fig:vacbox} 
\end{figure}

The second leak test 
involved pulling vacuum around localised sections of welding lines and testing them one by one. 
The test measures the leak rate from the helium-rich atmosphere, through the welds of the primary membrane, to a local volume evacuated by means of a vacuum box (Figure~\ref{fig:vacbox}). 
The vacuum boxes come in a variety of shapes  
to fit corrugations, corners, and flat welded regions.  Overall, about 80\% of the total length of welding lines on the inner membrane was tested and no leaks were found. 
It is worth mentioning that a similar double-test procedure 
was performed on the ProtoDUNE-DP cryostat's inner membrane and yielded the same result. This gives confidence that the helium leak-check performed by the vendor can be fully trusted when it comes time to verify the cryostat inner membranes for the far detector, and that performing the second test will be unnecessary. Retesting single welded lines on a DUNE-scale cryostat would require months.

\subsubsection*{Roof penetrations}
The leak-tightness of all feedthrough flanges was verified. This was done to avoid the potential formation of argon gas (GAr) pockets on the cryostat roof, a working area during operation, and also to prevent air leaks that could pollute the LAr volume.
Each penetration was tested after closing the corresponding flanged connections. 

The test procedure changed slightly depending on the stage of the detector preparation: the helium was injected into the penetration under test either from the inside (cryostat volume still accessible) or using the argon gas exhaust lines present in every penetration (detector installation completed).

All flanges except one were certified to be leak-tight; the chimney of one temperature profiler (see Section~\ref{sec:ci:t-pro}) was found to be faulty. None of the actions taken fixed the leak entirely, therefore an enclosure was constructed and installed around the leaky flange, defining a buffer volume in which argon gas circulated continuously.

\subsubsection{Cryostat Mechanical Performance and Test Campaign}
\label{sec:mechtests}

The cryostat inner and outer structures must satisfy European regulations and standards, as the \pdsp{} cryostat was installed and operated in Europe. In addition, since the design and the construction methods and technologies are prototypes for the larger DUNE cryostats, compliance with U.S. standards and regulations is also required. 

Rigorous quality assurance and control procedures were carried out on all materials and techniques used for the construction of the \pdsp{} cryostat. The cryostat outer structure is instrumented with strain gauges and displacement sensors for making the required measurements.  
To fully validate the mechanical performance of the cryostat, a careful stress analysis of each structural component was done, together with pressure tests before and after filling with LAr. 

\begin{figure}
\centering 
\includegraphics[width=0.8\columnwidth]{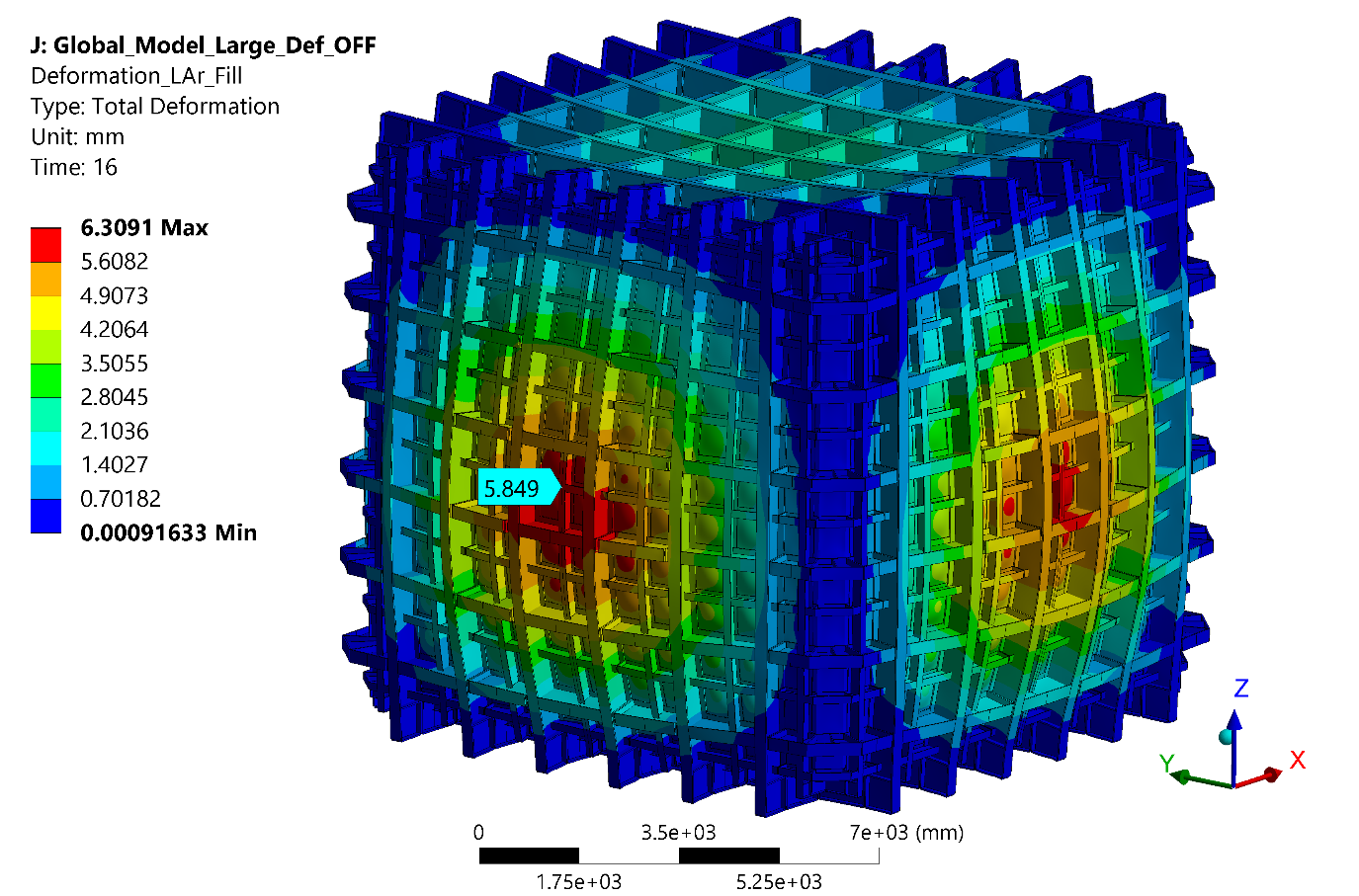}
\caption[Expected deformation of the cryostat]{Expected deformation (using an exaggerated scale) of the cryostat during standard operations (liquid argon level of 7.4\,m and $\Delta P = +57 $ mbarg. Predictions are obtained using the \emph{shell} FEA model (see text).}
\label{fig:global_deform} 
\end{figure}

Two Finite Element Analysis (FEA) models were developed to evaluate the cryostat design against the required codes and regulations. Predictions from the models were compared to the experimental data collected during the commissioning phase and operation. A relatively simple \textit{beam model} includes simulations only of the cryostat outer structure and is used to study the moments and forces acting on the cryostat. This model was developed for the relative ease of obtaining reliable verification studies. The \textit{shell model} is a more complex simulation that includes the entire inner vessel, which contributes to the overall stiffness of the structure. This model was developed to provide detailed predictions of structural deformations as well as stresses on the frame and outer membrane. Its results are compared against the measured stresses to validate the cryostat behaviour for the sizing case (i.e., the cryostat  full of LAr and at 350\,mbarg over-pressure). It can predict strain and deformation during the three main phases: 
commissioning, filling with LAr, and operation.  Figure~\ref{fig:global_deform} illustrates a possible deformation during standard operation. Note that the image vastly exaggerates the actual deformation.

Strain gauges and displacement sensors are installed 
at the positions where the simulation predicts the highest stresses.
All sides of the cryostat are instrumented in similar locations to check the symmetrical behaviour of the structure. 

Four displacement sensors, commonly known as \textit{Linear Variable Differential Transformers} (LVDTs), are installed to monitor the symmetry and magnitude of the cryostat's global deformation. A total of 55 strain gauges (SGs) are used to measure local deformation of the structure under stress. 
A detailed description of the instrumentation can be found in~\cite{straingages}.

The sensors are read out continuously during cryogenics commissioning, in order to  $(i)$ check whether the maximum stresses and strains are exceeded in the most loaded areas and whether the structure demonstrates  elastic behaviour; $(ii)$ monitor the maximum deformations and their symmetry; $(iii)$ serve as a secondary safety indicator of abnormal cryostat behaviour during commissioning. 


The \pdsp{} monitoring system is complemented by a pressure gauge to measure the pressure difference between the cryostat and the local atmospheric pressure, and four temperature sensors to monitor variations in the pit where the cryostat is installed and on the surface of its outer structure. 

Pressure tests validate the safety of the cryostat under pressure so that work activities can proceed in and around it, and they provide a final leak check on the cryostat roof  penetrations. The success of these tests is a fundamental prerequisite for the cryogenics safety permit approval, which is required for the start of the cryogenics commissioning.    

In a first pressure test, with the cryostat still empty, the over-pressure is increased to 200~mbarg,  then decreased in steps of 50~mbarg.  The maximum value is that which allows testing of the overall structure while keeping the most stressed areas within the elastic range. This value must also be far below  safety valve maximum pressure, which is set at 350~mbarg. A second pressure test is performed at the end of the cryogenics commissioning, before starting operations, with the cryostat 96\% full of LAr. The cryostat pressure is increased up to an over-pressure of 280 mbarg, again in steps of 50~mbarg. When performed on \pdsp{} each test lasted a few hours and  the maximal over-pressure remained stable for about an hour.  

\begin{figure}
\centering
\begin{subfigure}{.48\textwidth}
  \centering
  \includegraphics[width=.9\linewidth]{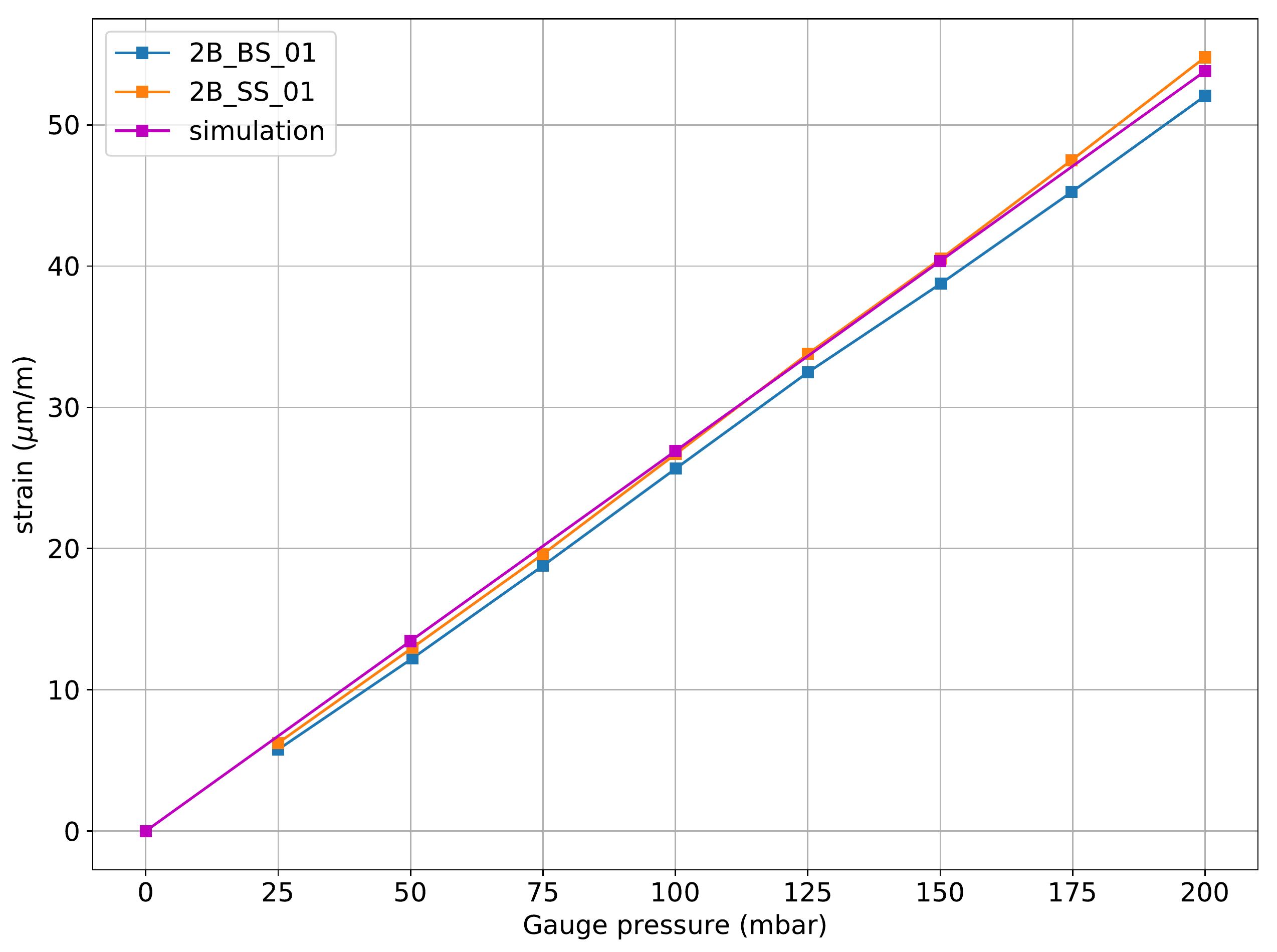}
  \caption{ \label{fig:sensorsvspressure} }
\end{subfigure}%
 \vspace{.4cm}
\begin{subfigure}{.48\textwidth}
  \centering
   \includegraphics[width=.92\linewidth]{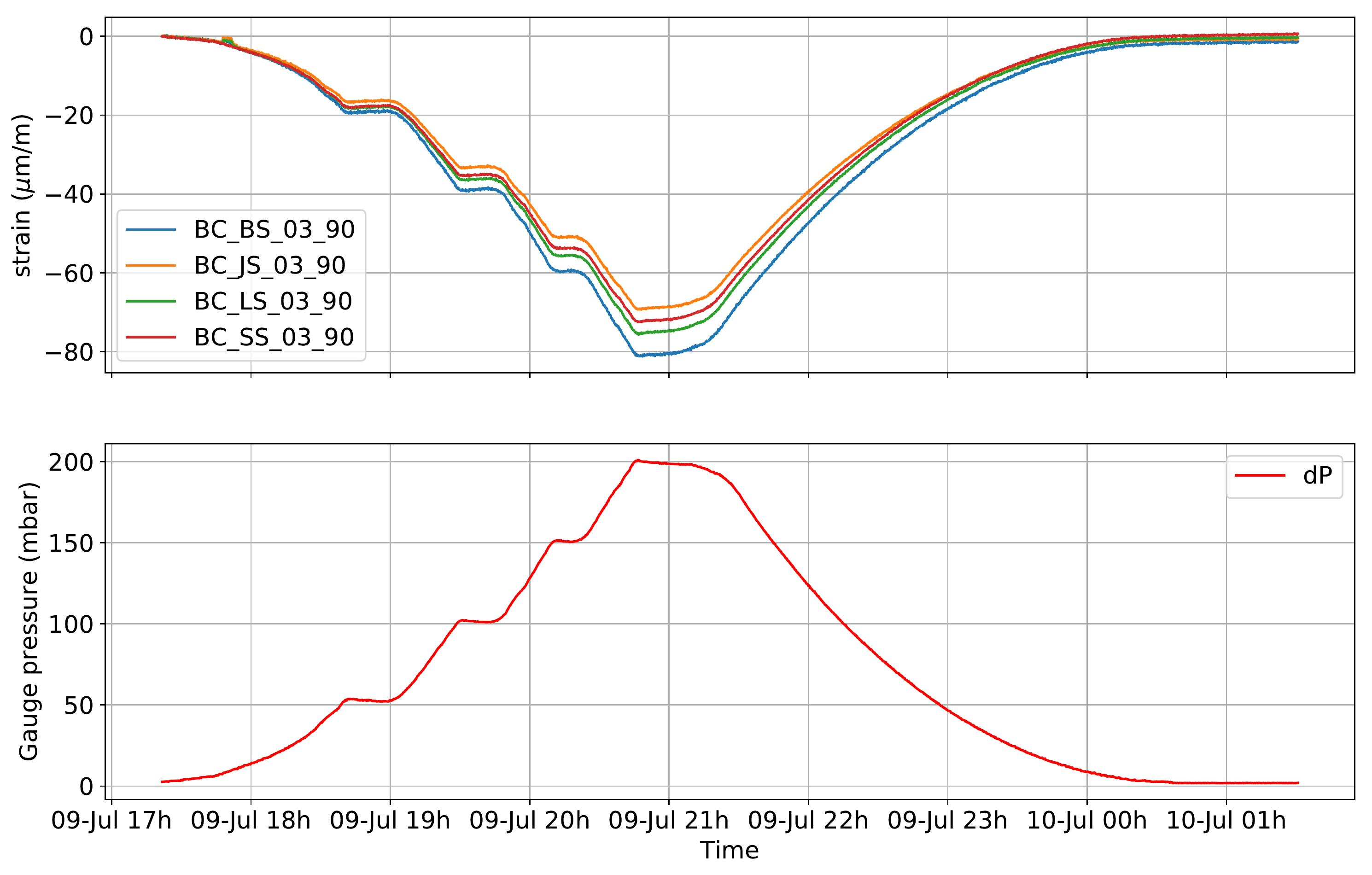}
  \caption{\label{fig:sensPvsTime} }
  \end{subfigure}
\caption[First pressure test]{\label{fig:first_ptest} Sampled results from the first pressure test. Left: strain-vs-pressure plot of two uni-axial gauges on the I-beams compared against the FEA simulation. Right: four bi-axial gauges in the same positions on the different detector walls (top) are compared to the time-trend for the pressure (bottom).}
\end{figure}

All tests performed during the validation campaign were successful and the FEA model  reproduced the cryostat behaviour in all cases. The pressure tests demonstrated the elastic behaviour of the cryostat structure: strain gauges showed a clear linear relation with the pressure and returned to zero at the end of the cycle (see Figure~\ref{fig:first_ptest}). Identically instrumented locations on the different walls showed symmetric readings, with the exception of the wall with the closed TCO, where differences are expected. 

The values of the displacement and strain gauges measured during the fill were compared to the FEA predictions and found to fulfill the safety requirements. This comparison also validated the FEA model itself, which was then used to predict the cryostat behaviour at the sizing case. The outcome of the simulation was compatible with both EU and U.S. standards, thus providing the final qualification of the inner and outer cryostat mechanical structures.   



\subsection{Cryogenics, Cooling and Purification System}
\label{sec:cryogenics}

The cryogenics system provides the equipment and controls for receiving the argon, for purging, cooling down, and filling the cryostat, for maintaining the LAr in the cryostat at the desired temperature, pressure and level, and for purifying the argon and maintaining the required operational purity. 
This system has been developed jointly by Fermilab and CERN, building on experience at Fermilab from the Liquid Argon Purity Demonstrator (LAPD)~\cite{Adamowski_2014}, the 35t prototype, and MicroBooNE, and at CERN from the WA105 1$\times$1$\times$3 Dual Phase prototype and the design and operation of large-volume noble liquid detectors. 
 The designed system described in this section is or will be applied to both ProtoDUNEs, the Short-Baseline Neutrino Near Detector (SBND), and the 
DUNE Far Detector~\cite{icalepcs2019:wepha049}. 
The Far Detector implementation will differ in that it will generate the liquid nitrogen 
locally in the underground area. 

\subsubsection{Overview} 
\label{sec:cryoOverview}

\begin{figure}
\centering 
\includegraphics[width=1.0\columnwidth]{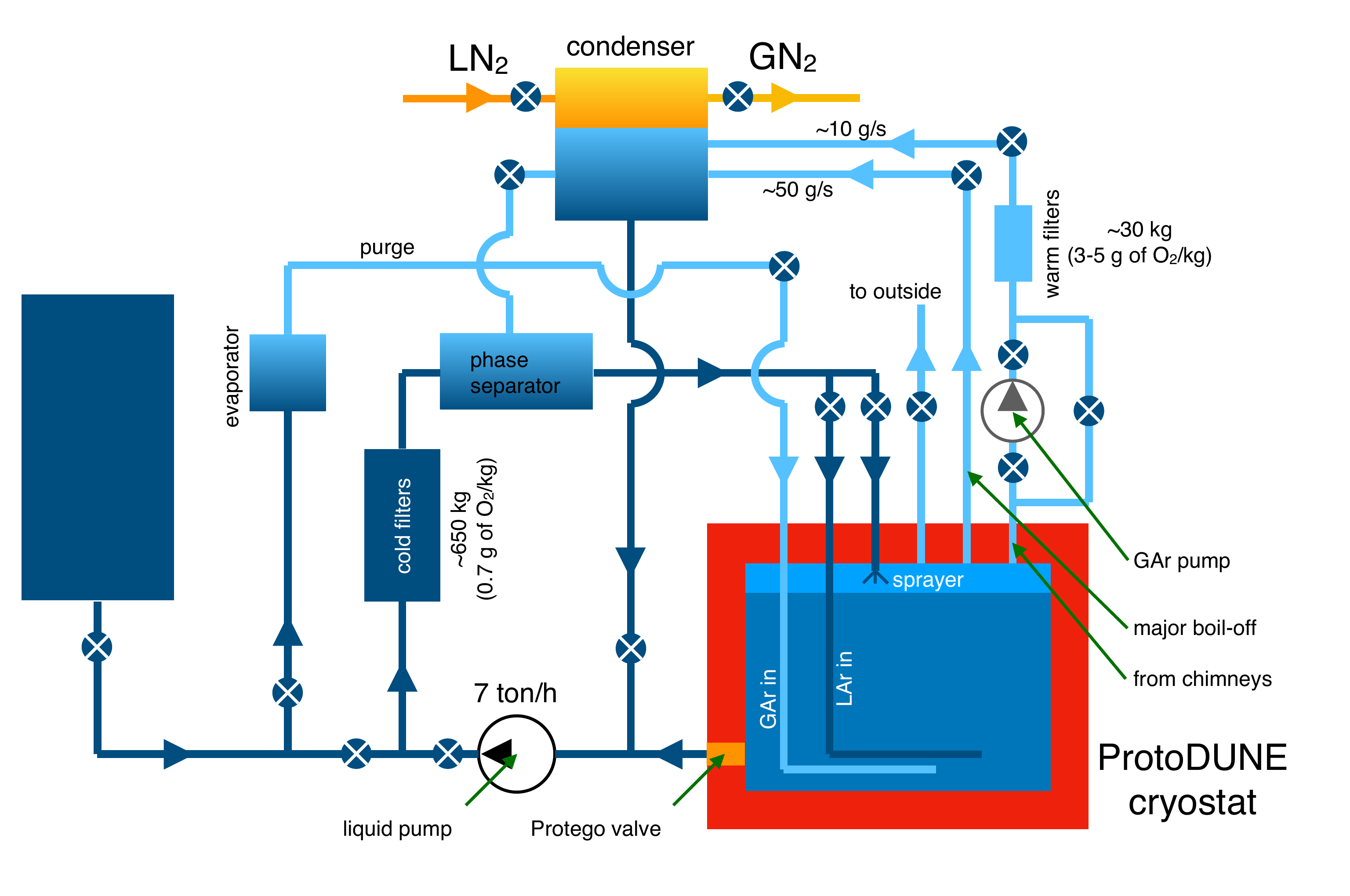}
\caption[Cryogenics process flow diagram]{Cryogenics process flow diagram. Light and dark blue lines show the path of the argon gas and liquid, respectively. The yellow line shows the nitrogen line used to re-condense the gas argon via a condenser before being injected into the purification chain. }
\label{fig:cryo-process-flow} 
\end{figure}

The Process Flow Diagram for the ProtoDUNE cryogenics system is shown in Figure~\ref{fig:cryo-process-flow}. 
The cryogenics infrastructure is divided into three main parts: the external cryogenics located outside the building housing the detector, the proximity cryogenics located next to the cryostat, and the internal cryogenics located inside the cryostat.
The external cryogenics facility includes the systems used for the receipt and storage of the cryogens used in the cryogenics system. 
Designed to serve both ProtoDUNE detectors, separate lines take LAr, GAr, and LN$_2$ to both installations. 
A 50 m$^3$ (70\,t LAr capacity) vertical dewar was used to receive the LAr deliveries for the \pdsp{} initial filling period. 
The specifications for the oxygen, water, and nitrogen contamination in the delivered argon are 2, 1, and 2\,ppm, respectively.  An analyzer rack near the dewar was used to check the levels of these impurities in the delivered LAr batches and
a 55\,kW vaporizer was used to deliver the gaseous argon to the cryostat. LN$_2$ deliveries were received and stored in a second 50\,m$^3$ vertical dewar (40\,t LN$_2$ capacity); the LN$_2$ is used in cool-down and normal operations, as explained below.
A $0.5 \,\mu$m mechanical filter is located on the LAr feed line to prevent any impurities in the LAr supply from entering the purification system and the cryostat. 
Since a GAr/H$_2$ mixture (2\% H$_2$) is used to regenerate the LAr and GAr purification filters, a separate 10\,m$^3$ storage dewar dedicated to this function connects to a cylinder of hydrogen (H$_2$) and a GAr line.

To fulfill the LAr purity requirements, the LAr and the liquid boil-off needs to be collected, purified, and reintroduced into the system. The proximity cryogenics takes care of all actions during the recirculation of both in liquid and gas phase. The system comprises the argon condenser, the purification system for the LAr and GAr, the LAr circulation pumps, and the LAr/LN$_2$ phase separators. 

The recirculation done at the liquid phase represents the major flow.  During normal operations, the continuous re-circulation rate was 7.0\,t/h, giving a full volume turnover time of 4.6 days, which kept the impurity concentration below $20$\,ppt oxygen equivalent. The LAr is transferred to the liquid purification system via external pumps located on one side of the cryostat more than 5\,m below the liquid level to avoid cavitation or vapor-entrapment. This purification system further reduces the initial impurity levels. The system has two pumps available to ensure continuous operation during maintenance. Only one pump is in service at a time. The purification system consists of three filter vessels; the first contains molecular sieve (4~\AA) to remove water, and the others contain alumina porous granules covered by highly active metallic copper for catalytic removal of O$_2$ by Cu oxidization. In addition, $15\, \mu$m mechanical filters are installed at the exit of the chemical filters.
 Saturation of the chemical filter occurs when the trapped/reacted impurity budget exceeds the removal capacity of the filter material. At this point the LAr flow is brought directly to the mechanical filter and the regeneration of the saturated chemical filter can start. Filter regeneration typically takes two days, after which the system returns to nominal operation.

The second main flow is given by the recirculation of the gas boil-off. The LAr in the cryostat continuously evaporates at a slow rate due to unavoidable heat ingress.  Therefore the boil-off is collected, condensed, purified, and reintroduced into the system. The main part of the argon vapour, roughly corresponding to  75\% of the total boil-off, is collected via a dedicated penetration in the cryostat roof. The vapor is then directed towards the argon condenser, which is a heat exchanger that uses the vaporization of LN$_2$ to provide the cooling power to condense the GAr. The newly condensed argon is then injected into the main liquid flow where it follows the liquid purification path carried out by means of the cold filters. 
About 15\% of the argon boil-off is removed through purge pipes connected to each penetration on the roof. This portion of gas is driven towards a purification system at RT that is composed of chemical filters similar to those used for the liquid. The design has a diaphragm pump to raise the pressure of the gas and a pressure-control valve to continuously monitor the gas flow to the condenser and maintain the pressure within the cryostat. 
The standard operating pressure is 1050\,mbar (absolute); on occasion the pressure was regulated on gauge from a minimum of 10\,mbar above ambient to a maximum of 250\,mbar. Under standard conditions the pressure regulation remains better than $\pm 1$~mbar.


The diaphragm pump failed at some point during operations. While this pump was bypassed,
the flow of the argon gas was instead controlled by adjusting the pressure.
This demonstrated that the diaphragm pump was not essential and highlights the
importance of designing flexibilty into the cryogenics system.

\begin{figure}
\centering 
\includegraphics[width=0.8\columnwidth]{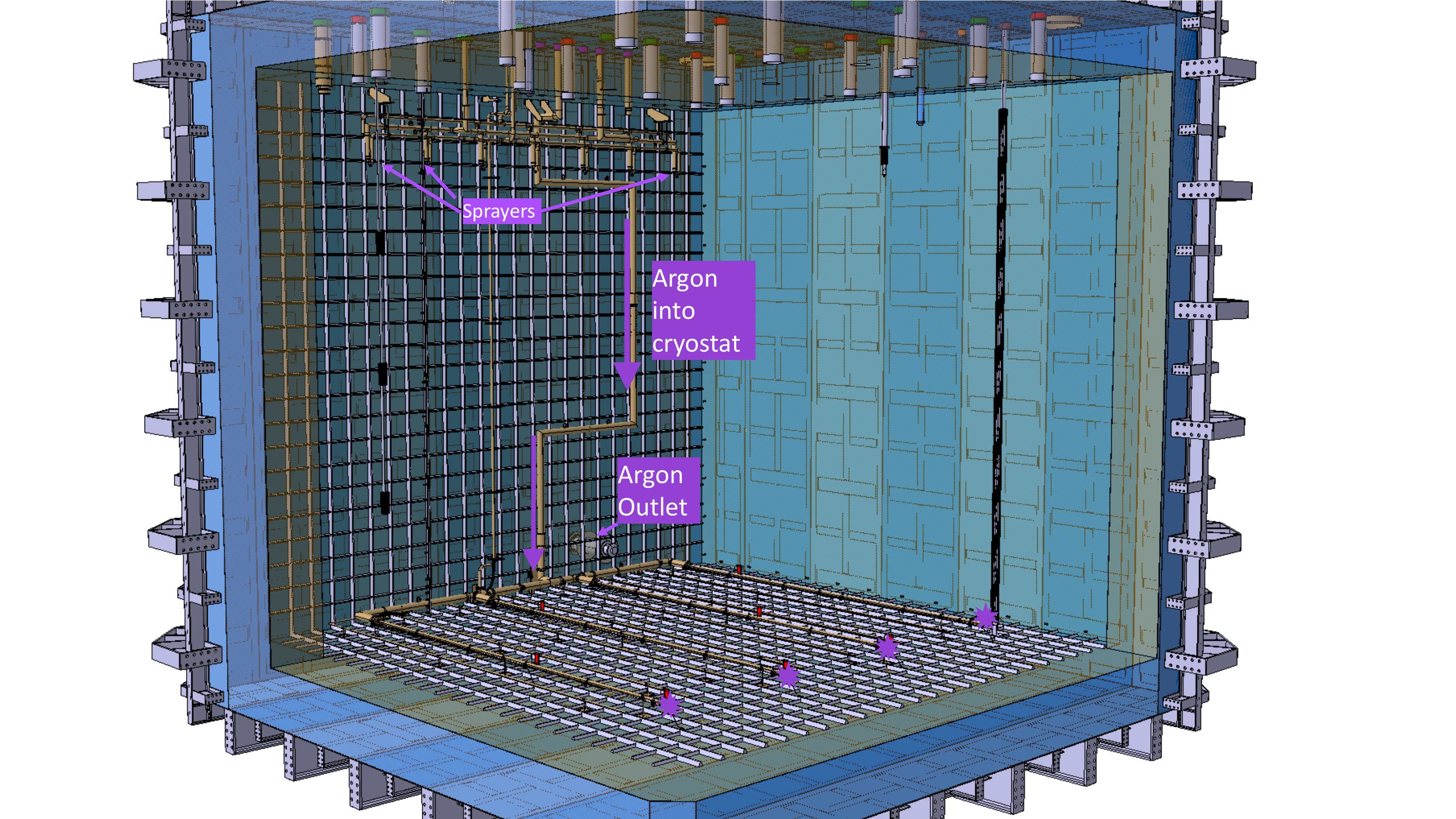}
\caption[Cryogenics internal pipes ]{Cryogenics internal pipes and services. }
\label{fig:cryo-internal} 
\end{figure}

The internal cryogenics encompasses all the cryogenics equipment located inside the cryostat, including manifolds for the cool-down operation and for the distribution of the liquid and gaseous argon. 
The pipes for LAr distribution are positioned on the bottom of the cryostat and the outlets are at the end of the pipes, roughly opposite to the side penetration from which the purification system extracts the LAr (see Figure~\ref{fig:cryo-internal}).
The LAr entering the tank is warmer than the ambient by 0.4\,K. This is a consequence of the isentropic efficiency of the LAr pump and the total heat load going to the transfer lines and valve boxes involved in the LAr circulation and purification processes. This temperature difference is crucial in that it induces an upward flow of the newly purified argon from its entry point at the bottom of the cryostat.

\subsubsection{Phases of Operation}
\label{sec:cryoOp}

Once the cryostat construction and the installation of all scientific equipment is complete, the cryostat is cleaned (dust removal) in preparation for cool-down and fill. The first step is the purge in open loop (also called piston-purge) in which the atmosphere in the cryostat is replaced with argon gas. To ensure the purity of the input argon, the argon pipings are isolated, evacuated to less than 0.1\,mbar absolute pressure, and back-filled with high-purity argon gas. This cycle is repeated several times to reduce contamination levels in the piping to the ppm level. The argon gas is then injected into the cryostat through a set of pipes at the bottom of the cryostat. The flow nozzles are directed downward to spread the gas across the bottom of the tank and produce a stable, upwardly advancing argon wave front. The exhaust is removed from the top using the main GAr outlet and vented outside of the building. A control valve regulates the pressure in the cryostat during this entire operation. The side purge lines located on each roof penetration are also used to evacuate the exhaust gas, ensuring that all volumes (especially trapped volumes) are properly purged. The vertical flow velocity of the advancing GAr is set to 1.2\,m/h, which is twice the diffusion rate of the air downward. This causes the advancing pure argon-gas wave front to displace the air rather than just dilute it.  After 44\,hours (seven volume changes), the purge process is complete with residual air reduced to a few ppm. 

The second stage of the purge process is done in ``closed loop" for one week. During this stage the GAr is recirculated through the GAr purifier and sent back to the bottom of the cryostat.
The gas purification system further removed the water-vapor outgassing from the FR4 circuit-board material and the plastic-jacketed power and signal cables present inside the cryostat. 
This closed-loop purge process further reduced the oxygen and water contamination inside the cryostat to sub-ppm and ppm-levels,  respectively, at which point the cool-down could start. It also allowed assessment of the residual leak rate to atmosphere and the nitrogen content before cool-down.

The cool-down of the cryostat and detector is performed by flowing LAr and GAr into the cryostat through a manifold located near the top. Sprayers deliver a mix of LAr and GAr in atomised form that is distributed inside the cryostat by another set of GAr-only sprayers. During this operation the gas is exhausted using the pressure control valve, keeping the cryostat pressure at 70\,mbarg.
The sprayers guarantee a flat profile of the fluid (LAr and GAr) coming out, so as to meet the cool-down requirements of the cryostat and in particular those of the detector, which are more stringent. The maximum cool-down rate for the TPC is 40\,K/h with a maximum temperature difference between any two points in the detector volume of $\sim$50\,K.

\begin{figure}
\centering 
\includegraphics[width=0.9\columnwidth]{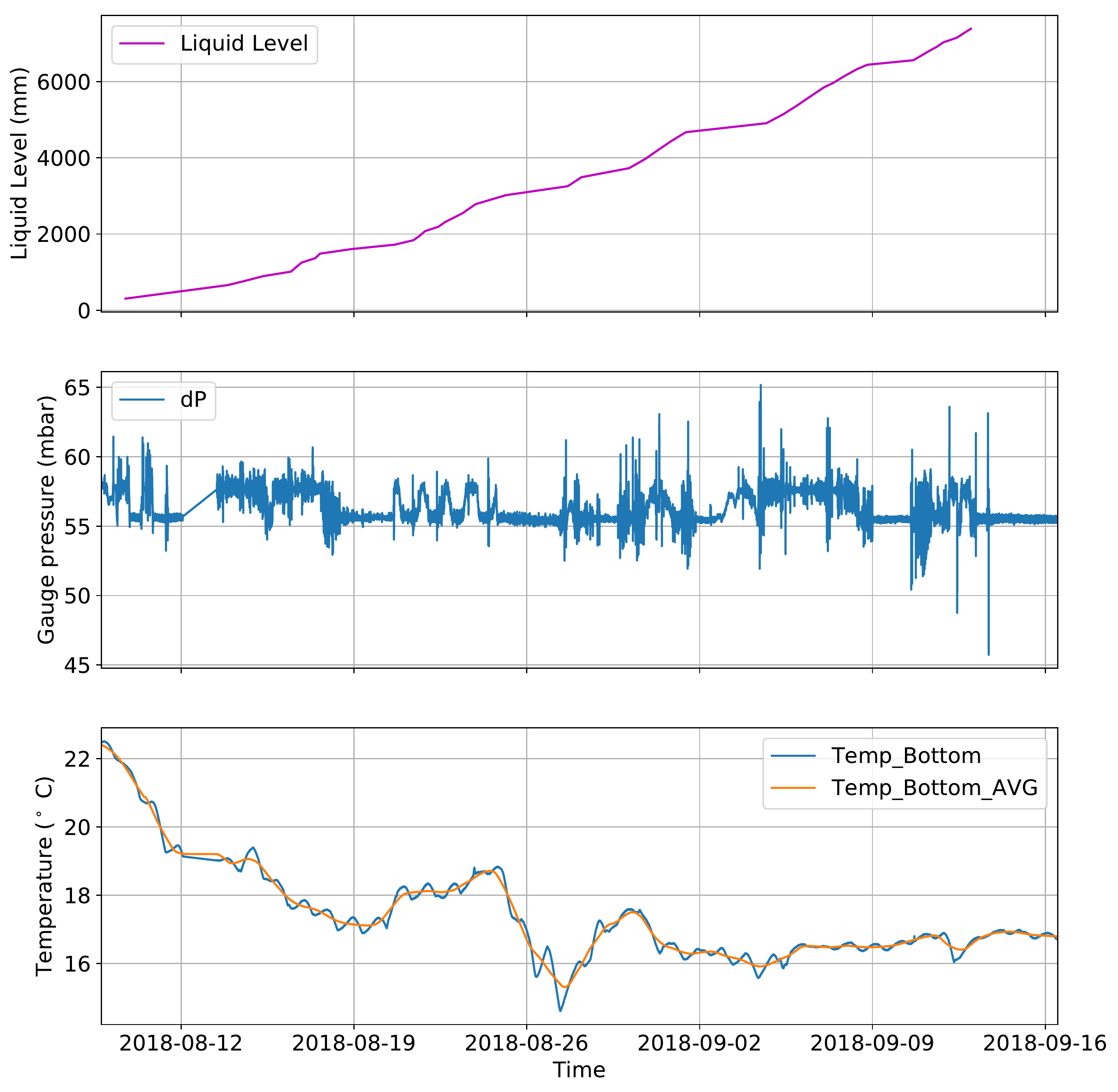}
\caption[Filling trends]{LAr level (top), pressure (middle) and cryostat outer structure temperature (bottom) trends during the filling of the NP04 cryostat.}
\label{fig:filling} 
\end{figure}

Once the cryostat and the TPC are cold, the fill starts. LAr is transferred through the cryostat-filling pipework from the 70\,t LAr storage tank via the filtration system for purification and the LAr phase separator, which allow for the injection in the cryostat of argon in both liquid and gas phases. The \pdsp{} filling process took about six weeks. Figure~\ref{fig:filling} shows the liquid level, pressure, and temperature trends observed during the filling of the cryostat. 

Once the cryostat is filled, the system can enter steady-state operation. 
During operation, as shown in Figure~\ref{fig:cryo-process-flow}, the boil-off that passes through the purge pipes on the signal feedthroughs is filtered directly at room temperature and then condensed with the main boil off. 
Before being reintroduced in the cryostat as liquid, it is purified and mixed with the bulk of the LAr coming from the cryostat.

\begin{figure}[tb]
\centering 
\includegraphics[width=0.9\columnwidth]{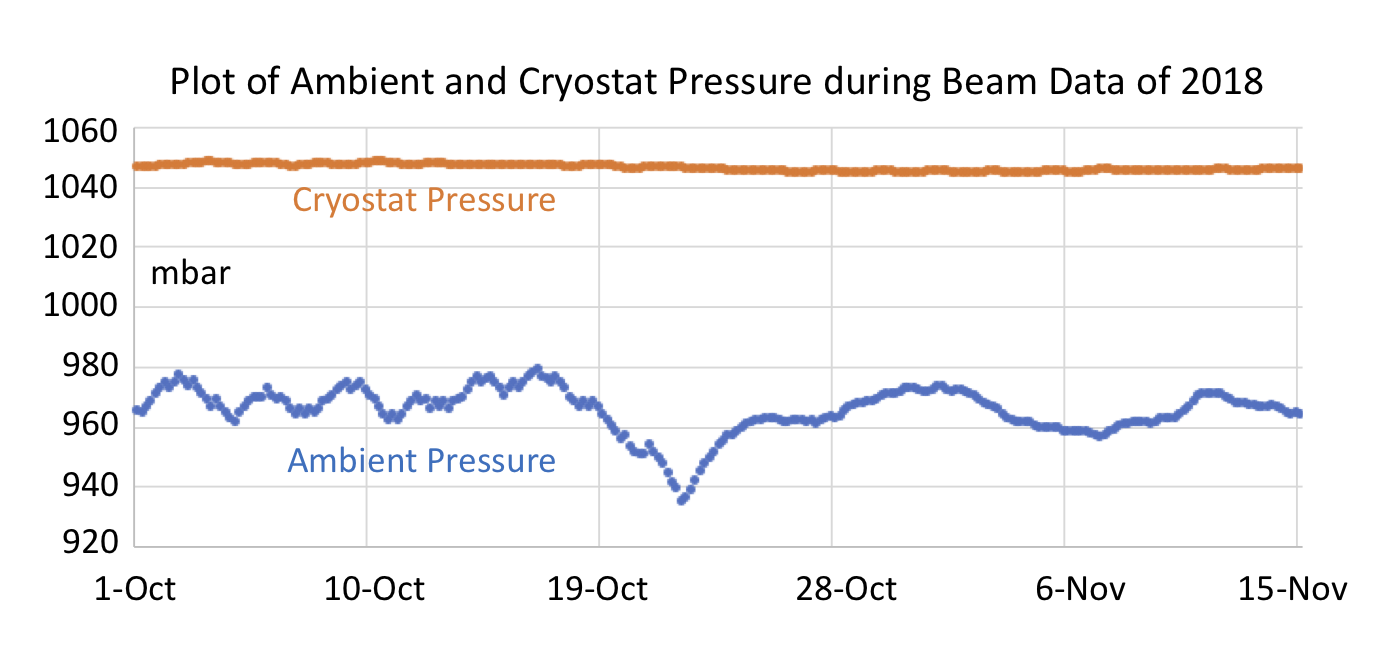}
\caption[Pressure during beam period]{Stability of the pressure inside the cryostat versus atmospheric pressure variations during the beam period.}
\label{fig:pressure_beatime} 
\end{figure}

Table~\ref{tab:cryoparam} lists the final parameters of the \pdsp{} cryogenics, Figure~\ref{fig:pressure_beatime} shows the stability of the cryostat pressure against ambient pressure variations during the beam run period (October - November 2018). 
Table~\ref{tab:cryoheat} shows the heat leak balance of the cryogenics system as measured during steady-state operations after the cryogenics cold commissioning.

\begin{table}
\caption{\label{tab:cryoparam} \pdsp{} cryogenics parameters after commissioning }
\centering
\begin{tabular}{ l c  }
\hline
Liquid level  & 7.45\,m \\
Liquid volume & 540\,m$^3$ \\
Liquid mass & 752.76\,t \\
Normal operating pressure & 1050.00\,mbar \\
\hline
\end{tabular}

\end{table}

\begin{table}
\caption{\label{tab:cryoheat} Heat leak balance as measured during steady-state operations after the cryogenics cold commissioning }
\centering
\begin{tabular}{ l c  }
\hline
Circulation/purification  process  & 2150\,W \\
Proximity cryogenics & 1200\,W \\
Cryostat including top cap & 5170\,W \\
Miscellaneous (cold electronic, cables, penetrations, rods, etc..)
 & 1980\,W \\
\hline 
TOTAL heat load & 10500\,W \\
\hline
\end{tabular}

\end{table}

At the end of operations the tank will be emptied and the LAr will be returned to the storage tank outside the building, and from there unloaded back to LAr trucks.

\subsubsection{Cryostat Pressure Control}
\label{sec:cryoPcontrol}

The pressure inside the cryostat is maintained within a very narrow range by a set of active controls. Pressure-control valves can increase or decrease the cooling power in the condenser by controlling the amount of LN$_2$ flowing to the heat exchanger and the amount of GN$_{2}$ that is vented. Other pressure-control valves can be used to vent GAr to atmosphere and/or introduce clean GAr from the storage, as needed.  

During normal operation the pressure-control valves are set to control the internal cryostat pressure to $1050 \pm1$\,mbar absolute. Excursions of a few percent from this value set off warnings to alert the operator to intervene, but more severe deviations (i.e., pressure exceeding 200\,mbarg or going below 30\,mbarg) trigger automatic actions: the system isolates the cryostat from any potential source of pressure, first closing the valves connecting the cryostat to the circuits used for gas make-up, purge, or cool-down, and the valve to the phase separator. Then the LAr circulation pump is stopped and the side penetration valves are closed.  

If the pressure is too high, the system increases the LN$_2$ flow through the heat exchanger inside the condenser
and powers down any heat sources within the cryostat (e.g., detector electronics). 
At this point some of the GAr is vented to reduce the pressure in a controlled way. On the other hand if the pressure is too low, fresh GAr can be introduced into the cryostat through the GAr make-up line, a line dedicated to taking in argon directly from the outside supply. 

The ability of the control system to maintain a set pressure is dependent on the size of pressure fluctuations (due to changes in flow, heat load, temperature, atmospheric pressure, etc.) and the volume of gas in the system. During normal operation \pdsp{} has 0.45\,m of gas ullage at the top of the cryostat. This is 5\% of the total argon volume and it is the typical vapour fraction used for cryogenic storage vessels. Reaction times to changes in the heat load are slow, typically on the order of one hour.

The cryostat is equipped with an additional high-integrity, mechanical, fail-safe over-pressure and under-pressure protection system capable of preventing catastrophic structural failure of the cryostat in any condition. The system is composed of a diverter valve connected to a set of two devices,  combination pressure  and vacuum safety valves (PSV/VSVs), 
located on the cryostat roof.  Functions for both over- and under-pressure are combined in a single device for efficiency. One of the two identical devices in the system is installed in service mode and the other in stand-by, to guarantee that one is always active in case, for example, maintenance is required. 

The device in service monitors the differential pressure between the inside and the outside of the cryostat and opens rapidly when the differential pressure is outside a preset range. In case of excessive pressure, the PSV/VSV opens and argon is released. The pressure within the cryostat falls and argon gas discharges into the argon vent riser. The valve is designed to close when the pressure returns below the preset level. 
In case the pressure detected is too low, the PSV/VSV opens and allows air to enter the cryostat to restore a safe positive pressure.

\subsubsection{Cryogenics Control System}

The cryogenics control system is a fully automated system that enables continuous cryogenic operation through the various modes of operation described in Section~\ref{sec:cryoOp}, while allowing manual actions by the operator as needed. It is a modular system, both in terms of the electrical and control hardware and the process control programming.

This system's combination of automated and manual operation allowed its implementation at an early stage of the project, i.e., before the project inputs were fully defined, and enabled adjustments as the process and instrumentation evolved. It also made it possible to commission the cryogenics system in batches, i.e., to operate in one mode while implementing the control programming for the next mode. 


The electrical and control architecture is based on a functional analysis of the cryogenics system and the subsequent Product Breakdown Structure (PBS), Piping and Instrumentation Diagram (P\&ID), parts list (detailing the characteristics of the instrumentation), and process logics specification. The hardware architecture consists of three different hardware installations, each controlled by a dedicated Programmable Logic Controller (PLC) Siemens s317, remote input/output modules (I/O), and control cabinets. Each hardware installation corresponds to one of the three main cryogenics subsystems, namely, the external infrastructure for the storage and the proximity system for the two ProtoDUNE cryostats. Each control cabinet is dedicated to a specific function, e.g., LAr/nitrogen management, purification, phase separation, condensing, argon circulation. Each cabinet includes industrial cryogenics instrumentation: actuators (valves, pumps) and sensors. The infrastructure for the \pdsp{} hardware installation includes a total of 10 control cabinets and 630 signals.

The control system software was developed using the UNIfied Industrial COntrol System (UNICOS)~\cite{UnicosSoft}.
The software architecture relies on a Process Control Object (PCO) breakdown structure. The process logic specification used as basis for the programming was designed with a modular structure and relies on associated option modes, sequencers, and interlocks. 
The option modes of the master PCO correspond to the sequence of operations of the cryogenics system, namely, default (which covers the fall-back situation with the cryostat pressure protection always active), open-loop purge, closed-loop purge, cool-down, fill, steady-state, and emptying. 
These option modes are used either to switch on a set of dependant units fulfilling the function for a given stage of operation or to set a collection of states of the controlled objects, e.g., fixed position or regulation mode for a valve, or a set-point. 

In case of abnormal behaviour, software interlocks prevent any further automatic actions. 
The specifications and programming were completed and implemented sequentially, in batches corresponding to the operation modes. This allowed starting safely with an incomplete system,  continuing to program while the system was already in operation, and  adjusting the program as needed during commissioning. 

 The system is equipped with several features to ensure safe and continuous cryogenics operation, maintainability, and to provide flexibility for system evolution. For example, all equipment essential to the functioning of the system is powered from redundant electrical power supplies, including Uninterruptable Power Supplies (UPS) and diesel generators. Vacuum enclosures placed in the cryostat pit are constantly monitored. A degraded vacuum could signal a potential argon leak, therefore an alarm would be raised to indicate this condition.

\subsubsection{Cryogenics Commissioning}
\label{sec:cryoCommiss}

Quality assurance and quality control were performed during the design, construction, installation and commissioning phases. During the construction and installation phases, non-destructive tests (X-rays, He leak tests and pressure tests) were successfully executed. During cold commissioning another series of tests was completed successfully, the main two of which were a check of the I/O signals and a functional test of all valves and equipment. The I/O check consisted of verifying all connections and the synchronisation of the 630 signals, testing the response to actions like opening/closing valves and to sensor readings, and verifying the sensor calibration or run calibration sequence for each control valve. In addition, tests were run on the Ethernet or hardwired signal lines dedicated to the exchange of information between the cryogenics system and the detector system, the safety system, and the CERN Central Control room.

A `mirror' station (not connected to the actual field equipment) provided a functional test environment for the process control logic. The tests were subsequently carried out on the real system to verify first alarms, interlocks, and the sequential function charts. Operation modes were tested afterwards, during the system commissioning.

         \cleardoublepage

\section{Detector Components}  
\label{sec:detcomponents}
      \subsection{Inner Detector: High Voltage}
\label{sec:detcomp:inner:hv}

A liquid argon time projection chamber (LArTPC) requires an equipotential cathode plane at high voltage (HV) and a precisely regulated interior electric field (E field) to drift electrons from particle interactions to sensor planes. The \pdsp{} LArTPC consists of a vertical cathode plane assembly (CPA), vertical anode plane assemblies (APAs), and sets of conductors surrounding the drift volume that are collectively called the field cage (FC). The FC provides a graded voltage profile between the CPA and APA in order to produce a uniform E field in the drift volume. 

\begin{figure}[h]
\centering
\includegraphics[width=0.75\textwidth]{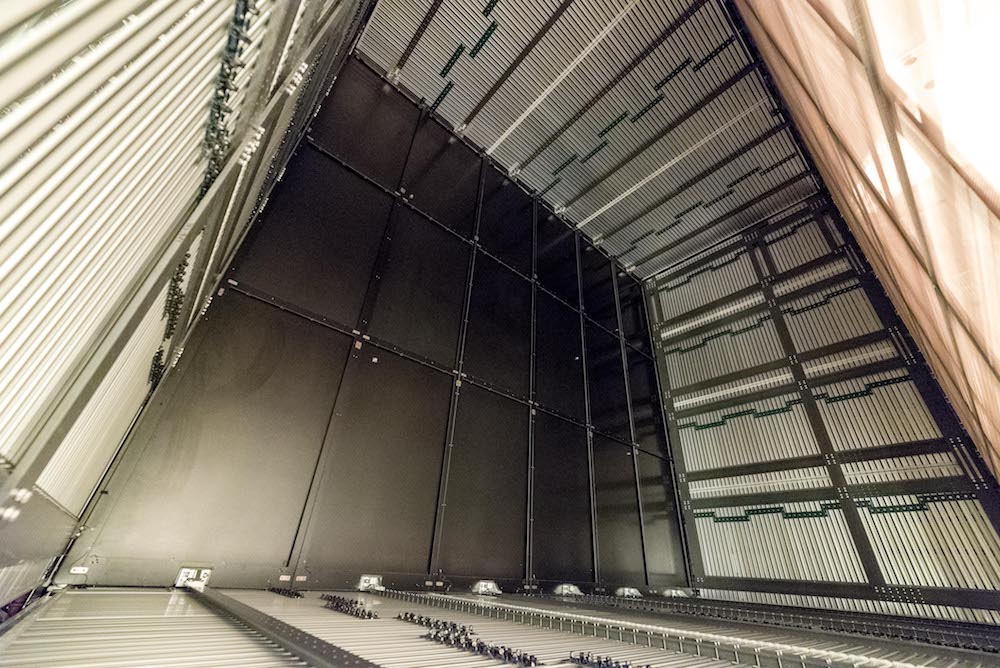}
\caption{One of the two drift volumes of \pdsp{}. The FC modules enclose the drift volume between the CPA array (at the centre of the image) and the APA (upper right). The endwall FCs are oriented vertically; the top and bottom units are horizontal. The staggered printed circuit boards connecting the endwall FC profiles are the voltage divider boards.}
 \label{fig:drift-volume}
\end{figure}

Figure \ref{fig:drift-volume} shows the TPC configuration. Six top and six bottom FC modules connect the horizontal edges of the CPA and APA arrays, and four endwalls connect the vertical edges (two per drift volume). Each endwall is composed of four endwall modules. A Heinzinger -300\,kV 0.5\,mA HV power supply delivers voltage to the cathode. Two HV filters in series between the power supply and HV feedthrough filter out high-frequency fluctuations upstream of the cathode.

       \subsubsection{Cathode Plane Assembly (CPA)}
\label{sec:detcomp:inner:hv:cpa}


The CPA is located in the middle of the TPC, dividing the detector into two equal-distance drift volumes. The CPA's 
7\,m $\times$ 6\,m area is made up of six panels, each of which is constructed of three vertically stacked modules. The same modular structure and materials will be used in the far detector design.
The scope of the \pdsp{ }CPA includes:

\begin{itemize}
\item 18 CPA modules, each with a frame and resistive  sheet, 
\item HV bus connecting the resistive sheets and modules, and
\item HV cup for receiving input from the power supply.
\end{itemize}

Several requirements are placed on the HV system. Electrically, the CPA must provide an equipotential surface at $-$180\,kV nominal bias voltage that remains stable for data taking. The resistive sheet must provide slow discharge of accumulated charge in the event of an unexpected HV breakdown,
to prevent damage to the readout electronics.  To ensure a uniform drift field, the flatness must remain to within 1 cm while submerged in LAr. 
Mechanically, the CPA needs to support the full weight of the four connected top and bottom FC modules, as well as that of a person during the installation.  It must allow for cryostat roof contraction during the cool-down to LAr temperature.  Furthermore, it must be constructable within the cryostat, and its materials must 
inhibit trapped LAr volumes.

\subsubsection*{CPA design}
The cathode plane surface is made of a resistive material, which, in the event of HV breakdown at a point on a CPA module, will restrict the sudden change in voltage to  a relatively localised area, thus preventing discharges from potentially damaging either the cold electronics (CE), the cryostat, or the capacitively coupled anode planes.  The rest of the CPA maintains its original bias voltage, and gradually discharges to ground through the high resistivity of the cathode material. 


The 
18 CPA modules are constructed of strong 6 cm thick FR4 (the fire-retardant version of G10) frames. The frames hold 3 mm thick FR4 sheets laminated on both sides with a commercial resistive Kapton film of type D11261075 \cite{dupont}. Each CPA module is 1.16\,m wide and 2\,m high, and they stack three high to form a  CPA panel of height 6\,m.  The CPA plane consists of six panels placed side-by-side and has the same dimensions as each of the two APA planes.

The surface of the frame facing the APAs is covered by a set of resistive FR4 strips with a bias voltage different from that of the CPA resistive sheets, chosen such that the frame itself causes no distortion in the drift field.

\begin{figure}[htbp]
\centering
\includegraphics[width=0.75\textwidth]{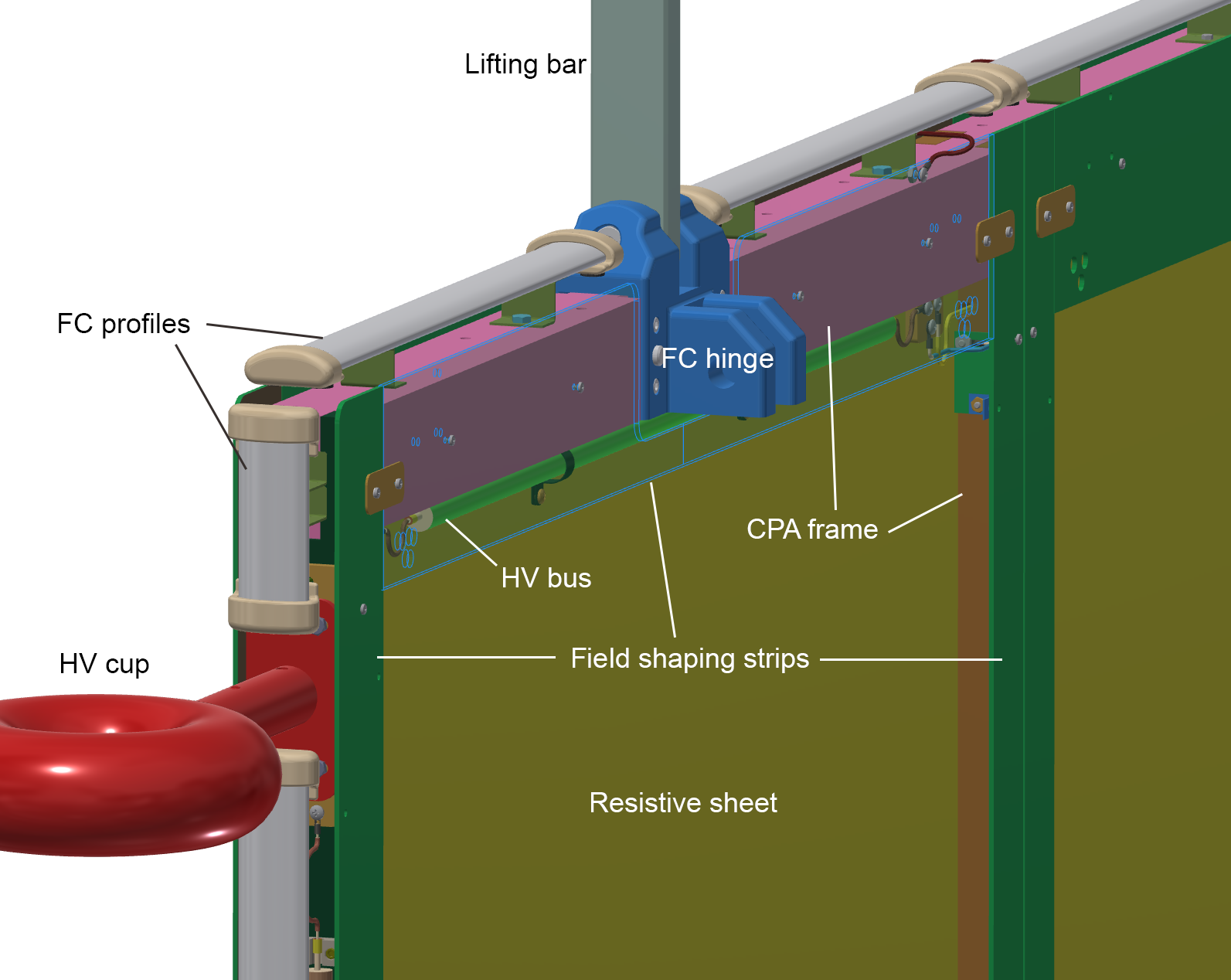}
\caption{HV input cup connection to CPA array. The system of electrically interconnected HV buses allows biasing of the entire CPA at -180~kV and provides the required voltage and current to all the FC modules.}
 \label{fig:donut_cpa}
\end{figure}

The outer edges of the cathode plane facing the cryostat wall are covered with the same metal profile assemblies as used in the field cage (FC), described in Section~\ref{sec:detcomp:inner:hv:fc}. This limits electric field strength to below 30\,kV/cm, as required, in the areas around the CPA frame and eliminates the need for a special design of these most crucial regions of the cathode plane. The edges of the CPA are effectively a continuation of the FC, as shown in Figure \ref{fig:donut_cpa}.  Since the FC 
profiles are the only objects facing grounded surfaces, they are the most likely candidates for HV discharges to ground. To limit peak current flow, these edge profiles are connected to the field-shaping strips by means of a wire jumper. 

\begin{figure}[htbp]
\centering
\includegraphics[width=0.75\textwidth]{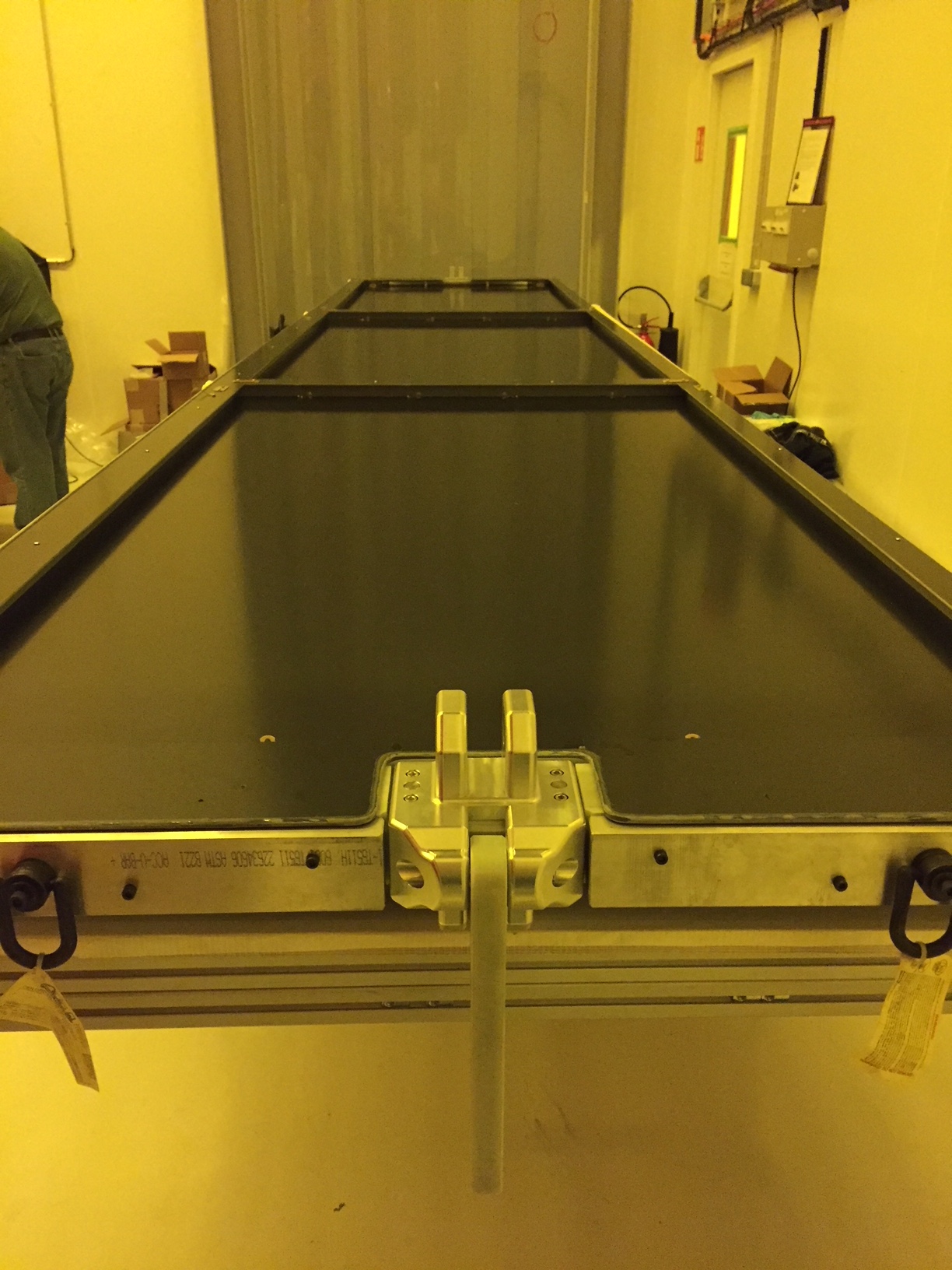}
\caption{Completed 6~m long \pdsp{} CPA panel on the production table. }
 \label{fig:cpa-panel}
\end{figure}

The CPA is connected to the HV feedthrough through a receptacle, called the \textit{HV cup} as shown in Figure \ref{fig:donut_cpa}, at the downstream side of the cryostat (with respect to the beam entrance) and biased at $-$180\,kV. It provides the voltage and the required current to all the FC modules (top, bottom, and end walls) through electrical interconnects (Section~\ref{sec:detcomp:inner:hv:fc}).


The design takes into account deformation and stress due to the pressure from the circulating LAr as well as shrinkage due to the temperature change. 
For example, to ensure contact between the CPA modules after cool-down, a gap of 0.7\,mm, corresponding to the calculated amount of separation that contraction will remove as they cool down, has been introduced. The joints between the FC and the CPA are also designed to accommodate an estimated shrinkage of 5.2\,mm of the steel supporting beam between the CPA and APAs.

\subsubsection*{Mechanical and electrical interconnections between modules}

Three modules are stacked vertically to form the 6\,m height of a CPA panel as shown in Figure~\ref{fig:cpa-panel}. 
The frames of these modules are bolted together using tongue-and-groove connections at the ends. The resistive cathode sheets and the field-shaping strips are connected using metallic tabs to ensure redundant electrical contact between the CPAs. 

Each CPA panel is suspended from the cathode rail using a central lifting bar.  Due to the roof contraction 
as the cryostat is cooled, it was calculated that each CPA 
would move $\sim$2\,mm relative to its neighbours. Several pin-and-slot connections are implemented at the long edges of the CPA panels to ensure the co-planarity of the modules while allowing for a small vertical displacement.

The electrical connectivity of the resistive sheets within a CPA panel is maintained by several tabs through the edge frames. The voltage is passed from one CPA panel to another through embedded cables in the panels, referred to collectively as the HV bus, as shown in Figure \ref{fig:donut_cpa}. Redundant connections in the HV bus between CPA panels are used to ensure reliability. The HV bus also provides a low-resistance path for the voltage needed to feed the FC resistive divider chains. The required connections to the FC modules are made at the edges of the CPA. Along its perimeter, 
the HV bus cables are hidden between the field-shaping strip overhang and the main cathode resistive sheet.  The cables are capable of withstanding the full cathode bias voltage to prevent direct arcing to (and as a result, the recharging of) a CPA  that discharges to ground. Connections are flexible in order to allow for FC deployment, thermal contraction, and motion between separately supported CPA components.

      \subsubsection{Field Cage (FC)}
\label{sec:detcomp:inner:hv:fc}
The FC covers the top, bottom, and endwalls of all the drift volumes, thus providing the necessary boundary conditions to ensure a uniform electric field and shielding it from the cryostat walls. The FC is made of adjacent extruded aluminum profiles running perpendicular to the drift field and set at increasing potentials along the 3.6\,m drift distance from the CPA HV (-180kV) to ground potential at the APA planes. 
Other elements of the FC are the ground planes (GP) sitting above the top and below the bottom FC modules. They confine the electric field in the liquid phase, avoiding high field both in the gas phase at the top of the cryostat and close to the piping at the bottom. The structures holding the profiles are made of insulating fibreglass-reinforced plastic (FRP). FRP has good mechanical strength at cryogenic temperatures and low coefficient of thermal expansion. The FC modules come in two distinct types: the identical top and bottom modules, which run the full length of the detector, and the endwall FC modules, which are installed vertically to close the detector drift volume at either end.

The FC is divided into mechanically and electrically independent modules, which comes with several advantages. As a consequence of electrically subdividing the modules, the stored energy and therefore the risk of detector damage is limited. In addition, the division acts as a protection from transient surges. FC modules have their own, independent voltage divider network providing the necessary linear voltage gradient. In case of a resistor failure in a divider chain field, distortions would be restricted to the FC module concerned.
The mechanical subdivision simplifies construction and assembly of the FC.

\subsubsection*{Top/Bottom Field Cage and Ground Planes}
\begin{figure}[h]
\centering
\includegraphics[width=1.0\textwidth]{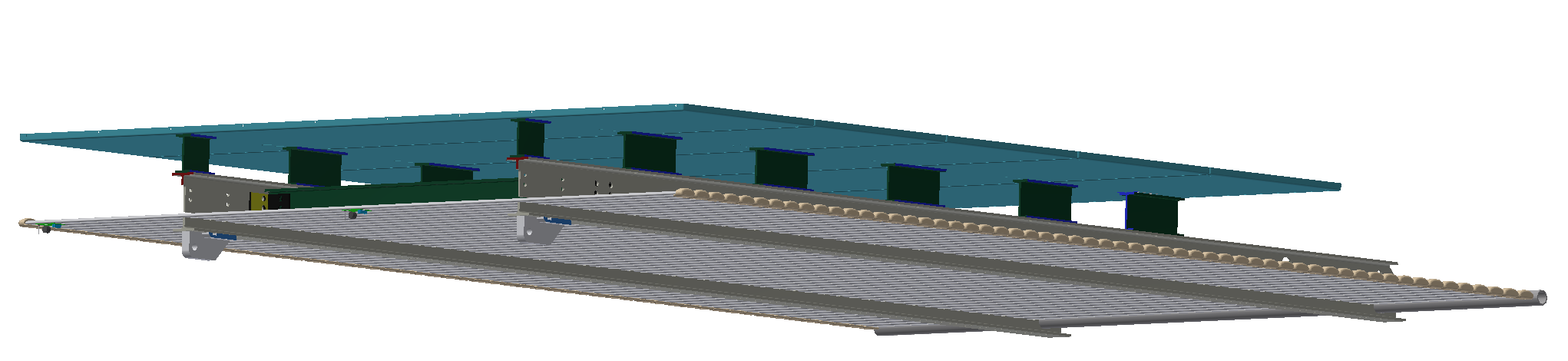}
\caption{Drawings of a Top/Bottom FC module. The lower gray section illustrates the FC profiles and the upper blue section is the ground plane (GP).}
 \label{fig:top-bot-fc}
\end{figure}

There are six top and six bottom FC assemblies, all like the one shown in Figure~\ref{fig:top-bot-fc}. The assemblies are constructed from pultruded FRP I-beams and box beams that support the aforementioned aluminum profiles. The length and width are 2.3\,m and 3.5\,m, respectively, each assembly comprising 57 aluminum profiles. A GP consisting of modular perforated stainless steel sheets runs along the outside surface of each top and bottom FC with a 20\,cm clearance. The gas region at the top of the volume, also referred to as the ullage, which is necessary for safe and stable operation of the LAr cryogenics system, contains many grounded conducting components with sharp features near which the electric field could easily exceed the breakdown strength of gaseous argon if directly exposed to the energised FC. The GP protects against this. The bottom FCs are equipped with GPs to shield from cryogenic piping and other sensors with sharp features on the cryostat floor. 

\begin{figure}[h]
\centering
\includegraphics[width=0.48\textwidth]{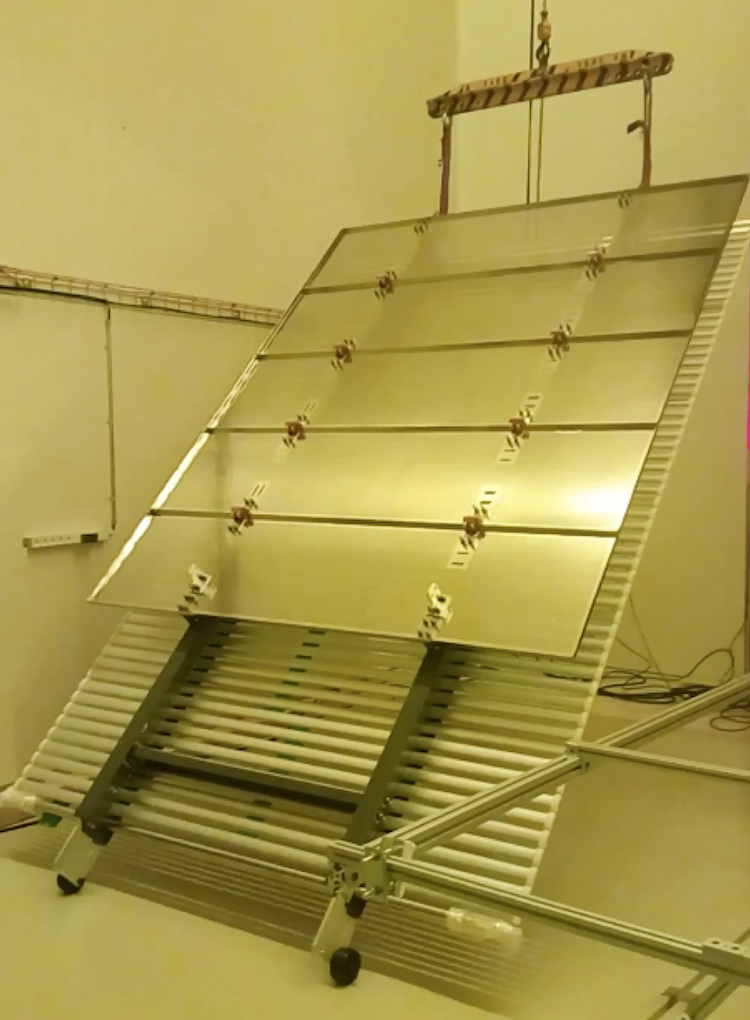}
\caption{Lifting of a Top FC module for subsequent installation on the CPA panel. The side of the FC module that is lifted will attach to the CPA while the other side (bottom side) is connected with latches to the APA.}
 \label{fig:fc-lifting}
\end{figure}

The connections between the top and bottom FC modules and the CPAs are made with aluminum hinges 2.54\,cm in thickness that allow the modules to be folded in on the CPA during installation. The hinges are electrically connected to the second profile from the CPA. The connections to the APAs are made with stainless steel latches that are engaged once the top and bottom FC modules are unfolded and fully extended towards the APA. A top FC module being lifted for installation on the CPA panel is shown in Figure \ref{fig:fc-lifting}.

\subsubsection*{Endwall Field Cage}
\begin{figure}[h]
\centering
\includegraphics[width=0.6\textwidth]{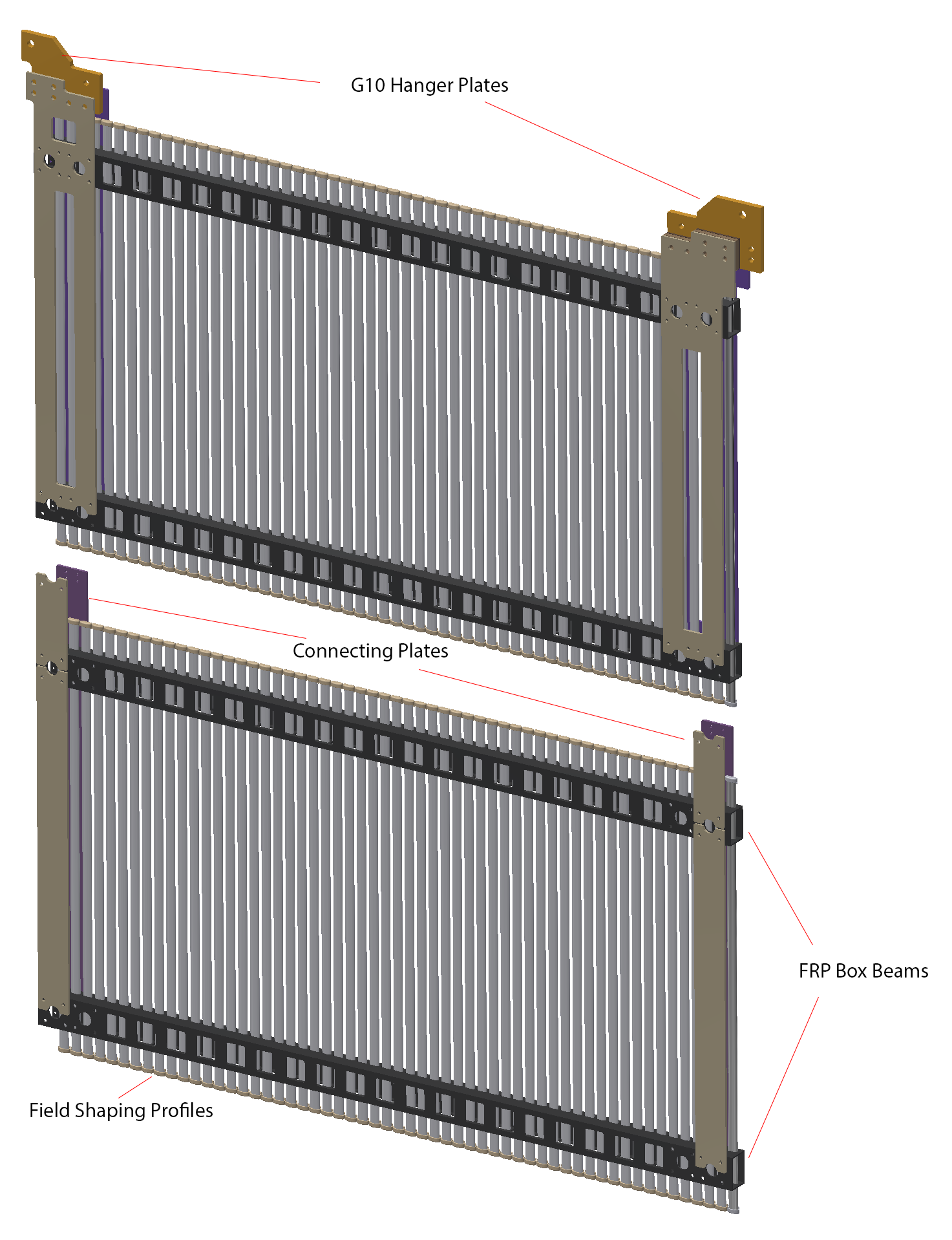}
\caption{Top: uppermost module of the endwall FC. The two G10 hanger plates connect the endwall FC to the detector support system (DSS) beams above the APAs and CPAs. Bottom: regular endwall FC module. Three of those modules are stacked vertically below a top module, to form the total height of the drift volume.}
 \label{fig:endwall-fc}
\end{figure}

\begin{figure}[h]
\centering
\includegraphics[width=0.48\textwidth]{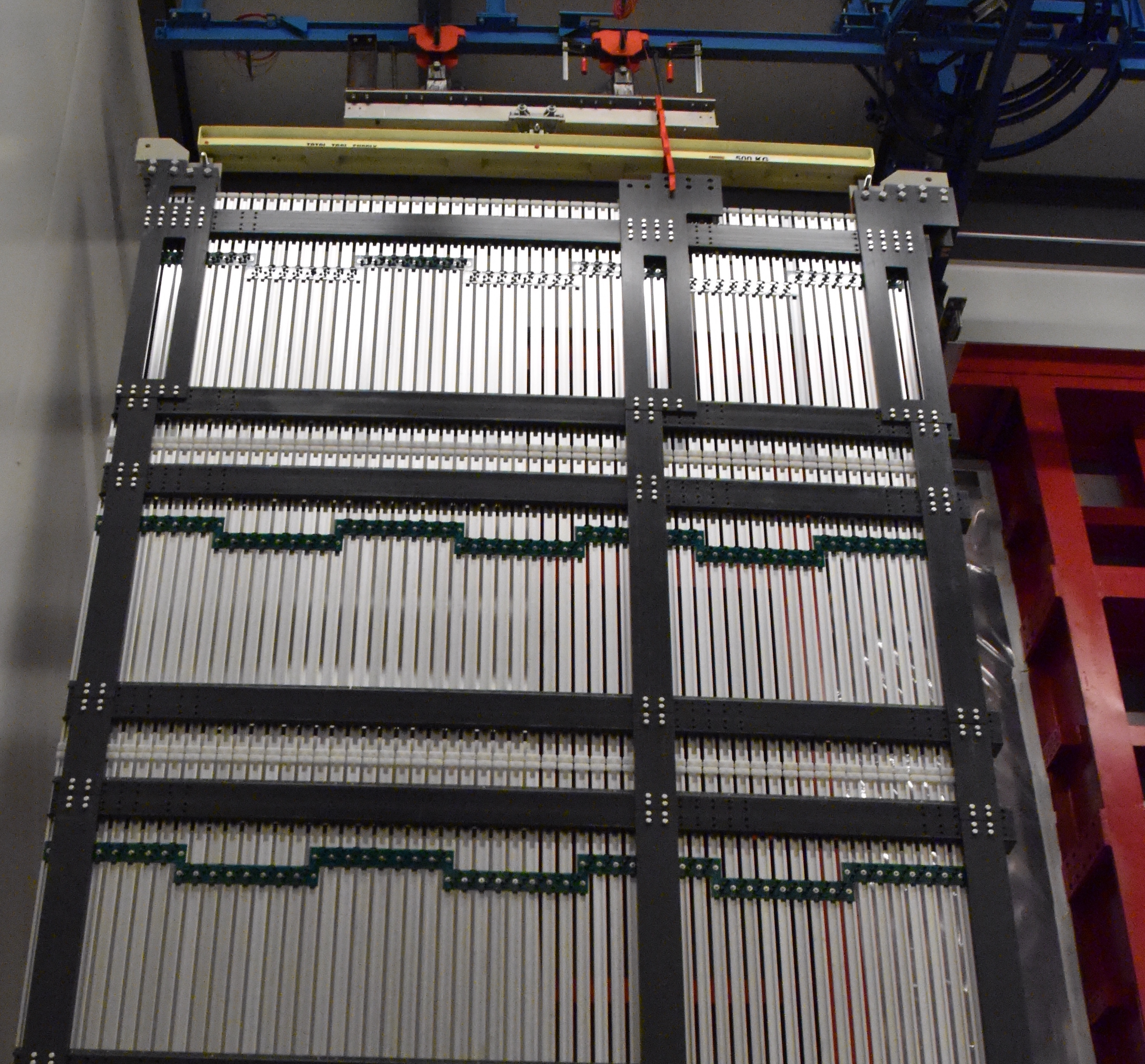}
\caption{Endwall FC assembly, the stacked modules are visible. The endwall is hung in front of the cryostat before insertion through the TCO. The visible face is the one that will face the inside of the detector active volume. The dark grey FRP plates are the support structure of the endwall and interconnected between the volumes.}
 \label{fig:endwall-fc-foto}
\end{figure}

Each of the two drift volumes has two endwall FCs one on each end. Each endwall FC is in turn composed of four stacked endwall FC modules, the topmost equipped with hanger plates. Each endwall FC module is constructed of two FRP box beams each 3.5\,m long as shown in Figure \ref{fig:endwall-fc} (dark grey) and Figure \ref{fig:endwall-fc-foto}. The endwalls are not equipped with GPs as there is enough side clearance between the cryostat wall and FC, thus avoiding high E-fields. In ProtoDUNE-SP the endwall at the beam entry point is customized to hold the beam plug,  described in Section~\ref{sec:detcomp:inner:hv:bp}.

\subsubsection*{Field Cage Profiles}
The 
FC modules consist of extruded aluminum field-shaping-profiles. The profile shape minimises the electric field strength between a given profile and its neighbours and between a profile and other surrounding parts. The profile ends have a higher surface electric field, especially those at the corners of the FC (boundary with APA or CPA). To prevent HV breakdowns in the LAr, the ends of the profiles are encapsulated by custom UHMWPE (Ultra-High-Molecular-Weight Polyethylene) caps. The caps are designed and experimentally verified to withstand the full voltage across their thickness.

\subsubsection*{Voltage Divider Boards and Terminations}
A resistive divider chain interconnects all the metal profiles of each FC module to provide a linear voltage gradient between the cathode and anode planes.

\begin{figure}[h]
\centering
\includegraphics[width=0.6\textwidth]{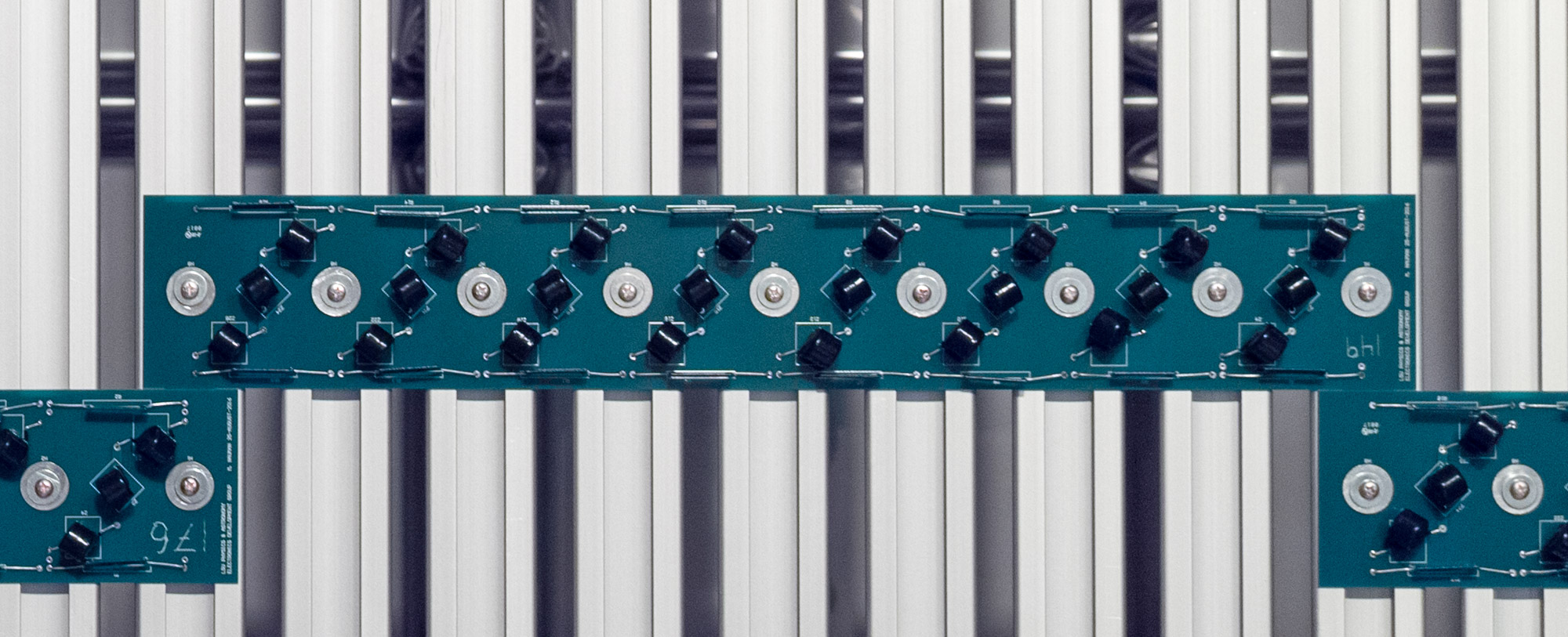} \\
\includegraphics[width=0.6\textwidth]{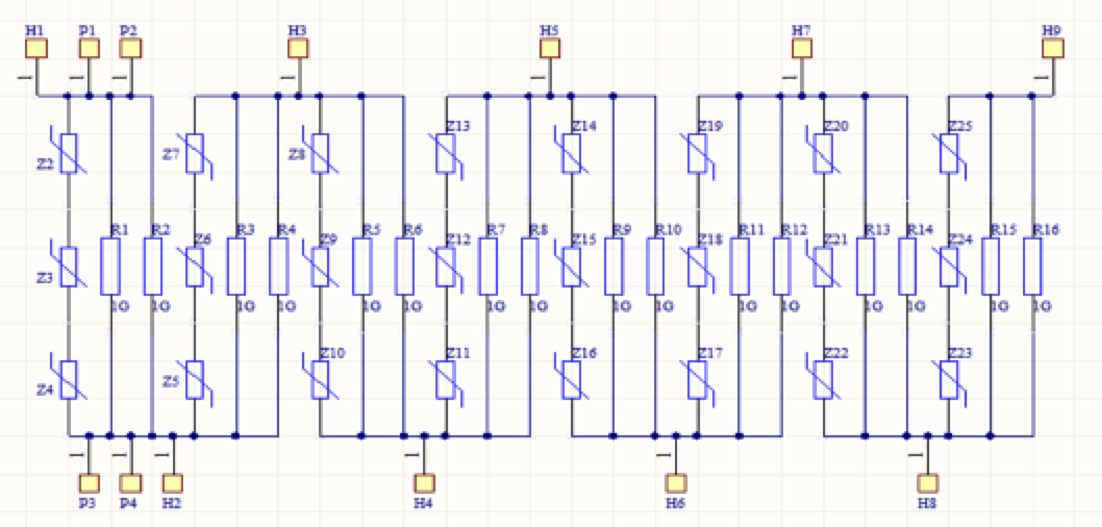}
\caption{Resistor divider board photograph (top) and schematic (bottom).}
 \label{fig:divider_board}
\end{figure}

The resistive divider chain is a chain of resistor divider boards each with eight resistive stages in series. Each stage (corresponding to a 6\,cm distance between FC profile centers) consists of two 1\,G$\Omega$ resistors in parallel yielding a parallel resistance of 0.5\,G$\Omega$ per stage to hold a nominal voltage difference of 3\,kV. In the event of a HV breakdown, each stage is protected against HV discharge by varistors. Three varistors (with 1.8\,kV clamping voltage each) are wired in series and placed in parallel with the associated resistors. A photo and schematic of the resistor divider board are shown in Figure~\ref{fig:divider_board}.
Each FC divider chain connects to an FC termination board in parallel with a grounded fail-safe circuit at the APA end. The FC termination boards are mounted on the top of the upper APAs and the bottom of the lower APAs. Each termination board provides a default termination resistance, and an SHV cable connection to the outside of the cryostat, via the CE signal feedthrough flange, through which it is possible to either supply a different termination voltage to the FC or monitor the current flowing through the divider chain, or both.
 
      \subsubsection{Beam Plug } 
\label{sec:detcomp:inner:hv:bp}
\begin{figure}[h]
\centering
\includegraphics[width=0.48\textwidth]{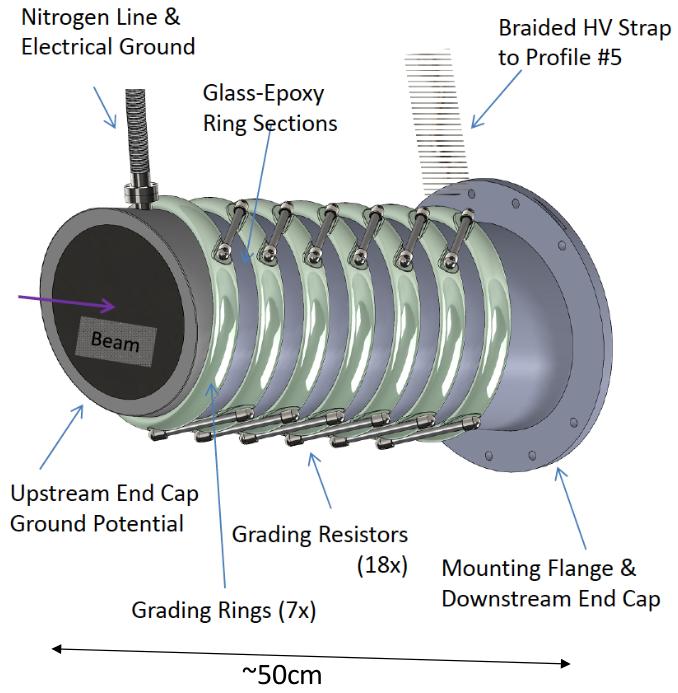}
\includegraphics[width=0.48\textwidth]{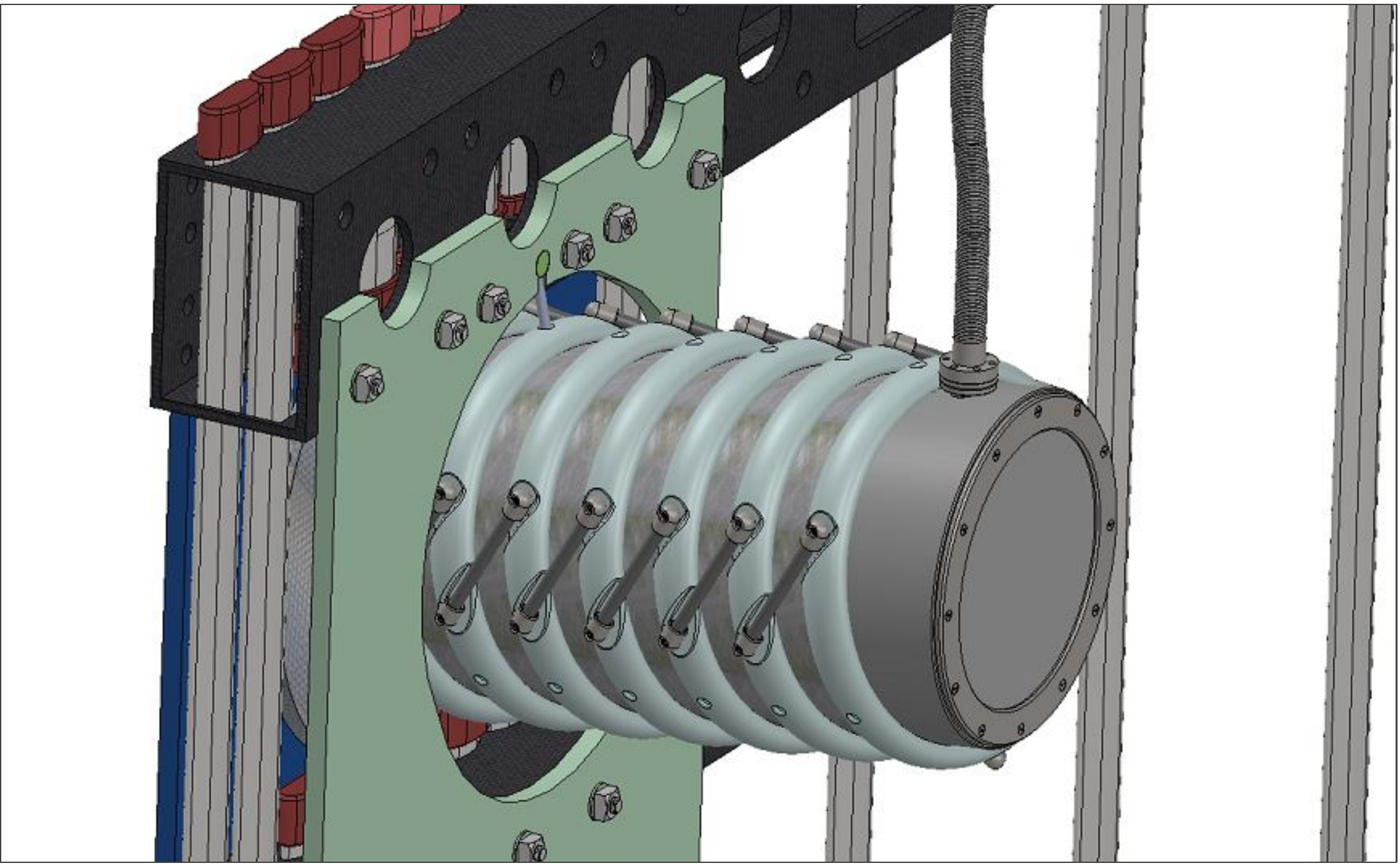}
\caption{The beam plug is a  composite pressure vessel filled with dry nitrogen gas. Left: The vessel is about 50\,cm in length and about 22\,cm in diameter. The pressure vessel is divided into sections with each section bonded to a stainless steel grading ring. The grading rings are connected by three parallel paths of resistor chain. Right: Beam plug to FC interface.}
 \label{fig:beamplug}
\end{figure}

\begin{figure}[h]
\centering
\includegraphics[width=0.75\textwidth]{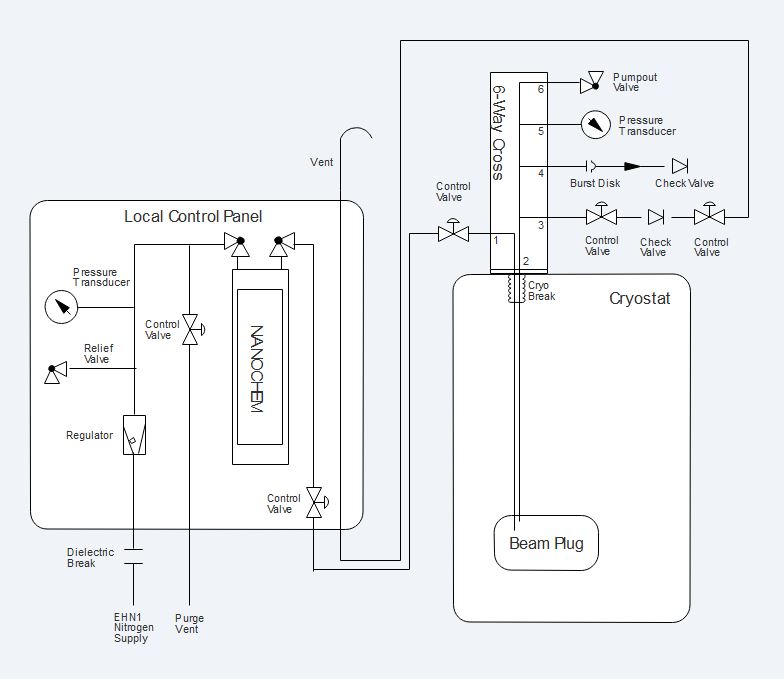}
\caption{Beam plug nitrogen gas system schematics. The Local Control Panel is mounted on top of the cryostat near the DN160 flange feedthrough. The nitrogen line enters the cryostat via the six-way flange, which also has a burst disk for emergency pressure relief, and temperature/pressure sensors.}
 \label{fig:beamplugN2}
\end{figure}

To minimise the material interactions of the particle beam in the cryostat upstream of the TPC, a volume of LAr along the beam path (between the cryostat inner membrane and the FC) is displaced, and replaced by a less dense volume of dry nitrogen gas. The gas is contained within the \textit{beam plug}, a cylindrical glass-fiber composite pressure vessel, about 50\,cm in length and 22\,cm in diameter. It is illustrated in Figure~\ref{fig:beamplug}.
A pressure relief valve and a burst disk are installed on the nitrogen fill line on the top of the cryostat (externally) to ensure the pressure inside the beam plug does not exceed the safety level of about 22\,psi. The nitrogen system schematic is shown in Figure~\ref{fig:beamplugN2}. 

The beam plug is secured to the endwall FC support structure as illustrated in Figure~\ref{fig:beamplug}. Beam is fed into one of the two available drift volumes of the TPC, the drift volume not receiving beam is used for cosmic ray studies. The front portion of the beam plug extends  5\,cm into the active region of the TPC  through an opening in the FC. The FC support is designed with sufficient strength and stiffness to support its weight. 
The total internal volume of the beam plug is about 16 liters. 

The requirements on the acceptable leak rate is between $7.8\times 10^{-5}$ scc/s and $15.6\times 10^{-5}$ scc/s. This is very conservative and is roughly equivalent to leaking  15\% of the nitrogen in the beam plug over the course of a year. In the worst-case scenario in which all the nitrogen in the beam plug leaks into the LAr cryostat, the increase in concentration is about 0.1\,ppm, which is still a factor of 10 below the maximum acceptable level, as specified by light-detection requirements. Over the course of about 1.5 years of beam plug operations in LAr at \pdsp{}, no detectable leak was observed. 

At nominal operation, the voltage difference across the beam plug (between the first and the last grading ring) is 165\,kV. 
To minimise risk of electrical discharges, the beam plug is divided into sections, each of which is bonded to stainless steel conductive grading rings. The seven grading rings are connected in series with three parallel paths of resistor chains. The ring closest to the FC is electrically connected to one of the FC profiles. 
  The electrode ring nearest the cryostat wall is grounded to the cryostat (detector) ground. 
The maximum total power dissipated by the resistor chain is about 0.6\,W.

      \subsubsection{High Voltage (HV) Components}
\label{sec:detcomp:inner:hv:hvgp}

The TPC high voltage (HV) components include the HV power supply, cables, filter circuit, HV feedthrough,  and monitoring instrumentation for currents and voltages (both steady state and transient).

\begin{figure}[h]
\centering
\begin{minipage}{\textwidth}
  \centering
 $\vcenter{\hbox{\includegraphics[width=0.2\textwidth]{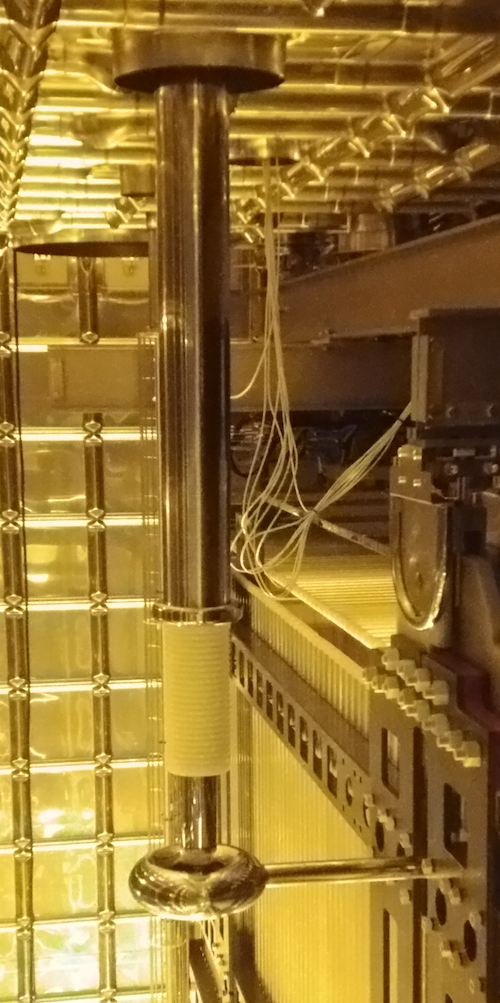}}}$
 \hspace*{0.001\textwidth}  $\vcenter{\hbox{\includegraphics[width=0.75\textwidth]{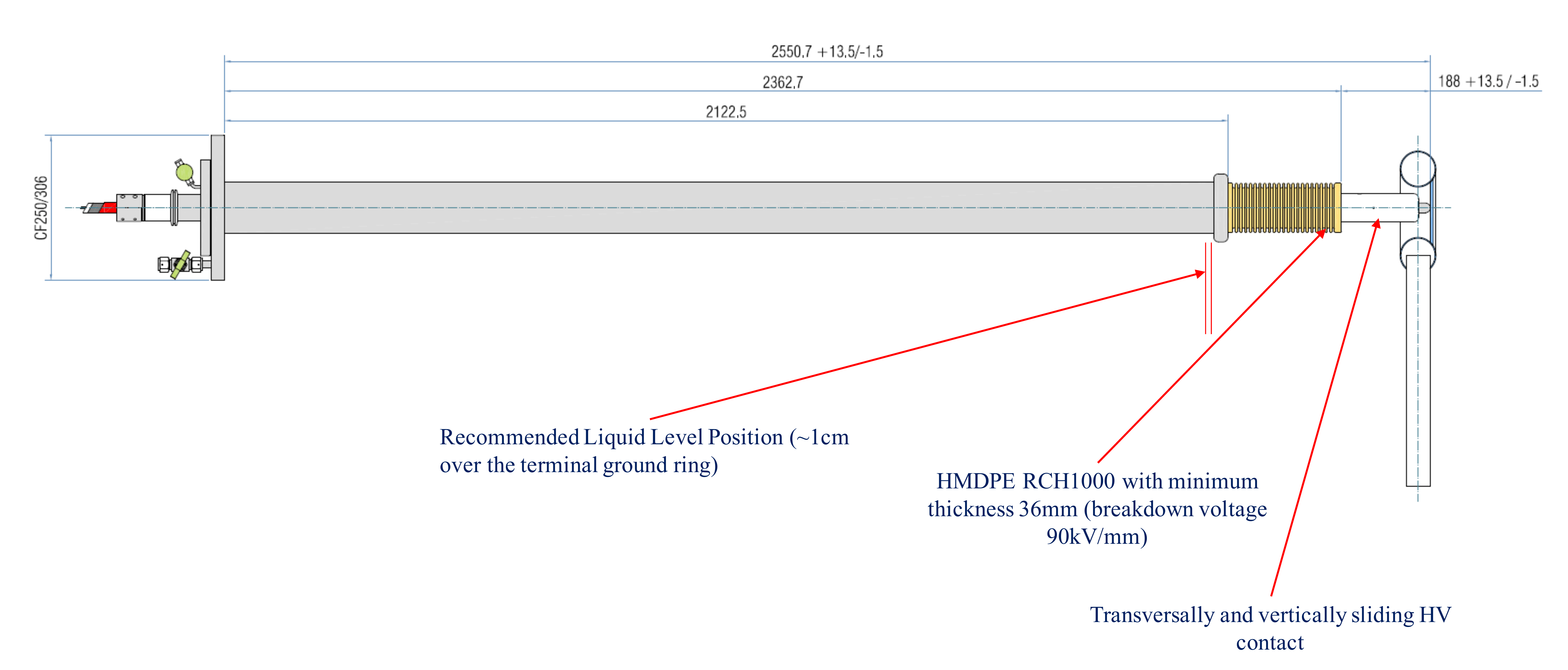}}}$
\end{minipage}
\caption{Photograph after installation and drawing of the HV feedthrough. The distance from the cup to the top surface is approximately 1.3\,m.}
 \label{fig:feedthrough}
\end{figure}

The $-$180\,kV voltage necessary to produce the required electric field of 500\,V/cm is delivered by the power supply through an RC filter and a HV feedthrough to the CPA. The design of the HV feedthrough is based on the reliable construction technique adopted for the ICARUS HV feedthrough \cite{Amerio:2004ze} and shown in Figure \ref{fig:feedthrough}. Before installation, the feedthrough was successfully operated for several days in a test stand at voltages up to 300\,kV. 

The feedthrough design has a coaxial geometry, with an inner conductor (HV) and an outer conductor (ground) insulated by UHMWPE,  as illustrated in Figure \ref{fig:feedthrough}. 
The outer conductor,  a stainless-steel tube, surrounds the insulator, extending down through the cryostat into the LAr.  In this geometry, the E field is confined within regions occupied by high-dielectric-strength media (UHMW PE and LAr).  The inner conductor is made of a thin-walled stainless steel tube to minimise the heat input and to avoid the creation of argon gas bubbles around the lower end of the feedthrough. A contact, welded at the upper end for the connection to the HV cable, and a round-shaped elastic contact for the connection to the cathode, screwed at the lower end, completes the inner electrode. Special care has been taken in the assembly to ensure complete filling  of the space between the inner and outer conductors with the (polyethylene) PE dielectric, and to guarantee leak-tightness at ultra-high vacuum levels.

Filter resistors are placed between the power supply and the feedthrough.  Along with the cables, these resistors reduce the discharge impact by partitioning the stored energy in the system.  The resistors and cables together also serve as a low-pass filter reducing the 30\,kHz voltage ripple on the output of the power supply.

The filter resistors are of a cylindrical design. Each end of an HV resistor is electrically connected to a cable receptacle. A cylindrical insulator is placed around the resistor, and a grounded stainless steel tube surrounds the insulator.

The instrumentation in \pdsp{} provided useful information on HV stability. Outside the cryostat, the HV power supply and cable-mounted toroids monitor the HV. The power supply has capabilities down to tens of nA in current read-back and is able to sample the current and voltage every 300\,ms.  The cable-mounted toroid is sensitive to fast changes in current;  the polarity of a toroid's signal indicates the location of the current-drawing feature as either upstream or downstream of it. 

Inside the cryostat, pick-off points near the anode monitor the current in each resistor chain. Additionally, the voltage of the ground planes (GPs) above and below each drift region can diagnose problems via a high-value resistor connecting the GP to the cryostat.

\subsubsection*{HV Commissioning, Beam Time Operation and Stability Runs}

During cool-down and LAr filling, a power supply was used to supply $-$1\,kV to the cathode and monitor the current draw of the system. As the system cooled from room temperature to LAr temperature, the resistance increased by $\sim$10\%, consistent with expectations. Once the LAr level had exceeded the height of the top GP, the voltage was ramped up to the nominal voltage.

Two types of instabilities emerged in the cold side of the HV system. The first type was a so-called current blip, during which the system drew a small excessive current that persisted for no more than a few seconds. The magnitude of the excess current during current blips increased over the subsequent three weeks from 1\% to 20\%. The second type of instability, called a ``current streamer,'' exhibited persistent excessive current draw from the HV power supply with accompanying excessive current detected on a GP and on the beam plug. These two types of instabilities occurred periodically throughout 
the beam run. The frequency of both types increased over time after the system was powered on, until a steady state of about ten current blips/day and one current streamer  every four hours was reached. These effects are consistent with a slow charging-up process of the insulating components of the FC supports, which then experience partial discharges that are recorded as HV instabilities. This process restarts after every long HV-off period. 

\begin{figure}[h]
\centering
\includegraphics[width=1.0\textwidth]{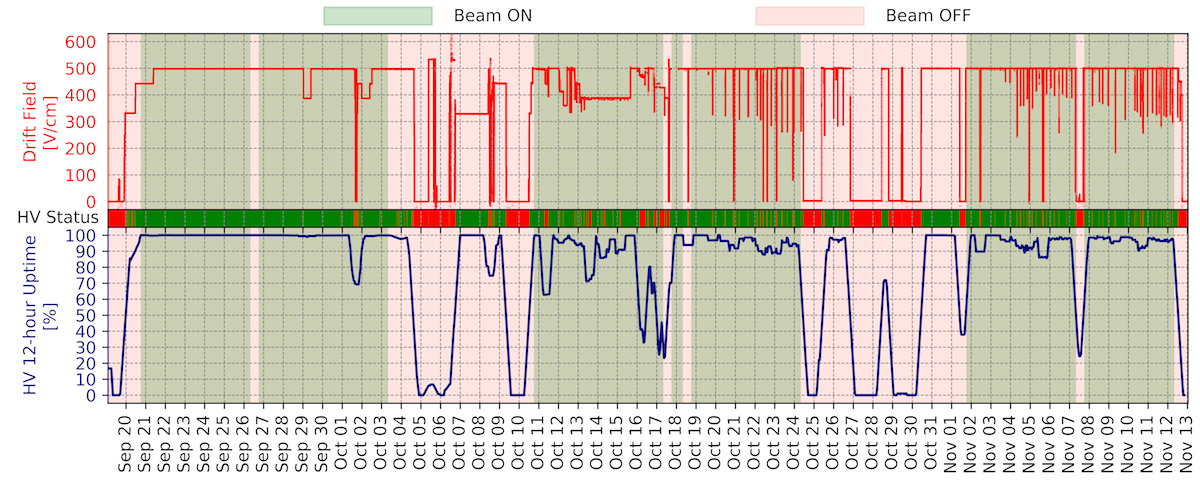}
\caption{The performance of the HV system across the test beam period, Sep-Nov 2018. The top panel shows the drift field delivered to the TPC; the middle panel indicates HV cuts during periods when the system was not nominal (some periods not visible due to their short timescale); and the bottom panel shows the moving 12-hour uptime of the HV system based on these HV cuts.}
 \label{fig:beam-run-summary}
\end{figure}

\begin{figure}
    \centering
     \includegraphics[width=.7\textwidth]{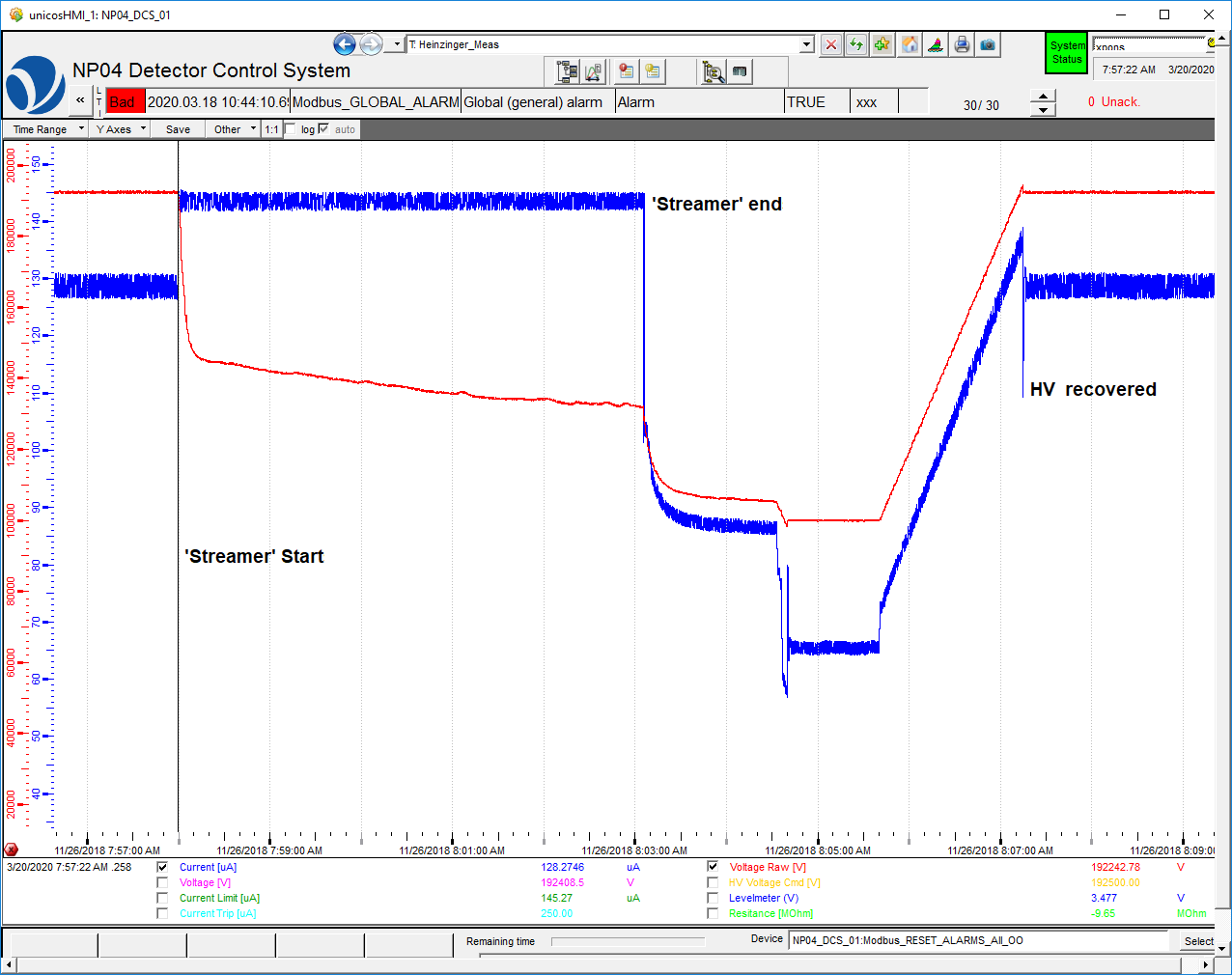}
     \caption{Automatic Recovery of a `streamer' event recorded on 26/11/2018 by DCS. The HV current (in blue) is limited at 142 µA; the voltage (in red) decreases until the discharge is finished, returning to the nominal voltage after a predefined time.}
    \label{fig:HV_streamer} 
    \end{figure}

In addition, these processes seemed to be enhanced by the LAr bulk high purity, which allowed the electric current to develop. 
At low purity electronegative impurities acted as quenchers, blocking the development of the leakage current. During the 2018 beam run periods, priority was given to operating the \pdsp{} detector with maximal uptime in order to collect as much beam data as possible at the nominal HV conditions \cite{Stocker:2021}. In some cases, mostly outside of the beam run period,  the HV system was turned off momentarily to allow the HV system components to discharge. This is reflected as larger dips in the uptime plot shown in Figure \ref{fig:beam-run-summary} \cite{Wood:2021}. During periods 
when the rest of the subsystems (including the beam) were stable, the moving 12-hour HV uptime fluctuated between 96\% and 98\%.  Automated controls to quench the current streamers were then successfully implemented in an auto-recovery mode. These helped to increase the uptime significantly, by optimising the ramping down and up of the HV power supply voltage, which was performed in less than four minutes. The process of the auto-recovery mode is shown in Figure \ref{fig:HV_streamer}.

Investigating the long-term behaviour of the HV instabilities and understanding their origin became goals of the long-term operation of \pdsp{} in 2019. As mentioned above, it appeared that the current-streamer effect is a charging-up process with its frequency increasing with time after a long HV-off period. This behaviour was  
observed repeatedly and this cause was confirmed in 2019. The current-streamer rate stabilised at 4-6 events per day, and the location 
remained on the same single ground plane (GP\#6). 
The rate and location were approximately independent of the HV applied on the CPA in the 90\,kV to 180\,kV range.

More recently, after 
changing the LAr re-circulation pump  in April 2019, the detector was operated for several months in very stable cryogenic conditions and with very high and stable LAr purity (as measured by purity monitors and cosmic rays). During this period, a significant evolution was observed. The HV system was set and operated at the nominal value of 180\,kV at the CPA  for several weeks without interruption. 


To better understand the current-streamer phenomenon, the HV system was operated for about fifty days without the auto-recovery script, and the current streamers were left to evolve naturally. They typically lasted six to 12 hours, exhibiting steady current and voltage drawn from the HV power supply, and they eventually self-quenched without any intervention. During this period, the repetition rate was significantly reduced to about one current streamer every 10-14 days; this rate can be compared to the 4-6 per day in the previous periods with auto-recovery on.

The auto-recovery script was then re-enabled and the current-streamer rate stabilised at about one in every 20 hours; in addition, the intensity of the current streamer on the GP was reduced with respect to the previous periods. As in the previous runs, the current streamers always occurred on the same GP (GP\#6) with a small leakage current on the beam plug hose, which is close to GP\#6. 

This behaviour is a further indication that the current streamers were, in fact, a slow discharge process of charged-up insulating materials present in the high-field region outside of the FC. The auto-recovery mode did not allow a full discharge, so the charging up was faster, and the streamer repetition rate was shorter.

The LAr purity loss experienced at the end of July 2019 was accompanied by the complete disappearance of any HV instabilities. 
They gradually reappeared when the electron lifetime exceeded 200 microseconds, and their intensity constantly increased as purity improved. This behaviour replicated that observed after the initial filling, and is consistent with what has been observed on other similar systems \cite{Bromberg_2015}, thus supporting the hypothesis that the HV instabilities are enhanced by the absence of electronegative impurities in high-purity LAr.

The effects of the current streamers on the front-end electronic noise and the PD background rate were investigated. No effect  
of the current streamers on the FE electronics was observed. On the other hand, recent analysis of the data collected by the PDS during active current streamers has indicated a high single photon rate on the upper upstream part of the TPC. This is consistent with the activities recorded on GP\#6, which is located exactly at this upper upstream area.

         \cleardoublepage

\subsection{Inner Detector: Anode Plane Assemblies and Front-end Electronics} 
\label{sec:APA}
      \subsubsection{Anode Plane Assemblies (APA)}
\label{sec:detcomp:inner:apafe:apa}
\begin{figure}[htb]
\centering
\includegraphics[width=0.85\textwidth]{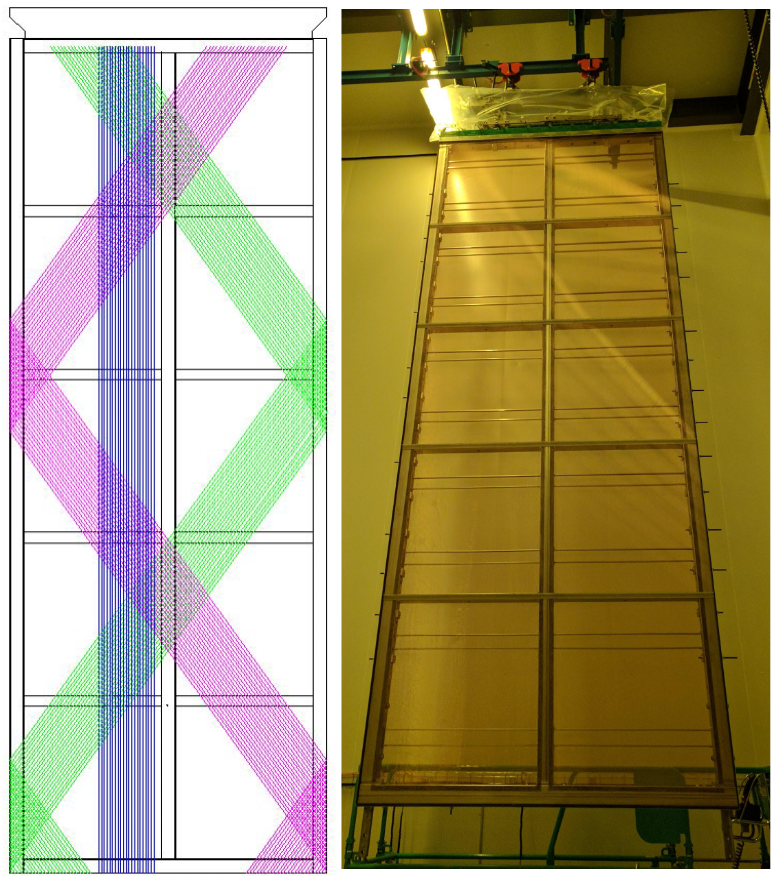} 
\caption{\label{fig:tpc_apa1}Left: Sketch of a \pdsp{} APA showing portions of the wire layers: the induction layers U (green) and V (magenta); and the collection layer X (blue). Only portions are shown to accentuate their angular relationships to the frame an d to each other.  The induction layers are connected electrically across both sides of the APA.  The grid layer (G) wires (not shown), run vertically, parallel to the X layer wires;  separate sets of G and X wires are strung on the two sides of the APA; they are not wrapped.  The mesh is not shown. Right: Assembled APA hanging inside the clean room next to the \pdsp{} cryostat at CERN.}
\end{figure}

Anode Plane Assemblies (APAs) are the detector elements used to 
detect ionization electrons created by charged particles traversing the LAr volume inside the \pdsp{} TPC. There are two APA arrays, one on the outer side of each of the two drift volumes. Each array comprises three APAs (6.3\,m tall, 2.3\,m wide, 0.12\,m thick) hung vertically and  adjacent to each other. Each APA has layers of sense and shielding wires wrapped around a framework of lightweight, rectangular stainless steel tubing, as shown in Figure~\ref{fig:tpc_apa1}. The sense wire readout is performed by cold electronics (CE) attached at the upper end (head) of each APA.

The APAs are designed and built to address the physics performance specifications listed in Table~\ref{tab:physicsrequirements}, which are defined to ensure  high-efficiency event reconstruction throughout the entire active volume of the LArTPC.  
\begin{table}[h!]
\centering
\begin{tabular}{l c}
\hline
Specifications & Value \\
\hline
MIP identification     &  100\% efficiency \\
Charge reconstruction & >\,90\% efficiency for >\,100\,MeV \\
Vertex resolution ($x$, $y$, $z$) & 1.5\,cm, 1.5\,cm, 1.5\,cm \\
\hline
Particle identification & \\
\hline
Muon momentum resolution & <\,18\% for non-contained, < 5\% for contained \\
Muon angular resolution & <\,1$^{\circ}$ \\
Stopping hadron energy resolution & 1-5\% \\
\hline
Shower identification & \\
\hline
Electron efficiency & >\,90\% \\
Photon mis-identification & <\,1\% \\
Electron angular resolution & <\,1$^{\circ}$ \\
Electron energy scale uncertainty & <\,5\% \\
\hline
\end{tabular}    
\caption{Physics requirements that motivate APA design parameters.}
\label{tab:physicsrequirements}
\end{table}

Identifying minimum-ionizing particles (MIPs) is a function of several detector parameters, including argon purity, drift distance, diffusion, APA sense-wire pitch, and equivalent noise charge (ENC).  It is required that MIPs originating anywhere inside the active volume of the detector 
be reconstructed with 100$\%$ efficiency.   The choice of wire pitch combined with 
the other high-level parameter values, listed in Table~\ref{tab:apaparameters},  is expected to enable  this  efficiency and provide good tracking resolution and good granularity for particle identification.

The specified vertex resolution of 1.5\,cm along each coordinate direction follows from a requirement that it be possible to determine the fiducial volume to 1\%.  
The chosen wire pitch achieves this for the $y$ and $z$ coordinates. The resolution on $x$, the drift coordinate, will be higher than in the $y$-$z$ plane due to the combination of drift velocity and electronics sampling rate.

In the construction of an APA, the frame is first covered on both sides with a fine mesh that defines a uniform ground across the frame. Along the length of the frame and around it, over the mesh layer, layers of sense and shielding wires are strung at carefully selected angles relative to each other. The wires are terminated on  boards that both anchor them and provide the connections to the CE.

\begin{table}[htbp]
\centering
\begin{tabular}{l c}
\hline
Paramater & Value \\
\hline
Active Height & 5.984 m\\
Active Width & 2.300 m\\ 
Wire Pitch (U,V) & 4.669 mm\\ 
Wire Pitch (X,G) & 4.790 mm\\ 
Wire Position Tolerance & 0.5 mm \\ 
Wire Plane Spacing & 4.75 mm\\ 
Wire Angle (w.r.t. vertical) (U,V) & 35.7$^{\circ}$\\ 
Wire Angle (w.r.t. vertical) (X,G) & 0$^{\circ}$\\ 
Number Wires / APA & 960 (X), 960 (G), 800 (U), 800 (V) \\ 
Number Electronic Channels / APA & 2560 \\
Wire Tension & 5.0 N \\
Wire Material & Beryllium Copper \\
Wire Diameter & 150 $\mu$m \\ 
Wire Resistivity & 7.68 $\mu\Omega$-cm $@$ 20$^{\circ}$ C \\
Wire Resistance/m & 4.4 $\Omega$/m $@$ 20$^{\circ}$ C \\
Frame Planarity & 5 mm \\
Photon Detector Slots & 10 \\
\hline
\end{tabular}    
\caption{APA design parameters.}
\label{tab:apaparameters}
\end{table}



The grid plane wires (G) are not connected to the electronic readout.  Separate sets of grid wires run along each side of the APA, parallel to its long edge. 
 The two planes of induction wires (U and V) wrap continuously around both sides of the APA in a helical fashion. 
 The collection plane wires (X) run vertically, parallel to the grid wires.   The ordering of the layers, from the outside in, is G-U-V-X, with the mesh layer at the centre. All wire layers span the entire face on both sides of the APA frame.
 
The angle of the induction planes in the APA ($\pm$35.7$^{\circ}$) was chosen such that each induction wire only crosses a given collection wire one time, reducing the ambiguities that the reconstruction must address.  
Coupled with the wire pitch, this angle makes it possible for an integer multiple of electronics boards to read out one APA.

The operating voltages of the APA layers are listed in Table~\ref{tab:bias}.  When operated at these voltages, the drifting ionization follows trajectories around the grid and induction wires, ultimately terminating on a collection plane wire; i.e., the grid and induction layers are completely transparent to drifting ionization, and the collection plane is completely opaque.  The grid layer is present for pulse-shaping purposes, effectively shielding the first induction plane from the drifting charge and removing the long leading edge from the signals on that layer. 
The mesh layer serves to shield the sense planes from pickup from the PDS.

\begin{table}[htbp]
\centering
\begin{tabular}{l c}
\hline
APA layer & Bias voltage \\
\hline
Grid (G) & $-$665\,V\\ 
Induction (U) & $-$370\,V\\ 
Induction (V) & 0 V\\ 
Collection (X) & 820\,V\\ 
Mesh (M) & 0\,V\\
\hline
\end{tabular}    
\caption{Bias voltages for APA wire planes and mesh.}
\label{tab:bias}
\end{table}

The wrapped 
configuration allows the APA array to fully cover the active area of the LArTPC, minimizing the amount of dead space between the APAs that would otherwise be occupied by electronics and associated cabling.   

The current design of the DUNE-SP far detector module implements three APA arrays that run the 60\,m length of the TPC. One runs down the centre of the detector, the other two are installed along the outer walls, and two CPA arrays are interleaved between them, creating a set of four drift fields. The central APA array, flanked by drift fields, requires sensitivity on both sides, which the wrapped 
induction-plane wire design enables. 
Whereas this double-sided feature is not strictly necessary for the \pdsp{} arrangement, in which 
only the inner side of each APA faces a drift field,  it is compatible with this setup, as the grid layer on the wall side effectively blocks any ionization generated outside the TPC from drifting in to the sense wires. 

The APAs are wound with 150$\mu$m (.006\,in) diameter beryllium copper (CuBe) wire ($\sim$98\% copper, $\sim$1.9\% beryllium), used for its high durability and yield strength.  The X- and G-plane wires extend the full 6\,m length of the APA and are not wrapped. The diagonal, wrapped wires (U and V planes) extend 3.9\,m across each face of the APA.  To prevent deflection from gravity, electrostatic forces, and liquid drag from any moving LAr, a set of \textit{combs} supports these wires at regular intervals along the length of the APA, keeping the longest unsupported wire length under 1.6\,m. The combs are slotted pieces of 0.5\,mm thick G10, mounted on each of the frame's four cross braces. The wires are held at a tension of 5\,N.

The wire tension and wire placement accuracy specifications ensure that the wires are held taut in place with no sag.  Wire sag can impact both the precision of reconstruction and the transparency of the TPC.  The tension of 5\,N is low enough that as the wires contract during cool-down, they stay safely below the vendor-tested 25.6\,N breaking tension value.

\begin{figure}[htb]
\centering
\includegraphics[width=0.45\textwidth]{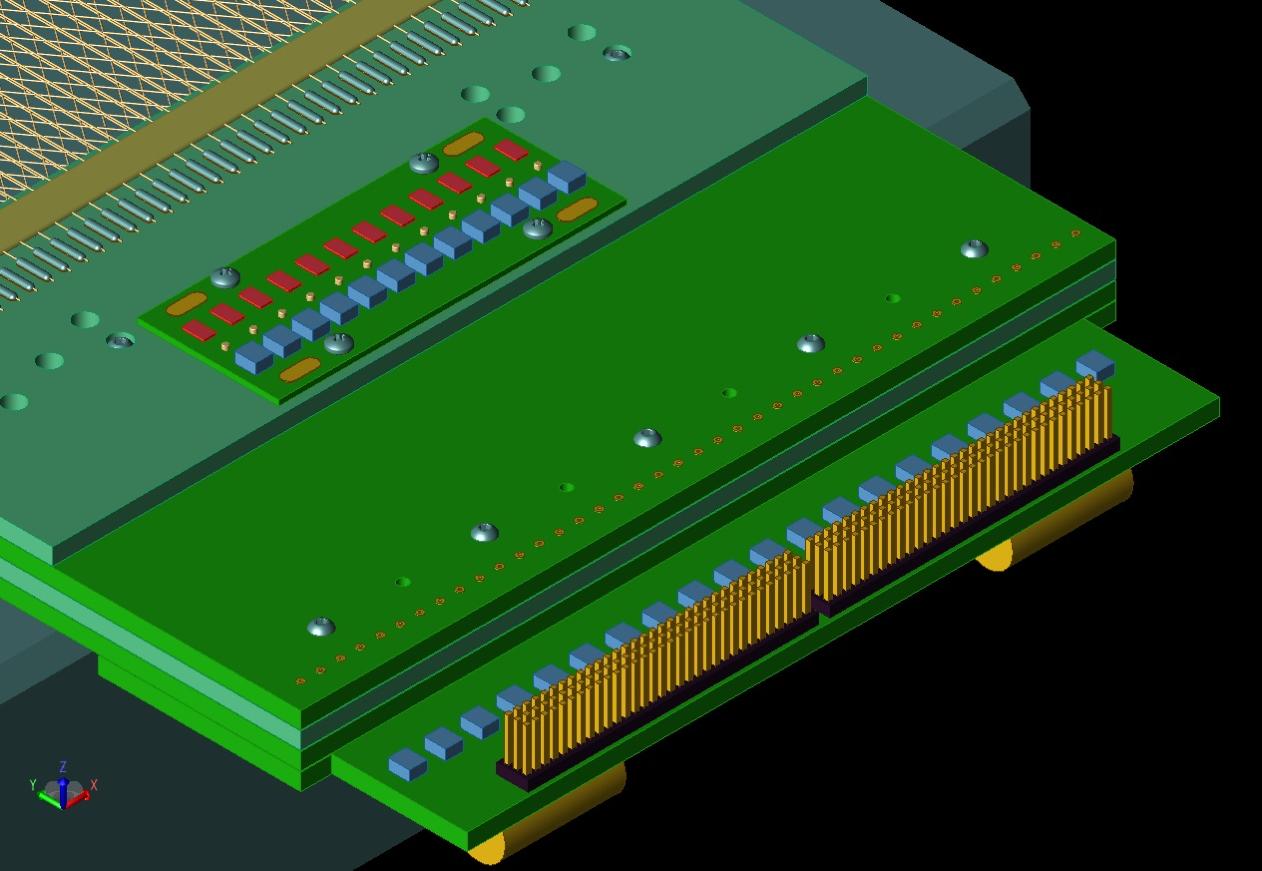}
\includegraphics[width=0.45\textwidth]{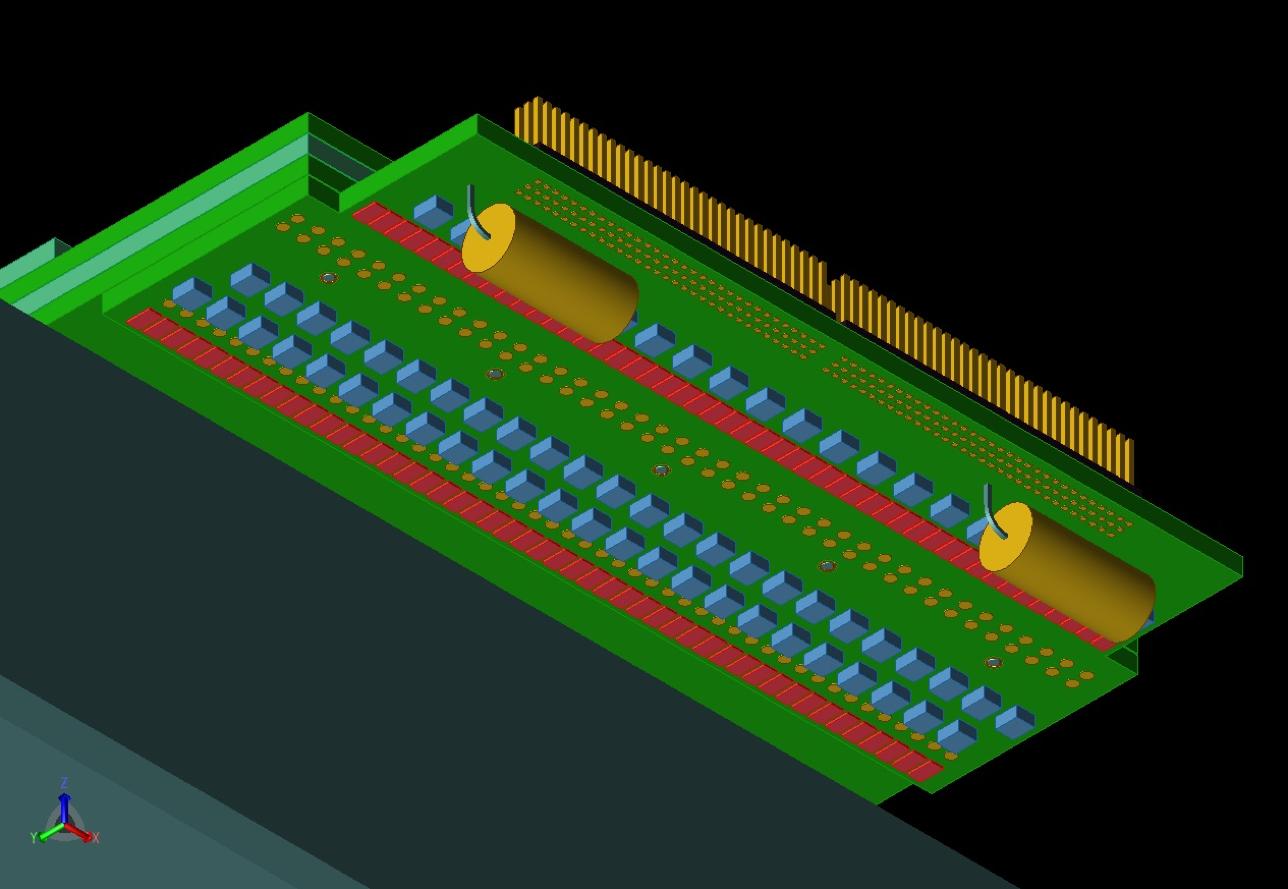}
\caption{\label{fig:tpc_apa_boardstack} Views of an APA wire-board stack from above (left) and below (right). The X plane wire board is the lowest in the stack. The left image shows the main capacitive-resistive (CR) board connected to the X plane board as it extends beyond the stack. V and U plane boards are attached to the X plane board using Mill-Max pins. The G plane board (outermost and extending less far) is glued on to the board stack of U, V, X, planes. The G plane has its own CR board that is attached 
as shown in the left figure.}
\end{figure}

All APA wires are terminated on stacked wire boards, installed at the head end of each APA to provide the connections to the CE, as shown in  Figure~\ref{fig:tpc_apa_boardstack}. 
The APA has ten adjacent sets of these board stacks on each side rather than one long stack spanning the entire width.
Within each stack, one board corresponds to one wire plane. 
Attachment of the wire boards begins with the (innermost) X wire plane to the lowest board in the stack. These wires are strung top to bottom along each side of the APA frame, soldered, epoxied to connections on the board, and trimmed. As each subsequent layer is applied, its wires are attached to the next highest wire board layer. Mill-Max pins and sockets provide the electrical connections between the circuit boards within a stack.

\begin{figure}[htb]
\centering
\includegraphics[width=0.7\textwidth]{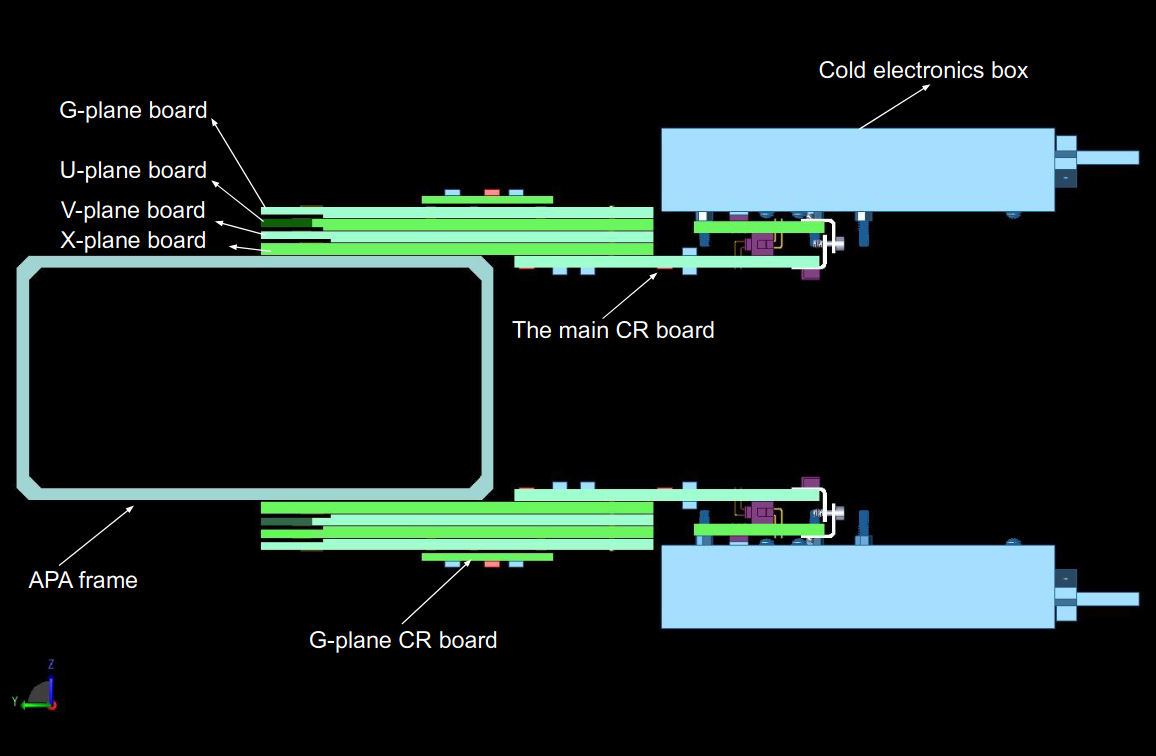}
\caption{\label{fig:tpc_apa_electronics_connectiondiagram}Connections between the APA wires, viewed from the APA edge. The symmetric set of wire boards within a stack can be seen on both sides of the APA, with the CR board extending further to the right. The CR board provides a connection to the CE housed in the boxes at the far right of the figure. }
\end{figure}

The wire boards for the X and G planes accommodate 48 wires each, and those for the U and V planes accommodate 40 each.  Each board stack, therefore, has 176 wires, but only 128 signal channels since the G wires are not read out.  
The total of 20 stacks per APA results in 2,560 signal channels and a total of 3,520 wires starting at the top of the APA and ending at the bottom.  Each APA holds a total of $\sim$\,23.4 km of wire. 
Figure~\ref{fig:tpc_apa_electronics_connectiondiagram} depicts the connections between the different elements of the APA electrical circuit.

At the head end of the APA, the 4.75~mm wire-plane spacing is set by the thickness of these wire boards. 
The first plane's wires solder to the surface of the first board, the second plane's wires to the surface of the second board, and so on.  For installation, temporary toothed-edge boards beyond these wire boards are used to align and hold the wires until they are soldered to pads on the wire boards.  After soldering, the extra wire is snipped off. 

Attached to the wire board stack, capacitive-resistive (CR) boards provide DC bias and AC coupling to the wires. The CR boards carry a bias resistor (51 M$\Omega$) and a DC-blocking capacitor (3900 pF) for each wire in the X and U planes. 
Connections from the CR boards to the CE are made through a pair of 96-pin Samtec connectors. In the case of the outermost G plane, 
wire boards connect adjacent groups of four wires together  and bias each group through an RC filter whose components are placed on special G plane CR boards.  
All CR boards are attached to the board stacks after fabrication of all wire planes.   

\begin{figure}[htb]
\centering
\includegraphics[width=0.9\textwidth]{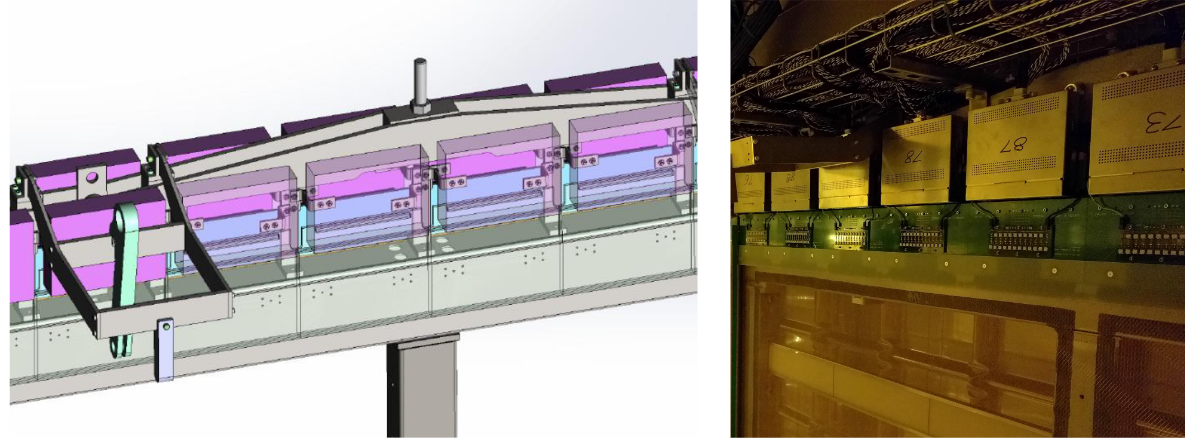} 
\caption{\label{fig:tpc_apa_electronicsmountingdiagram} 3D model of 
modular CE boxes (left). A set of CE boxes installed on an APA in \pdsp{} (right).}
\end{figure}

Pins extending outward from the CR boards provide connections from the APA to the modularly designed CE. Each board stack has one CE module connected to it. Each CE module is housed in a small metallic enclosure called a CE box that provides electrical shielding and mechanical support, 
simplifies installation and replacement, and 
helps with the dissipation
of argon gas generated by the warm electronic components.  The CE modules are mounted in such a way that any of them can be removed from 
the inner side of the APA after APA installation. 
Figure~\ref{fig:tpc_apa_electronicsmountingdiagram} illustrates the CE boxes and their installation in \pdsp{}. 

      \subsubsection{TPC Front-end Cold Electronics (CE)}
\label{sec:detcomp:inner:apafe:fe-ce}
The \pdsp{} TPC read-out electronics are referred to as ``cold electronics (CE)'' because the system resides in the LAr,
mounted directly on the APA, 
thus reducing channel capacitance and noise by minimising the length of the connection between an anode wire
and its corresponding electronics input.


The CE signal processing is implemented in ASIC chips using CMOS technology,
which has been demonstrated to perform well at cryogenic temperatures,
and includes amplification, shaping, digitisation, buffering, and multiplexing (MUX) of the signals.
The CE is continuously read out,
resulting in a digitised ADC sampling from each APA channel (wire) at a rate of up to every 500\,ns (2\,MHz maximum sampling rate).

\begin{figure}[htb]
\centering
\includegraphics[width=0.4\linewidth]{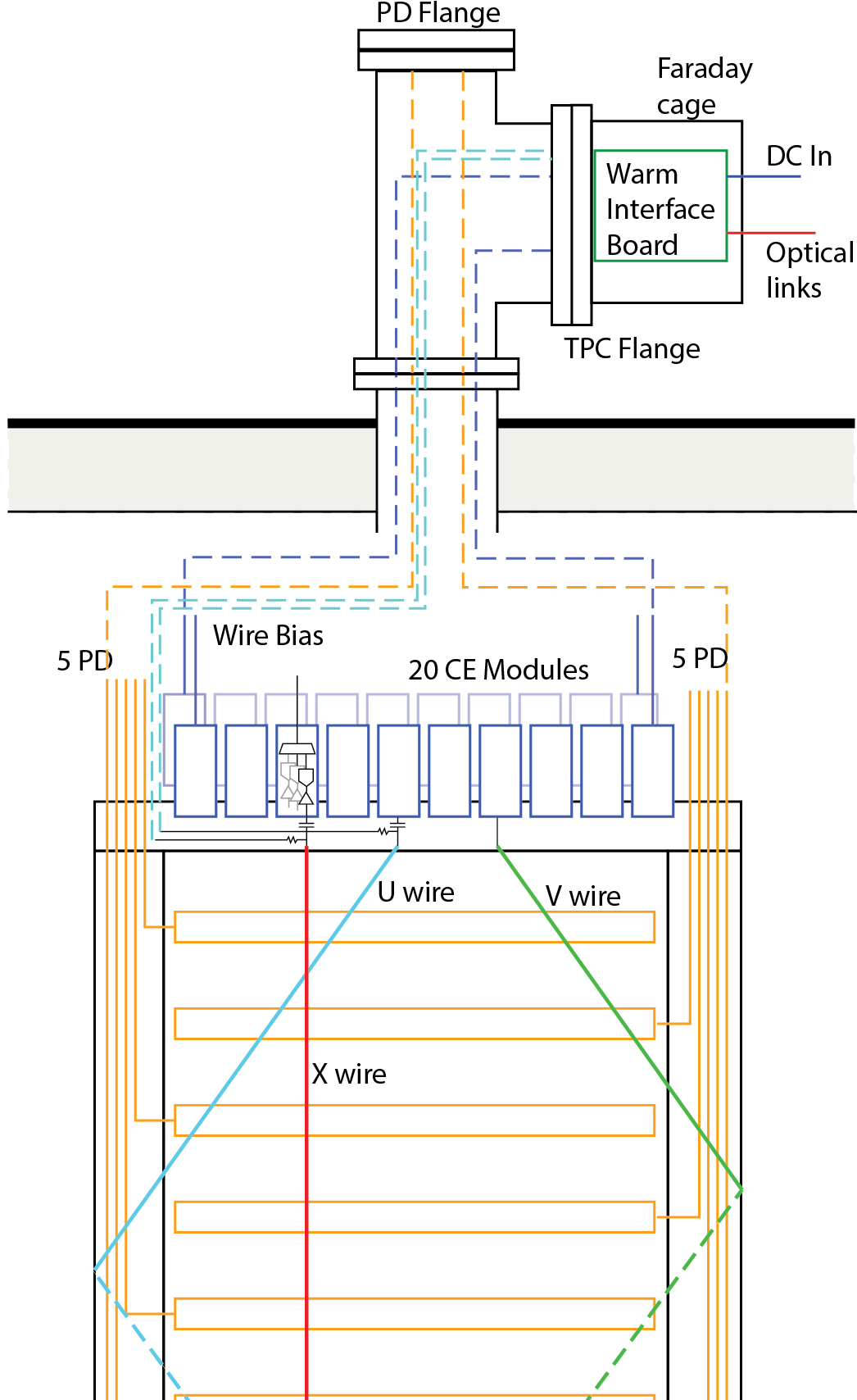}
\caption{\label{fig:tpcce_apa_flange}Power and readout cable connection between the APA and signal flange on the cryostat. The orange cables represent photon detector connections, the turquoise lines represent wire biases and field cage (FC) terminations and blue lines represent power and readout cables for the CE. }
\end{figure}

The 2,560 channels from each APA are read out by 20 front-end motherboards (FEMBs), each providing digitised wire readout from 128 channels. One cable bundle connects each FEMB to the outside of the cryostat via a CE feedthrough in the signal cable flange at the top of the cryostat, where a single flange services each APA. Each cable bundle contains wires for low-voltage (LV) power, high-speed data readout, and clock/digital-control signal distribution. In addition to the CE cables, eight separate cables carry the TPC wire-bias and FC termination voltages from the signal flange to the APA wire-bias boards as
shown schematically in Figure~\ref{fig:tpcce_apa_flange}.

\begin{figure}[htb]
\centering
\includegraphics[width=0.9\linewidth]{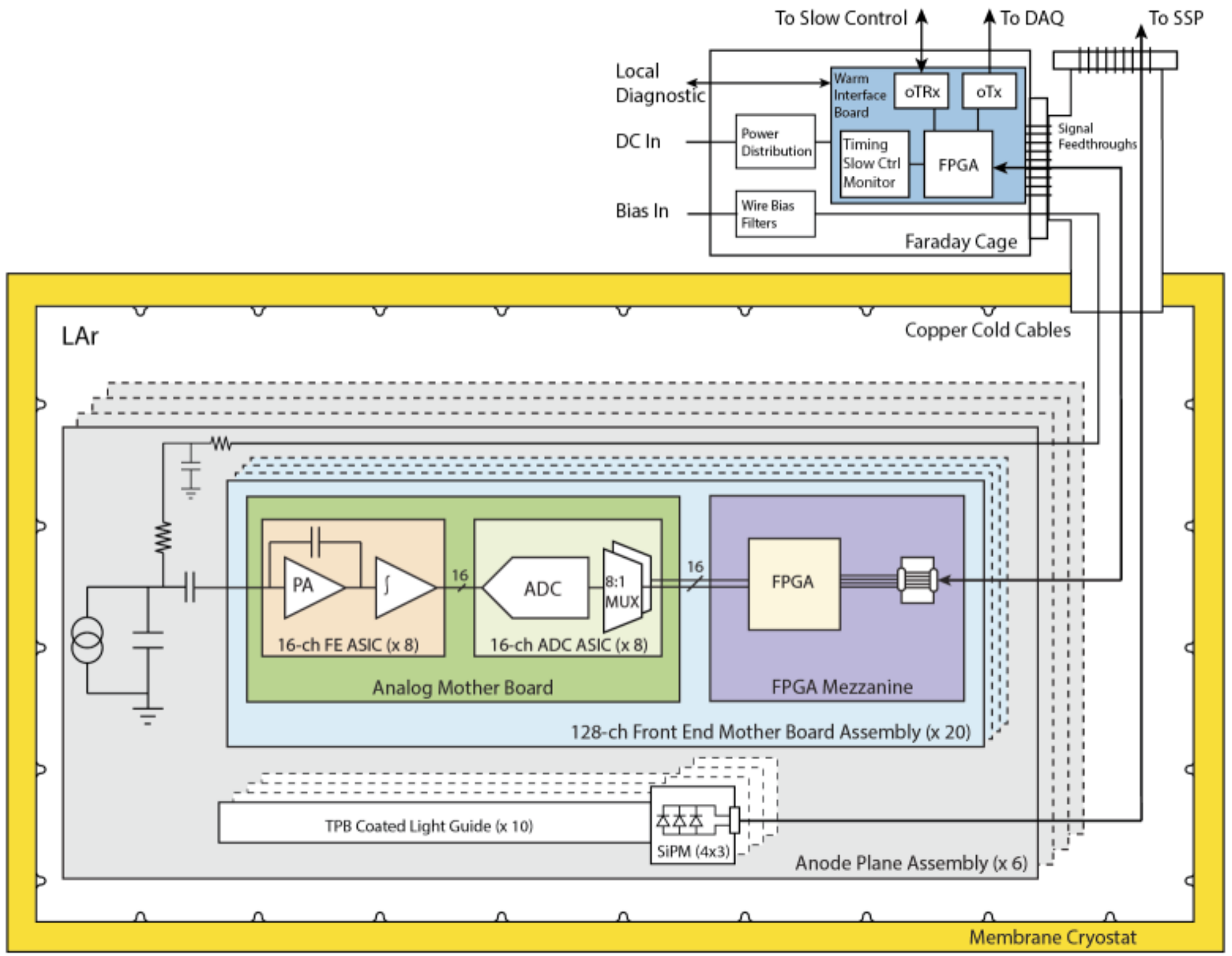}
\caption{\label{fig:tpcce_schem}{}The CE architecture. The basic unit is the 128-channel FEMB.}
\end{figure}

The main component of the CE architecture, illustrated in Figure~\ref{fig:tpcce_schem}, is the 
128-channel FEMB, which itself consists of an analogue motherboard and an attached FPGA 
mezzanine card for processing the digital outputs.
Each APA is instrumented with 20 FEMBs, for a total of 2,560 channels per APA.
The FEMBs plug directly into the APA CR boards, making the connections from the 
U, V, and X plane wires to the charge amplifier circuits as short as possible.

\begin{figure}[htb]
\centering
\includegraphics[width=0.7\linewidth]{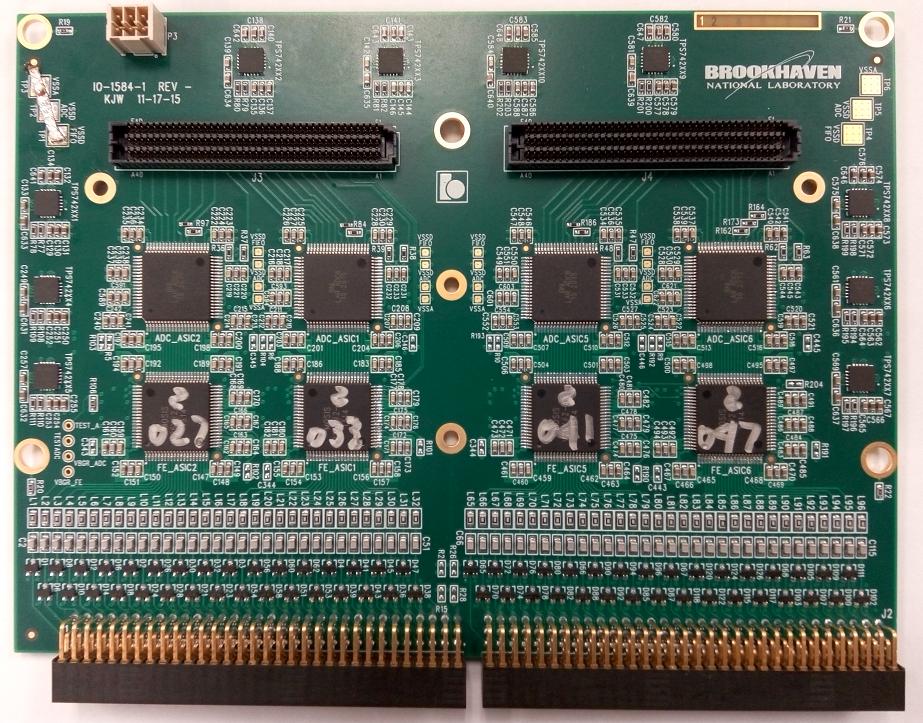}
\includegraphics[width=0.7\linewidth]{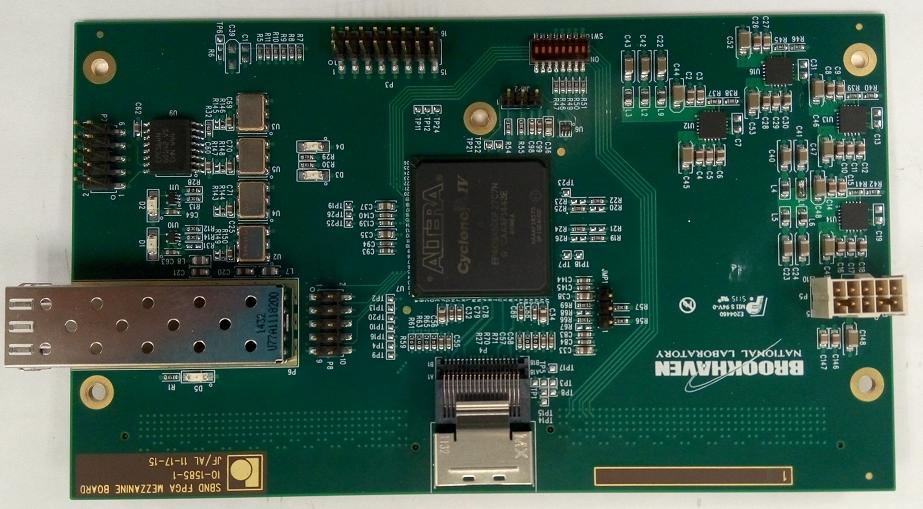}
\includegraphics[width=0.7\linewidth]{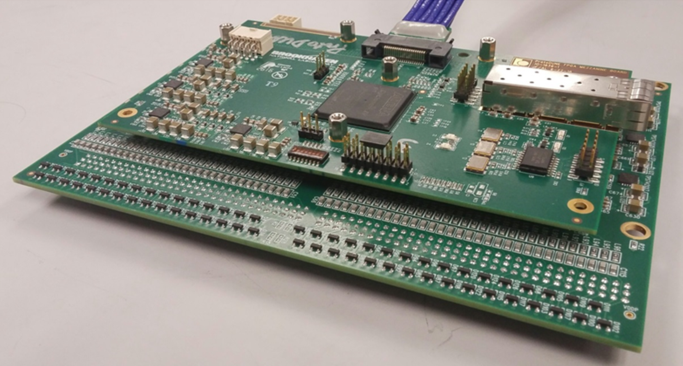}
\caption{\label{fig:tpcce_CMBpix}The Front End Mother Board (FEMB).
  Top: The analogue mother board, showing four ADC ASICs and four FE ASICs, surface mounted.
  The other side of the board has another four ADC and FE ASICs.
  Middle: The FPGA mezzanine.
  Bottom: The complete FEMB assembly.
  The cable shown is the high-speed data, clock, and control cable.}
\end{figure}

The analogue mother board is instrumented with eight 16-channel FE ASICs,
eight 16-channel ADC ASICs, LV power regulators, and input-signal protection circuits.
The 16-channel FE ASIC provides amplification and pulse shaping.
The 16-channel ADC ASIC comprises 12-bit digitisers performing at speeds up to 2 MS/s, local buffering,
and an 8:1 MUX stage with two pairs of serial readout lines in parallel. Figure~\ref{fig:tpcce_CMBpix} shows the 
analog motherboard, the FPGA mezzanine, and the complete FEMB assembly.

Each FE ASIC channel has a charge amplifier circuit with a programmable gain, selectable from 
4.7, 7.8, 14, and 25\,mV/fC
(full-scale charge of 55, 100, 180 and 300\,fC),
a high-order anti-aliasing filter with programmable time
constant (peaking time 0.5, 1, 2, and 3 $\mathrm{\mu}$s),
an option to enable AC coupling,
and a baseline adjustment for operation with either the collecting (200\,mV) or the non-collecting (900\,mV) wires.
Shared among the 16 channels in the FE ASIC are the bias circuits, programming registers,
a temperature monitor, an analog buffer for signal monitoring, and the digital interface.
The FE ASIC's estimated power dissipation is about 6\,mW per channel at 1.8\,V supply.

The FE ASIC was implemented using the commercial CMOS process (0.18\,$\mu$m and 1.8\,V), which 
is expected to be available for at least another ten~years. 
The charge amplifier input MOSFET is a p-channel biased at 2\,mA with a L/W (channel length/width) ratio
of 0.27\,$\mu$m / 10\,$\mu$m, followed by dual cascade stages.


Each FE ASIC channel is equipped with an injection capacitor that can be used
for testing and calibration and is 
enabled or disabled through a
dedicated register. The injection capacitance has been measured using 
a calibrated external capacitor. The measurements show
that the calibration capacitance is extremely stable, changing from
184\,fF at room temperature to 183\,fF at 77\,K. This result and the measured
stability of the peaking time demonstrate the high stability of the
passive components as a function of temperature. Channel-to-channel and chip-to-chip
variation in the calibration capacitor is typically less than 1\%. 


The ADC ASIC design is also implemented using the CMOS process (0.18\,$\mu$m and 1.8\,V).
The ADC ASIC is a complex design with 320,000 transistors, while the FE ASIC has 16,000.
The transistor design work has been done following the rules for long cryogenic lifetime.
Shared among the 16 channels in the ADC ASIC are the bias circuits, programming registers,
an 8:1 MUX, and the digital interface.
The estimated power dissipation of ADC ASIC is below 5\,mW per channel at 1.8\,V supply.

The ADC ASIC data are passed to the FPGA mezzanine board for transmission to the warm electronics
located on the outside of the signal flange.
The FPGA has four 4:1 MUX circuits that combine the 16 serial lines from the eight ADC
channels into four serial lines of 32 channels each, and 
four $\sim$\,1.2\,Gigabit-per-second (Gbps) serial drivers that drive the data in each
line over cold cables to the warm interface electronics (WIBs). The FPGA on the mezzanine card is also responsible for communicating with the
control and timing systems from the WIB and providing the clock and control signals required by the FE and ADC ASICs.

\begin{figure}[htb]
\centering
\includegraphics[width=0.7\linewidth]{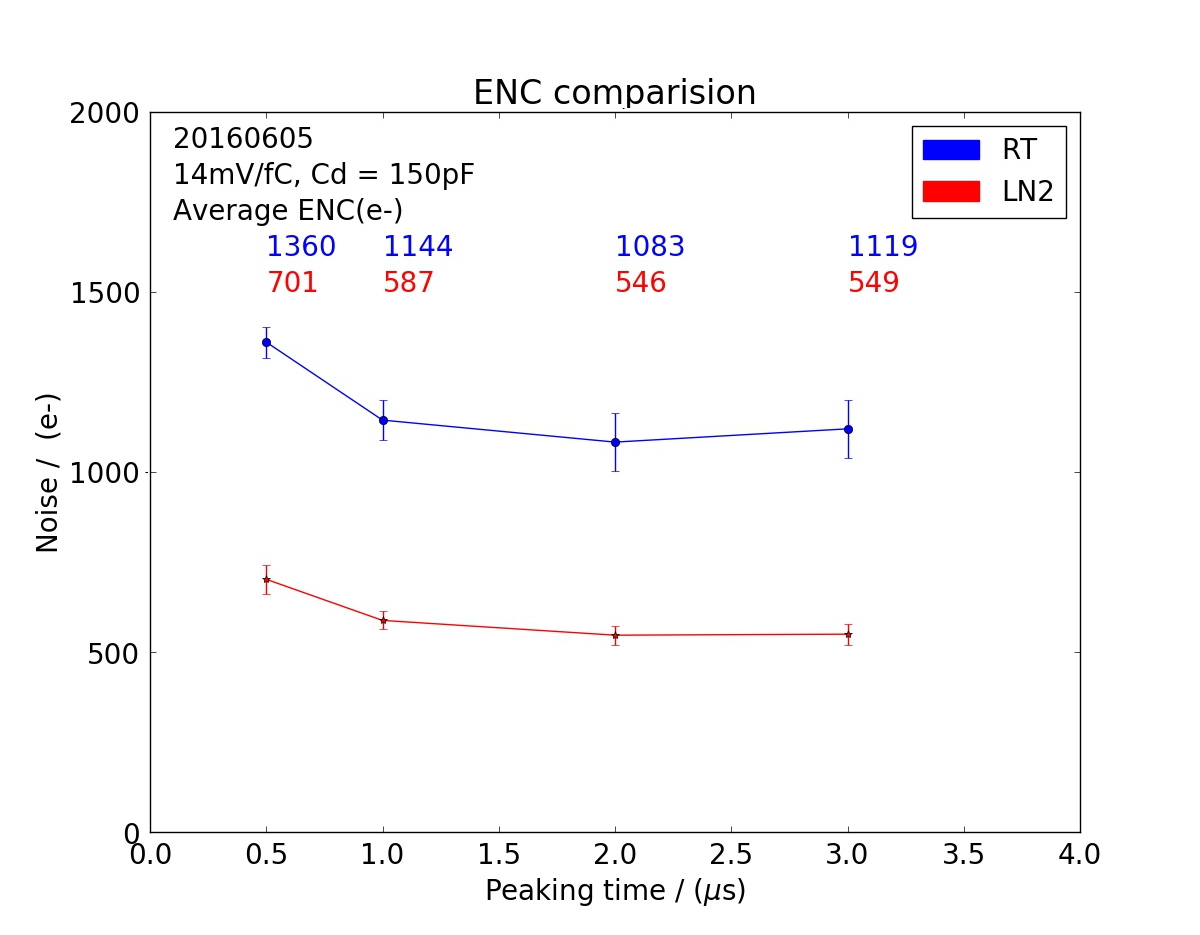}
\includegraphics[width=0.7\linewidth]{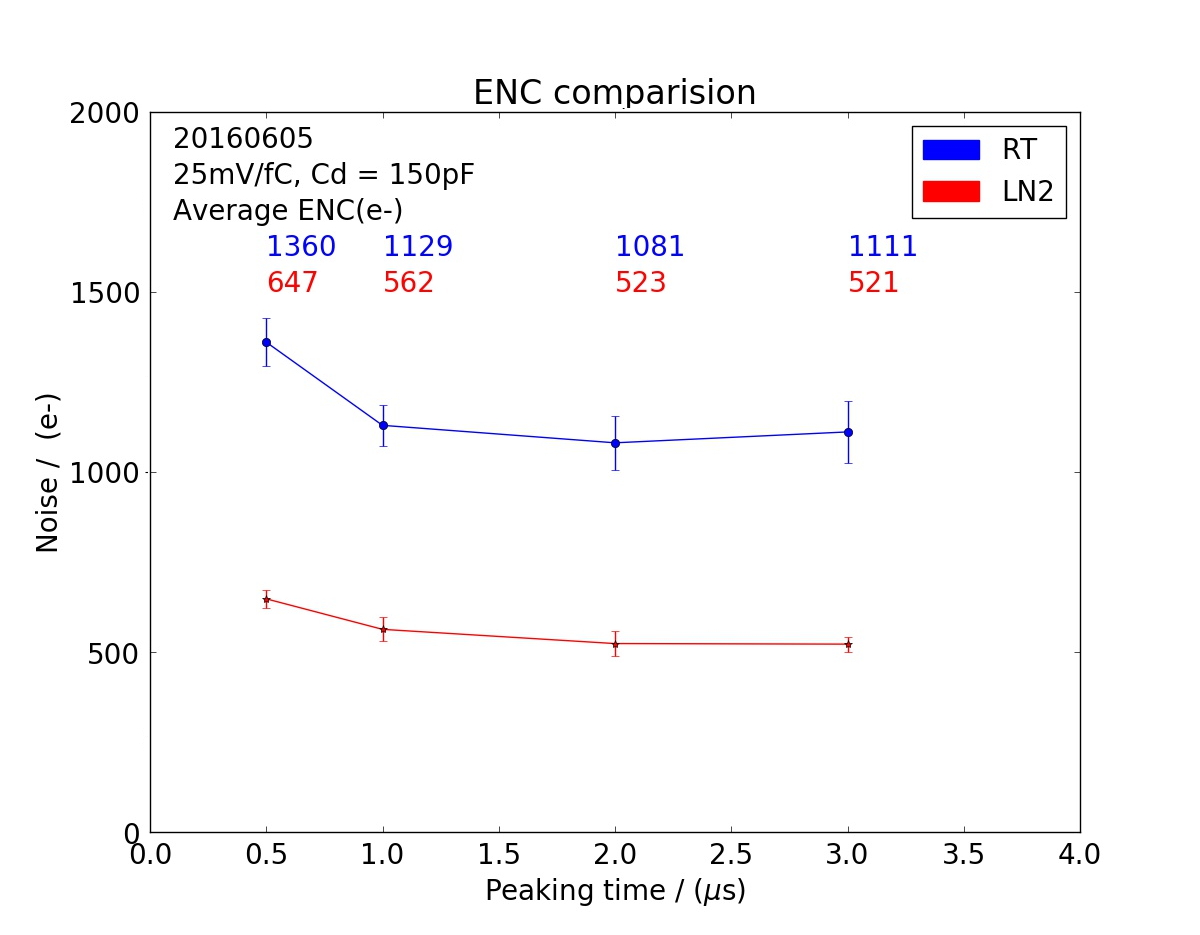}
\caption{\label{fig:tpcce_enc}Measured ENC vs filter time constant from the latest prototype version of the FEMB
for two different gains, 14\,mV/fC and 25\,mV/fC. In the legend RT refers to room temperature measurements and LN2 refers to measurements made in liquid nitrogen.}
\end{figure}

The FPGA and all other electrical components on the FEMB
assembly has been evaluated and characterised at RT (300\,K) and LN2 (77\,K) temperature. During these tests the FEMB has been temperature-cycled multiple times. In addition, power-cycle tests at cryogenic temperature have been performed. Figure~\ref{fig:tpcce_enc} shows the measured ENC as a function of 
filter-time constant (peaking time) for two different gains as measured on a prototype FEMB. ENC is the value of charge 
(in electrons) injected across the detector capacitance that would produce at the output of the 
shaping amplifier a signal whose amplitude equals the output RMS noise. These measurements
were made with the prototype FEMB at both RT and submerged in LN2 with a wire-simulating input capacitance of $C_f~=~150$\,pF
(equivalent to an approximately 7\,m sense wire load).
In LN2, for peaking times $>$1\,$\mu$s, less than 600\,e$^{-}$ was measured. For comparison,
a MIP travelling perpendicularly to the wire plane in the direction of wire spacing is
expected to deposit at least $\sim$\,10,000\,e$^{-}$ on the collection wires, for a worst-case
SNR is about 16 to 1\footnote{For the minimum performance requirements of 500\,V/cm and a 3\,ms electron lifetime}. One of the key ingredients to achieving a low-level noise performance is the grounding of the experimental setup. The details of the \pdsp{} grounding scheme and methods to ensure a low noise level are discussed in Section~\ref{sec:assy:gr}

\begin{figure}[htb]
\centering
\includegraphics[width=0.45\linewidth]{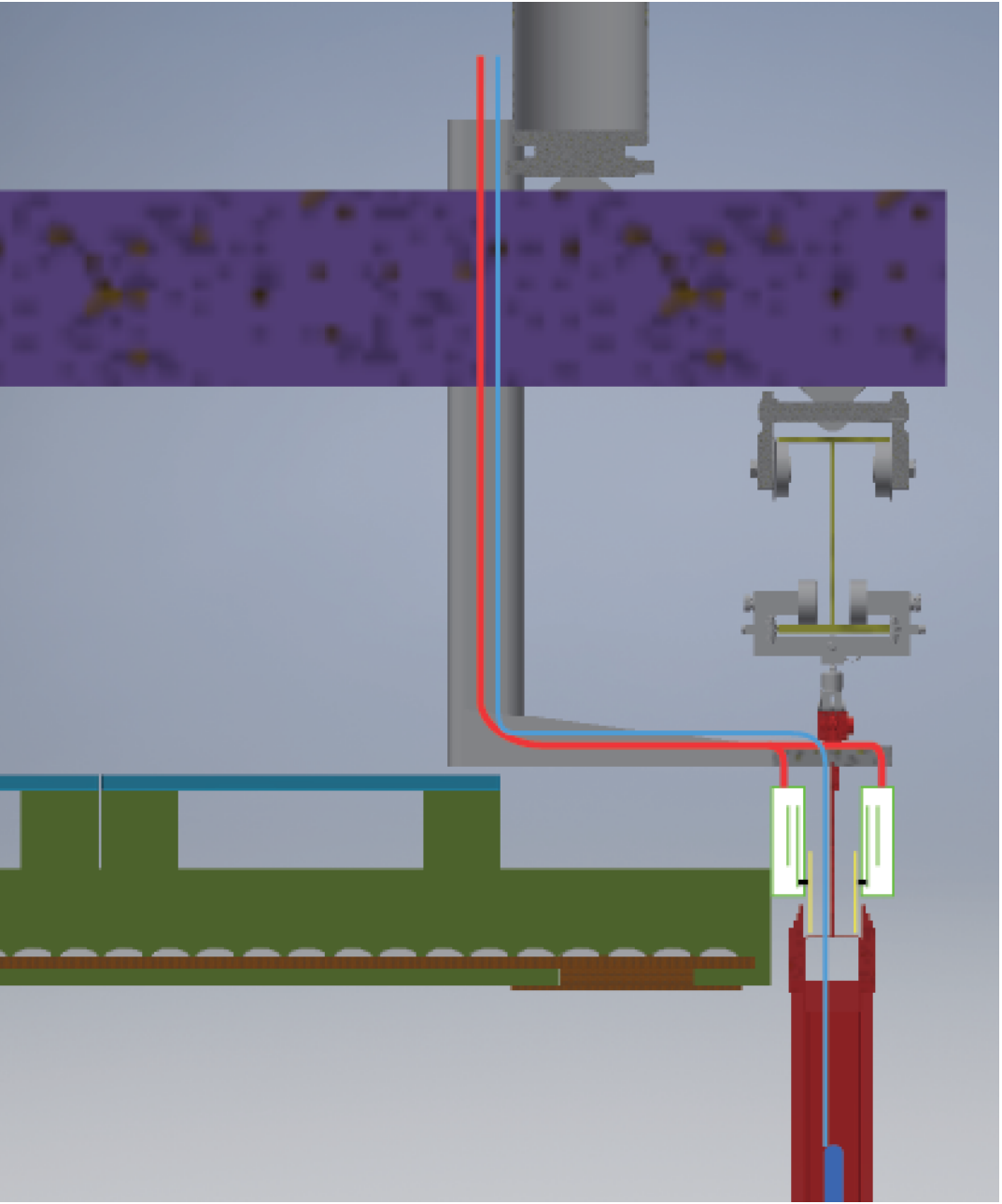}
\includegraphics[width=0.45\linewidth]{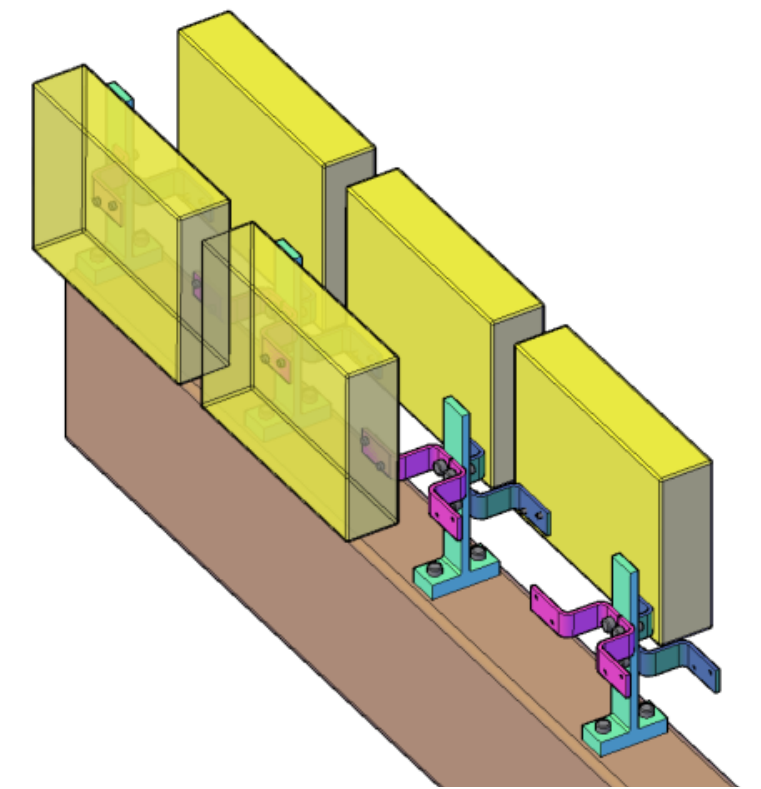}
\caption{\label{fig:tpcce_box}Left: cable routing from CE box to signal flange on the cryostat. Right: CE box (yellow) for the FEMB.}
\end{figure}

As mentioned in Section~\ref{sec:detcomp:inner:apafe:apa}, each FEMB is enclosed in a Faraday box, called a CE box, that provides shielding from noise. As shown in Figure~\ref{fig:tpcce_box}, it 
provides the electrical connection between the FEMB and the APA frame. Mounting hardware inside the CE box connects the ground plane of the FEMB to the box casing, which 
is electrically connected to the APA frame via twisted conducting wire. This is the only point of contact between the FEMB and
APA, except for the input amplifier circuits connected to the CR board, which also terminate to
ground at the APA frame.
      \subsubsection{TPC Front-end Warm Electronics}
\label{sec:detcomp:inner:apafe:fe-we}
\begin{figure}[htb]
\centering
\includegraphics[width=0.9\linewidth]{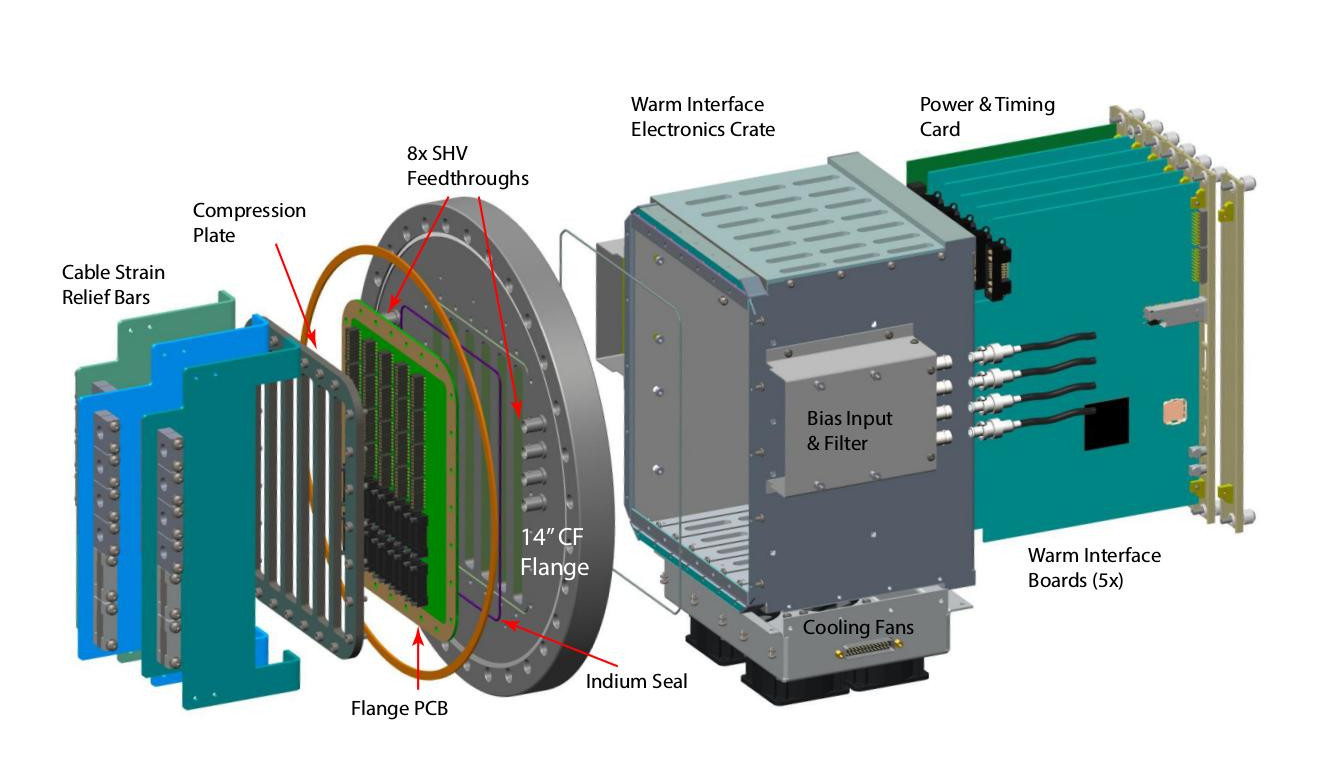}
\caption{\label{fig:tpcce_ceflange_dune}Exploded view of the \pdsp{} signal flange.}
\end{figure}

The warm interface electronics are housed in warm interface electronics crates (WIECs)
attached directly to the signal flange.  The WIEC shown in Figure~\ref{fig:tpcce_ceflange_dune}
contains one power and timing card (PTC), up to five warm interface boards (WIBs) and a passive
power and timing backplane (PTB), which fans out signals and LV power from the PTC to the WIBs.

The WIB is the interface between the DAQ system and up to four FEMBs. It receives the system clock and control signals from the timing system and provides for processing and fan-out of those signals to the four FEMBs. The WIB also receives the high-speed data signals from the four  FEMBs and transmits them to the DAQ system over optical fibres.  The WIBs are attached directly to the TPC CE feedthrough on the signal flange. The feedthrough board is a PCB with connectors to the cold signal and LV power cables fitted between the compression plate on the cold side, and sockets for the WIB on the warm side. 
Strain relief for the cold cables is supported from the back end of the feedthrough.


The PTC provides a bidirectional fibre interface to the timing system.  The clock and data
streams are separately fanned-out to the five WIBs. 
The PTC fans the clocks out to the WIB over the PTB, which is a passive backplane attached directly to the PTC and WIBs. The received clock on the WIB is separated into clock and
data using a clock/data separator.
\begin{figure}[htb]
\centering
\includegraphics[width=0.7\linewidth]{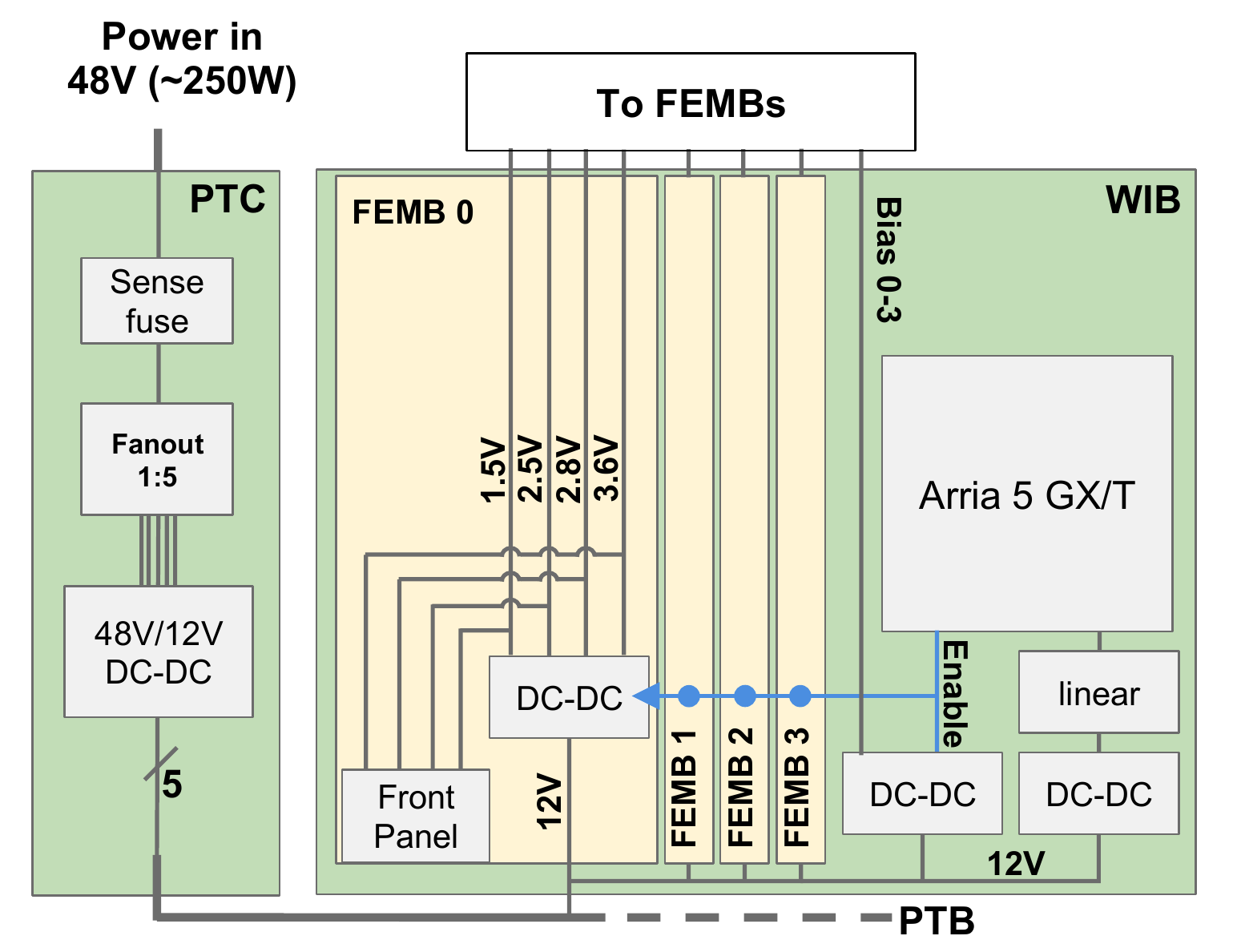}
\caption{\label{fig:tpcce_wib_power}LV power distribution 
to the WIB and FEMBs. Power of 250\,W is for a fully-loaded crate 
with the majority of the power dissipated by the 20 cold FEMBs in the LAr.}
\end{figure}
The PTC also receives LV power for all CE connected through the signal flange, approximately 250\,W at 48\,V for a fully-loaded flange (one~PTC, five~WIB, and 20~FEMB). The LV power is then stepped down to 12\,V via a DC/DC converter on-board the PTC and fanned out on the PTB to each WIB, which provides the necessary 12\,V DC/DC conversions and fans the LV power out to each of the cold FEMBs supplied by that WIB, 
as shown in Figure~\ref{fig:tpcce_wib_power}. 
The majority of the 250\,W drawn by a full flange is dissipated in the LAr by the cold FEMB.

Each WIB contains a unique IP address for its UDP (User Datagram Protocol) slow control interface. In addition, the WIB is capable of receiving the encoded system-timing signals over bi-directional optical fibres on the front panel, and processing these using either the on-board FPGA or clock synthesizer chip to provide the 50\,MHz clock required by the CE.

The FPGA on the WIB is an Altera Arria V GT variant, which requires a 125\,MHz clock for its state machine that is provided by an on-board crystal oscillator. It can drive the high-speed data to the DAQ system up to 10.3125\,Gbps per link,  implying that all data from two FEMB (2$\times$5\,Gbps) could be transmitted on a single link. On top of that, the FPGA has an additional Gbps Ethernet transceiver I/O based on the 125\,MHz clock, which provides real-time digital data readout to the slow control system as well.

         \cleardoublepage

\subsection{Inner Detector: Photon Detection System} 
\label{sec:detcomp:inner:pds}
      The Photon Detection System (PDS) in the DUNE FD is critical for the DUNE physics program and the ProtoDUNE PDS serves as a prototype for the DUNE FD PDS system.
The topology and particle trajectories for events produced by accelerator neutrinos can be reconstructed in the TPC using charge signals with drift times referenced to the accelerator clock. 
However, the observation of potential nucleon decay or neutrinos from supernova bursts and other cosmological sources require additional timing information. 
To make these observations, the PDS must have large acceptance for light in each of the 12\,m $\times$ 14\,m $\times$ 58.2\,m 17\,kt FD modules and a timing resolution that localises beam and self-triggered TPC events to mm-scale resolution; it must exist within the design constraints of the TPC without affecting its operation; and the entire system must be implemented so as to minimise cost while maximising light detection efficiency.
These requirements translate to a minimum photon yield of at least $0.5$\,PE/MeV at a maximum drift distance of 3.6\,m and a minimum timing resolution of 100\,ns. 
The PDS in \pdsp{} demonstrates that current designs satisfy all of these requirements and, with some revisions, will satisfy the physics goals of the DUNE FD.

The PDS obtains event and timing information from the photons produced by LAr scintillation as particles traverse the detector. 
LAr is highly transparent to its scintillated light, producing photons at 128\,nm with a Rayleigh scattering length of about 90\,cm and an absorption length that is entirely dependent on the presence of impurities~\cite{Babicz:2018gqv,Babicz:2020den}.
Minimum ionising particles generate about 50,000~photons per MeV of deposited energy, but fewer photons are produced in the presence of electric fields due to the reduction of electron-ion recombination. 
At the nominal 500\,V/cm field strength in \pdsp{}, the photon yield is about 24,000~photons per MeV.
Roughly a quarter of the photons produced are promptly emitted through singlet configuration decay of excited argon dimers; the remaining light is emitted with a lifetime of approximately 1.3\,$\mu$s from the triplet configuration \cite{heindl:2011}. 
The measurement of prompt photons produced in LAr scintillation serves to temporally and spatially localise relevant events, by  setting a ``$t_{0}$'' for each event.

\begin{figure}[!htbp]
    \centering
    \includegraphics[height=9.5cm,width=1.05\textwidth]{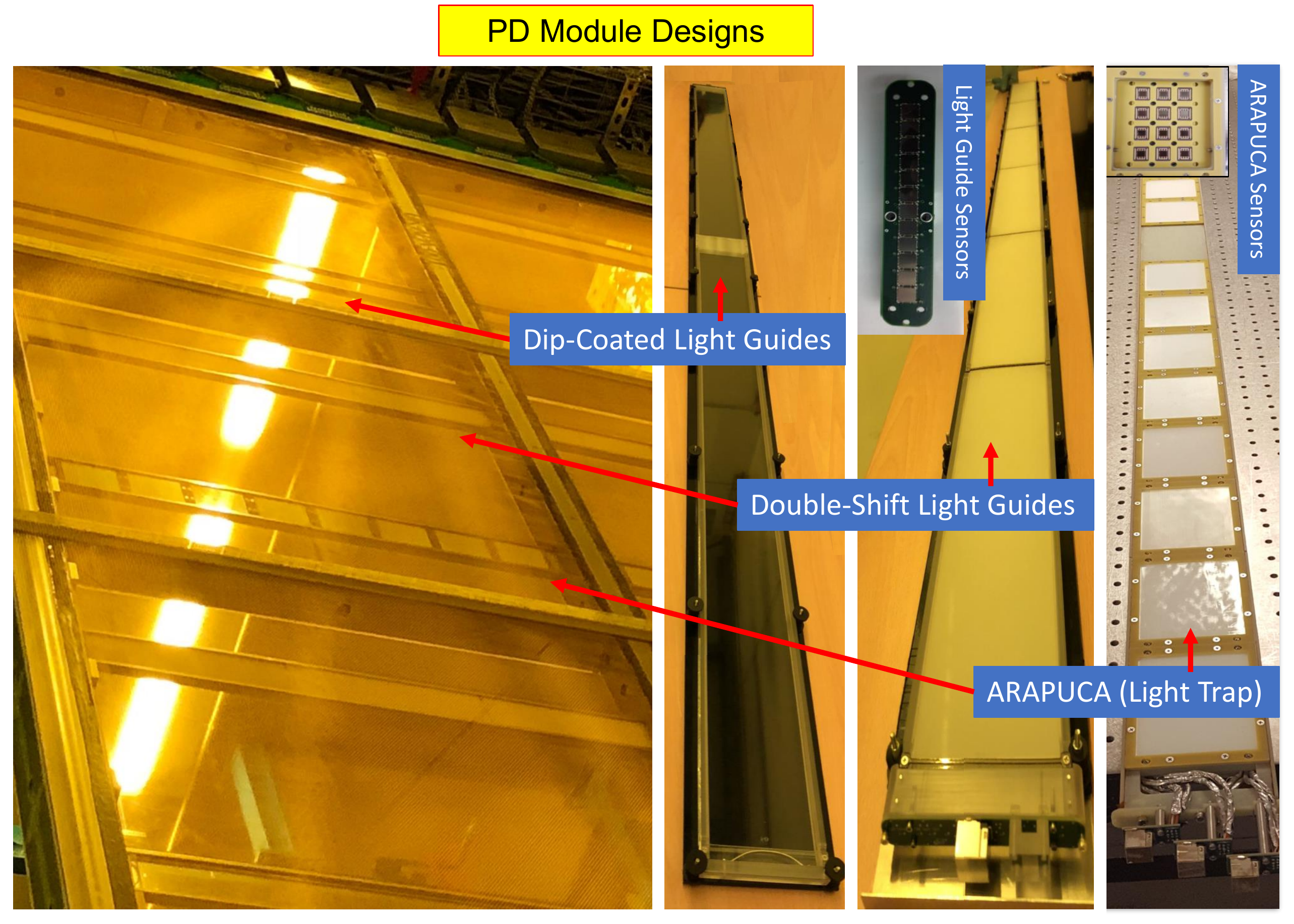} 
    \caption{Photon Detectors in a \pdsp{} APA. The three types of photon collector technologies are shown in the APA (left) and before installation (right), along with the arrangement of photosensors. Before installation and from left to right, the modules are: a dip-coated light-guide module, a double-shift light-guide module, and an ARAPUCA (light trap). The right-most collection of photosensors shows the arrangement of a single cell in the ARAPUCA and the sensor collections just to its left show the arrangement of photosensors in the dip-coated and double-shift modules.   }
    \label{fig:photondetectorsapa}
\end{figure}

The PDS is made up of modules, each combining a photon collector and a photon sensor. 
\pdsp{} implemented 60 modules using three different module designs, 
“double-shift light guides” (29),  “dip-coated light guides” (29), and “ARAPUCA” light traps (2).
The designs share a common low-profile elongated rectangular shape for insertion into slots in the APA frames, to sit between the APA's two sets of wire planes.
All modules have the same external dimensions and mounting system. 
Ten PD modules were inserted into each APA, regularly spaced in the vertical dimension of each of the six APAs.
This configuration maximises the detection efficiency of the PDS without affecting the operation of the TPC. 
The module dimensions, $2.3 \text{~cm} \times 11.8 \text{~cm} \times 209.7 \text{~cm}$, are dictated by the need to mount them in the APA frames without impairing  the mechanical stability or useful fiducial volume of the TPC.
The active area of each PD module is  1744\,cm$^2$ for both types of light guides, and  1223\,cm$^2$  for the ARAPUCA bars, for a total coverage of  about 12.5\% of the APA surface.
PD placement in an APA frame along with examples of all three types of detector technologies are shown in Figure~\ref{fig:photondetectorsapa}.


Signals for the PDS system are routed directly to 24 instances of an electronic readout system, called a SiPM Signal Processor (SSP), all of which independently synchronize and interface directly with the \pdsp data acquisition (DAQ) system.
A pair of standalone UV LED calibration systems, called Light Calibration Modules (LCMs), are used to precisely determine photosensor calibration, timing resolution, and to observe the long-term stability in the PDS.
      \label{sec:detcomp:inner:pds:coll}
\subsubsection{Photon Collector: Dip-coated  Wavelength-shifting Light Guides }
\label{sec:detcom:inner:pds:coll:wlsDC}
The dip-coated collector design uses diamond-polished acrylic, cut from cast LUCITE UTRAN manufactured by Palskolite, Inc., and supplied by EMCO Industrial Plastics as the light-guide ``bar''. 
The bar has an index of refraction of 1.49 and is dipped in a solution of tetraphenyl-butadiene (1,1,4,4-Tetraphenyl-1,3-Butadiene, TPB) and other solvents (including a surfactant) to produce a wavelength-shifting (WLS) layer on the outside surface.
When a particle passes through the active volume, VUV light produced from scintillation at 128\,nm interacts with the surface coating on the bar, where it is shifted to blue light at 425\,nm, and isotropically emitted. 
Once the blue light enters the light guide, it travels via total internal reflection and is absorbed by a collection of photosensors at one end. 
A conceptual example of the dip-coated WLS light guide is shown in Figure~\ref{fig:wls}.

\begin{figure}[!htbp]
    \centering
    \begin{minipage}[t]{0.5\textwidth}
        \includegraphics[width=1.0\textwidth]{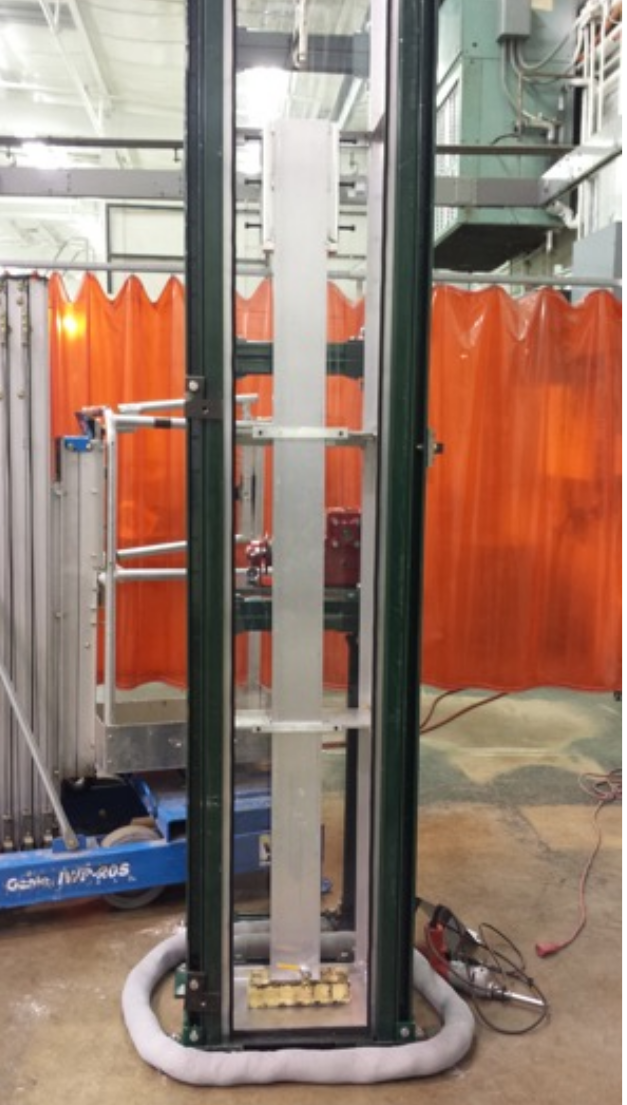} 
    \end{minipage}
    \caption{Mechanized dipping setup at Fermilab, used for the production of dip-coated light guides.}
     \label{fig:dippingsetup}
 \end{figure}

Bar construction began with the industrial manufacture of the acrylic, which was shipped pre-polished. 
The first quality control step was a visual inspection of the acrylic which ensured that bars used for production were relatively free from scratches or chips and matched pre-production bars in terms of polish quality. 
Production bars were then measured at five points of contact using calipers to verify their required dimensions and tolerances: ($209.270 \pm 0.127 \text{\,cm})\,\times\,(8.471\,\pm 0.127 \text{\,cm})\,\times\,(0.599\,\pm 0.060 \text{\,cm})$.
Once a bar was approved for use, a unique serial number was etched on it. 
The next steps were to anneal (bake and then slowly cool) the bars then dip them in the WLS solution.
Annealing was done to strengthen the bar material and as a form of quality control.
Bar annealing took place in a large oven in Lab 3 at Fermilab that was set at about $76.7^\circ$C.
Baking lasted approximately four hours. 
Annealed bars were required to bear no visual signs of crazing (mechanical stress breaks that make the bar appear cloudy).
Accepted bars were cleaned with ethanol and dipped in a coating solution, consisting of  50\,mL toluene, 12\,ml ethanol, 0.1\,g acrylic pellets, and 0.1\,g TPB for five minutes, where the acrylic pellets ensure that the coating index of refraction matches that of the bar \cite{Moss:2016yhb}.
A mechanised dipping vessel was prepared to automate the process and standardise bar quality. 
Finally, successfully prepared, annealed, and dipped bars were hung in air under a fume hood to dry for at least 30 minutes. 
The dipping and drying processes were sensitive to moisture and required a low-humidity environment (<15\% relative humidity).
The dipping vessel at Fermilab is shown in Figure~\ref{fig:dippingsetup}.

Quality control for fully dipped and dried bars consisted of observing the attenuation of 200\,nm light in the bars in both warm and LAr temperature regimes.
The warm measurements used full-size bars whereas measurements at LAr temperatures took place on shorter bars that were cut to fit within the TallBo facility at the Proton Assembly Building (PAB) at Fermilab. 
Both measurement campaigns yielded an attenuation length in excess of 2\,m.

\subsubsection{Photon Collector: Double-shift Wavelength-shifting Light Guides}
\label{sec:detcom:inner:pds:coll:wlsDS}
\begin{figure}[!htbp]
    \centering
    \begin{minipage}[t]{1.0\textwidth}
        \includegraphics[width=1.0\textwidth]{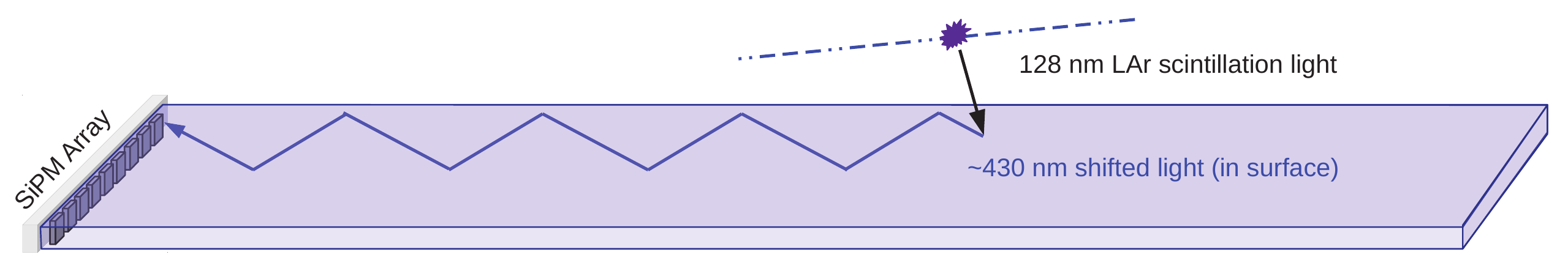} 
    \end{minipage}
    \begin{minipage}[t]{1.0\textwidth}
        \includegraphics[width=1.0\textwidth]{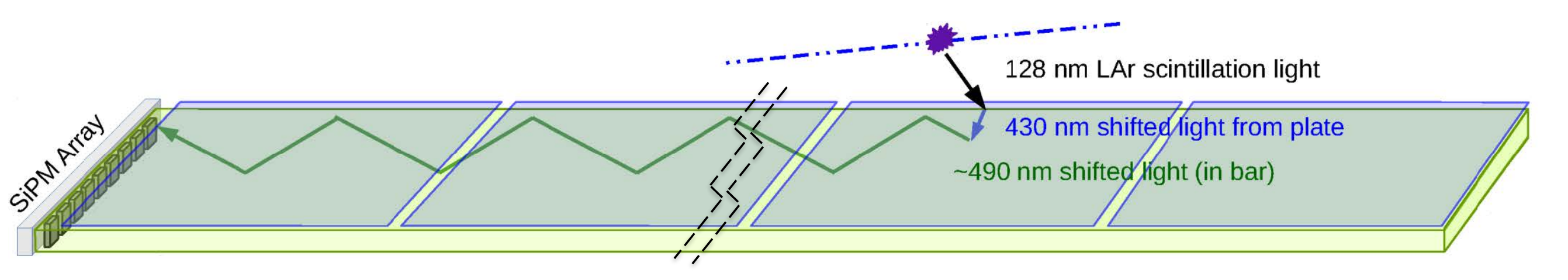}
    \end{minipage}
    \caption{A conceptual example of the dip-coated and double-shift WLS light guides. The top design shows conceptually how the dip-coated light guide collects scintillation light at 128\,nm and converts it to 430\,nm. The bottom design shows the operation of the double-shift WLS light guide. It converts from 128\,nm to 430\,nm and then again to 490\,nm. }
    \label{fig:wls}
\end{figure}

The double-shift light-guide collector~\cite{Howard:2017dqb,Howardthesis} combines the use of WLS TPB-coated radiator plates with a WLS light-guide bar. 
In this design, LAr scintillated light at 128\,nm undergoes two wavelength-shifting steps, once to 425\,nm blue light in the plate and then to 490\,nm green light in the bar. 
A conceptual example of the double-shift WLS light guide is shown in Figure~\ref{fig:wls}.
The benefits of this design are in the increased efficiency of photosensor detection for light at 490\,nm and in the production and quality control of the technology.
During and after construction, bars are separable and components can be independently studied and optimised.

The outer WLS material is a set of twelve separate radiator plates that completely cover the light guide, six on each side,  placed end-to-end.
The plates are made from 1/16 inch acrylic (purchased in one order from McMaster Carr) that were laser-cut in pairs at Fort Collins Plastics in Fort Collins, CO.  
The plates are coated by hand in a WLS solution consisting of 5\,gm of scintillation grade (>99\%) TPB to 1,000\,gm of dichloromethane,  using a Binks high-volume lower pressure sprayer system. 
Two plates, cut from a single piece of acrylic, along with three breakout tabs for quality control, are coated with the WLS at the same time (tab-plate-tab-plate-tab).

Once the plates are coated, they are annealed in a vacuum oven at 80$^\circ$C, a temperature just below the melting point of the acrylic. 
The annealing incorporates the TPB into the acrylic, ensuring that it will not separate or precipitate due to aging or cryocycling in LAr. 
Metrology tests are carried out to ensure that no changes in bar dimension occurred during annealing.
The plates are then separated and the breakout tabs are tested for brightness using a McPherson VUV monochromator with an H$_2$ lamp source, and calibrated with a NIST-calibrated photodiode.
Each plate is evaluated with respect to a standard tab, averaging the response from the two tabs on either side of it \cite{Macias:2021}. 
The evaluation is relative; plates are accepted only if they are brighter than baseline plates that meet efficiency goals derived from DUNE physics requirements.
The measurement of baseline plate efficiency was conducted on a LAr test stand at Fermilab, on two prototype paddles using a cosmic ray trigger \cite{Howard:2017dqb}.

The second WLS material and primary light guide for the double-shift collectors is made from special commercial Eljen EJ-280PSM crosslinked polystyrene.
The polysterene was cut into bars that were required to meet the following dimensions within tolerances: ($209.15 \pm 0.05 \text{~cm}) \times (8.6 \pm 0.05 \text{~cm}) \times  (0.6 \pm 0.05 \text{~cm}$).
Bars that passed were next accepted or rejected on the basis of internal attenuation length, which was measured in a dark box with a 435\,nm LED at room temperature. 
The correlation between attenuation length at room temperature and at LAr temperature was established at Indiana University on a set of shorter Eljen bars. 
The results of these combined studies showed that a room temperature attenuation length measurement greater than 6.4\,m in the dark box ensured an attenuation length in LAr greater than 2\,m. 
An attenuation length greater than 6.80\,m as measured in the dark box at Indiana University was therefore required for acceptance \cite{Macias:2021}. 


\subsubsection{Photon Collector: ARAPUCA Light Trap}
\begin{figure}[!htbp]
    \centering
        \includegraphics[width=0.75\textwidth]{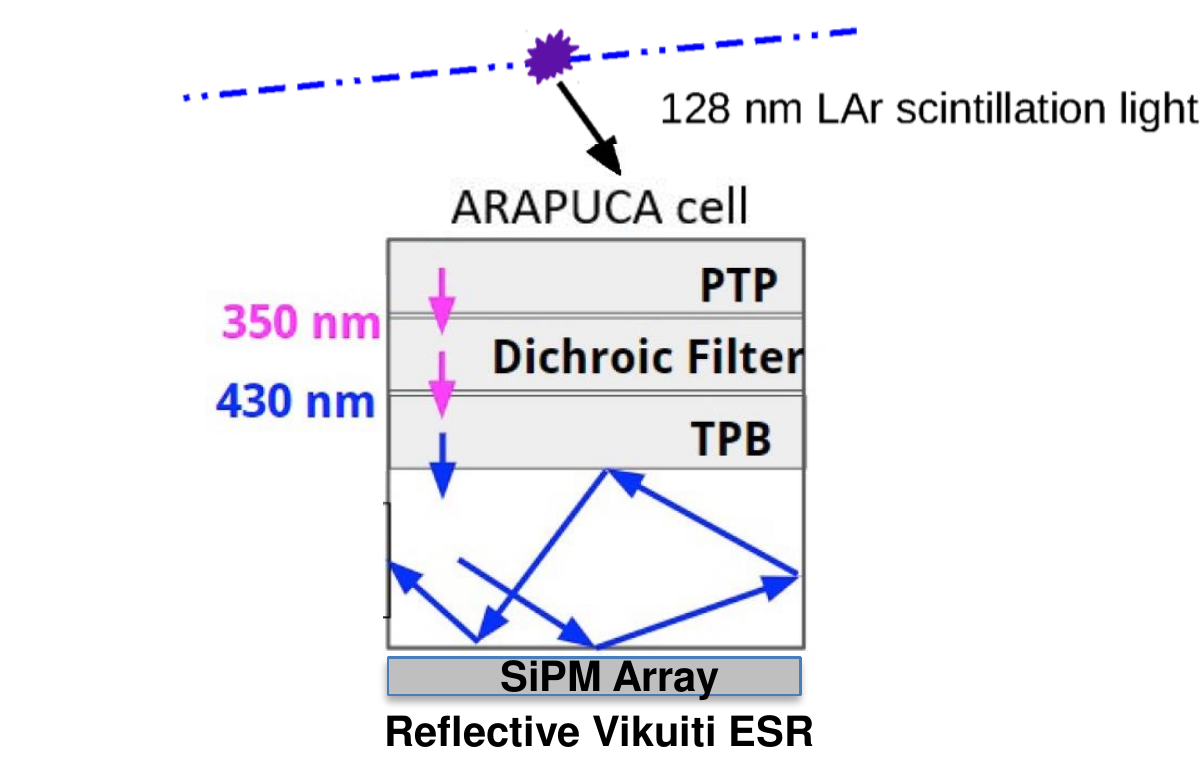} 
    \caption{A schematic diagram of the ARAPUCA light trap. Scintillation light is first wavelength shifted to 350\,nm to pass through a dichroic filter, then again to 430\,nm after the filter, at which point it can no longer return through the acceptance window. It internally reflects until absorbed by the photosensor array. }
    \label{fig:arapuca}
\end{figure}

\begin{figure}[!htbp]
    \centering
        \includegraphics[width=0.75\textwidth]{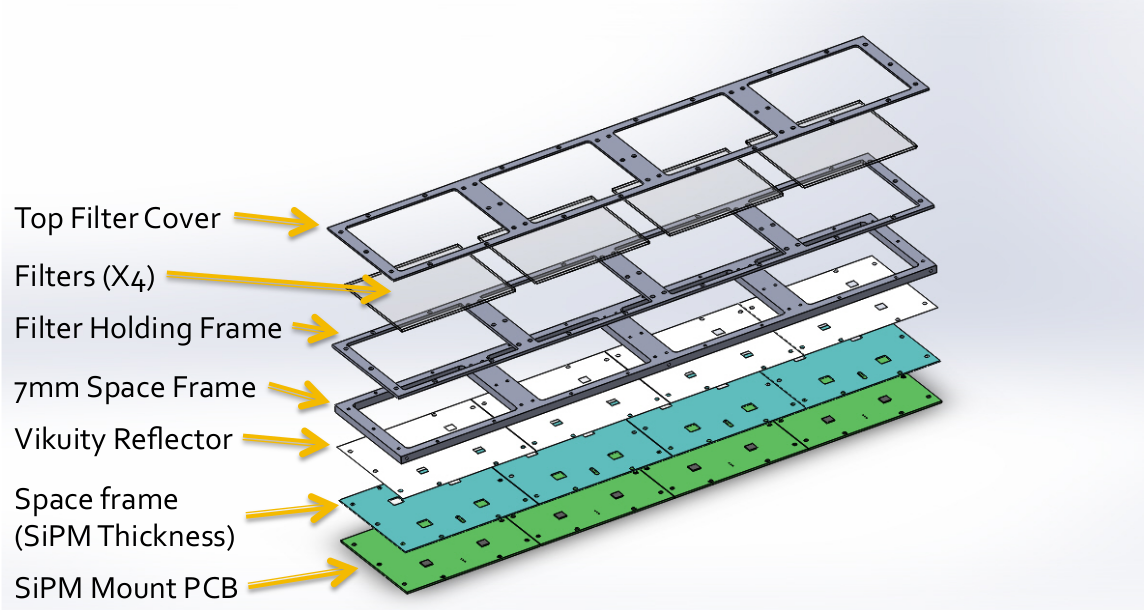} 
    \caption{Exploded view of one ARAPUCA supercell. The filter cover, filters, and filter holding frame are all embedded in the space frame. MPPCs photosensors are mounted in the SIPM Mount PCB and supported by the second space frame. }
    \label{fig:cartucho}
\end{figure}

The operating concept of the ARAPUCA is the trapping of incident photons inside a highly reflective chamber until they reach a sensor, yielding high detection efficiency despite a limited active coverage \cite{Machado_2016}.
A single ARAPUCA cell consists of a highly reflective chamber with a single-direction acceptance window.
The window consists of a dichroic optical filter sandwiched between thin layers of wavelength shifting materials.
A conceptual demonstration of the ARAPUCA concept is shown in Figure \ref{fig:arapuca} and its construction is shown in Figure \ref{fig:cartucho}.
%
 

For the collector to act as a photon trap, the external face of the dichroic filter has a WLS coating with an emission wavelength less than the cutoff wavelength of the filter. 
The transmitted photons pass through the filter where they encounter a second WLS-coated surface. 
This second coating has emission spectra that exceed the cutoff wavelength, thus trapping the photon inside the box. 
Trapped photons reflect off the inner walls and the filter surface(s) (of reflectivity typically greater than 98\%) and have a very high probability of impinging on a photosensor before being lost to absorption.
%
Dichroic filters with cutoff at 400\,nm were acquired from Omega Optical.
The internal surfaces of the filter are lined with VIKUITI\texttrademark 3M specular reflector coated with TPB, and P-Terphenyl (1,4-Diphenylbenzene, PTP) is applied to the external side of the filter.   
PTP and TPB coatings were applied using a resistive evaporation technique under a clean vacuum of at least $2.0 \times 10^{-6}$ torr. 
A quartz microbalance was used to monitor rate and thickness during coating and was consistently placed in the same position relative to the substrates and target material. 
PTP and TPB surface densities are about 200 and 300$\mu g/\mathrm{cm}^2$, respectively.

Quality control for each batch of evaporated coating consisted of the random testing of one of four pieces in liquid argon. 
The test pieces were cooled in cold argon vapor for two hours before being fully immersed in LAr for up to 20 hours. 
When a piece did not pass this test, the entire batch was thoroughly cleaned of the coating, recoated, and a piece retested. 
There were some early failures that required recoating, but after establishing  a pre-shipping cleaning and packaging protocol from the vendor and a cleaning protocol before coating, no batches were rejected.
Before installation in \pdsp{}, several ARAPUCA prototypes were tested in LAr with $\alpha$ sources and cosmic ray muons\,\cite{Segreto:2018jdx}.


      \subsubsection{Photosensors}
\label{sec:detcomp:inner:pds:sensors}
The \pdsp{} PDS uses three different silicon photon sensor (SiPM) models commercially manufactured by two different companies, SensL Technologies, Ltd. (now, ON Technologies) and Hamamatsu Photonics.
All sensors have the same active area, $6\times6$\,mm$^2$. 
Samples of the different sensors types were tested and characterised at LN$_2$ temperature before use. 

SensL Technologies, Ltd provided (MicroFC-60035-SMT) sensors with pixel size 35\,$\mu$m. 
Two types of series models were used, A-Series and C-Series, with the switch due to a change in availabilty from the manufacturer.
SensL devices equip most of the WLS bar-type modules, 
i.e., 21 dip-coated modules and 22 double-shift modules. 
Each module array contains 12 SiPMs passively ganged in groups of three and read out as four independent channels.
A total of 172 channels are equipped with 516 SensL SiPMs.

Hamamatsu Photonics provided two versions of its Multi-Pixel Photon Counter (MPPC) of model number S13360-6050, a CQ ``Quartz Window'' type and a VE ``Through Silicon Via'' type for use in the WLS bar detectors and the ARAPUCAs. 
Both have a pixel  of 50\,$\mu$m. 
The VE MPPC is coated with epoxy resin and uses through-hole electrodes called TSV (through-silicon via). 
In VE MPPCs, the space around the active area is reduced with respect to standard wire-bonding techniques, allowing for better packaging. 
The Quartz Window MPPC is an uncoated MPPC in a ceramic package with a quartz window, aimed to better resist thermal stresses. 
All ARAPUCA modules are equipped with Hamamatsu CQ MPPCs, while eight dip-coated and seven double-shift modules are equipped with VE MPPCs.
All photosensors were tested at warm temperature before mounting them on the boards, to verify compliance with producer specifications.  
The equipped boards were thermal-cycled and tested again before being coupled to the lightguides.




      \subsubsection{Photon Detector Readout Electronics}
\label{sec:detcomp:inner:pds:readout}
For each photon detector, signals from one of the three different ganging schemes are summed together and connected, in the cold volume, to a 20\,m long multi-conductor Cat6 cable that feeds through the detector flange. 
No front-end electronics are necessary in the cold volume for the operation of the \pdsp{} PDS.
The readout system transmits unamplified signals from the photosensors in the LAr volume to the outside of the cryostat and performs processing and digitisation using a SiPM Signal Processor (SSP), a readout developed and manufactured by Argonne National Laboratory.
Signal cables from each of the six detector flanges connect directly to four of 24 SSP modules. 
An SSP module consists of 12 individual readout channels packaged in a self-contained 1U module. 
Each channel contains a fully-differential voltage amplifier and a 14\,bit, 150\,MHz analogue-to-digital converter (ADC) that digitises the waveforms received from the SiPM arrays. 
The front-end amplifier is configured as fully-differential with a high common-mode rejection, and receives the SiPM signals into a termination resistor that matches the characteristic impedance of the signal cable. 
The SSP module can operate three separate WLS light guide modules or one ARAPUCA module.

\begin{figure}[!htbp]
\centering
\includegraphics[height=7.5cm,width=1.0\textwidth]{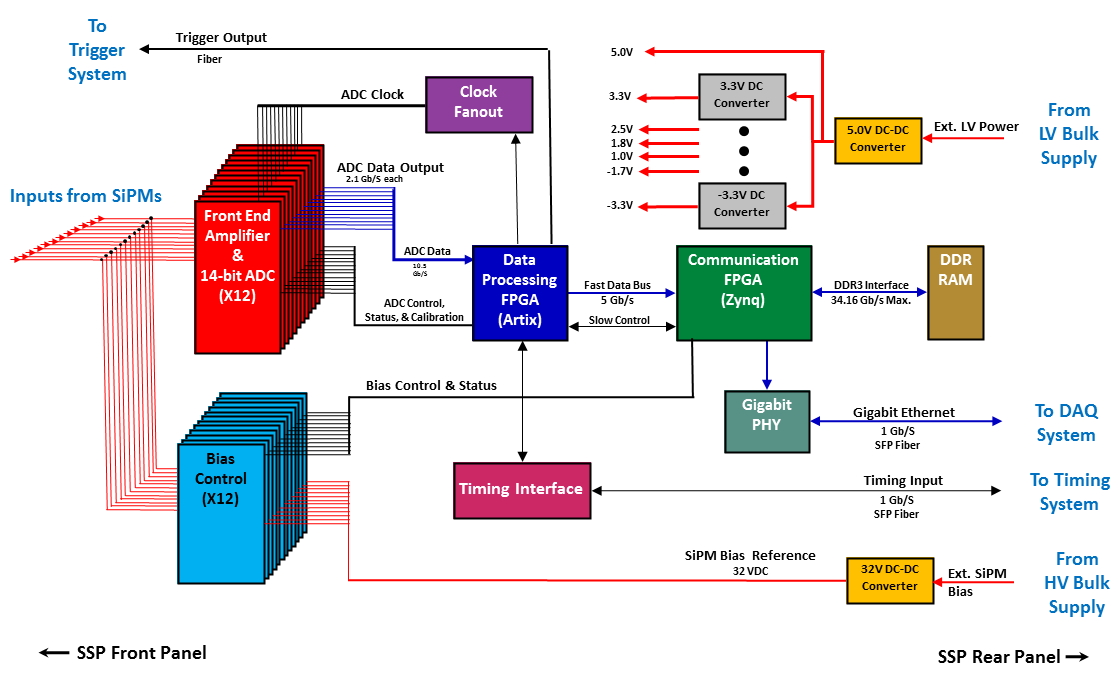}    
\caption{Operational schematic of the \pdsp{} SSP. This diagram shows how the SSP manages the timing, readout, and bias voltage of 12 channels of ganged photosensors. The SSP also interfaces with the trigger, DAQ, and timing systems.}
\label{fig:ssp}
\end{figure}

The digitised data is stored in pipelines in the SSP, for up to 13\,$\mu$s for a single output, per channel. 
The processing is pipelined, and performed by a Xilinx Artix-7 Field-Programmable Gate Array (FPGA). The FPGA implements an independent Data Processor (DP) for each channel. 
The processing incorporates a leading edge discriminator for detecting events and a constant fraction discriminator (CFD) for sub-clock timing resolution. 
An operational schematic of the SSP is shown in Figure~\ref{fig:ssp}.
Each channel can be individually triggered by any of the following: a periodic timestamp trigger, a SSP-internal trigger based on a leading-edge discriminator local to the individual channel, or a SSP-external trigger from the timing system. 
If a SSP trigger is present, the channel will produce a data packet consisting of a header and a waveform of predefined length, which comprises a series of ADC values. 
The header contains bookkeeping information (e.g., module and channel numbers, timestamps, trigger type) and some calculated integral values for the waveform. Because the SSP trigger system is distinct from the global trigger, the unit of data produced when an SSP channel triggers is called  a ``packet,'' and the term ``trigger''  refers to the global triggers only. 
The SSP boardreader generates a fragment when a trigger produced by the timing system is observed, and this fragment will contain all packets received from the SSP with timestamps in a window $\pm$\,2.5\,ms from the timestamp of the trigger. 

In general, an SSP fragment will contain a fixed 12 packets with identical timestamps corresponding to the trigger time, one for each channel, and also an arbitrary number of additional packets generated when a channel's discriminator fires. 
It should be expected that a different number of packets will be observed for each channel within a given fragment. 
In the case that only the SSP-external trigger is enabled, exactly 12 packets should be present in a single fragment.
In such a case the packets received by the SSP that do not fall within the window around a timing system trigger will be dropped and never included in a fragment. 
All 24 SSPs are individually synchronised to the dedicated \pdsp{} timing system time-stamp. 
\label{sec:detcomp:inner:pds:installtestcern}
\subsubsection{PDS Quality Control and Installation at CERN}
\begin{figure}[tbh]
    \centering
    
    \includegraphics[height=9.5cm,width=.65\textwidth]{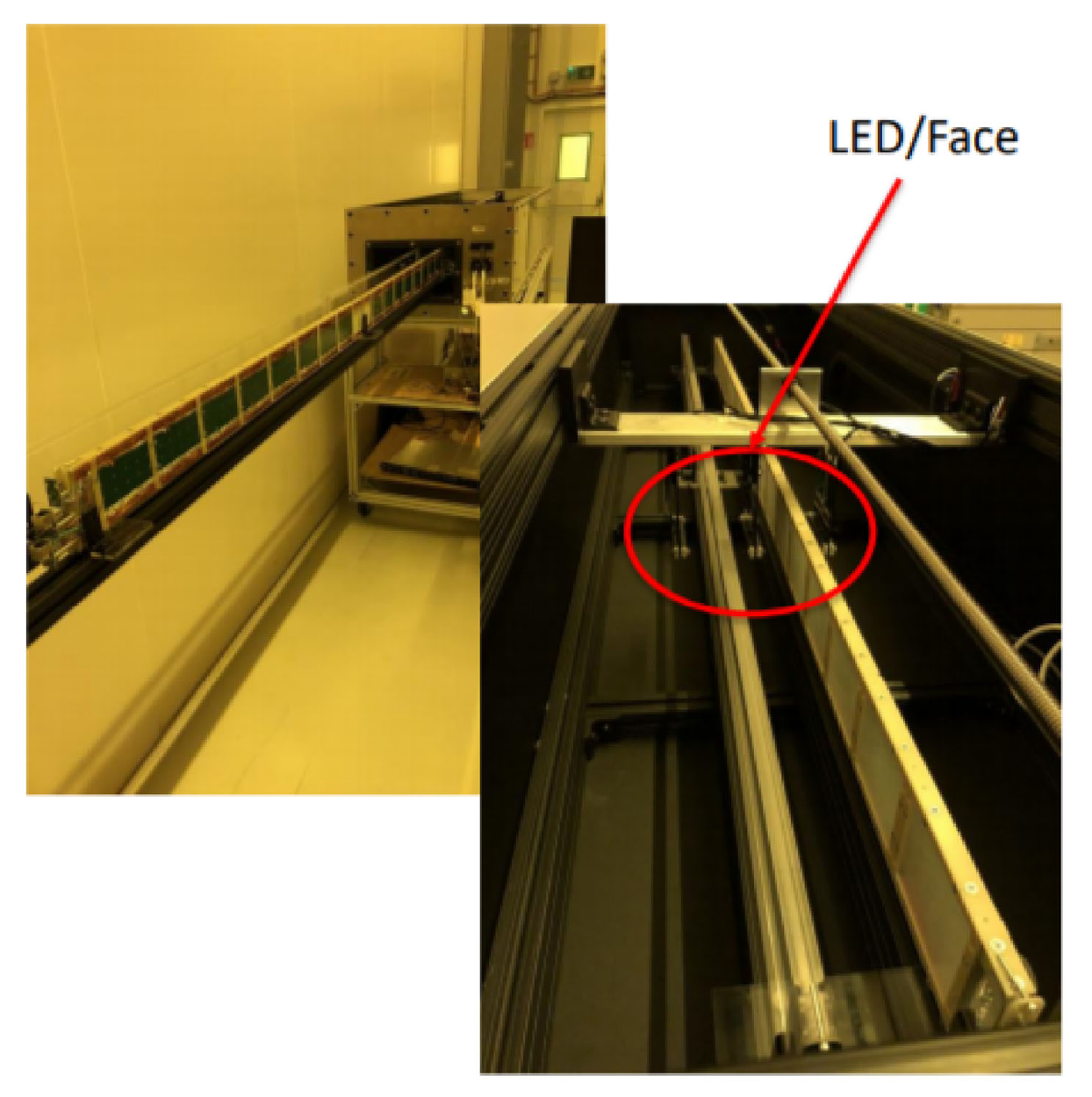}
    
    \caption{PDS Darkbox Scanner and Sample. A photon detector was light sealed in the box and exposed to incident light along its entire length provided by an LED. }
    \label{fig:Darkbox_Scanner}
\end{figure}

All PD modules for \pdsp{} arrived at CERN ready for APA installation. 
PD modules were shipped from Colorado State University (CSU), via a special PD-Crate, which contained 12 PD modules, individually packaged in an anti-static bag and placed between anti-static foam.
The final quality control steps before installation were a visual inspection of each PD module and an assessment of each module using a PD darkbox scanner, shown in Figure~\ref{fig:Darkbox_Scanner}.

In the standard procedure for the darkbox scanning measurements, both the module photosensors (SiPMs/MPPCs) and the darkbox-LED were allowed to warm up for 30 minutes, using nominal voltage. 
This assured stability within each scan and reliability for module-to-module comparisons. 
The PDS group used LBNEWare, custom made software which interfaces with the SSPs directly, as the DAQ for these tests. 
For each module scan, the LED was set to pulse and read out 5000 times at multiple locations along each side of the PD module. 
This allowed for sufficient signal observation and determination of the mean response of the module along each position.
The mean response was determined by fitting a Gaussian to the 5000 integrated waveforms. 
Prior to arrival at CERN an attenuation length had been calculated for each PD module using the relative brightness (mean signal) along each position. 
At CERN, a cross-check in pseudo-attenuation length was made and the module with the longest pseudo-attenuation length was selected for installation into the APA.  
Modules were placed facing the TPC for the best photon yield response. 
Spare modules were also sent to CERN and were sorted for use in the detector through the same module-to-module comparisons using the same pseudo-attenuation measurement.
Figure~\ref{fig:Darkbox_Scanner} shows the result of a sample scan \cite{Macias:2021}.

Each APA frame held ten PDS modules, inserted on rails, equally spaced along the full length of the APA frame. 
The spacing between modules along the $y$ direction was approximately 60\,cm.
PDS modules were inserted into the frames of completed APAs, between the sets of wire layers.
Two ARAPUCA devices were mounted  in \pdsp{}, one in APA~3 directly in front of the beam plug for the observation of photons from beam interactions, and the other in the middle of APA~6 to observe photons from cosmic particles.
The two light-guide designs filled the remaining modules in alternating positions in the APAs. 
Once the PD modules were installed into an APA, the entire APA and its components were tested in the cold box. 
The full PDS chain, including photosensors, cold cables, warm cables, SSP, DAQ, and the connection to the slow control system was then tested. 
The full test included tests for functionality, stability, trigger rates, threshold calibrations, and PD module response and comparison, as well as  thermal-cycling at $\sim150$K.
With respect to the slow control and DAQ, nominal configuration settings were implemented in GUI format, with additional configurations that could be uploaded at run time. 

\begin{figure}[tbh]
    \centering
    
    \includegraphics[height=9.5cm,width=.85\textwidth]{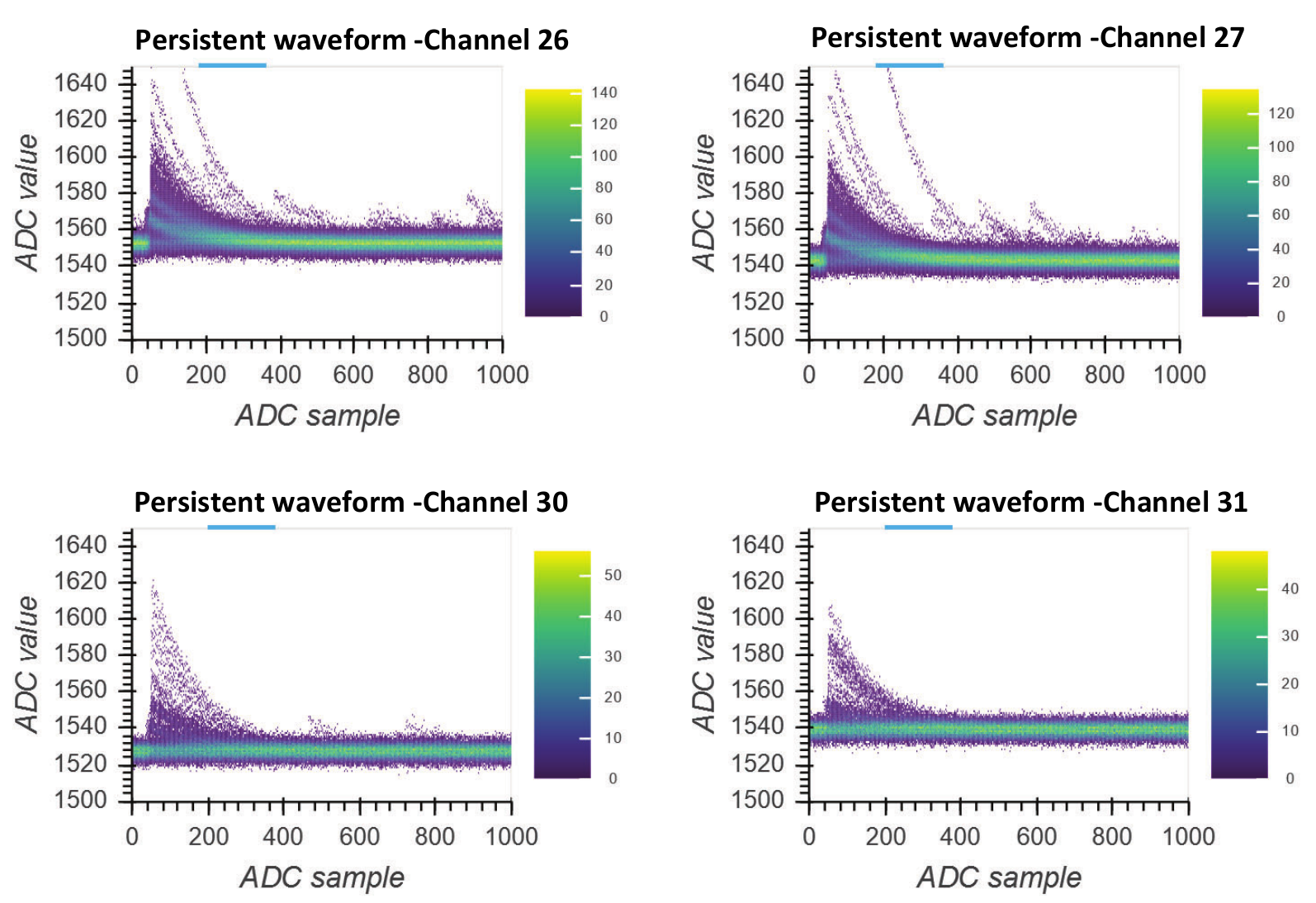}
    \caption{Photon detector cold box waveform example. Each plot shows the raw response of the photon detector channel under quality control testing in the CERN cold box before installation into ProtoDUNE-SP.}
    \label{fig:PDcoldbox_wvfm_ex}
\end{figure}

Data rates were also tested using the DUNE Light Calibration Modules (LCMs) and the Detector Control System data limits were established and optimised for data taking. 
The cold box helped determine if live data were sufficient by using the data to prepare for online monitoring diagnostics, such as persistent traces, leading-edge amplitude histograms, integral waveforms, average waveforms, and internal/external triggering for every PDS channel.
The functionality of each PDS channel was verified by inspecting waveforms using persistent traces in the online monitoring system, as shown in Figure~\ref{fig:PDcoldbox_wvfm_ex}. 
The cold box tests and online monitoring allowed the observation of overall clean waveforms in the data. 
They also provided an opportunity to observe faults with the grounding or disconnects in the SSPs, channels, and cables.
The cold box tests also allowed preliminary module-to-module comparisons. 

\begin{figure}[!htbp]
    \centering
   \begin{minipage}[t]{1.0\textwidth}
        \includegraphics[width=0.9\textwidth]{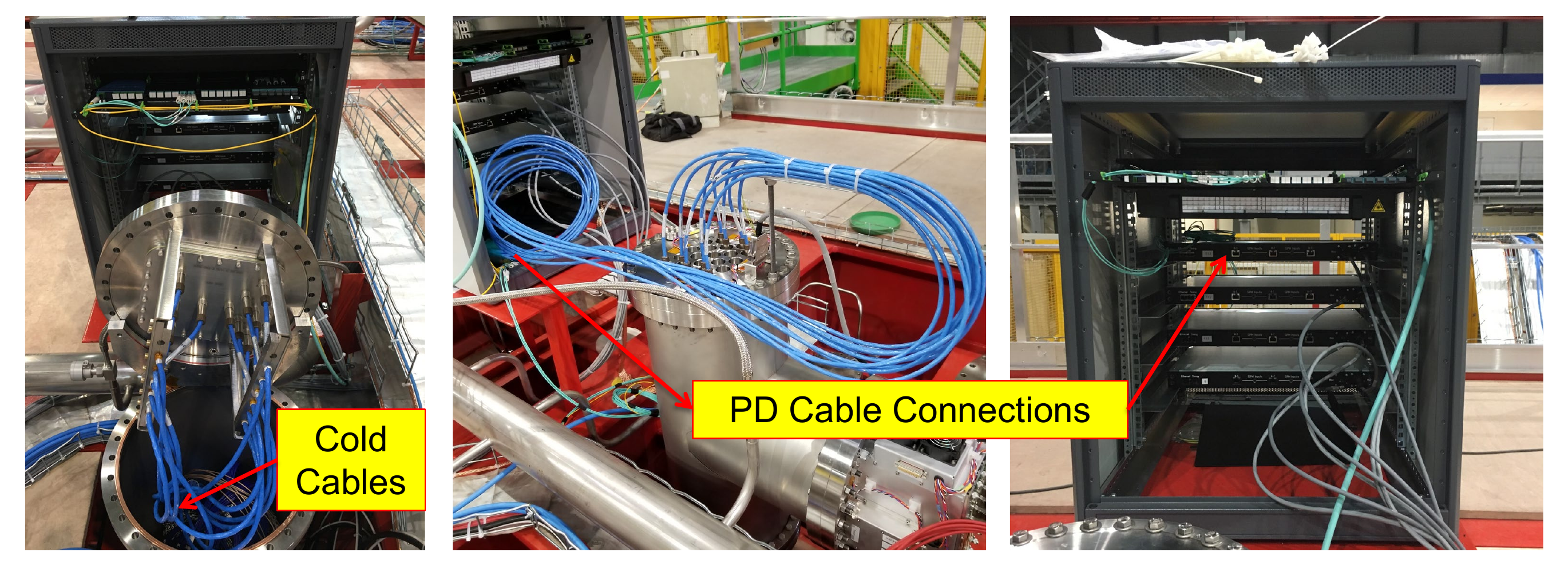} 
   \end{minipage} 
   \caption{PDS Cable Routing. From left to right, the three pictures show cold cables connecting to the flange from inside the cryostat, warm cables connecting from the warm side of the flange, and warm cables connecting from the flange warm side to the SSPs.   }
    \label{fig:pds-flange}
\end{figure}

\begin{figure}[!htbp] 
    \centering
    \includegraphics[width=0.9\textwidth]{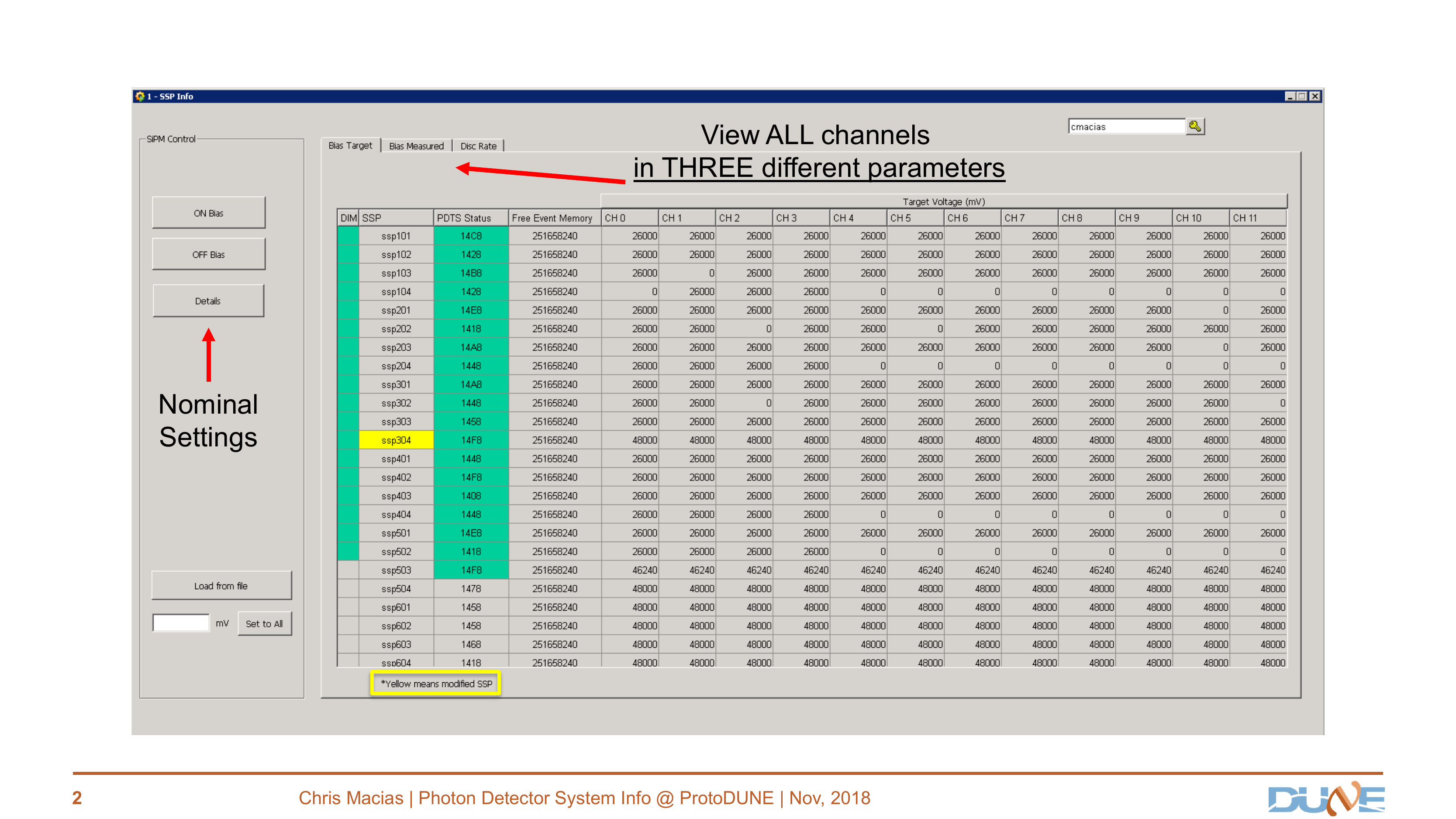} 
    \caption{SSP Setup and Readout GUI in the DCS. All channels for all SSPs can be biased and readout from the DCS software. }
    \label{fig:PDS-DCS}
 \end{figure}
The cryogenic PD cables were installed at the same time as the PD modules. 
Each APA has ten Cat6 PD cables, one per module, with alternating installation sides. 
During installation of the APAs in the cryostat, cables were routed through feedthroughs in the cryostat roof.
From there, the cables were connected to the SSP modules (see Figure~\ref{fig:pds-flange}).
During the cold box tests and after installation, the PDS was fully interfaced with both the Detector Control System (DCS) and the monitoring system.
Photosensors can  be biased and unbiased from the DCS, their status can be displayed, and the bias voltage can be modified manually or  according to different preset conditions, as shown in Figure~\ref{fig:PDS-DCS}.


\subsubsection{Photon Detector UV-Light Calibration System}
\label{sec:detcomp:inner:pds:monitor}
\begin{figure}
    \centering
    \includegraphics[width=1.0\textwidth]{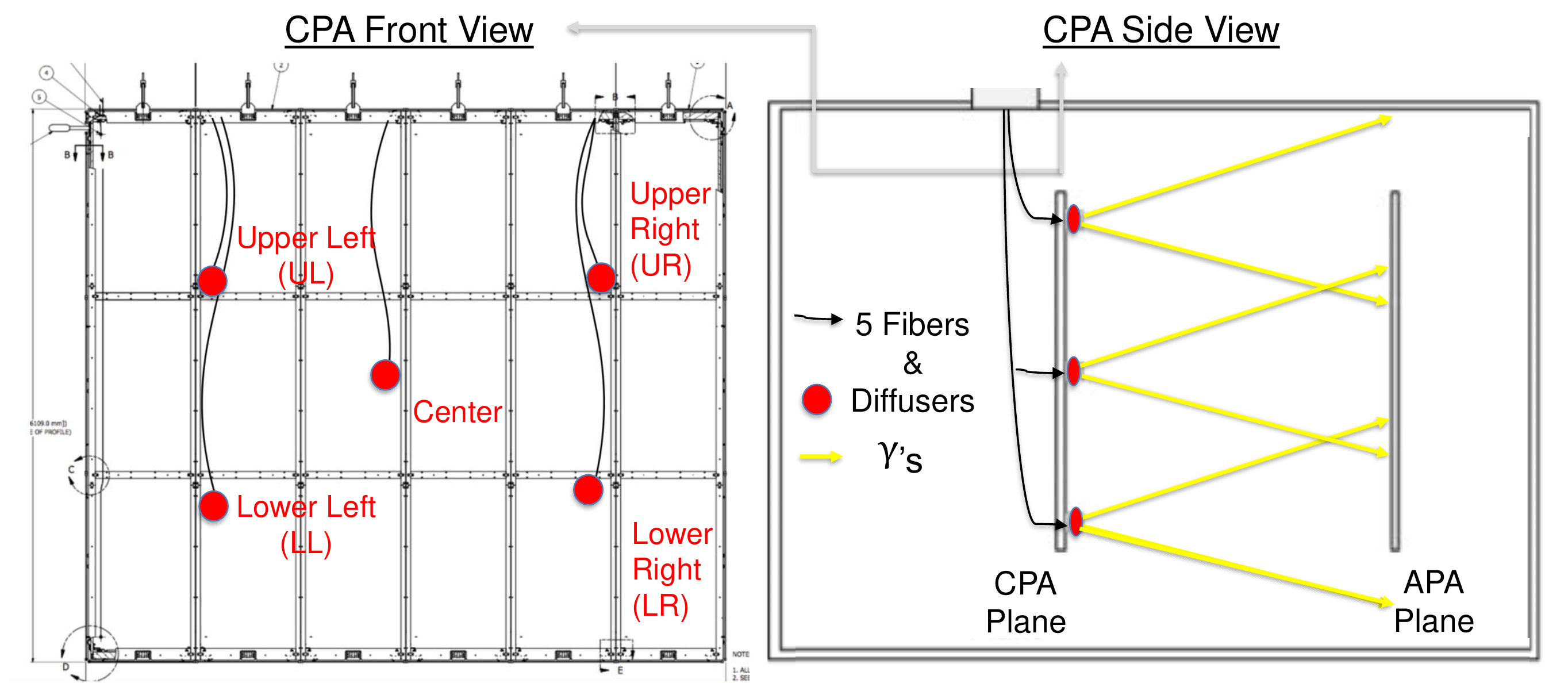}
    \caption{Schematic view of the Light Calibration Module system}
    \label{fig:DCMs}
\end{figure}

A UV-light-based calibration and monitoring system, designed and fabricated by Argonne National Laboratory, is used to calibrate SiPM gain and cross-talk, and to monitor linearity, time resolution, and long-term stability of the system. 
The system hardware consists of both warm and cold components.
The system has no active components within the cryostat and in no way alters the operation of the PDS or the HV system. 
The active system component consists of an external 1U rack-mount Light Calibration Module (LCM). 
The LCM generates 275\,nm UV-LED light pulses that propagate through a quartz fiber-optic cable to diffusers located at the CPA. 
The calibration module consists of an FPGA-based control logic unit coupled to an internal LED Pulser Module (LPM) and an additional bulk power supply. 
Light diffusers located on the CPA surface uniformly illuminate the APA surface and hence the PDS light collection modules. 
Five light diffusers on each face of the CPA plane are used: one in the center and four in each of the four  CPA corners, as shown in Figure~\ref{fig:DCMs}. 

The calibration system produces UV light flashes with a predefined pulse amplitude, pulse width, repetition rate, and pulse duration.  
It also provided an external trigger for the light collection system.
Pulse multiplicity control offers the option to produce two pulses at a fixed time difference to study timing properties of the photon system as well as trigger delays. 
The UV light system was used as a complement to  cosmic-ray muons for calibration.

\subsubsection{Full PDS Performance}
\label{sec:detcomp:inner:pds:perform}
The performance of the full PDS 
as measured during \pdsp{} operations is reported in detail in the \pdsp{} performance paper~\cite{Abi:2020mwi}. 
A summary is included here for completeness.
All channels exhibit a linearity of gain response to varying bias voltage ($\mathrm{V_B}$). 
The actual breakdown voltage ($\mathrm{V_{bd}}$) of the multi-sensor channel at LAr temperature, however, shows a relatively large spread, particularly for the 12-H-MPPC channels of the ARAPUCA modules.
For these 12-H-MPPC channels, the signal-to-noise (SNR) values are around 6, while for the 3-S-SiPM channels of the double-shift and dip-coated bar modules the SNR is in the range 10 to 12.
Within the uncertainties of the calibration  measurements taken during operations, neither the gain nor the other parameters were found to drift significantly over time for any of the sensors used in the \pdsp{} PDS. 
The photon detection efficiency was evaluated through eight independent measurements using muon and electron data at four different beam momenta, supplemented by simulation. 
Photon detection efficiencies average 2\% for the single area ARAPUCA cell, 0.21\% for the double-shift module, and 0.08\% for the dip-coated modules (see Table 4 in~\cite{Abi:2020mwi}).
An extrapolation of the performance to a PDS system consisting entirely of ARAPUCA modules indicates that it can perform calorimetric energy reconstruction with an expected light yield of 1.9 photons/MeV at 3.3 m from the anode. 
This performance exceeds the specifications of the DUNE far detector by almost a factor of four.
         \cleardoublepage

\subsection{Cosmic Ray Tagger (CRT)} 
\label{sec:CRT}
As \pdsp{} sits on the surface, it experiences 20\,kHz of cosmic-ray muons entering the detector. To provide external reconstruction to a sample of these cosmic-ray muons and beam halo-muons, a system of scintillation counters external to the cryostat, called the CRT, 
covers almost the entire upstream and downstream faces of the TPC. The fact that the CRT only covers the front and back faces of the TPC means that many cosmic-ray muons do not cross any part of the CRT.  
Both tagged and untagged muons provide important calibration data and performance indicators.

The scintillation counters used in the CRT were originally built and deployed for the outer veto of the Double Chooz experiment~\cite{DoubleChooz}. The CRT is composed of 32 modules, each of active area 1.6\,m $\times$ 3.2\,m, arranged into mechanically independent super-modules of four modules each. 

\begin{figure}
\centering
\includegraphics[width=.9\textwidth]{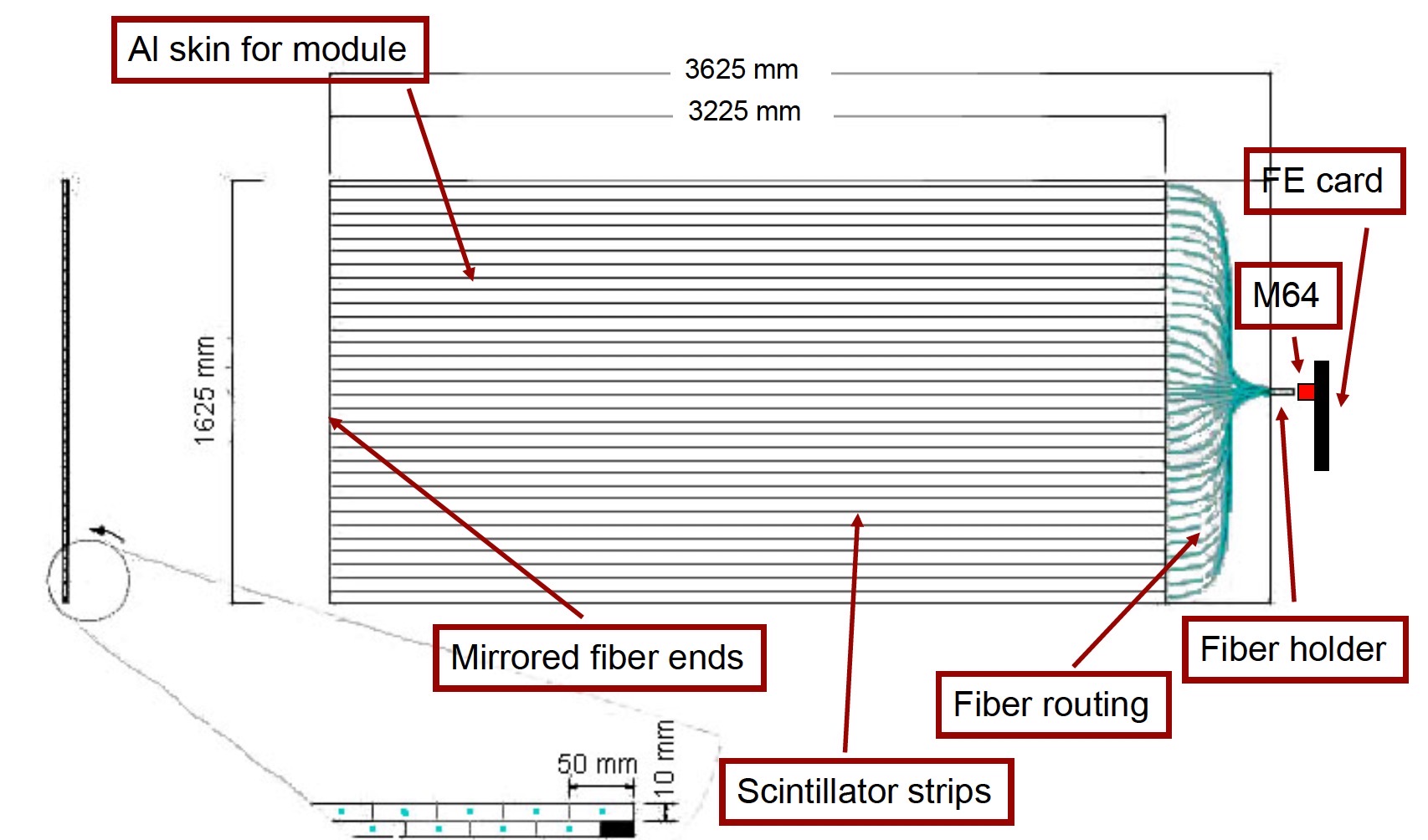} 
\caption{Labeled illustration of a CRT module.}
\label{fig:CRT_module_details}
\end{figure}

Each module is instrumented with 64 scintillator strips 5\,cm wide and 365\,cm long, arranged in two parallel planes of 32 each. A schematic of a CRT module is shown in Figure~\ref{fig:CRT_module_details}.

\begin{figure}
\centering
\includegraphics[width=.7\textwidth]{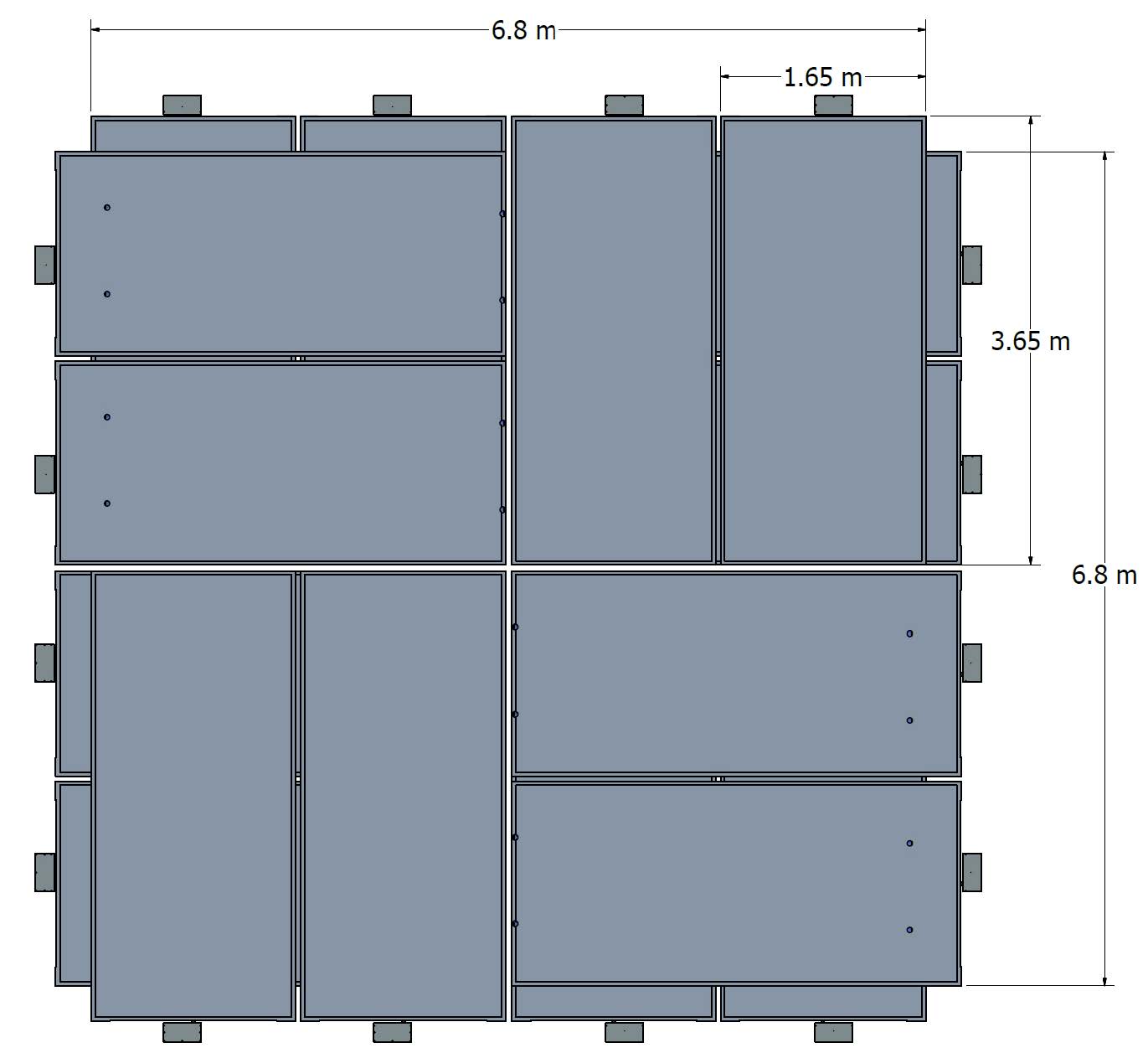} 
\caption{Drawing of the downstream CRT module assembly, showing the four super-modules positioned in a square. Each super-module consists of two modules edge-to-edge with scintillator strips running in the $x$ direction, and either behind or in front of them, two modules with strips running in the $y$ direction. The active area of the entire assembly is (6.8\,m)$^2$, as indicated. For the upstream portion, the assembly is split in two, left and right.}
\label{fig:CRT_assembly}
\end{figure}

\begin{figure}
\centering
\includegraphics[width=.8\textwidth]{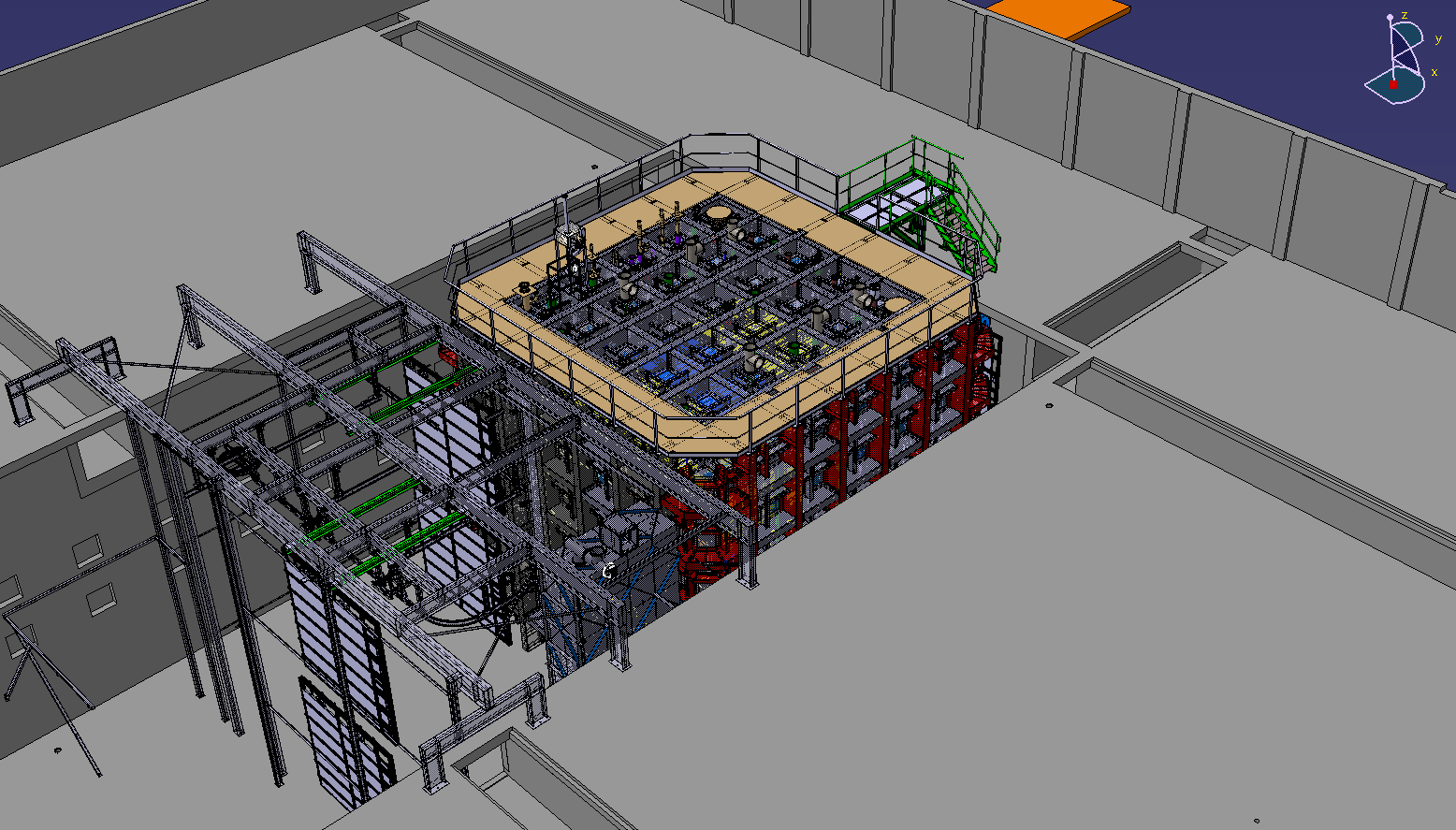} 
\caption{3D CAD view showing the upstream CRT planes split left and right to accommodate the beam pipe.} 
\label{fig:upstr_CRT_planes}
\end{figure}

Each strip has a wavelength-shifting scintillating fiber that transports the light to an individual pixel of a 64 multi-anode photo-multiplier tube (Hamamatsu M64).
 The two layers of strips are offset by half a strip width to maximise coverage. The extra centimeters of strip, compared to the length of a module (320\,cm), are outside the active area. 
 Each module measures a one-dimensional spatial position from its strip number and measures its given position along $z$ from its super-module's placement. To reconstruct a CRT hit in three-dimensions, two modules are placed edge-to-edge with their strips all parallel, and two more placed behind them in $z$ with their strips rotated by 90 degrees. This composes one super-module of dimensions 3.65\,m by 3.65\,m, as illustrated in Figure~\ref{fig:CRT_assembly}. The coordinate system is shown in Figure~\ref{fig:upstr_CRT_planes}.

For the \pdsp{} downstream face, four super-modules are arranged edge-to-edge in a square roughly the size of a \pdsp{} face, also shown in Figure~\ref{fig:CRT_assembly}. This assembly of CRT modules is centred with respect to the centre of the TPC in $x$ and placed 10.5\,m from the upstream face of the TPC in the cryostat. 

\begin{figure}
\centering
\includegraphics[width=.6\textwidth]{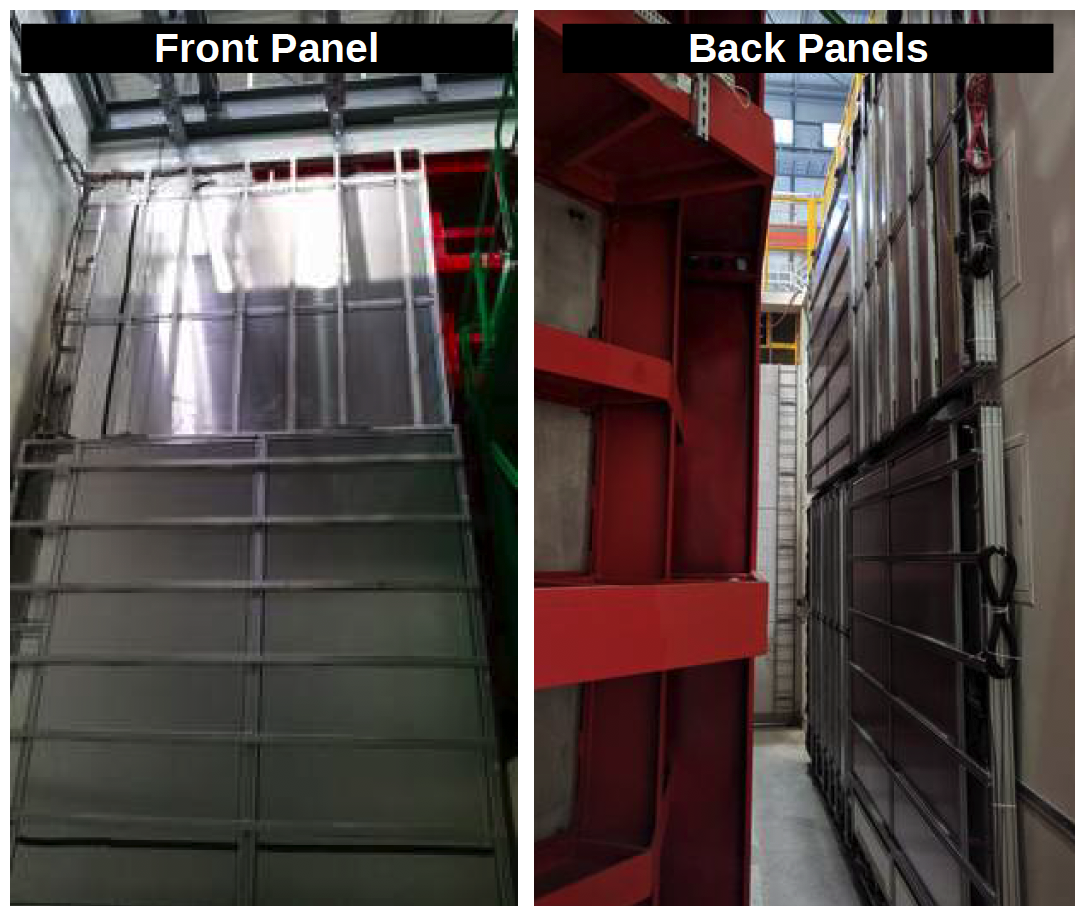} 
\caption{Photos of the beam-left upstream CRT assembly of two super-modules (left) and the downstream assembly of four super-modules. } 
\label{fig:CRT_front_back}
\end{figure}

The positioning of the upstream super-modules of the CRT system is complicated by the presence of the beampipe requiring that this portion of the CRT system be split, with the right and left halves offset from each other in $z$. One set of the two vertically stacked super-modules is placed 2.5\,m from the front face of TPC (left of beam) and the other (right of beam) is placed 9.5\,m upstream of it, as shown in Figure~\ref{fig:upstr_CRT_planes} and in the photographs in Figure~\ref{fig:CRT_front_back}. In each stack, the upper and lower super-modules are actually offset slightly from each other in $z$ (the upstream to downstream direction) so that they can both hang from the same bar.

Scintillation light from the strips is measured by the M64 photomultipliers. Customised ADC boards then digitise the signals of all strips in the module. A CRT module will trigger if any ADC signal is above threshold and provide the Central Trigger Board or CTB (see Section~\ref{sec:readout:daq}) the digitised readings of all 64 strips in a single CRT module with a timestamp that has a 20\,ns resolution. 

\begin{figure}
    \centering
    \includegraphics[scale=0.35]{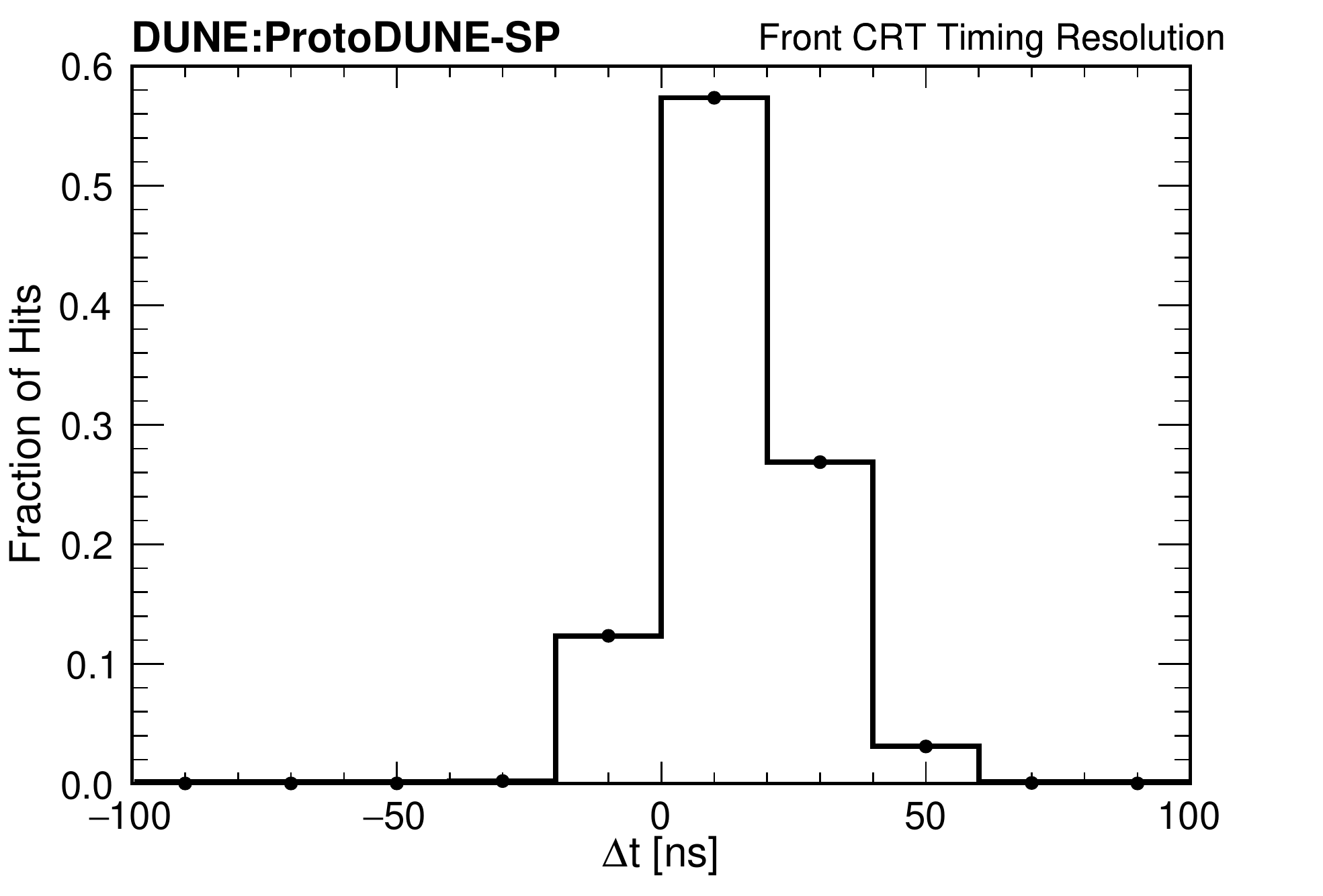}
        \includegraphics[scale=0.35]{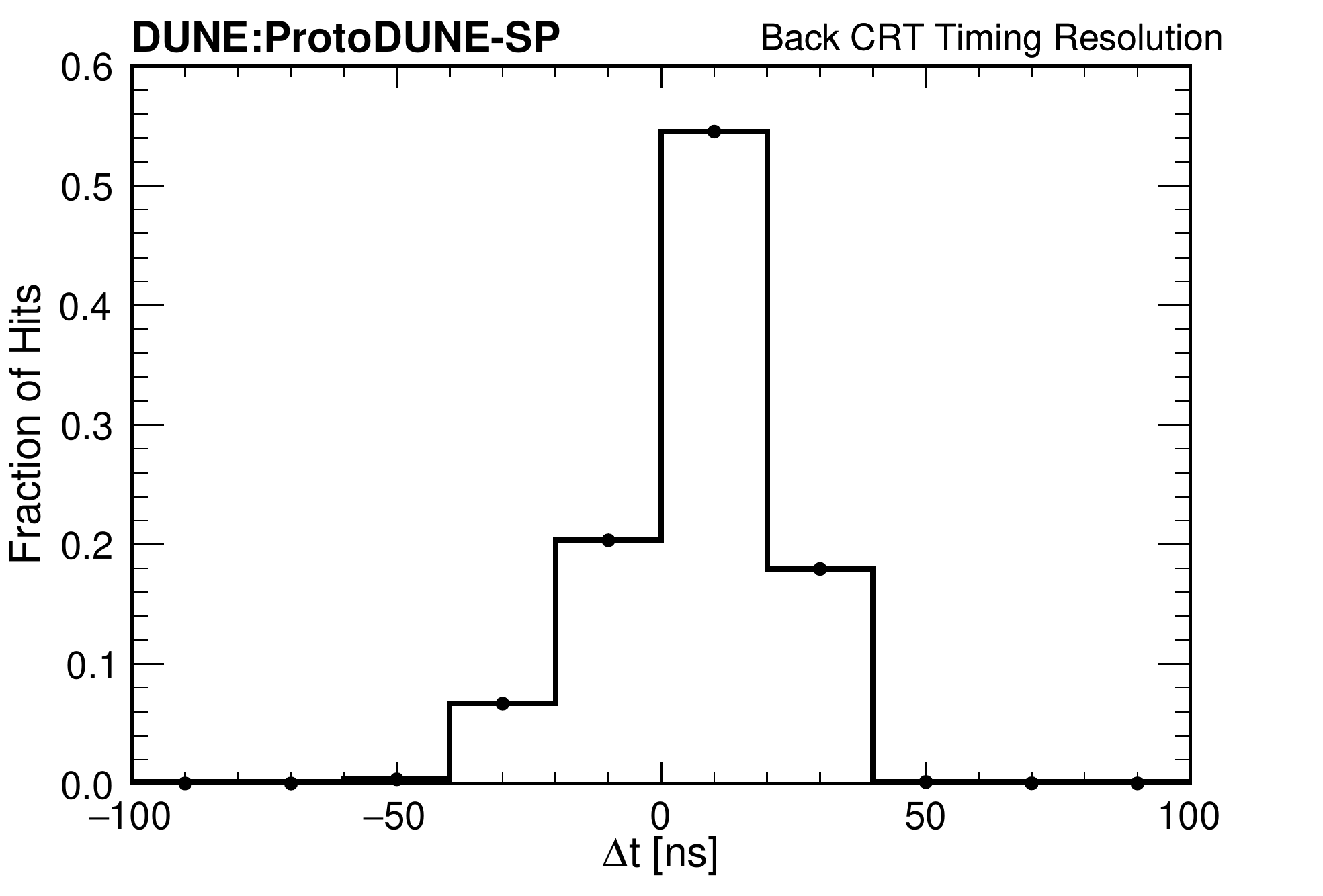}
    \caption{Timing resolution for CRT reconstructed hits on the front CRT modules (left) and on the back CRT modules (right) from data taken in November 2018.}
    \label{fig:crtHitResolution}
\end{figure}

Offline reconstruction then assembles the CRT signals stored for an event into three-dimensional hits. First, strip signals below the dark count threshold are rejected and then the pulse magnitude, CRT module number, the strip's identification number, and a timestamp for each strip is collected from the raw data. These primitively act as ``one-dimenstional'' hits and are then sorted between upstream and downstream CRT modules based on their CRT module channel number. Reconstructed CRT hits are then constructed by combining strip signals of overlapping CRT modules that occur within a coincidence window. The coincidence window is set at 80\,ns or 4 time ticks of the CTB and was decided upon using information on the timing offsets between CRT modules obtained during commissioning. The timing resolution of a CRT hit is measured as the difference between the timestamps on two overlapping CRT modules, measured to be less than 60~ns as seen in Figure~\ref{fig:crtHitResolution}.


        \cleardoublepage

\subsection{Cryogenics Instrumentation} 
\label{sec:CryoInst}

\begin{figure}
\begin{center}
\includegraphics[width=0.9\textwidth]{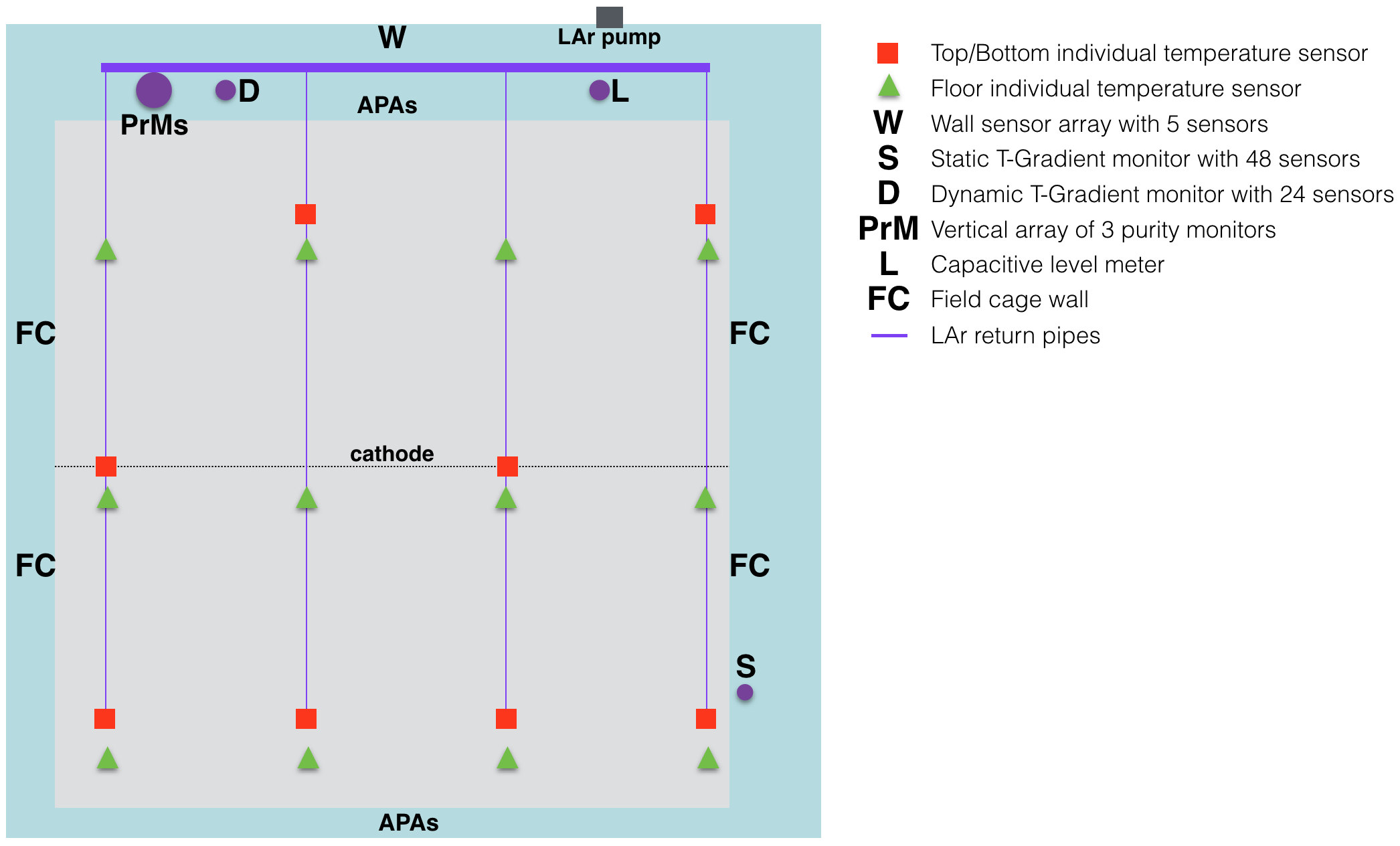}
\caption{Top view of the cryostat with the distribution of LAr instrumentation devices. The upstream side of the detector where the beam enters the cryostat is on the left hand side of the image. \label{fig:inst_map}}
\end{center}
\end{figure}

\pdsp{} includes instrumentation 
designed to monitor the quality and stability of the overall detector environment, to  ensure that the LAr quality is adequate for operation of the detector, and to help diagnose the source of any changes in detector operations (see Figure~\ref{fig:inst_map}). 
Monitoring instrumentation placed inside the cryostat includes thermometry to monitor the cryostat cool-down and fill, level meters to monitor the height of the LAr surface, purity monitors to provide a rapid assessment of the electron-drift lifetime independent of the TPC, and a system of internal cameras. Gas analysers used to monitor the cryostat purge process are located outside the cryostat. 

\subsubsection{Purity Monitor}
\label{sec:ci:purity}
Achieving an electron lifetime that is long enough so that electrons can drift sufficient distances is a challenging aspect of LArTPCs. The electron (e-) lifetime in a LArTPC is inversely proportional to, and determined by, the electronegative impurity concentration in the LAr, making the LAr purity an essential concern for successful operation and physics reach of the detector. 
Electron loss due to electronegative impurities in a LArTPC can be parameterised as $N(t)=N(0)e^{-t/\tau}$, where $N(0)$ is the number of electrons generated by ionization and not recombined with argon ions, $N(t)$ is the number of electrons after drift time $t$, and $\tau$ is the electron lifetime. 
\pdsp{} was designed to have the same 3.6\,m drift distance and 500~V/cm electric field as  planned for the DUNE-SP far detector modules. Given the drift velocity of approximately 1.5~mm/$\mathrm{\mu s}$ in this field, the drift time from cathode to anode is roughly 2.3~ms. 
Thus, if the electron lifetime is $\mathrm{2.3 ms}/[-\ln(0.8)] \simeq \mathrm{10~ms}$, then the LArTPC signal attenuation, \([N(0)-N(t)]/N(0)\), remains less than 20\% over the entire drift distance.


Purity monitors are used to independently infer the effective free electron lifetime in a LArTPC.  
It works by illuminating a photocathode with deep UV light to generate an electron current, then collects the drifted current at an anode a known distance away. Attenuation of the
current is related to the electron lifetime. In the \pdsp{} purity monitors the UV light is generated by an external xenon light source and delivered by quartz fibres to the inside of the cryostat. The fraction of photoelectrons generated at the purity monitor cathode that arrives at the anode ($Q_A/Q_C$) after the drift time $t$ is a measure of the electron lifetime $\tau$: $Q_A/Q_C=e^{-t/\tau}$, where $Q_C$ is the combined charge of the electrons generated at the photocathode, $Q_A$ is the combined charge of the electrons collected by the anode after drift time $t$, and $\tau$ is the electron lifetime.

The purity monitors are placed inside the cryostat but outside of the TPC volume due to their size. 

Although \pdsp{} receives ample cosmic ray data to perform electron lifetime measurements, the purity monitor system was found to be essential for providing quick, reliable, real-time information. It enabled operators to catch purity-related changes caused by LAr recirculation issues in time to correct them. 

In addition, since the purity monitors have much smaller volumes than the LArTPC, the measurements from this system are affected to a much smaller degree by the space charge caused by cosmic rays. Since purity monitors provide run-by-run electron lifetime measurement in liquid argon, they have unique importance for DUNE's deep-underground far detector charge calibration, where the cosmic-ray-based calibration is very challenging due to the low cosmic statistics.


\subsubsection*{Purity Monitor Design}

\begin{figure}[h]
\centering
\includegraphics[width=0.7\textwidth]{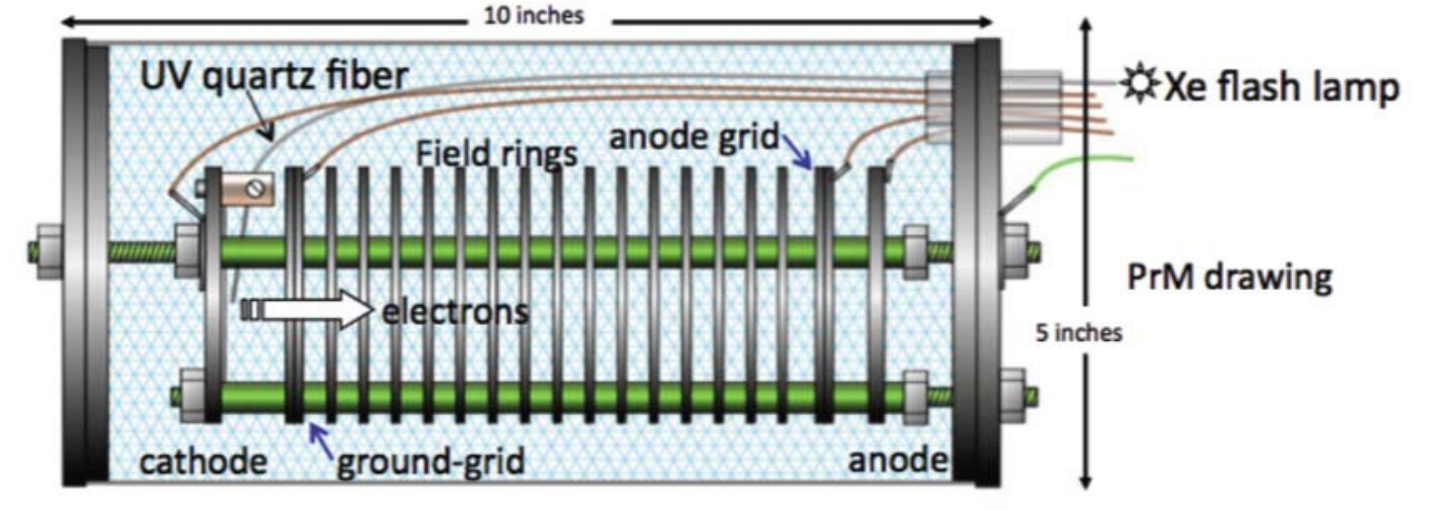}
\caption{Schematic diagram of the 
purity monitor design \cite{Amerio:2004ze}.}
 \label{fig:prm}
\end{figure}

The purity monitor design follows that of the monitors used in the ICARUS experiment~\cite{Amerio:2004ze} (Figure~\ref{fig:prm}). It consists of a double-grid ion chamber with four parallel, circular electrodes: a disk holding a photocathode, two grid rings (one each in front of the anode and the cathode), and an anode disk. The cathode grid (labelled ``ground-grid'' in the figure) is held at ground potential. The cathode, anode grid, and anode can each be independently biased via modified vacuum-grade HV feedthroughs. The anode grid and the field-shaping rings are connected to the cathode grid by an internal chain of 50\,$\mathrm{M\Omega}$ resistors to ensure the uniformity of the electric field in the drift regions. A stainless steel mesh cylinder is used as a Faraday cage to isolate the purity monitor from external electrostatic background. 


Low signal strength has limited the precision and measuring ranges of purity monitors in previous LArTPC experiments \cite{Amerio:2004ze,Adamowski_2014}. To boost the signal strength for use in \pdsp{}, UV light was delivered to each purity monitor by eight fibres and an 8-channel feedthrough, resulting in signal magnitudes six times larger than with a single fibre. This increase came without penalty in timing resolution. 


The electron drift time is a key indicator of the measurement range and precision of the purity monitor. The electron drift time depends on the cathode--anode distance and the applied voltage within the monitor. During the commissioning and beam phases of \pdsp{} the electron drift time within the detector varied from 150$\mathrm{\mu s}$ to 3\,ms. The purity monitors were operated at different high voltage settings that  covered an electron drift time measurement range of 35$\mathrm{\mu s}$ to more than 10\,ms. The electron drift lifetime exceeded 3\,ms during the beam phase. 


\subsubsection*{Purity Monitor Data}
\begin{figure}[h]
\centering
\includegraphics[width=1.0\textwidth]{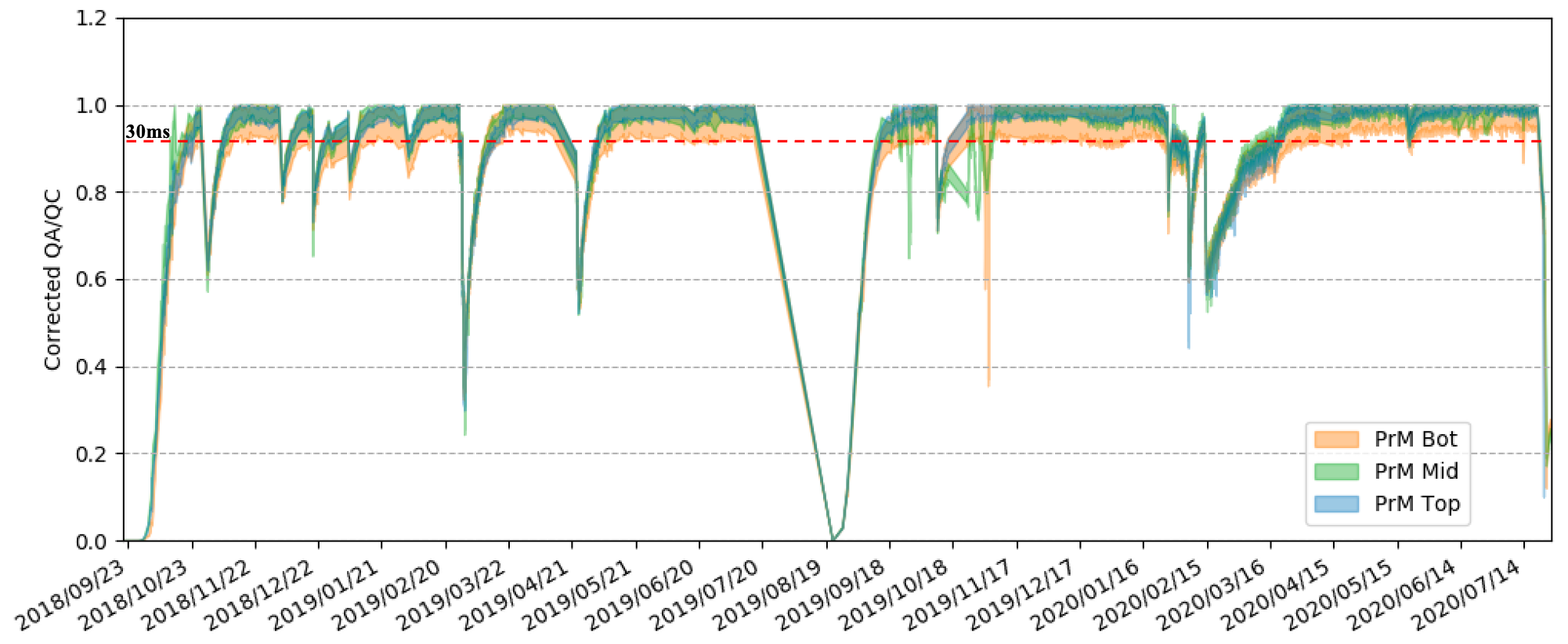}
\caption{
The anode-to-cathode signal ratios $Q_A/Q_C$ measured by three purity monitors in \pdsp{} as a function of time, September 2018 through February 2020. The purity is low prior to the start of circulation in October 2018. Later dips represent recirculation studies and recirculation pump stops. The shaded bands represent uncertainties of the measurements. }
 \label{fig:prm-qaqc}
\end{figure}

\begin{figure}[h]
\centering
\includegraphics[width=1.0\textwidth]{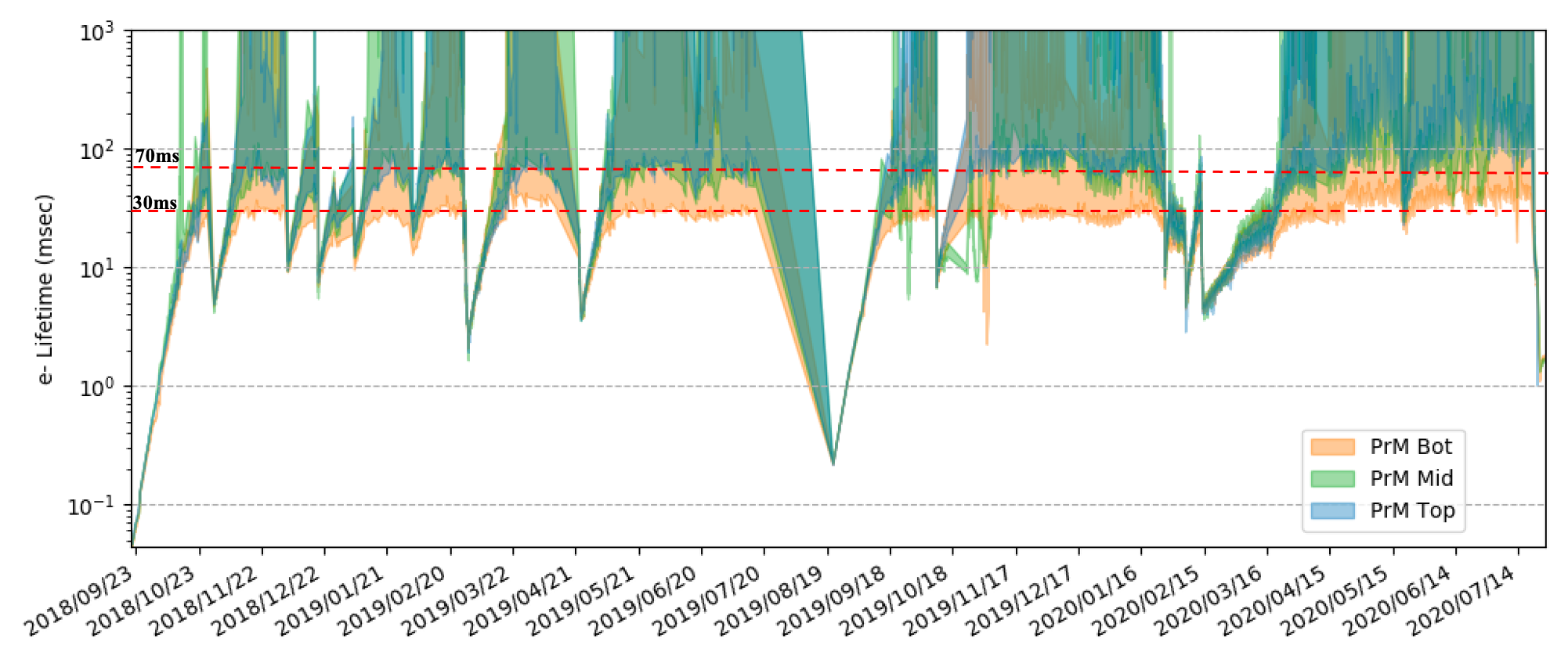}
\caption{The electron lifetimes measured by three purity monitors in \pdsp{} as a function of time, September 2018 through February 2020. The purity is low prior to the start of circulation in October 2018. Later dips represent recirculation studies and recirculation pump stoppages. The shaded bands represent uncertainties of the measurements.} 
 \label{fig:prm-lifetime}
\end{figure}

Three purity monitors were installed at heights of 1.8 m, 3.7 m, and 5.6 m from the bottom of the \pdsp{} cryostat  to  continuously monitor the LAr purity during all phases:  commissioning, beam test, and operation. Figures~\ref{fig:prm-qaqc} and~\ref{fig:prm-lifetime} show the anode-to-cathode signal ratios $Q_A/Q_C$ and the electron lifetime measured by each purity monitor from the commissioning phase, which started in September 2018, through the entire beam test, which ended in November 2018, and continued through the non-beam operations, which ended in February 2020. The shaded bands represent uncertainties in the measurements. All $Q_A/Q_C$ are normalised to a drift time of 2.3\,ms.

Each electron-lifetime measurement by the purity monitors is based on the signal ratios $Q_A/Q_C$ from 200 UV flashes at the same location occurring within a 40 second window. These measurements, taken regularly, were able to indicate incipient circulation-related issues on a quick timescale, mitigating potentially serious consequences for the detector. They caught a filter saturation during LAr filling and recirculation pump outages; these incidents show as sudden dips in Figures~\ref{fig:prm-qaqc} and~\ref{fig:prm-lifetime}.
The measurement uncertainties in Figure~\ref{fig:prm-qaqc} include statistical and time-dependent fluctuations as well as uncertainties in the baseline of the purity monitor signal waveform, the cathode and anode RC constants, the inefficiency of grid shielding, and the electrical transparency of the grids. The ``inefficiency of grid shielding'' refers to the inefficiency of the cathode and anode grids to shield against induced currents on the anode and cathode while electrons are drifting between the anode and cathode grids. The electrical transparency of a grid is the proportion of the electrons that pass the grid. 
Other uncertainties in such quantities as signal rise time and electron drift time were found to be small. The overall uncertainties in purity-monitor charge ratio measurement $\frac{\Delta(Q_A/Q_C)}{(Q_A/Q_C)}$ at 2.3\,ms drift time for the purity monitors, from highest position to lowest, are $1.9\%$, $2.2\%$, and $3.9\%$, respectively. 

Note that given $Q_A/Q_C=e^{-t/\tau}$, when $Q_A/Q_C=1$, the measured lifetime $\tau$ is infinity. As shown in \ref{fig:prm-lifetime}, when the LAr purity is stabilised after filtering, $Q_A/Q_C$ is close to one, which means that the electron loss (due to impurities) after drifting for 2.3\,ms is very small. The top and middle purity monitors measured the electron lifetime to be greater than 70\,ms, and the bottom monitor measured it as greater than 30\,ms. Considering the inverse relationship between the drift electron lifetime and the amount of oxygen equivalent impurity, the estimate predicts the impurity never went above 40\,ppt equivalent of oxygen in the week of data-taking. At the end of beam data-taking on November 11th, 2018, the impurity in the detector can be estimated to be approximately 3.4$\pm$0.7\,ppt oxygen equivalent \cite{bettini1991study, Abi:2020mwi}. These measurements indicate that \pdsp{} exceeded the high LAr purity required by the DUNE far detector. They also corroborate the high electron lifetime measurements ProtoDUNE-SP has previously reported using TPC tracks matched to the CRT \cite{Abi:2020mwi}, \cite{Diurba:2021}.

\begin{figure}[htbp]
\begin{center}

\includegraphics[width=1.0\textwidth]{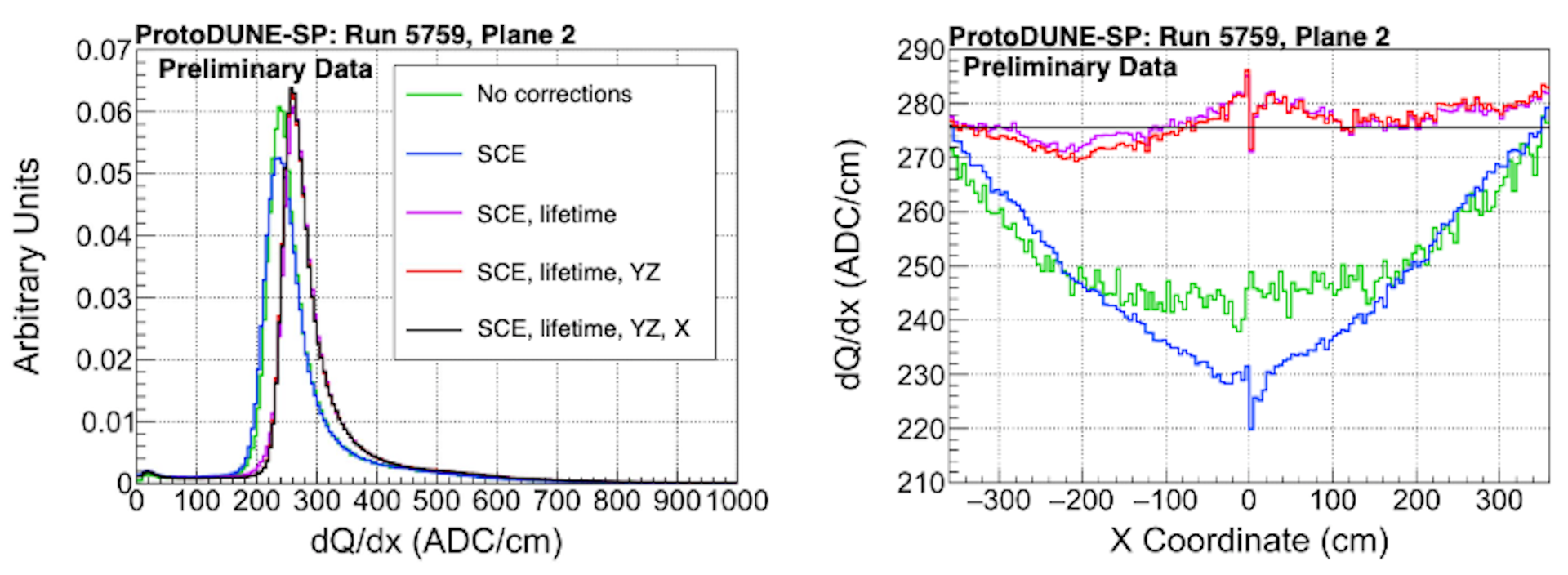}
\caption{Performance of the purity monitor based electron lifetime calibration for  cathode-crossing cosmic rays in ProtoDUNE-SP Phase-1: (Left) $dQ/dx$ distribution and (Right) $dQ/dx$ vs. electron drift distance $x$, comparing no calibration (green), space charge effect calibration (SCE) (blue), PrM electron lifetime calibration+SCE (violet), electron lifetime calibration+SCE+YZ correction, electron lifetime calibration+SCE+YZ+Y correction. After the purity monitor based lifetime calibration (violet), the 15\% charge loss on TPC signal (blue) is mostly corrected.} \label{fig:PrMCosmic}
\end{center}
\end{figure}

Again given  $Q_A/Q_C=e^{-t/\tau}$, the relative uncertainty in lifetime $\Delta\tau/\tau$ is propagated as $\Delta\tau/\tau=(\tau/t)\frac{\Delta(Q_A/Q_C)}{(Q_A/Q_C)}$. It follows then that for a given $Q_A/Q_C$ uncertainty of  $\frac{\Delta(Q_A/Q_C)}{(Q_A/Q_C)}$, the relative uncertainty in the measured electron lifetime is proportional to the electron lifetime $\tau$ and inversely proportional to the electron drift time $t$. For a long electron lifetime, the relative uncertainty in this lifetime is large, as shown in Figure~\ref{fig:prm-lifetime}. However, since the main sources of uncertainties in $Q_A/Q_C$ are not correlated with the lifetime, $\frac{\Delta(Q_A/Q_C)}{(Q_A/Q_C)}$ is not significantly correlated with the electron lifetime or the drift time, and therefore the large relative lifetime uncertainty at high purity does not indicate lower precision in the purity monitor's measurement of the charge loss during the electron drift time. The run-by-run purity monitor electron lifetime has been included in the ProtoDUNE-SP calibration database for TPC charge and energy calibration. The performance of the purity monitor based electron lifetime calibration tested with ProtoDUNE-SP cosmic rays is shown in Figure~\ref{fig:PrMCosmic}~\cite{ref:prmcalibUser2020}. After the purity-monitor-based lifetime calibration, the charge loss on the TPC signal due to LAr impurity is mostly corrected.

\subsubsection{Temperature Sensors}
\label{sec:ci:t-pro}

Since the purity monitors cannot be placed within the TPC volume, to monitor the cryogenics system and the LAr for homogeneous mixing, temperature monitors are used as a substitute. An extensive set of temperature measurements is recorded to create a detailed 3D temperature map of the detector volume that is used as input for the computational fluid dynamic (CFD) model validation.  Results from these CFD simulations can predict the LAr purity across the entire cryostat volume. The vertical coordinate is particularly important since it corresponds closely to the LAr recirculation and uniformity.
 

In \pdsp{} 92 high-precision temperature sensors are distributed near the TPC walls in two configurations (see Figure~\ref{fig:cisc_static_TG}):
i) a formation of two high-density vertical arrays (called T-gradient monitors), and 
ii) coarser 2D horizontal arrays at the top and bottom of the TPC. The bottom sensors are mounted on the LAr return pipes and the top sensors on the ground planes. 

Three elements are common to all systems: sensors, cables and readout. Lake Shore PT102 platinum sensors with 100\,$\Omega$  resistance (at 0°C) were chosen based on previous experience from the 35\,t prototype\cite{MONTANARI2015308}. For the inner readout cables, a custom cable made by Axon' was used. This 3.7\,mm diameter cable consists of four AWG 28 teflon-jacketed copper wires, forming two twisted pairs, with a metallic external shield and an outer teflon jacket . The four-wire configuration eliminates any influence of wire resistance on the measurements.

Finally, the readout system consists of an electronic circuit that includes: 
i) a precise 1\,mA current source for the excitation of the sensors,  
ii) a multiplexer reading out the different temperature sensors and forwarding the selected one to a single channel, and
iii) a readout system based on the National Instrument Compact RIO Device with a high-accuracy voltage signal readout NI9238 module that provides 24-bit resolution over a 1\,V range.  In addition, 12 standard temperature sensors are in contact with the bottom of the cryostat to detect the presence of LAr when filling starts, and five standard sensors are lined up vertically and epoxied onto one of the lateral walls to measure the temperature of the cryostat membrane at different heights during cool-down and filling. 

\subsubsection*{Static T-gradient monitor}

\begin{figure}[ht]
\begin{center}
\includegraphics[height=0.6\textwidth]{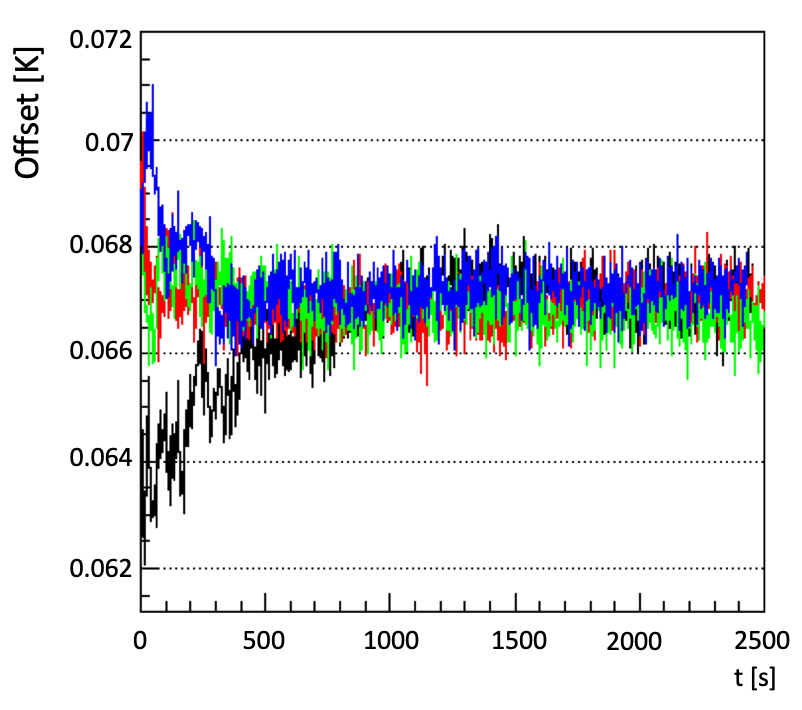}%
\caption{Temperature offset between the reference sensor and the three other sensors in a calibration set for the Static T-Gradient monitor. Offsets are shown as a function of time for four independent immersions in LAr with respect to the reference sensor. Within the time range used for the actual calibration (1000-2000\,s), both fluctuations for a single immersion and difference between immersions are below 1\,mK, demonstrating the performance and stability of the system and the robustness of the calibration procedure.  
\label{fig:cisc_static_TG}}
\end{center}
\end{figure}

In addition to the distributed temperature sensors described above, a vertical array of 48 sensors was installed 20\,cm away from the lateral field cage (see Figure~\ref{fig:cisc_static_TG}-left). 
Vertical spacing is 11.8\,cm for both the top and bottom 16 sensors and 23.6\,cm for the 16 in the middle, with the bottom (top-most) sensor 30.2 (738.5)\,cm from the bottom surface of the cryostat. 
Mechanical rigidity is provided by an 80\,$\times$\,25\,$\times$\,3\,mm$^3$ U-shaped fibre glass profile (FGP) that holds sensors and cables. Given the proximity to the field cage, the entire system is surrounded by a Faraday cage made up of 19 vertical 6-mm-diameter stainless steel rods that are arranged to form a cylinder 12.5\,cm in diameter. The rods and FGP are hung from the top of the cryostat and are  mechanically decoupled from each other to allow for their different contraction rates in the LAr. The rods are passed through FR4 rings attached to the FGP to maintain the cylindrical shape.

Sensors are mounted on a 52\,$\times$\,14\,mm$^2$ PCB with an IDC-4 connector, such that they can be plugged in at any time. Cables are housed in the inner part of the FGP and run vertically from the sensor to the flange, which has eight Sub-D 25-pin connectors. 

The temperature sensors are cross-calibrated in a controlled laboratory environment. During the calibration procedure, a set of four sensors (one of which is kept as a reference for all sets) are placed next to each other and submerged in LAr several times. The calibration procedure relies on the assumption of equal temperature for all sensors. Convection is minimised by placing the four sensors inside a 50 mm diameter, 1 mm thick cylindrical aluminum capsule (allowing fast cool-down in gas before immersion), which is placed in the centre of a 3D printed box with two independent concentric LAr volumes, and surrounded by 10 cm thick polystyrene walls. The stability of the calibration system, and the performance of the sensors and the readout system are shown in Figure~\ref{fig:cisc_static_TG}-right. The accuracy of the calibration was estimated to be 2.6\,mK \cite{Peris:2018}. This takes into account the uncertainty on the offset between any two sensors in the static T-Gradient monitor as well as differences between the four immersions in LAr.   

\subsubsection*{Dynamic T-gradient monitor}
\begin{figure}[ht]
\begin{center}
\includegraphics[height=0.77\textwidth]{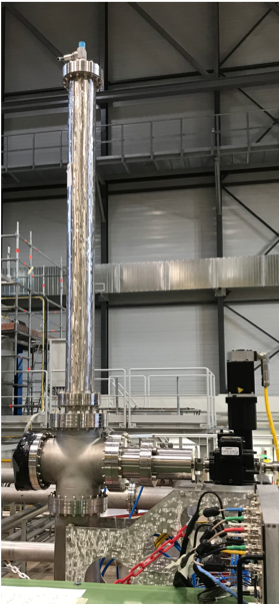}
\hspace{1cm}
\includegraphics[width=0.55\textwidth]{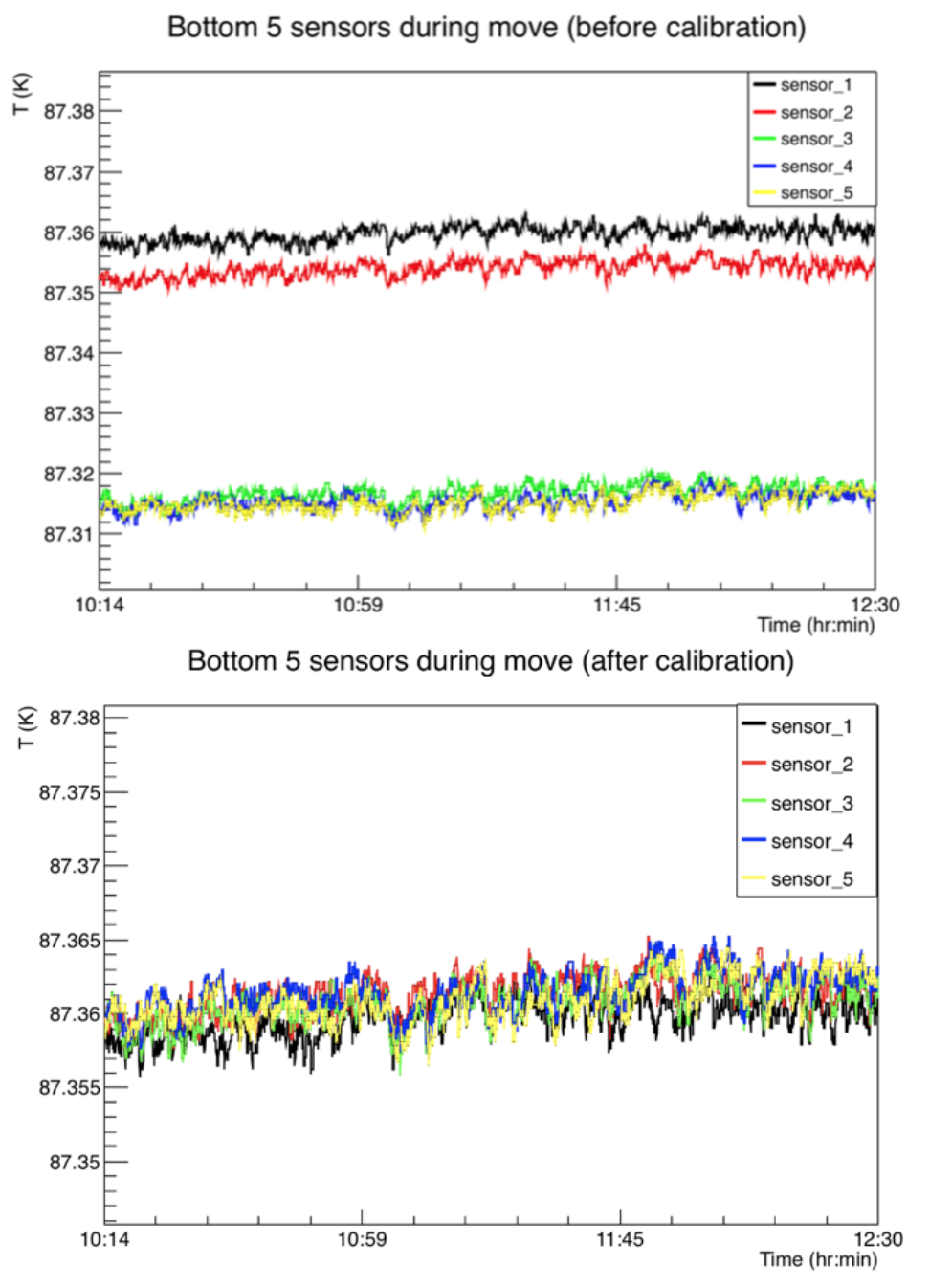}
\caption{Left: Dynamic T-gradient monitor enclosure showing stepper motor on the side and viewports to visually check the motion of the system. The top part of the enclosure houses the carrier rod when the system moves up. Right: Temperature measurement by the bottom-most five sensors 
before and after cross-calibration}.
\label{fig:cisc_dynamic_T}
\end{center}
\end{figure}

Finally, a movable system of temperature sensors, called the Dynamic T-gradient monitor, was also installed in \pdsp{}. 
The Dynamic T-gradient monitor is equipped with a stepper motor connected to a carrier rod on which 24 temperature sensors are mounted. 
The carrier rod with sensors is contained inside the gas-tight enclosure shown in Figure~\ref{fig:cisc_dynamic_T}-left. The stepper motor is mounted on the outside of the enclosure. Vertical spacing is 10\,cm for the five sensors at the top and bottom, and 50\,cm for the 14 sensors in the middle, where the bottom (top-most) sensor is 10 (750)\,cm from the bottom of the cryostat. Each sensor is soldered to a PCB. Cables are routed along the carrier rod and connected to a flange with Sub-D connectors on both sides.


The stepper motor engages with the carrier rod via a ferrofluidic dynamic seal and a pinion and gear that converts the motor's rotational motion into linear motion of the rod without jeopardising the integrity of the argon atmosphere inside the enclosure. 

The stepper motor can move the rod vertically, enabling temperature measurements at various heights in the TPC, and at the same heights by different sensors. Assuming a stable temperature profile over the calibration period of a couple of minutes, any offset in the temperature readings at a given height by different sensors can be attributed to the intrinsic offset between the two sensors, providing constants for their cross-calibration. By linking all adjacent sensors, it is possible to calibrate out all offsets with respect to a single one and achieve precise measurement of the relative vertical T-gradient.
See Figure~\ref{fig:cisc_dynamic_T}-right for the comparison of sensor readout before and after calibration. As will be shown in Section~\ref{sec:temp_analysis}, final confirmation of the robustness of the system was achieved during the ``pump-off'' detector period when the temperature became uniform over the detector height.

\subsubsection*{Temperature data analysis}
\label{sec:temp_analysis}

The systems described above have been collecting data for more than two years, resulting in a deeper 
understanding of LAr dynamics inside a cryostat. Some important outcomes are discussed below. 

Stability studies show that except near the LAr surface the difference in temperature between any two sensors in the same T-Gradient monitor remains constant to within 3\,mK, demonstrating the stability of the LAr 
system under standard operating conditions, and importantly, the reliability and longevity of the sensors. 

The slow and continuous process of progressively moving each sensor up (or down) to the location previously occupied by its neighbour is expected to yield a height profile that is effectively flat. Deviations from flatness would indicate a problem with one or more sensors. 

\begin{figure}
\begin{center}
\includegraphics[width=0.7\textwidth]{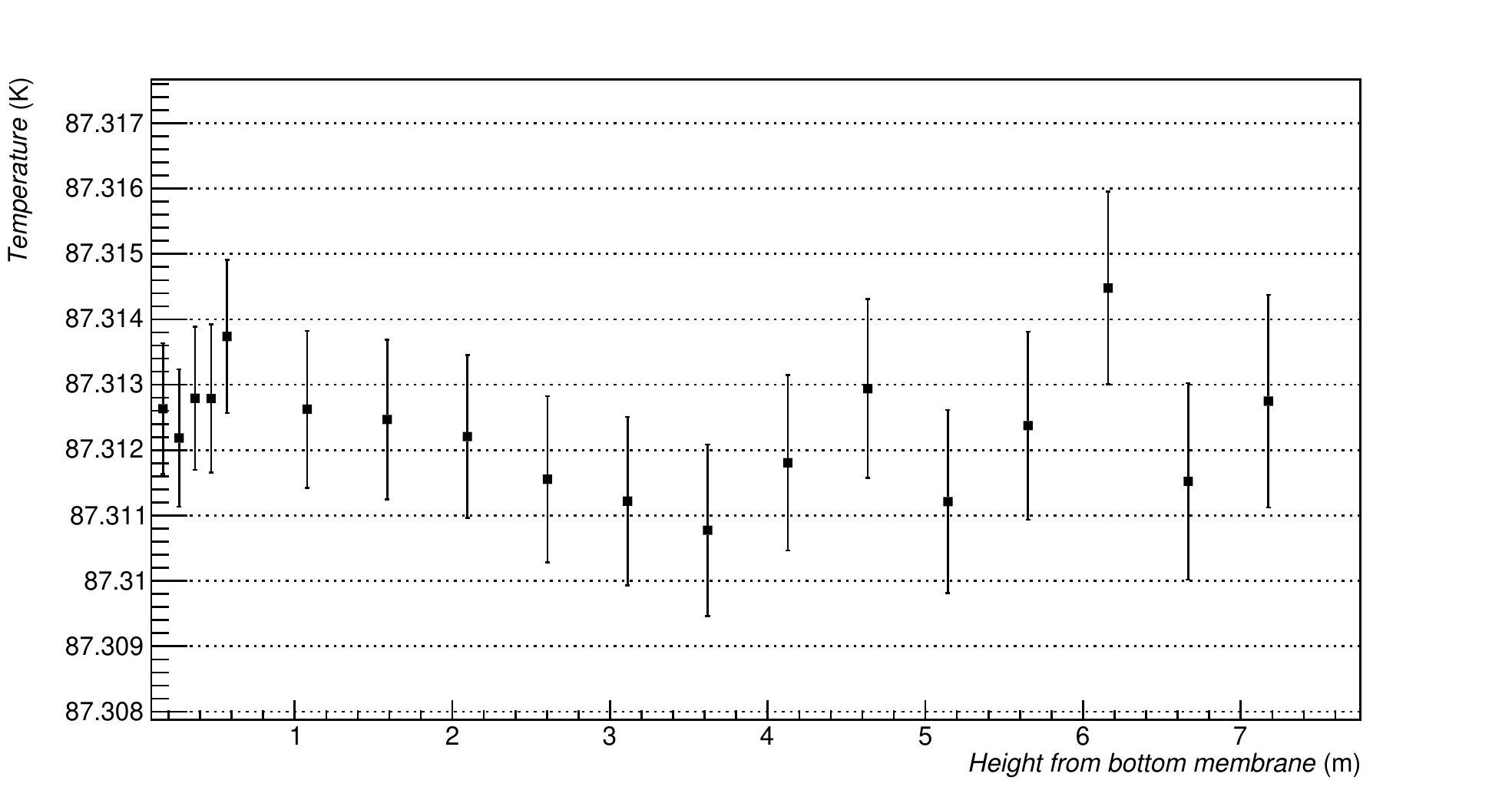} \\
\includegraphics[width=0.7\textwidth]{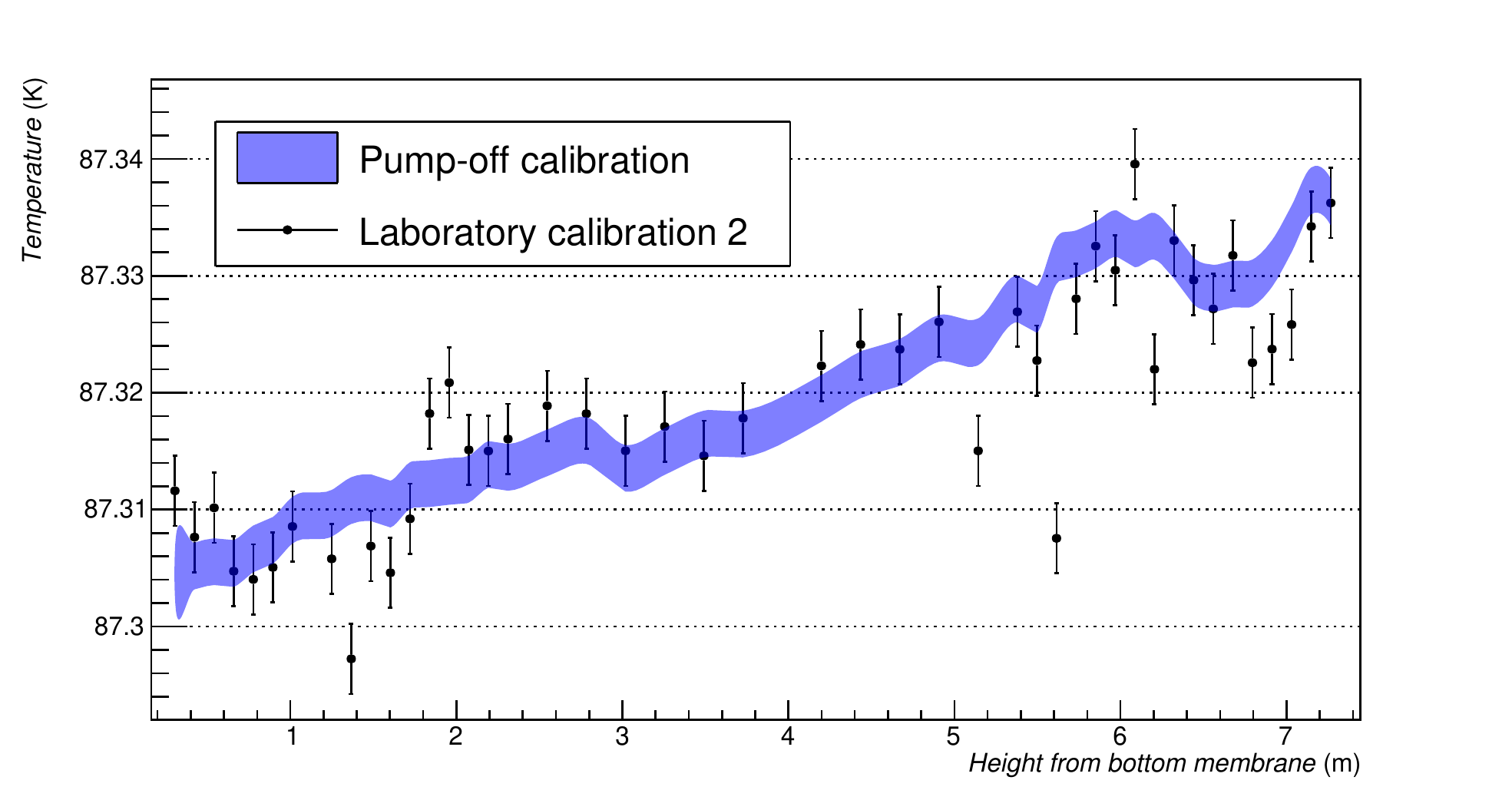}
\caption{Top: Dynamic T-Gradient with no recirculation shows a flat profile. Bottom: Static T-Gradient while gas pressure is raised; comparison of laboratory calibration to ``pumps-off'' calibration.
\label{fig:protodune_general_performance_pumpoff}}
\end{center}
\end{figure}

Vertical and horizontal cross-calibration profiles of the Dynamic T-Gradient monitor have been compiled under various conditions (e.g., stable, with recirculation system on and off, at different pressures). The pumps-off (i.e., no recirculation) measurements are particularly interesting since small temperature gradients are expected and were in fact observed by the Dynamic T-Gradient monitor (see Figure~\ref{fig:protodune_general_performance_pumpoff}-left). The flatness of this profile to within a few mK illustrates the homogeneity of LAr temperature. This measured homogeneity was used to re-calibrate the Static T-Gradient monitor (this is the so-called ``pumps-off'' calibration method) and the result shows good agreement with the laboratory calibration (see Figure~\ref{fig:protodune_general_performance_pumpoff}-right). 

\begin{figure}
\begin{center}
\includegraphics[width=0.98\textwidth]{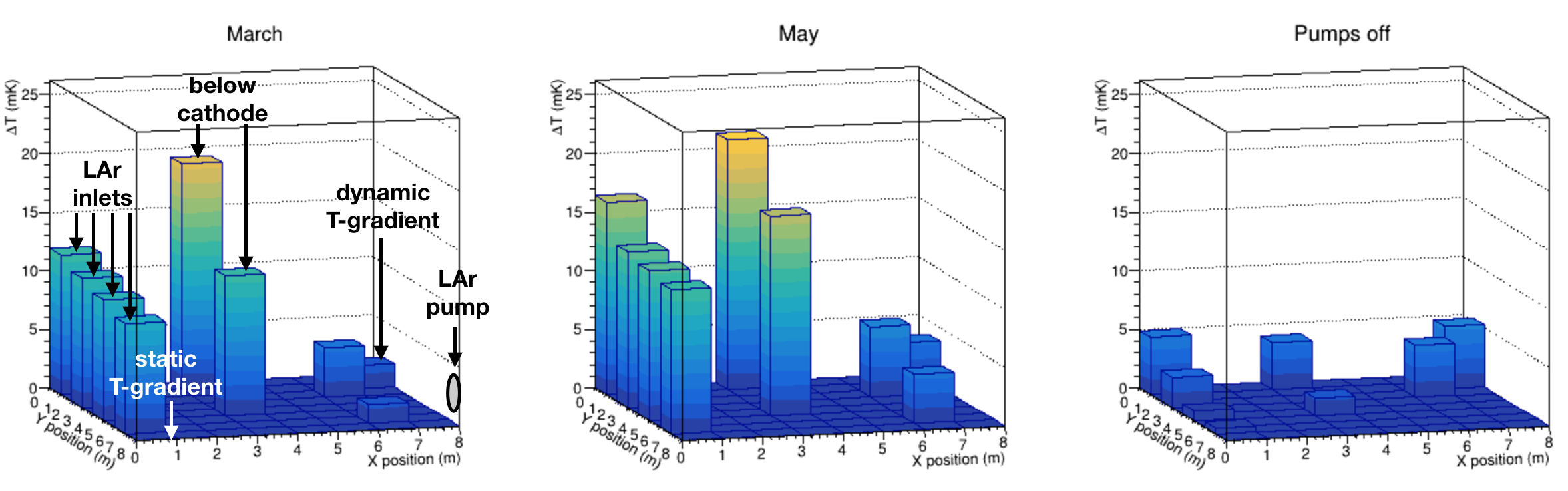}
\caption{Temperature difference between the Static T-Gradient bottom sensor and the horizontal grid of sensors placed 40\,cm apart over the bottom of the cryostat. Left and middle: temperature distribution with the recirculation system on. Right: temperature distribution when recirculating system is off.
\label{fig:protodune_general_performance_bottom}}
\end{center}
\end{figure}

These studies have demonstrated significant differences in temperature based on whether the recirculation system is on or off. These differences appear in both vertical and horizontal planes. The grid of sensors placed 40\,cm apart over the cryostat bottom shows a temperate peak (20 mK) below the cathode plane with a decrease towards the pump on the bottom beam left when the recirculation system is on (see Figure~\ref{fig:protodune_general_performance_bottom}-left), and a more homogeneous distribution when it is off (see Figure~\ref{fig:protodune_general_performance_bottom}-right).

\begin{figure}
\begin{center}
\includegraphics[width=0.7\textwidth]{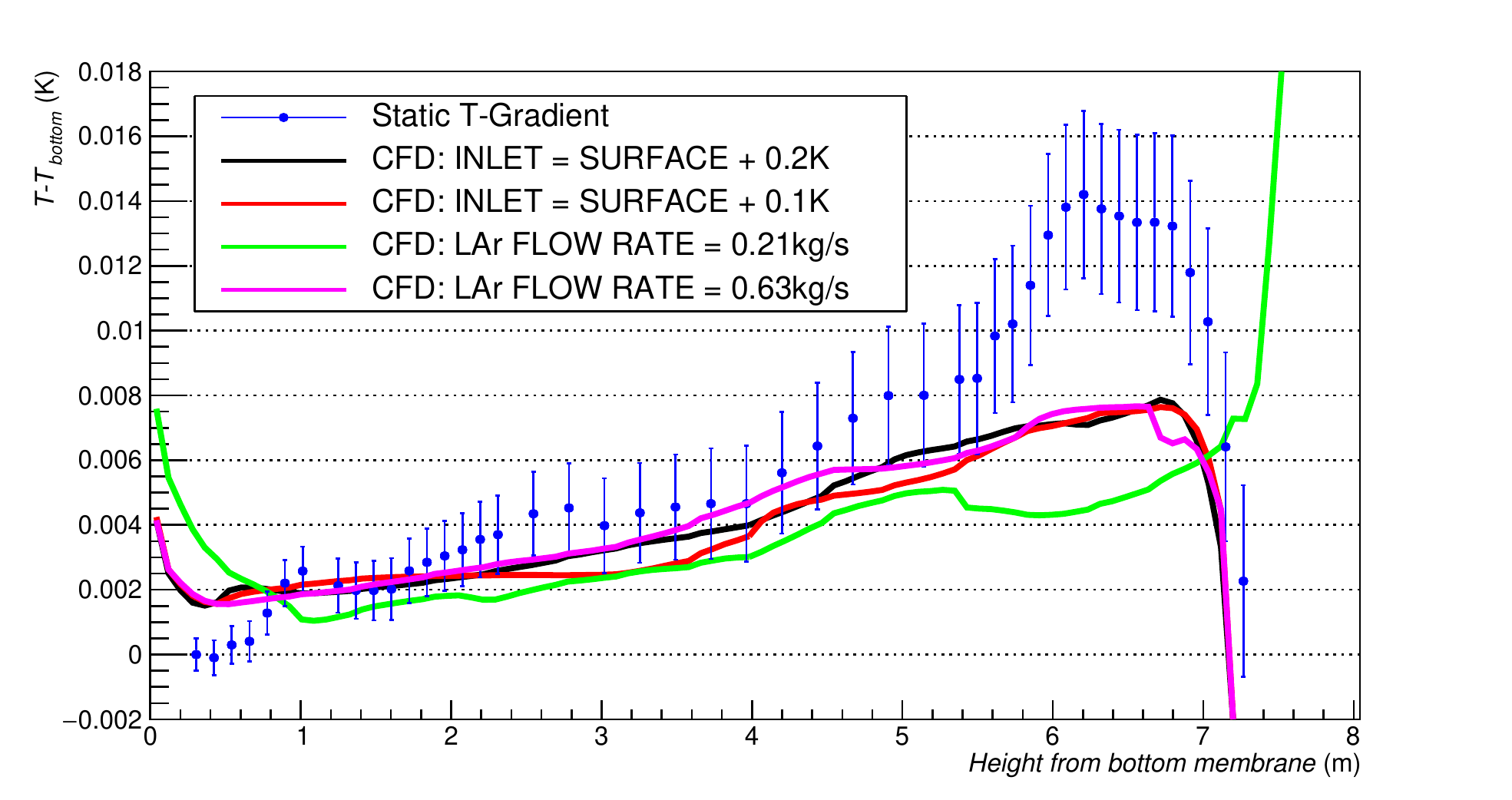} \\
\includegraphics[width=0.7\textwidth]{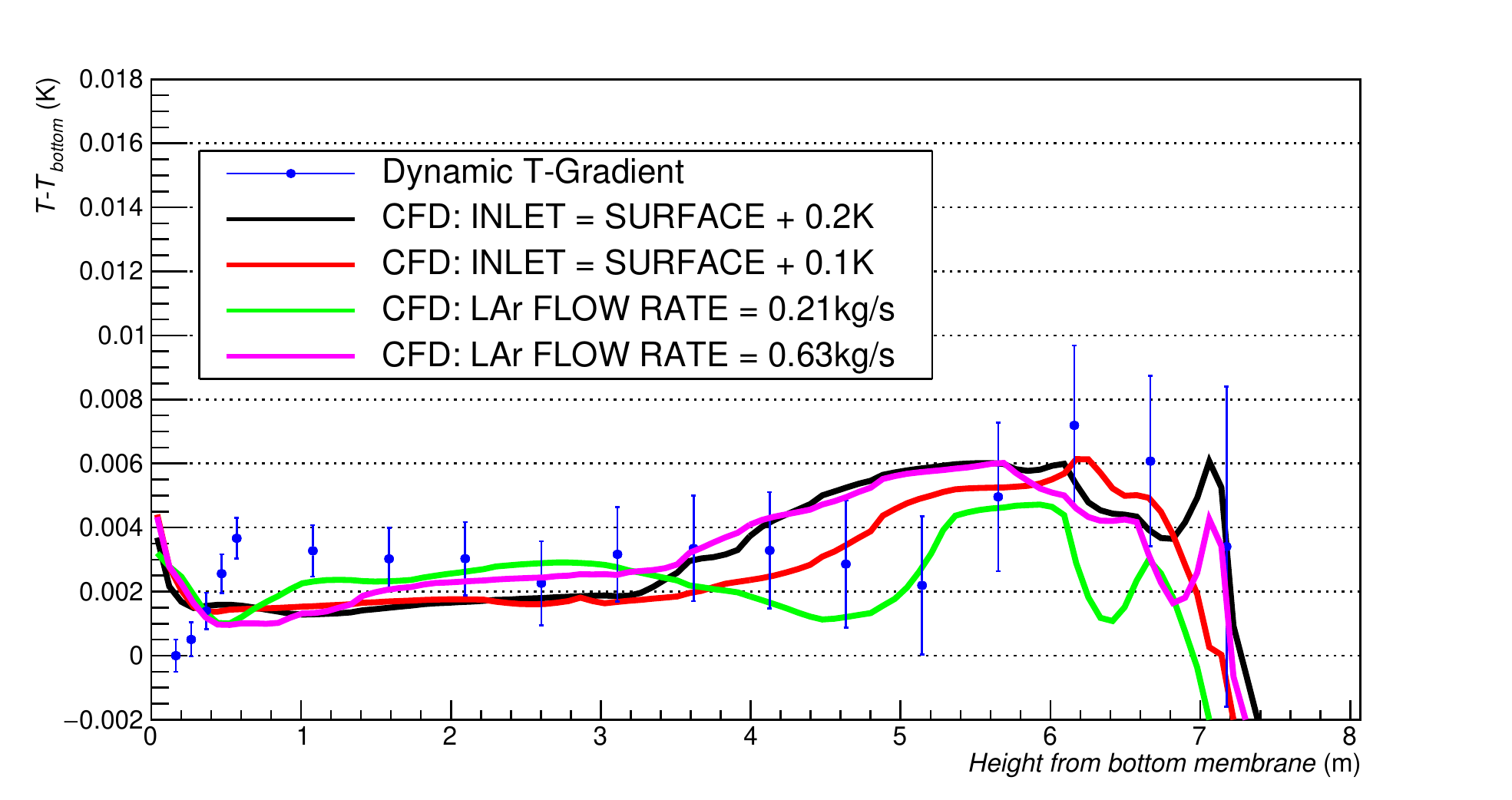}
\caption{Temperature profiles measured by the T-gradient monitors and comparison to the CFD model with different boundary conditions. Top: Static T-gradient monitor; Bottom: Dynamic T-gradient monitor. Unless specified, LAr flow rate is nominal, 0.42kg/s.
\label{fig:protodune_general_performance_cfd_comparative}}
\end{center}
\end{figure}

Finally, Figure~\ref{fig:protodune_general_performance_cfd_comparative} shows a comparison of the vertical profiles (under stable conditions) to CFD simulations for a variety of boundary conditions (e.g., varying flow rate and temperature of the incoming LAr). The CFD model reasonably predicts the main features of the data, but some details still need to be understood, e.g., the bump at 6.2\,m and the lower measured temperature at the bottom of the cryostat. More studies on CFD boundary conditions are needed. 

\subsubsection{Cameras}
\label{sec:ci:camera}
\begin{figure}[ht]
\begin{center}
\includegraphics[width=0.6\textwidth]{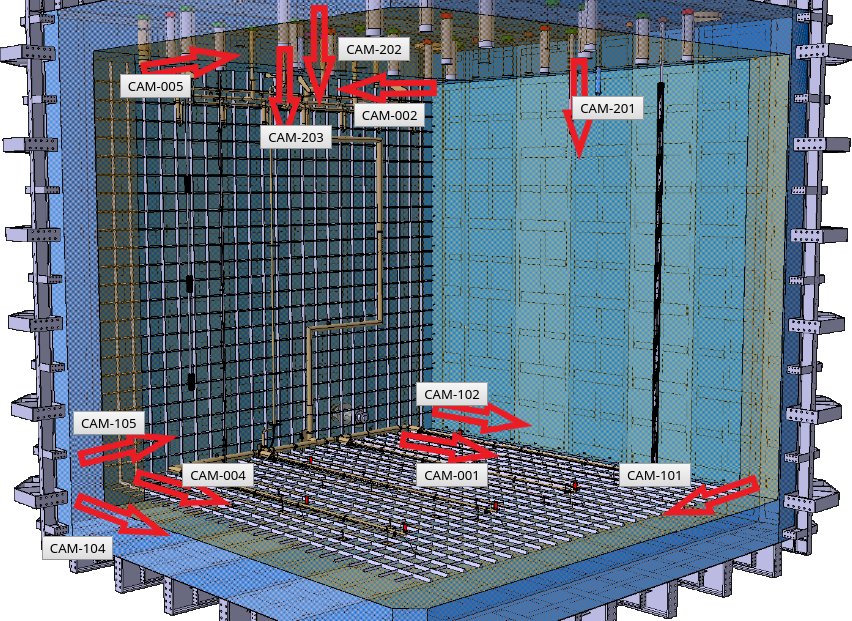}
\caption{A 3D view showing the locations of the 11 cameras in \pdsp{}.}.
\label{fig:pdsp-camera-locations}
\end{center}
\end{figure}

Cameras provide direct visual information about the state of the
detector during critical operations and when damage or unusual
conditions are suspected. Cameras can be used to verify stability, straightness, and alignment of the hanging TPC structures during cool-down and filling; ensure that no bubbling occurs near the ground planes; and inspect the state of movable parts in the detector (i.e., the dynamic temperature sensor).
Eleven cameras were deployed in \pdsp{} at the locations shown in Figure~\ref{fig:pdsp-camera-locations}. They successfully provided views of the detector during filling and throughout its operations. 
Two types of cameras, were deployed, ``cold'' for fixed, long-term use and ``warm'' for short-term inspections.

\subsubsection*{Fixed Cameras (cold)}
The cold fixed cameras
monitor the following items during filling:
\begin{itemize}
\item positions of the corners of each APA, CPA, FC, and GP (1~mm resolution);
\item relative straightness and alignment of the APAs, CPA, and FC (\(\lesssim 1\,mm\))
\item relative positions of profiles and endcaps (0.5\,mm resolution); and 
\item the LAr surface, specifically, the presence of bubbling or debris.
\end{itemize}

\begin{figure}[ht]
\begin{center}
  \includegraphics[width=0.4\textwidth]{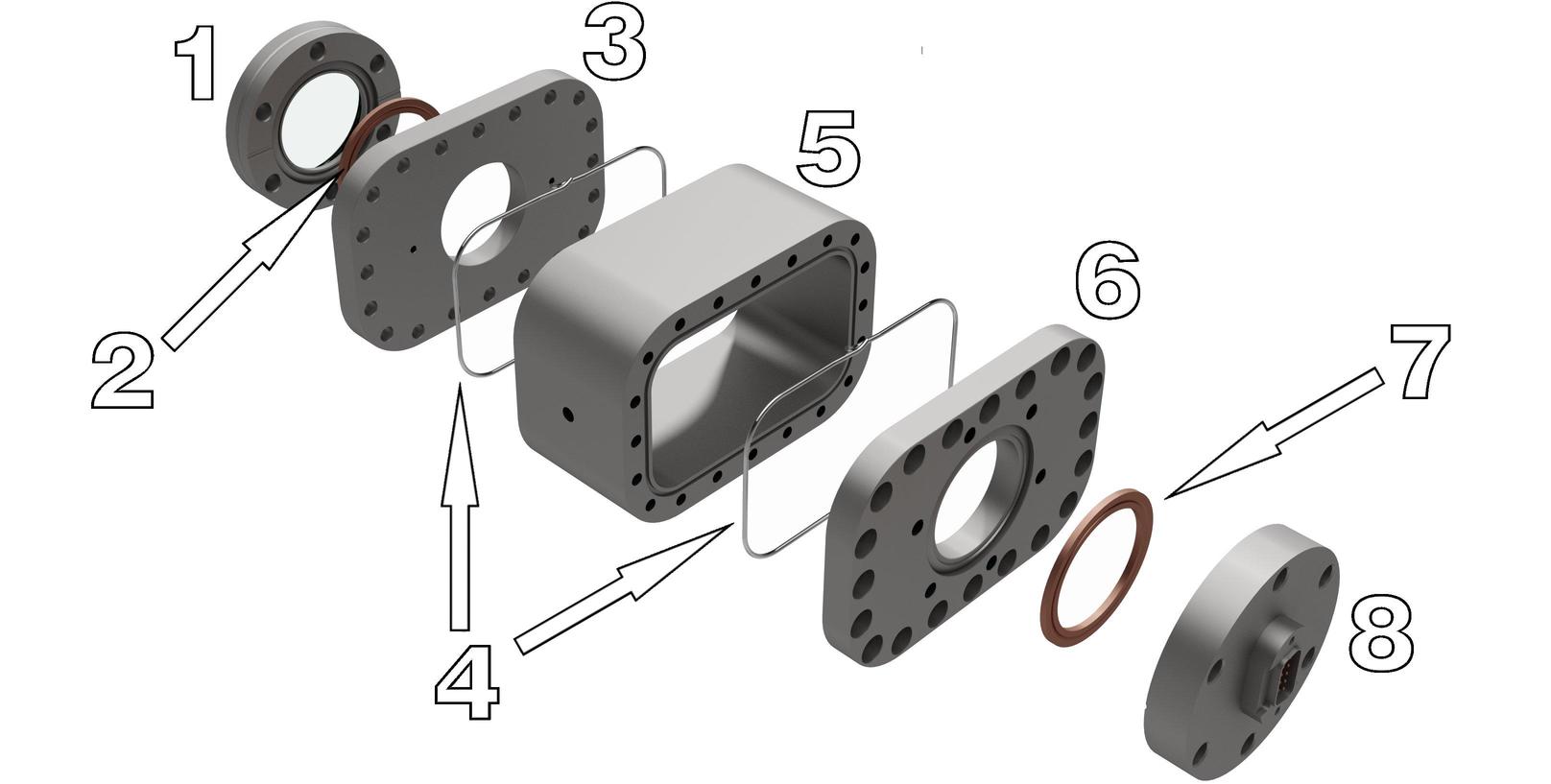}%
  \includegraphics[width=0.4\textwidth]{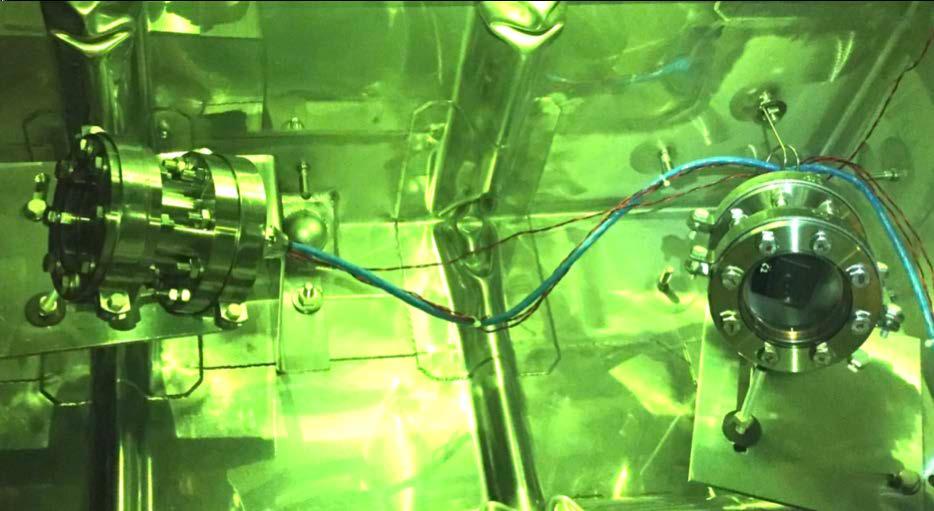}\\
  \hfill \includegraphics[width=0.4\textwidth]{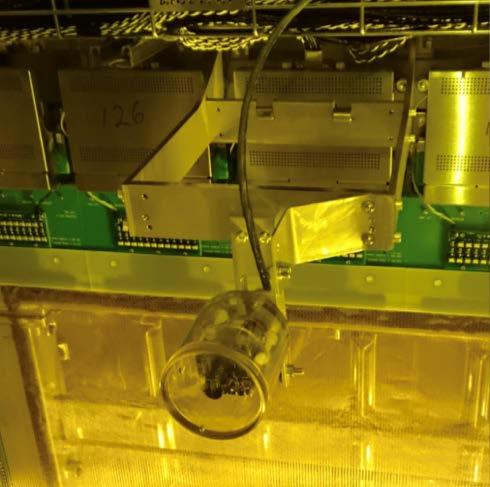}%
  \hfill \includegraphics[width=0.4\textwidth]{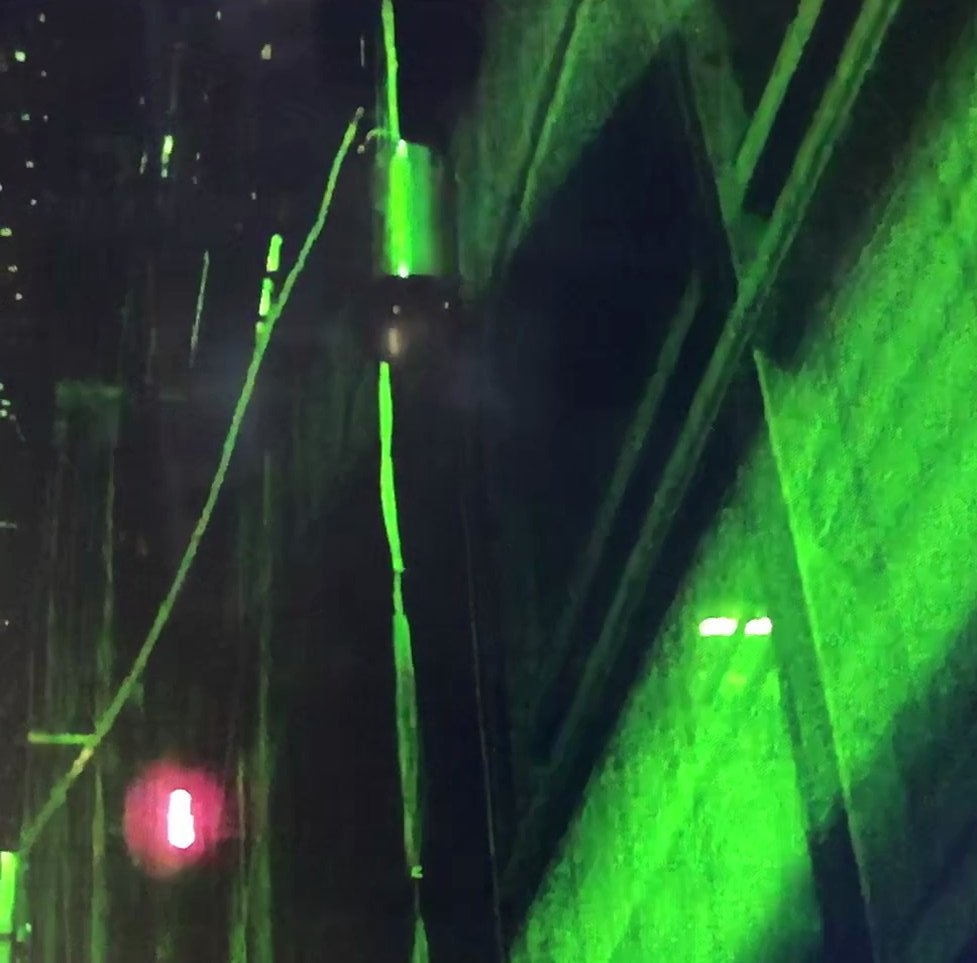}%
   \hfill
   \caption{Top left: a CAD exploded view of a vacuum-tight camera enclosure suitable for cryogenic applications \cite{Delaquis:2013hva}. Its numbered features are 
    (1) quartz window, (2 and 7) copper gasket, (3 and 6) flanges, (4) indium wires, (5) body piece, (8) signal \fdth.
    Top right: two of the \pdsp{} cameras in stainless steel enclosures similar to the CAD design on the top left. 
    Bottom left: one of the cameras in an acrylic enclosure.
    Bottom right: a portion of an image taken with camera 105 showing a purity monitor mounted outside the APA on the beam-left side. This photo was taken with \pdsp{} completely filled.}
  \label{fig:gen-fdgen-cameras-enclosure}
  \end{center}
\end{figure}

One design for the \pdsp{} fixed cameras uses an enclosure similar to the successful EXO-100 design \cite{Delaquis:2013hva}, see Figure~\ref{fig:gen-fdgen-cameras-enclosure}. Cameras 101, 102, 104, and 105, shown in Figure~\ref{fig:pdsp-camera-locations}, 
were installed in this type of enclosure.
A thermocouple in the enclosure allows temperature monitoring, and a heating element provides temperature control.  
SUB-D connectors are used at the cryostat flanges and the camera enclosure for signal, power, and control connections. An alternative successful design uses an acrylic enclosure, see Figure~\ref{fig:gen-fdgen-cameras-enclosure}, bottom left. Cameras 001, 002, 004, and 005, shown in Figure~\ref{fig:pdsp-camera-locations}, are placed in acrylic enclosures. All have operated successfully.

\subsubsection*{Inspection Cameras (Warm)}
The inspection cameras are selected to be as versatile as possible to cover the range of intended uses.
The following inspections have been done with the warm cameras:

\begin{itemize}
\item status of HV-feedthrough and cup,
\item status of FC profiles, endcaps (0.5\,mm resolution),
\item vertical deployment of calibration sources,
\item status of thermometers, especially dynamic thermometers,
\item HV discharge, corona, or streamers on HV feedthrough, cup, or FC,
\item relative positions of profiles and endcaps (0.5\,mm resolution), and
\item sense wires at the top of the outer APA wire planes (0.5\,mm resolution).
\end{itemize}

Unlike the fixed cameras, the inspection cameras operate only during an inspection. 
It is more practical to use commercial cameras for this purpose, which requires keeping the cameras warmer than -150$^{\circ}$C during deployment. Cameras of the same commercial model were used successfully to observe discharges
in LAr from a distance of 120\,cm~\cite{Auger:2015xlo}.

\begin{figure}[ht]
\begin{center}
  \includegraphics[height=0.3\textheight]{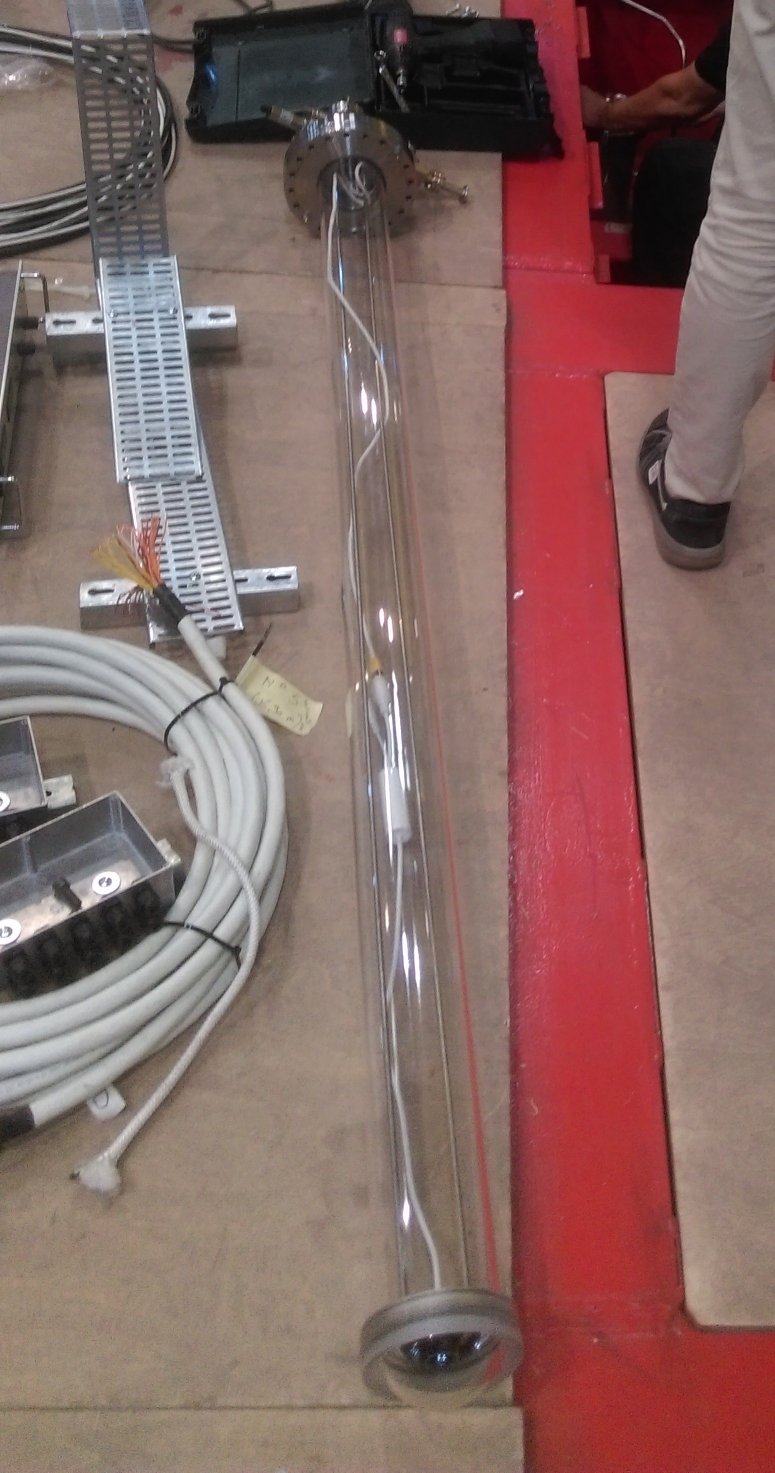}%
\caption{A photo of the \pdsp{} warm inspection camera acrylic tube immediately before installation; the acrylic tube is sealed with an acrylic dome at the bottom and can be opened at the top.}
\label{fig:gen-fdgen-cameras}
\end{center}
\end{figure}

In this design, the warm camera is contained inside a gas-tight acrylic tube inserted into the feedthrough, and can be removed for servicing, upgrade, or 
replacement at any time. Figure~\ref{fig:gen-fdgen-cameras} shows an acrylic tube enclosure and camera immediately before deployment. 
These acrylic tube enclosures 
were deployed at the positions marked 201, 202, and 203 in Figure~\ref{fig:pdsp-camera-locations}, 
and equipped with cameras with fish-eye lenses 
during initial operation.  One camera was then successfully removed without any evidence of having contaminated  the LAr. The other cameras were left in and used during post-beam operations.

\subsubsection*{Light-emitting system}
The light-emitting system uses LEDs to illuminate the parts of the detector in the cameras' fields of view with selected
wavelengths (IR and visible) that the cameras can detect.  Performance criteria for the light-emission system include the light-detection efficiency of the cameras and the constraint on
heat generation 
inside the cryostat. 
Very high-efficiency
LEDs help reduce heat generation; a 750\,nm LED~\cite{lumileds-DS144-pdf} with a specification equivalent to
33\% conversion of electrical input power to light was used.

\begin{figure}[ht]
\begin{center}
  \includegraphics[width=0.53\textwidth]{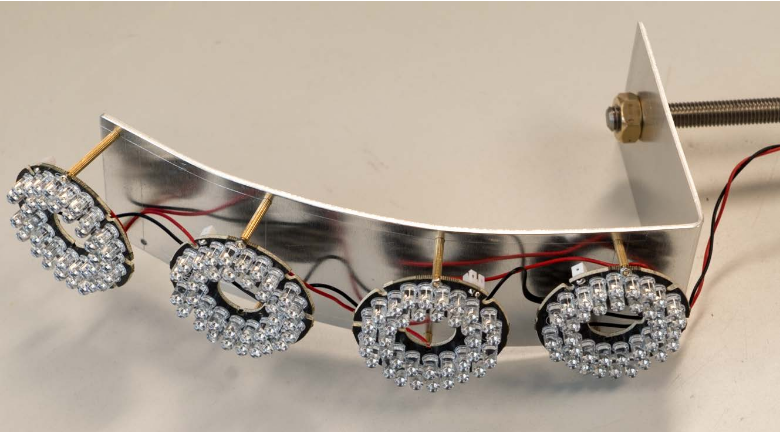}%
  \includegraphics[width=0.3\textwidth]{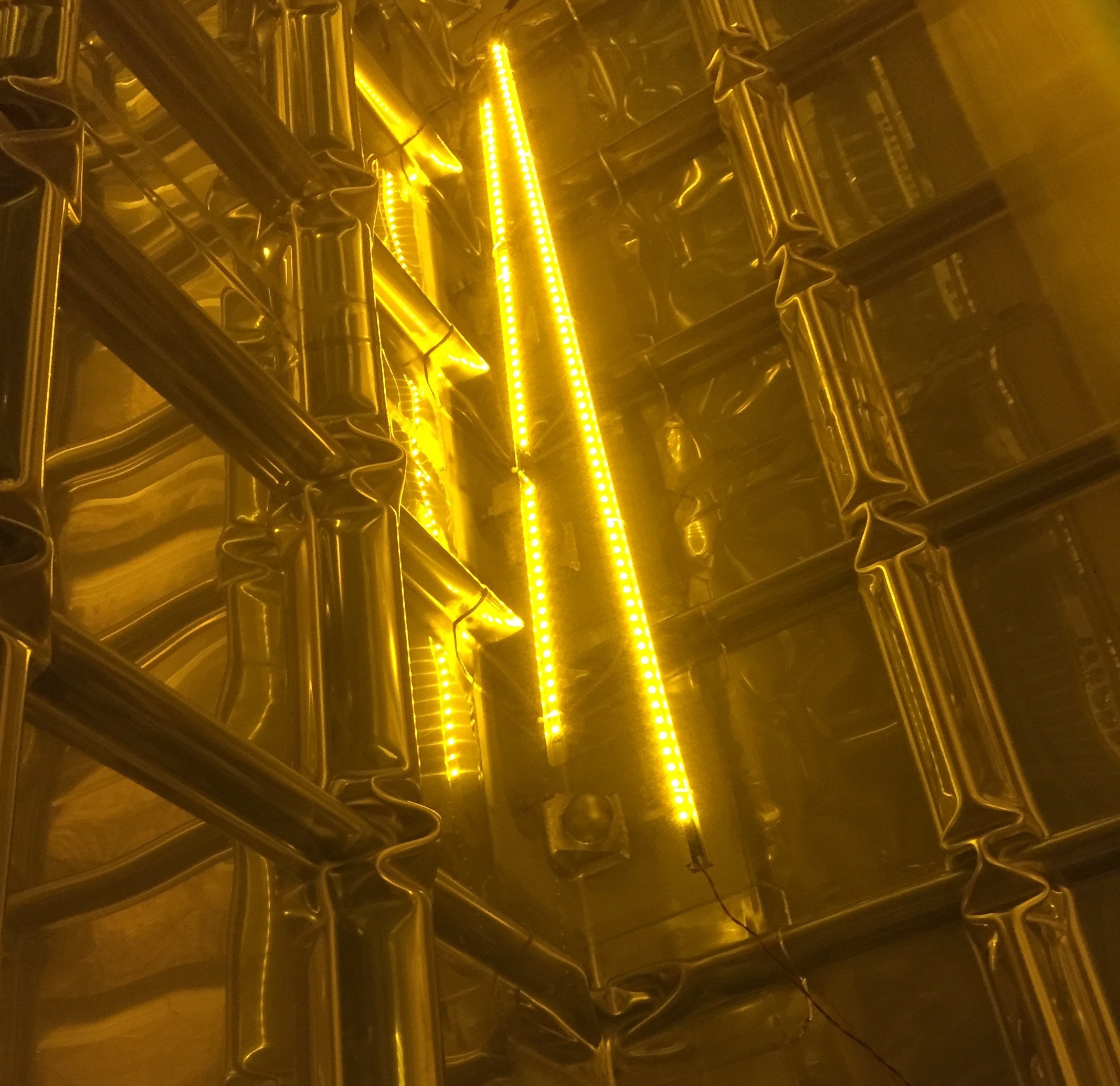}
  \hfill
  \caption{Left: Mounting of IR-LEDs to be placed directly under camera 002 near the beam-left manhole. Right: LED strip used for broad illumination}
     \label{fig:LED-cam002}
\end{center}
\end{figure}

One set of IR-LEDs is mounted directly under  camera 002 near the beam-left manhole, shown in Figure~\ref{fig:LED-cam002}. It is constructed from the IR LED modules that were removed after the initial operation.  Each LED module has 30 LEDs.  This set of LEDs draws about 2.4\,A of current at 12\,V at room temperature, and about 1.8\,A in LAr.  The assembly is installed vertically, with the LEDs aimed upstream of the beam entry behind the beam left APAs. 
When powered, the centres of the LEDs have a slight red glow.

\begin{figure}[ht]
\begin{center}
  \includegraphics[width=0.6\textwidth]{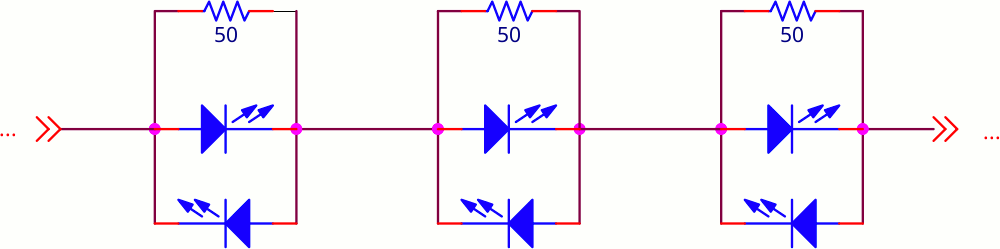}
  \caption{Example schematic for the LED chain, providing failure tolerance and two LED illumination spectra.}
\label{fig:cisc-LED}
\end{center}
\end{figure}

Additionally, chains of LEDs connected in series and driven by a constant current 
are used for broad illumination (see Figure~\ref{fig:LED-cam002} right), with each LED paired in parallel with an opposite polarity LED and a resistor (see Figure~\ref{fig:cisc-LED}).
This provides two different wavelengths of illumination by simply changing the direction of the drive current, and allows continued use of the chain even if an individual LED fails.

         \cleardoublepage

\section{Detector Assembly, Testing and Installation} 
\label{sec:Assembly}
\subsection{Detector Assembly}
\label{sec:assy:ship}
\begin{figure}[htb]
\centering
\includegraphics[width=0.9\linewidth]{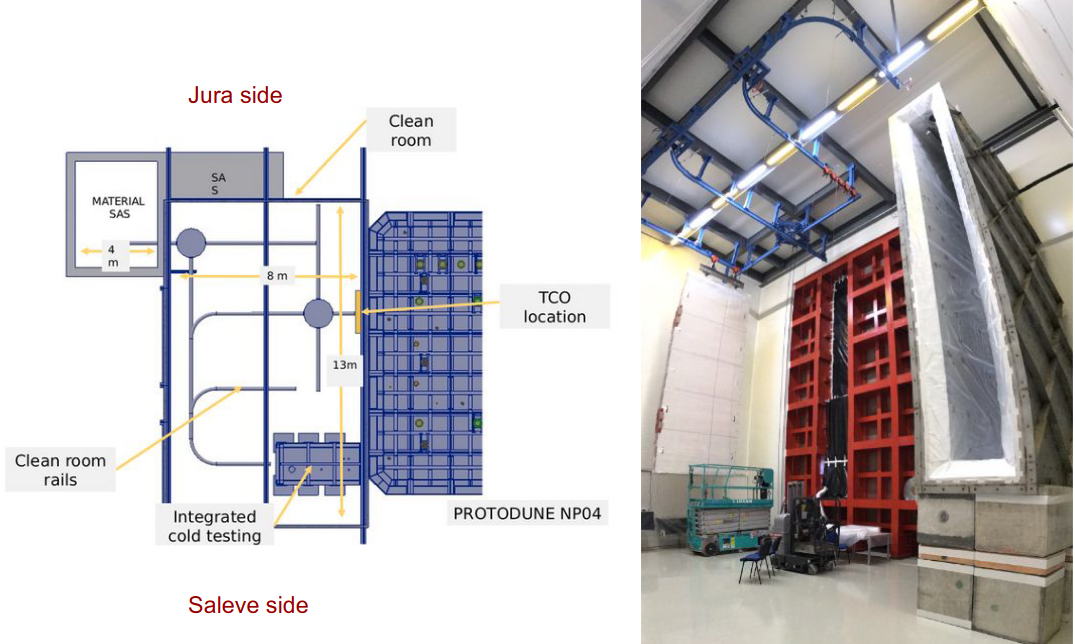}
\caption{ \label{fig:cleanroomwithAPA} Left: Schematic of the NP04 clean room showing the rail system, cold box, and its orientation with respect to the cryostat.  Right: A photograph taken inside the NP04 clean room. The blue structure attached to the roof is the rail system. It is shown supporting a (still covered) APA at left. The tall metal box on the right is the cold box.  
}
\end{figure}

The detector materials arrived at the EHN1 building in containers shipped from the production sites. The majority of the assembly and testing for \pdsp{} took place in the NP04 clean room inside EHN1. This space is equipped with a rail system to hang detector components, and move them into the integrated cold test stand, the cold box, and eventually  
into the cryostat for installation. Figure~\ref{fig:cleanroomwithAPA} shows the rail system and cold box in the clean room. The clean room satisfies the ISO-8 level of cleanliness 
and is equipped with filtered lights 
used to protect the photon detector coating.

As detector materials for \pdsp{} were brought into EHN1, they were passed into the material ``sas'' (the buffer zone between the EHN1 hall and the clean room) 
through its removable roof, unpacked and/or cleaned as necessary, hung from the rail system, then transported through a set of large doors into the clean room testing and assembly area where the following activities took place: 
\begin{itemize}
\item attachment of FC assemblies to CPA modules;
\item unpacking and testing of the PDS elements, and installation on the APA frames;
\item unpacking and testing of the CE elements, and mounting on the APAs; 
\item integrated testing of APA with PDS and CE; and
\item when ready, passage through the temporary construction opening (TCO) in the side of the cryostat for installation.
\end{itemize}

\subsubsection{APA Preparation and Integration}
\label{sec:assy:ship:apa}
\begin{figure}[htb]
\centering
\includegraphics[width=0.9\linewidth]{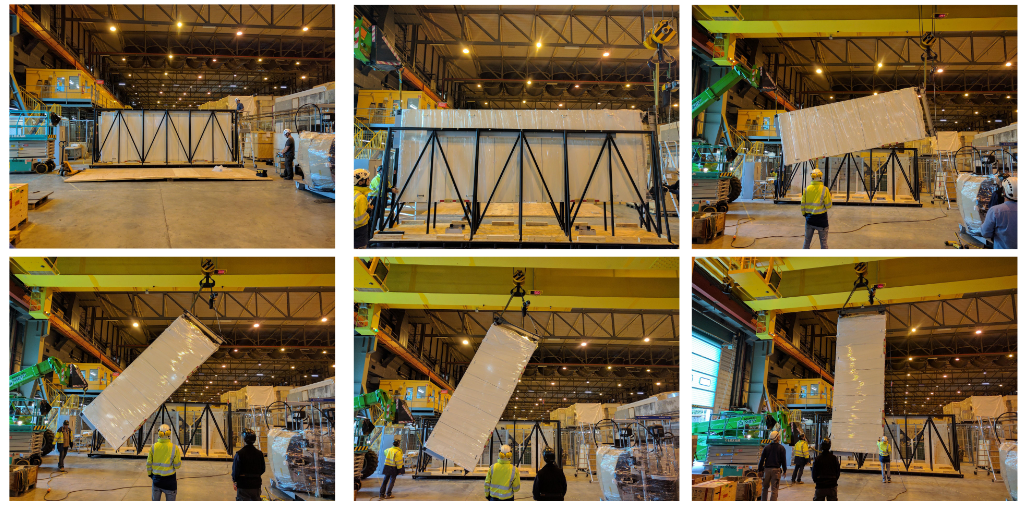}
\caption{ \label{fig:apa-tooling} Arrival of an APA in EHN1 and its positioning. 
The upper left image shows the orientation of the APA as delivered, the lower right shows its vertical orientation when it is lowered into the material sas.  
}
\end{figure}

As each arriving APA container is opened inside EHN1, special lifting fixtures are attached to each end of the APA, then   
attached to two conveyances 
that lift the APA out of the container, and rotate it 90$^\circ$ to vertical. The sequence is shown in Figure~\ref{fig:apa-tooling}.

Once an APA is 
properly oriented, the lifting strap and fixtures are removed from its lower edge, the roof hatch on the material sas is opened, and the APA is lowered through the hatch. It is then transferred to a rolling trolley attached to a series of rails, and moved into the clean room. The APA then goes through a series of acceptance tests for both electrical integrity and wire tension, as well as an 
inspection for broken wires or any other damage that could have resulted from shipment and handling. At this stage, survey of the APA geometry was also performed. 

\begin{figure}[htb]
\centering
\includegraphics[width=0.9\linewidth]{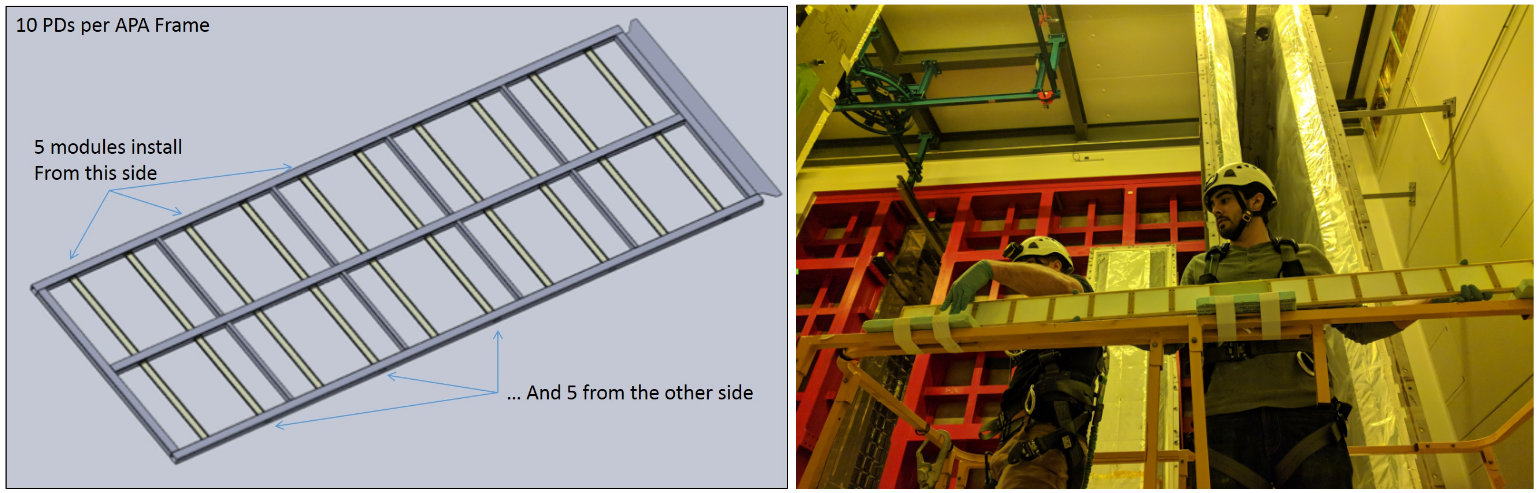}
\caption{\label{fig:pds-install}Left: PD slots (yellow) on the APA frame. Right: Installation of PDs on an APA.} 
\end{figure}

Next, 
ten photon detectors (PDs) are inserted into alternating sides of the APA frame, 
five from each direction.    After insertion, a PD is attached mechanically to the APA frame with fasteners, and a single electronics cable is attached and strain-relieved.  Each PD is tested immediately after installation to ensure proper operation and to verify the cable readout.  The design model and a photograph of an APA in the clean room during PD installation are shown in Figure~\ref{fig:pds-install}.

\begin{figure}[htb]
\centering
\includegraphics[width=0.45\linewidth]{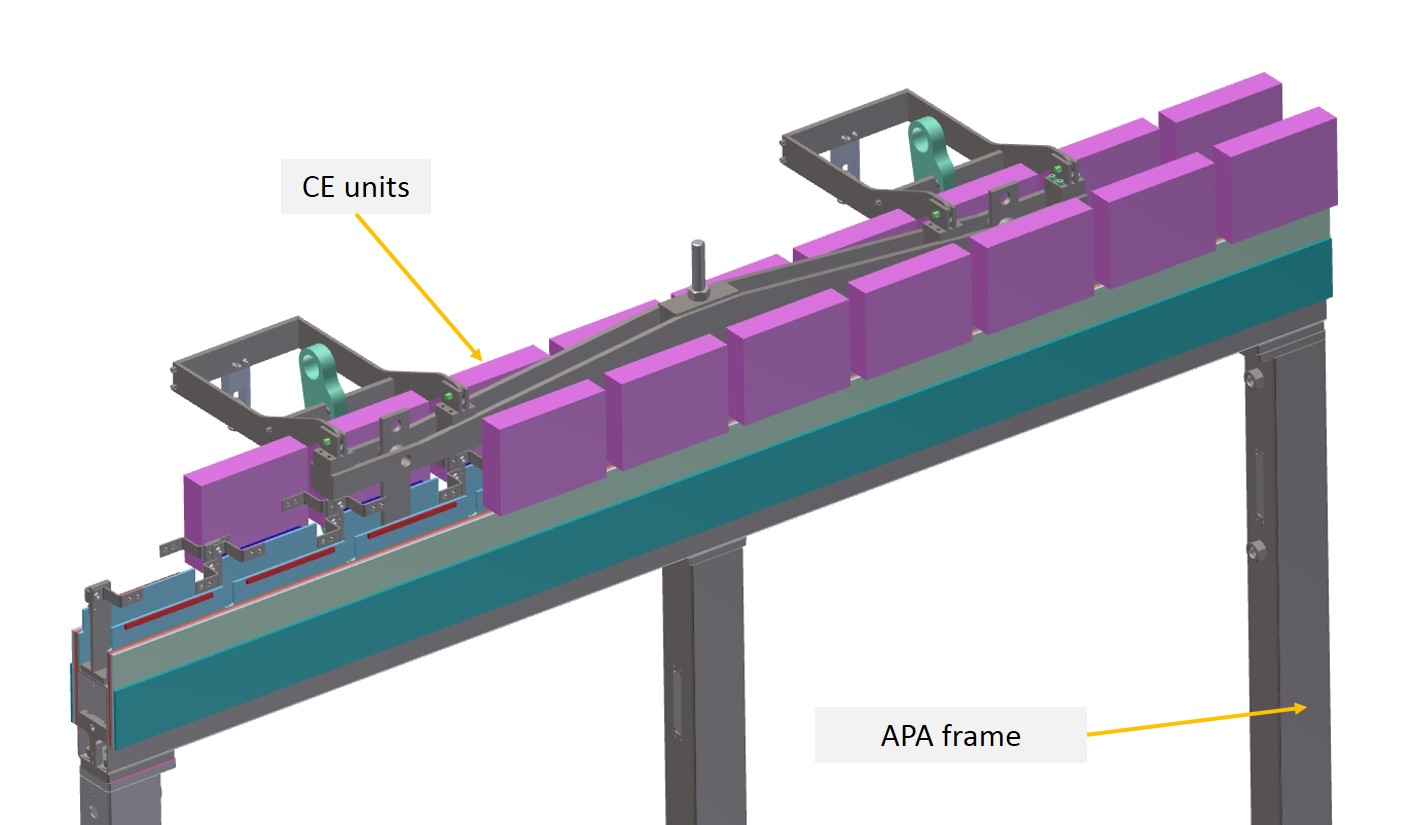}
\includegraphics[width=0.45\linewidth]{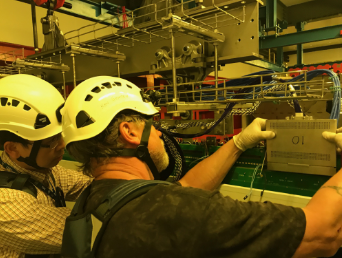}
\caption{\label{fig:ce-install}CE installation.}
\end{figure}

After PD installation, CR boards are mounted on the geometry boards then 20 cold electronics (CE) boxes are installed at the top of the APA frame. Figure~\ref{fig:ce-install} indicates the location of the CE boxes on the APA and shows them being installed.  These boxes are connected via matching electrical connectors on the FEMB. Mechanical fasteners affix the CE enclosure to brackets supported by the APA frame. Prior to their installation, the CE boxes undergo testing several times at room temperature for noise and channel response. Only those that are fully functional and have ENC levels typical for room temperature operations, 1000 -- 1500 e$^{-}$, are installed on the APA. 

The details of the warm tests and their results can be found in~\cite{bib:ce_installation_paper}. In addition to the TPC read-out electronics, each CE box includes a bundle of data and power cables that connect  the electronics to WIEC (warm interface electronics crate). 
Cable bundles were grouped together at appropriate lengths for the WIEC connections, and secured on the cable trays.

At this point, the APA is moved via the rails 
to the integrated cold test stand where the warm and cold tests are performed 
(see Section~\ref{sec:assy:box}). 

After completing the cold test procedure the APA 
is slowly warmed up back to room temperature. Then the cold box is opened, the cables are disconnected and secured, and the APA is extracted from the box on the rail system, 
ready to be moved into the cryostat and positioned for the final installation. Installation inside the cryostat is described later in this section.

\subsubsection{CPA, Field Cage Preparation and Integration}
\label{sec:assy:ship:cpa}
Upon arrival at EHN1, each individual CPA module, weighing roughly 24\,kg,  is lifted out of its shipping crate by hand. 

Three 1.16\,m wide, 2\,m tall 
CPA modules are placed on a flat table surface and screwed/pinned together end-to-end to form a 6\,m tall CPA column. Then a crane attaches at the top end of the CPA column with appropriate lifting straps and shackles, and rotates it to vertical. 
  
As each successive CPA column is ready, it is affixed adjacent to the previous one lengthwise with  1\,mm separation. 
Alignment is provided by two pins located on one side of each CPA that  fit into a vertical slot on the adjoining side of the next CPA. 

\begin{figure}[htb]
\centering
\includegraphics[width=.75\linewidth]{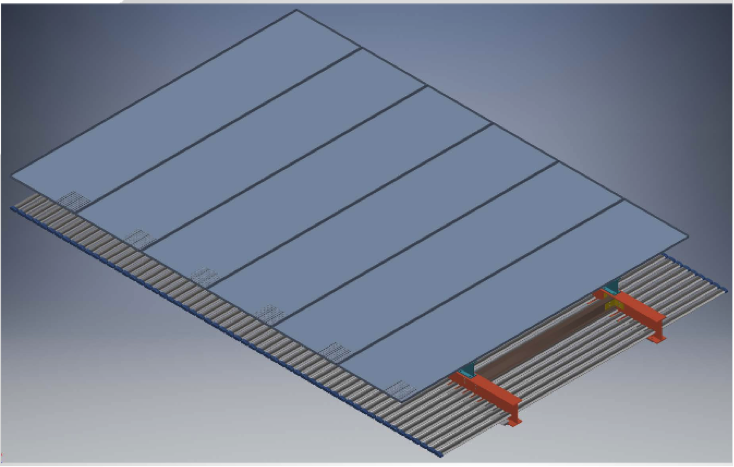}
\includegraphics[width=0.23\linewidth]{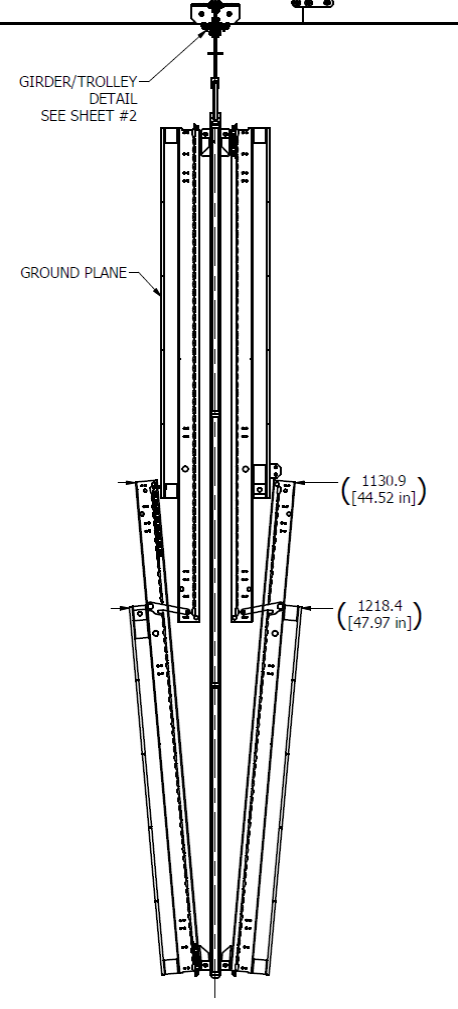}
\includegraphics[width=0.95\linewidth]{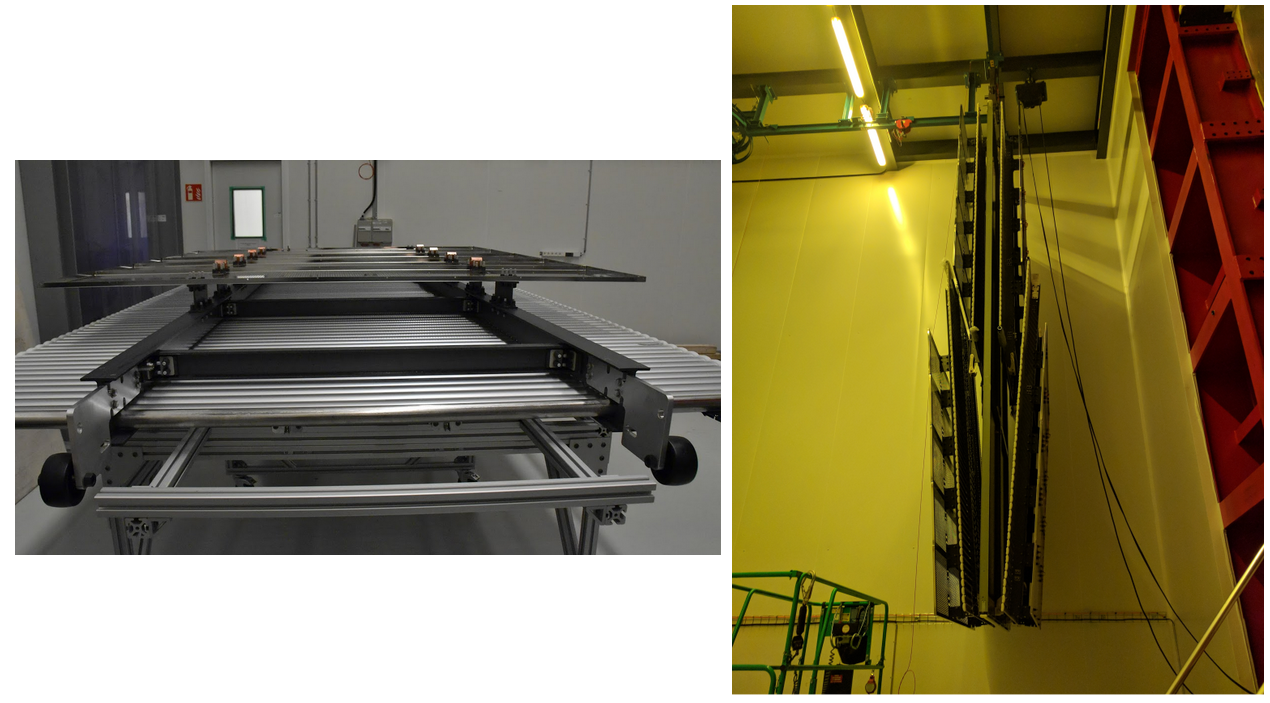}
\caption{\label{fig:fc-assy}Top Left: top or bottom FC assembly (they are symmetrical). Top Right: side view of a CPA pair with four FC assemblies (two top and two bottom) attached. Bottom: 
Photographs of an assembled pair of CPA columns with its four FC assemblies.}
\end{figure}

Three basic elements comprise the FC: the top, bottom and end-wall FC assemblies.  
The top and bottom FC assemblies are effectively mirror assemblies that are hinged from the top and bottom of the CPAs, respectively. 
Figure~\ref{fig:fc-assy} (upper left) is a schematic of a top/bottom FC assembly in which the ground plane covers one side of the field-shaping profiles.  
The upper right image in the figure illustrates a CPA pair with top and bottom FCs attached. The lower two images show units constructed inside the NP04 clean room.

\begin{figure}[htb]
\centering
\includegraphics[width=.48\linewidth]{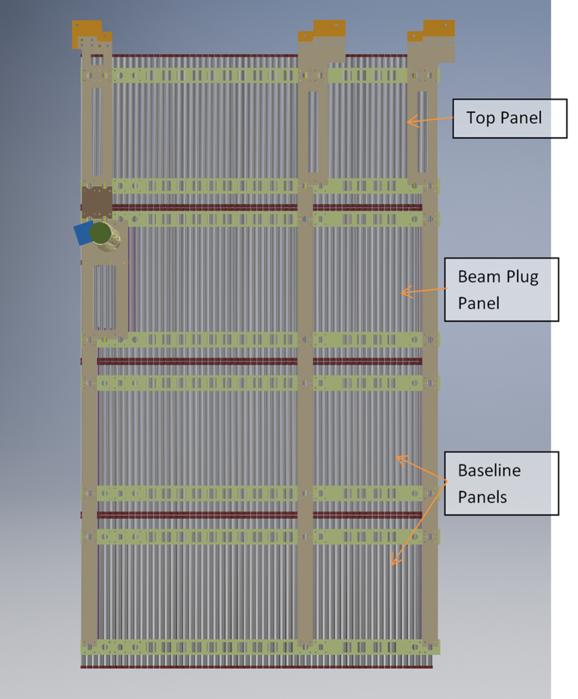}
\includegraphics[width=.46\linewidth]{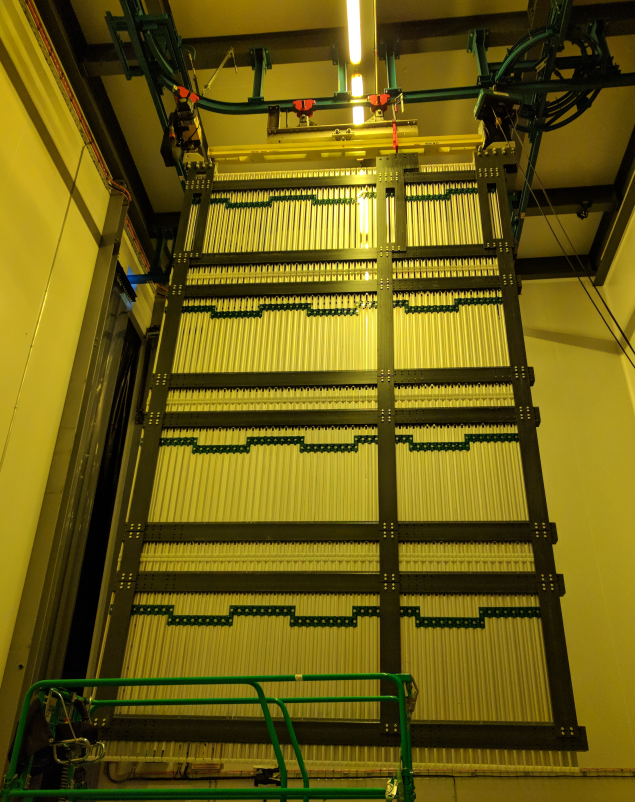}
\caption{\label{fig:fc-end-wall-panel}Left: Design model of the FC end wall panel. Right: Finalized assembly of one of the FC end wall panels. }
\end{figure}

An end-wall FC assembly includes four stacked end-wall FC modules, stacked via the overhead hoist near the TCO.  Figure~\ref{fig:fc-end-wall-panel} shows two images of an end-wall module, the design model and an assembled unit hanging on the rail system inside the clean room. Each of these assemblies was moved into the cryostat via the rails and positioned on the appropriate beam in the detector support system (DSS). 
The end-wall FC is supported by a spreader bar (hung from the same beam) that can swivel about the support point in order to allow proper positioning with respect to the APA and CPA.  

\subsection{Detector Installation}
\label{sec:assy:install}
\begin{figure}[htb]
\centering
\includegraphics[width=0.95\linewidth]{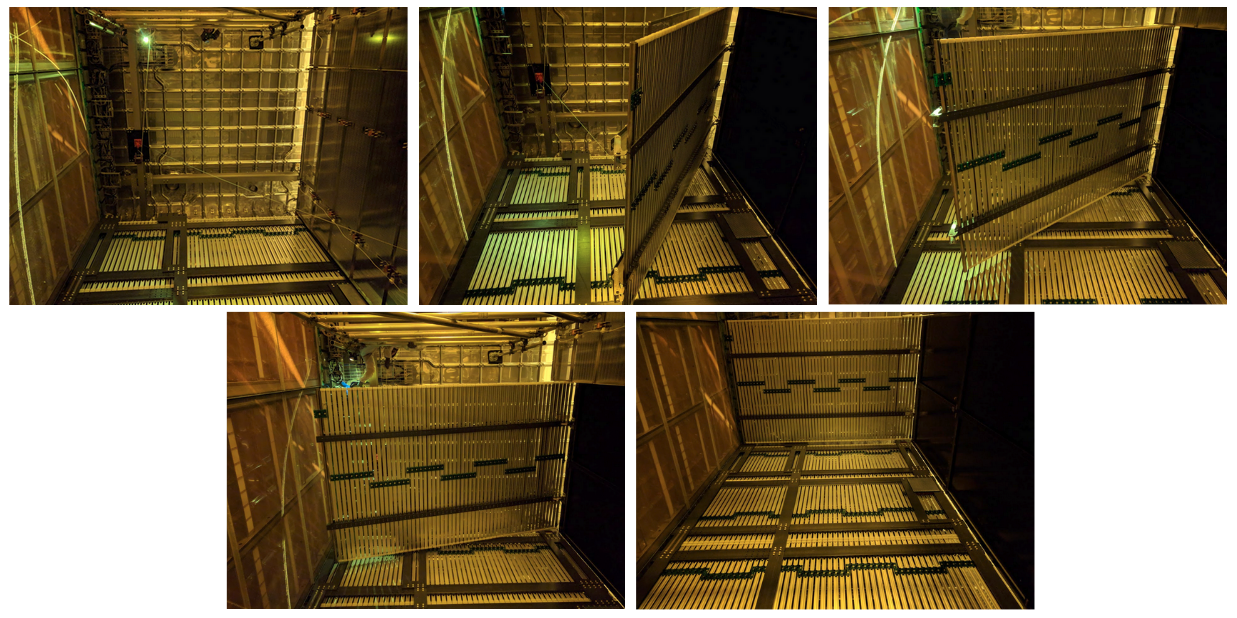}
\caption{\label{fig:deploy-fc-saleve-drift}Field cages deployed in the Sal\`{e}ve-side drift volume.}
\end{figure}

The detector installation steps as performed for ProtoDUNE-SP are outlined below.
\begin{itemize}
\item Each of the six instrumented and tested APAs was moved into the cryostat through the TCO and transferred onto the appropriate rail in the DSS .  They were then assembled into the two anode planes (Salève (south) side and Jura (north) side), of three APAs each. 
\item Signal cables from the TPC read-out electronics boards and from the PD modules were routed up to the cryostat roof 
and connected to the CE and PD flanges on the cryostat. 
Figure~\ref{fig:threeAPAs} shows three Sal\`{e}ve-side APAs hung next to the TCO opening, connected and ready to be pushed into their final installation position. 
\begin{figure}[htb]
\centering
\includegraphics[width=0.85\linewidth]{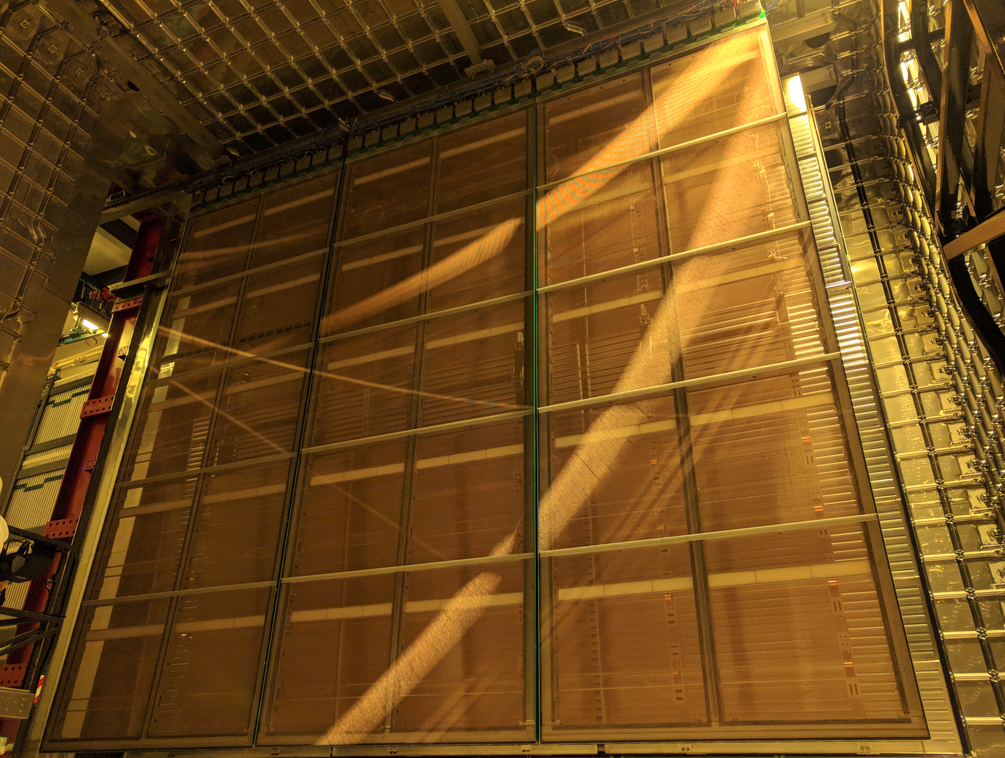}
\caption{\label{fig:threeAPAs}Three APAs installed and mechanically connected inside the cryostat.}
\end{figure}

\item After the first row of APAs was 
in place on the Sal\`{e}ve side, the bridge beam holding the APAs was bolted into place. Then the  two end walls for the Sal\`{e}ve-side drift volume were constructed and moved inside the cryostat, supported by another bridge beam. The left image in Figure~\ref{fig:twoEndWalls} shows the two end-wall units in the cryostat ready for final installation.
\begin{figure}[htb]
\centering
\includegraphics[width=.9\linewidth]{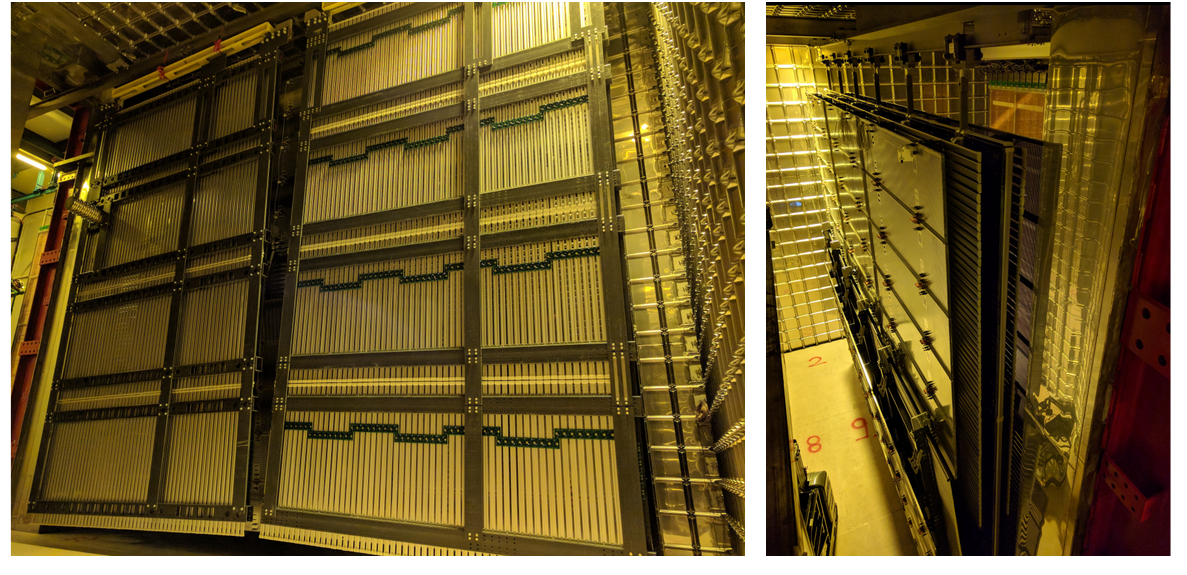}
\caption{\label{fig:twoEndWalls}. Left: Sal\`{e}ve-side drift FC end-walls are ready to be deployed inside the cryostat. Right: CPA and FC assemblies inside the cryostat ready to be pushed to their final position.}
\end{figure}

\item 
Each assembled FC and CPA column was then moved into the cryostat, supported by its own bridge beam. Figure~\ref{fig:twoEndWalls} shows the three CPA/FC units ready to be installed inside the cryostat.

\item Once the CPA/FC beam 
was bolted into position, the end-walls were mounted on the end-wall hangers and the spreader bars that were used to manipulate and move end-walls were removed.  
\item The FC units in the Sal\`{e}ve-side drift volume were deployed as shown in Figure~\ref{fig:deploy-fc-saleve-drift}.
\item After the Jura-side row of APAs were installed, 
the two end-walls for the Jura-side drift volume were constructed and moved inside the cryostat, supported by their own bridge beam. 
\item At this point 
all the TPC components were inside 
the cryostat and the TCO was closed.
\item 
Then the end wall in the Jura-side drift volume were placed into position on their hangers and the FC units were deployed.
\end{itemize}

      \subsection{Cold Box Tests}
\label{sec:assy:box}
\begin{figure}[htb]
\centering
\includegraphics[width=1\linewidth]{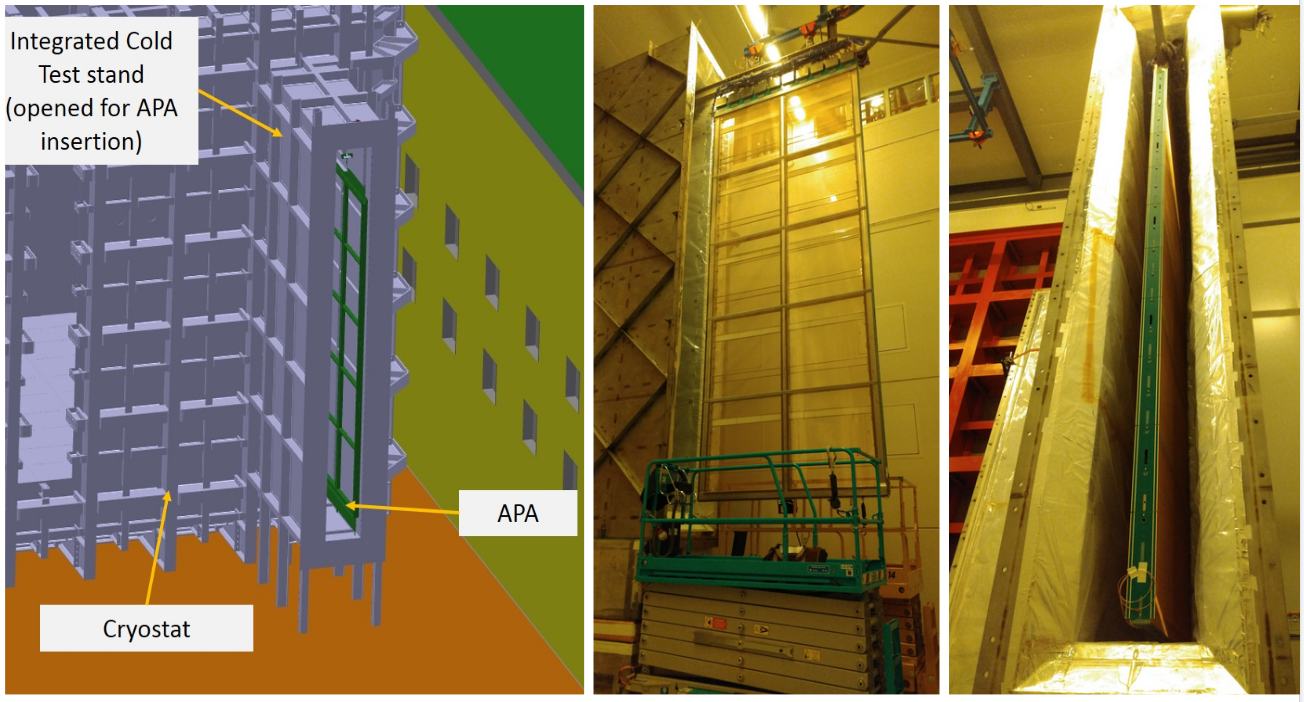}
\caption{\label{fig:cold-test-stand-open}Left: A model of the cold box 
in the \pdsp{} clean room. 
Middle: An APA in front of the cold box ready to be inserted for testing. 
Right: An APA inserted into the open cold box.} 
\end{figure}

The cold box is a large insulated container that    provides electrical isolation, light tightness, and an environment for conducting CE and PD tests on a single instrumented APA at both room and cryogenic temperatures.  Figure~\ref{fig:cold-test-stand-open} shows a model and photographs of the cold box with an APA.
At the top of the box, a crossing tube with a ConFlat fitting, similar to those in the cryostat,  connects to the warm-cold interface flange for the PD and CE cable connections.  Once the cables are routed and connected to their flanges, and an APA is moved inside, the end cap that completes the Faraday cage is put in place, closing the box.  

The cooling system is designed to first
purge the cold box with dry gas then cool the volume slowly to $\sim$150\,K using cold nitrogen gas, and 
maintain the inner volume at that temperature for approximately 48 hours.
The cool-down rate, the same as that used for the cryostat, was less than 10\,K/hr.

 The first set of tests on each APA in the cold box was performed at room temperature to evaluate the ENC performance of each channel with different gain and shaping times. After cool-down, a full set of tests was performed on each APA at low temperature  to assess functionality and electronic noise. Upon completion, the system slowly warmed the cold box volume back to room temperature. At this point, the cold box was opened, the cables were disconnected, and the APA was extracted from the box onto the rail system, and assuming it passed the tests, into the cryostat for installation.
 
\begin{figure}[htb]
\centering
\includegraphics[width=0.7\linewidth]{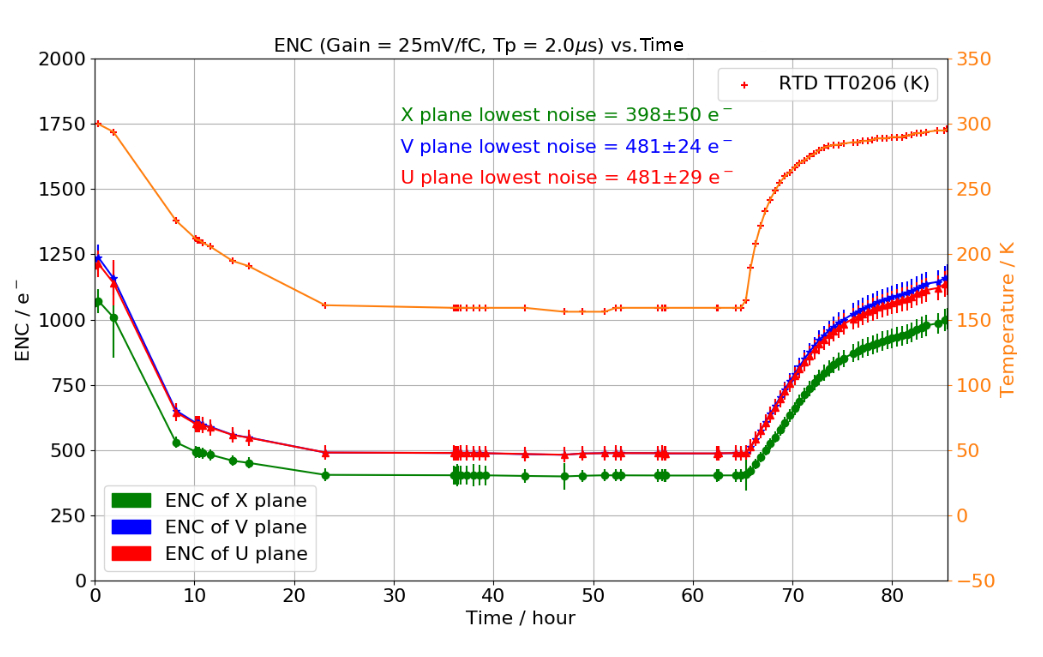}
\caption{\label{fig:cold-box-results}ENC performance in electrons of the APA+CE as a function of time  during the cold box test; red/blue are the induction (U/V) wire planes, and green is the collection (X) wire plane. The orange curve shows the temperature measurements taken nearest the CE Boxes.}

\end{figure}

The measured temperature dependence of the ENC noise  is shown in Figure~\ref{fig:cold-box-results}. The measured ENC was $\sim$1100 e$^{-}$ at room temperature for the collection plane and it dropped to $\sim$400 e$^{-}$ at cryogenic temperatures~\cite{bib:ce_installation_paper}. This agrees with the values measured in standalone tests.


      \subsection{Grounding and Shielding}
\label{sec:assy:gr}


To ensure adequate sensitivity of the \pdsp{} detector, two distinct grounding systems, a detector ground and a building ground, were put in place to isolate it from all other electrical systems and equipment, and to minimise the influence of inductive and capacitive coupling and ground loops. 

The detector ground consists of the steel cryostat outer vessel, the cryostat inner membrane, and all metal structures attached to or supported by the cryostat. The building ground encompasses the network of grounding bus bars and interconnected rebar previously installed in ENH1. A $\sim$1\,mm thick sheet of G10 is installed between the concrete floor and the bottom of the cryostat to provide an insulating barrier and to minimise low-frequency ground and noise currents between the detector and building grounds.

Detector readout racks are placed on or near the top of the cryostat to put them on the common detector ground and minimise the cable runs. All cryogenics and gas piping is connected to the building ground at regular intervals. Dielectric breaks are installed on these pipes near the top of the cryostat.

To avoid structural ground loops inside the cryostat, the APA frames 
are insulated from each other. Each frame is electrically connected to the cryostat at a single point on the CE feedthrough board in the signal flange where the cables exit the cryostat. Mechanical suspension of the APAs is accomplished using insulated supports. 

The analog portion of each 
FEMB contains eight front-end (FE) ASICs configured as 16-channel 
digitising charge amplifiers. 
The ground terminals on the ASICs' input amplifiers  are connected to the APA frame. All power-return leads and cable shields are connected to both the ground plane of the FEMB and to the signal flange.

Filtering circuits for the APA wire-bias voltages are locally referenced to the ground plane of the FEMBs through low-impedance electrical connections. This approach ensures a ground-return path in close proximity to the bias-voltage and signal paths. The close proximity of the current paths minimises the size of potential loops to further suppress noise pickup.

PD signals are carried directly on shielded, twisted-pair cables to the signal flange. The cable shields are connected to the 
cryostat at a second feedthrough, the PDS feedthrough, and to the PCB shield layer on the PDs. There is no electrical connection between the cable shields and the APA frame except at the signal flange.

The frequency domain of the TPC wire and PD signals are separate. The wire readout digitises at 2\,MHz with $<$~500\,kHz bandwidth at 1\,$\mathrm{\mu}$sec peaking time, while the PD readout operates at 150\,MHz with $>$10\,MHz bandwidth. They are separated from the clock frequency (50\,MHz) and common noise frequencies through the FE ASIC and cabling designs. All clock signals are transmitted separately with individual shields to avoid  interference from 
power lines. 

         \cleardoublepage

\section{Detector Readout and Control} 
\label{sec:detcomm}
      \subsection{Data Acquisition}
\label{sec:readout:daq}

The function of the DAQ System is to orchestrate the physics data taking. It collects data from the sub-detectors and the Cosmic Ray Tagger (CRT), and conveys the event data files and metadata to the offline computing system. 
An important aspect of the DAQ is its triggering system. The Central Trigger Board (CTB) receives signals from the Photon Detector readout electronics, the CRT and the beam instrumentation detectors, and forms trigger candidates. The timing system receives those trigger candidates, applies dead-time rules, and forms a global trigger signal associated with a unique timestamp. It then distributes the trigger to the detector electronics and to other DAQ components in such a way that all recorded data corresponding to a fixed time-window around the trigger timestamp can be captured.
Dedicated software applications take care of acquiring the data from the readout electronics and  packaging it appropriately into files that are kept in a local temporary storage (in the DAQ cluster) awaiting their transfer to  permanent storage, which is at both CERN and at Fermilab. 

\begin{figure}
\centering
\includegraphics[width=.7\textwidth]{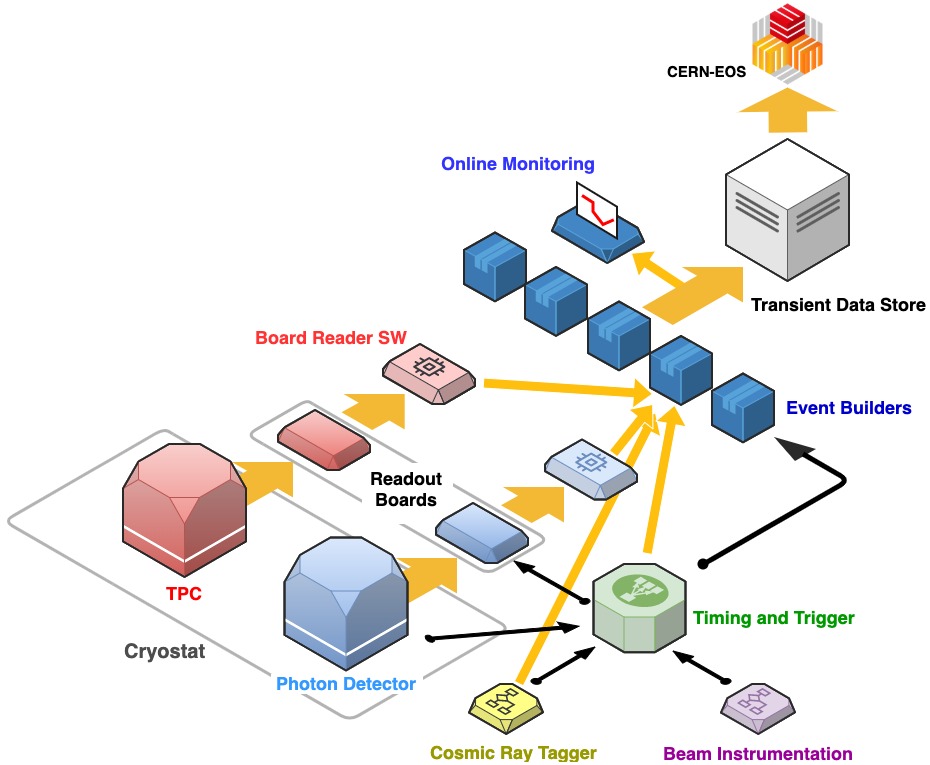} 
\caption{Overview of the DAQ system illustrating its interconnections, data flow, timing and trigger signals, and the interfaces to the front-end electronics and offline computing systems. The TPC readout systems (RCE and FELIX) receive data from the WIBs, which are then compressed and selected based on trigger information. Triggered data are then sent to the event-building farm and subsequently stored to disk. In parallel, a sample is prepared for online monitoring. Triggers are formed using inputs from the beam instrumentation, the PDS, and the CRT.}
\label{daq-overview}
\end{figure}

The DAQ software is also responsible for other functions, including the delivery of configuration parameters to the front-end electronics, the overall Run Control and the real-time monitoring of the data quality and performance of the DAQ system. Figure~\ref{daq-overview} illustrates the DAQ system and its interconnections.

Due to the extremely tight construction and commissioning schedule of \pdsp, the DAQ has been designed to use commercial off-the-shelf (COTS) components and readily 
available electronics boards:
the notable exceptions are the CTB and the timing system, which were custom developed for \pdsp. 
Most of the implemented solutions for the firmware and software are based on existing frameworks and generic technologies. 

The DAQ system design is driven by the need to support both the TPC and PDS front-end electronics, which have very different behaviors.  TPC data are streamed out at 2\,MHz without zero suppression. This allows for the acquisition of an unbiased dataset to be used offline to study data reduction methods for DUNE-SP. The photon detector electronics, on the other hand, applies a local threshold to the signals and sends to the DAQ only waveform data that are either above this threshold or that match an external trigger timestamp. 

The DAQ design is also constrained by the amount of data that can be transferred to the offline computing system for permanent storage. To down-scale the acquired data volume, the trigger system supports selection of data using inputs from the beam instrumentation detectors and the CRT.

Finally, the DAQ system is designed to support partitioning so that detector components can operate independently; this is particularly relevant for dedicated commissioning 
and calibration runs. 

\subsection*{Data rate}
At ~430 Gb/s, the TPC electronics data dominates the data volume. For the six APAs, data pass through a set of 30 Warm Interface Boards (WIBs), located on the cryostat flanges, on their way to the DAQ.

PDS data is collected by an array of 24 SSPs, and consists of a combination of externally triggered events and self-triggered events. The self-triggered data rate is tuned through the bias voltage settings to aim for a data rate into the DAQ of $\sim$\,3\,Gb/s. The externally triggered data contribute much less than 1\,Gb/s.

The DAQ was set up for a target trigger rate of 25\,Hz for the beam run, with a readout window of up to 5\,ms. These values were chosen to ensure that the bandwidth would stay well within the maximum of 20\,Gb/s for data transfer from the \pdsp online system to CERN's OpenSource Storage (EOS)~\cite{eos}. Using data compression, the typical beam data DAQ output bandwidth in \pdsp{} was on the order of 5 -- 8\,Gb/s.

\subsection*{The Timing System}
The timing system provides a stable and phase-aligned master clock for all the DAQ components. In \pdsp{}, a GPS-disciplined oscillator sends high-quality clock signals to an FPGA-based master unit, and the trigger system and SPS accelerator send external signals to it.  The master unit multiplexes the synchronisation and trigger commands into a single encoded data stream. This stream gets broadcast to all timing endpoints (the WIBs, the SSPs and the DAQ components) that require it, and is decoded into separate clock and data signals, ultimately providing synchronous triggers and  timing signals to all endpoints.


\subsection*{Trigger system} 
\label{trigger-section}
\begin{figure}
\centering
\includegraphics[width=.7\textwidth]{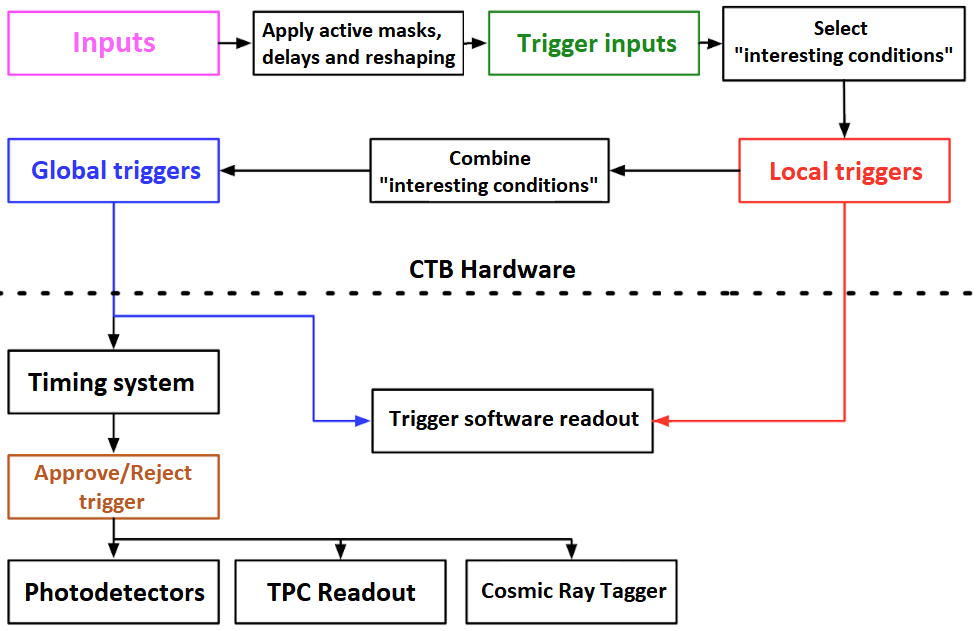} 
\caption{The Central Trigger Board (CTB) is designed to receive triggers from various subsystems and combine them into a global trigger based on a configurable input mask (or more sophisticated algorithm, if desired). The CTB provides functionality to time-stamp triggers globally, keep event counts, and provide information such as trigger type and error conditions.}
\label{fig:trigger}
\end{figure}

Figure~\ref{fig:trigger} illustrates the trigger system logic. The Central Trigger Board (CTB) receives information from the beam instrumentation, photon detectors, and the cosmic ray tagger (CRT), and forms trigger candidates that are passed on to the timing system. Due to the high rate of cosmic rays, the photon detector trigger inputs were 
not used in the trigger logic for \pdsp{}.
After applying any dead-time conditions, the timing system generates a global trigger signal and distributes it to the WIBs, the SSPs, and to the DAQ components, as needed. 
A simple dead-time logic was implemented for \pdsp{} that kept triggers at least 10\,ms apart from each other to ensure that the readout windows of two consecutive triggers would never overlap and to avoid rate spikes above 100\,Hz.

The CTB is designed around the MicroZed System-on-Chip (SoC) board, equipped with a Xilinx Zynq7020. The motherboard implements the hardware interface to the various systems, the FPGA implements trigger logic and interfaces with the timing, and the CPU and software elements manage the FPGA configuration and the communication with DAQ software. This solution supports up to 100 separate inputs that can be combined into a global trigger based on a configurable input mask (or more sophisticated algorithm, if desired). The firmware is designed to be suitably flexible such that the exact trigger selection can be decided at configuration time. Regardless of the trigger configuration, the decision is made in under $1~\mu$s.

\subsection*{Readout systems} 

\pdsp{} is equipped with three readout systems, two for the TPC and one for the PDS. 
The two TPC solutions, ATCA-based RCE and PCIe-based FELIX, are implemented for testing purposes.

The Reconfigurable Cluster Elements (RCE)~\cite{Herbst:2016prn} based readout, developed by the SLAC National Accelerator Laboratory, is a full meshed distributed architecture, based on networked SoC elements on the ATCA platform. The RCE was chosen as the baseline readout solution for \pdsp{} since it had already been successfully used in the 35\,ton prototype~\cite{Adams_2020:35ton}, and during \pdsp{}'s beam run it read out five of the six APAs (12,800 channels). 
RCE-based readout focuses on early data processing, with tightly coupled custom firmware and software implementations. The \pdsp{} version of the RCE accepts digitized raw data from the WIBs over optical fibre. 
The primary processing functions of the RCEs are compression and buffering of the raw data. The RCEs send the data to the back-end upon receipt of an external trigger. An output data rate of $\sim$\,1\,Gb/s per RCE can be sustained and a compression factor of four has been achieved~\cite{2019EPJWC.21401025T}. 

\begin{figure}
\centering  
\includegraphics[width=.45\textwidth]{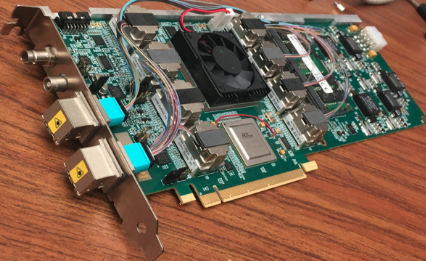} 
\caption{The FELIX BNL-711 PCIe card.}
\label{fig:felix}
\end{figure}

The Front-End LInk eXchange (FELIX) readout system, a project initially developed within the ATLAS Collaboration at CERN~\cite{Anderson:2016lfn}, is centered around Peripheral Component Interconnect Express (PCIe) technology.
The FELIX design aims to minimise the need for custom hardware and firmware development and rely instead on commodity servers, networking, and software. The FELIX PCIe card streams data arriving from the detector front-ends into a circular memory buffer in a host PC using a continuous direct memory access (DMA) transfer (with fixed 4\,kB block size). From this stage onward, all data processing is done by software executed on networked servers. Data compression can be offloaded from the CPUs through embedded hardware acceleration modules (Intel QuickAssist (QAT)~\cite{qat}) in order to achieve the required processing performance. Due to the potential of FELIX technology as a flexible and modular solution for DUNE, it was decided to use it to read out at least one APA in \pdsp{} as a proof of concept. A picture of the FELIX PCIe card is shown in Figure~\ref{fig:felix}.

The PDS electronics~\ref{sec:detcomp:inner:pds:readout} is based on a Silicon Photomultiplier (SiPM) Signal Processor (SSP) prototype module, which is a high-speed waveform digitizer with 12 channels per module. Each channel contains a fully differential voltage amplifier and a 14-bit, 150 megasamples-per-second (MSPS) analogue-to-digital converter (ADC) with 2.1\,Gbps data output. The timing is obtained by applying signal processing techniques to the leading edge of the SiPM signal, using the on-board Artix FPGA. It has deep data buffering ($ 13\,\mu s$), and operates with no dead-time up to 30\,KHz per channel. A total of 24 SSPs serve to read out the PDS modules in all six APAs. The DAQ receives data from the SSPs over 24$\times$1\,Gb/s fibers: TCP/IP is used as communication protocol, thus the PDS readout is fully software-based from a DAQ point of view.

For all readout systems, software-based applications called Board Readers read data from the TPC and PDS electronics to prepare event fragments corresponding to triggered time windows and send them to the event-building system.
Within the event-building system, several applications run in parallel. The Routing Master application assigns a specific event builder to each trigger and distributes this information to the Board Readers.
The Event Builders receive event fragments from the Board Readers, assemble them by timestamp into complete events, and store them into files, ready to be transferred to the offline computing. Another important function of the Routing Master is the handling of back-pressure from the downstream parts of the DAQ. If no event-builder application is ready to accept new events to build, the Routing Master informs the timing system, which starts vetoing triggers.

\subsection*{The DAQ software}

The DAQ software depends in large part on existing solutions and frameworks. Its role is to manage and monitor the data-flow, and provide a run control service and online monitoring \cite{Reynolds:2020} processes. 


The primary responsibility of the DAQ data-flow software is to pack and transport the data from the detector electronics acquisition (WIB and SSPs) to the temporary storage on the DAQ transient store. 

This software is based on artdaq~\cite{Biery:2013cda}, a data acquisition toolkit developed at Fermilab. Use of artdaq also gives the software functionality to deliver configuration parameters to the front-end electronics and to perform a real-time monitoring of the data quality (O(10) seconds) and of the DAQ system.

The Run Control is based on the Joint COntrols Project (JCOP)~\cite{jcop} extension for the WinCC-OA supervisory control framework by Siemens\textsuperscript{\textcopyright}. It is in common use at the LHC experiments and is officially supported by CERN. The system provides a number of services: it provides the operator interface for launching, executing, and terminating DAQ applications, it is used for selecting the desired  data-taking configuration, and it provides visualization of operational monitoring data and alerts the operator if any monitored values go out of range. 


\subsection*{Towards the DUNE Far Detector}
In addition to its being an important component for operating and validating the prototype as a whole, the \pdsp{} DAQ has served as a sandbox for validating data acquisition and triggering strategies for DUNE.

The FELIX-based readout has been chosen as the baseline readout system for both the TPC and PDS in DUNE, and the ATCA-based RCE in \pdsp{} has now been replaced so that the TPC readout is all FELIX-based. The DUNE system will be enhanced, adding the capability to carry out some data processing in the firmware (e.g., the extraction of trigger primitives from the detector's data stream).

The design of the timing system has also been confirmed. The hardware will be re-engineered for the DUNE far detector single phase module (DUNE-SP), retaining most of the features and the signalling protocol. 

Besides the scale of the system (\pdsp{} is 4\% of one DUNE module) and the different requirements on resilience, fault tolerance, and DAQ uptime, the the \pdsp{} and DUNE-SP DAQs differ mainly in how they are triggered. In DUNE-SP, the activity in the TPC and PDS, rather external signals, will provide the triggers. This requires that the DAQ be able to self-trigger based on the data received from the TPC and the PDS.

During the 2019 run, \pdsp{} achieved an important milestone demonstrating the feasibility of the proposed trigger approach for DUNE-SP. In this approach, raw waveforms from the TPC wires are sent to the FELIX system and buffered in the memory of high-end servers \cite{Vermeulen:2021}. A highly-optimised software runs on the CPUs of those servers and performs a hit-finding routine on each of the  2560 wires of an APA. These hits, called ``trigger primitives,'' are then processed to form ``trigger candidates'' for each APA. Trigger candidates are consumed in turn by the ``module-level trigger'' that issues trigger commands for the whole detector, for \pdsp{}, all six APAs.

\begin{figure}
\centering  
\includegraphics[width=.9\textwidth]{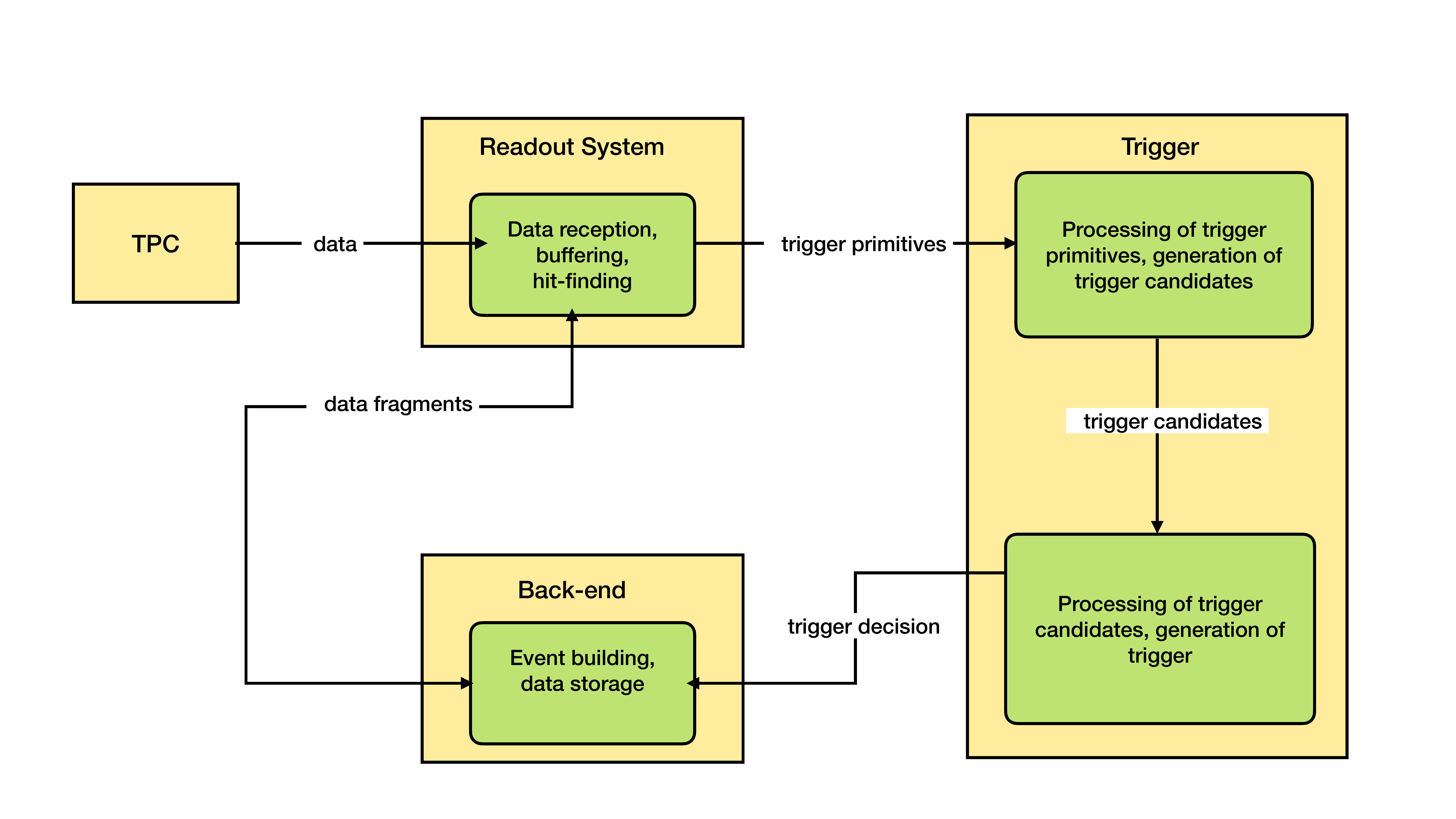} 
\caption{The flow of data in \pdsp{} in the self-triggering scenario.}
\label{fig:dataflow_selftrig}
\end{figure}

Once a trigger is issued, the back-end system collects the data corresponding to a window around the trigger timestamp and stores the data. The flow of data in the self-triggering scenario is shown in Figure~\ref{fig:dataflow_selftrig}. 

A single server can handle data reception, buffering, and hit-finding for a complete APA. To increase the capacity to two APAs in order to minimise space and power consumption for DUNE, further software bench-marking and optimisations are ongoing. Using an FPGA for hit-finding and trigger-primitive formation is also under study, as is the inclusion of PDS data into the self-triggering chain.

      \subsection{Detector Control System }
\label{sec:readout:dcs}

The \pdsp{} Detector Control System (DCS) (known historically as slow control) encompasses the hardware and software elements that ensure the safe and coherent operation of the detector. The DCS monitors all the incoming signals from the detector and provides a uniform interface for access and information exchange to all the detector subsystems. 
An alarm system is configured to report any abnormal conditions to the operators. Depending on the severity of the alarm, the DCS can alert designated experts via email or SMS and take assisted or automatic corrective actions, e.g., if it detects a dangerous situation it can  fire hardwired interlocks to protect the detector and its subsystems. 

Since \pdsp{} must thoroughly validate the detector design and performance, 
the DCS monitors and archives data for online and offline analysis at a much higher granularity and rate than would a conventional DCS system. It archives the sensor data as well as the power supply voltage, current, and temperature variations.

\subsubsection{DCS Overview}

The tight \pdsp{} schedule imposed a requirement that the DCS implement a proven and reliable solution.
The commercial Supervisory Control And Data Acquisition (SCADA) 
WinCC-OA~\cite{SiemensAGSCADAArchitecture} from Siemens, widely used at CERN, was selected as the basis.  The \pdsp{} DCS implements two CERN frameworks on top of WinCC-OA: the Joint Controls Project (JCOP) \cite{Holme2005TheFramework} and the Unified Industrial Control System (UNICOS)\,\cite{Milcent2009UNICOS:FRAMEWORK}. These frameworks provide guidelines, components, and tools designed to facilitate the implementation of a homogeneous control system. Their most salient features are:

\begin{itemize}
    \item automatic Programmable Logic Control (PLC) project generation;
    \item a number of sets of predefined widget tools, faceplates, a Finite State Machine (FSM) toolkit, and a generic configurable Human Machine Interface (HMI);
    \item an alarm and alert configuration system with automatic mail and SMS notifications, derived from the system used by the LHC experiments;
    \item an access control component providing detailed authorisation schemes to assign specific privileges to different user groups; 
    \item a run-time database, accessible for processing, visualisation, etc., in which to store the data coming from the devices;
    \item an archiver for long-term storage of the runtime database and offline use of that data;
    \item a C-compatible scripting language with a very large library of functions that can be used to implement all WinCC-OA functionality; 
    \item standard libraries for commonly used hardware (e.g., CAEN and Wiener power supplies) 
    that allow for centralised configuration of the implemented functions and parameters such as alarms, archiving parameters, operational values, and default panels, all at the crate, board, or channel level;
    \item support of a wide range of drivers and communication protocols such as Siemens S7, MODBUS, OPC DA, OPC UA, the CERN Distributed Information (DIM)\cite{Gaspar2001DIMCommunication}, and Data Interchange Protocol (DIP)\cite{Copy2018MonitoringSystem}. 
\end{itemize}

\begin{figure}
\centering
\includegraphics[width=.7\textwidth]{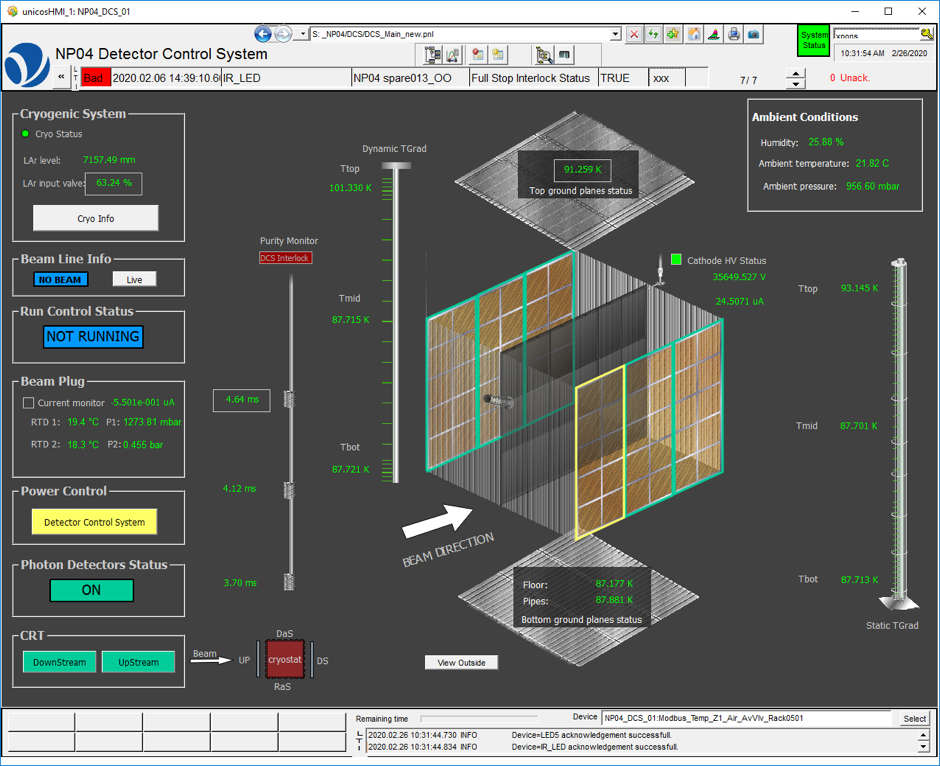}
\caption{The DCS overview display, showing the color-coded status of the detector components, the average temperatures on the two T-gradient rods, and the most recent electron lifetime measurements.}
\label{fig:dcs_advops} 
\end{figure}

To minimise the risk of unintended operator actions and maximise the stability of the system, the DCS interface presents panels that are designed to be as simple and straightforward as possible. The interface relies on standard color coding (see Figure~\ref{fig:dcs_advops}) and is based on two main concepts:

\begin{itemize}
    \item{} dynamic graphical objects that allow the operator to navigate to dedicated panels for each detector subsystem, and 
    \item{} data widgets, objects displayed as data values that allow operations to be performed on them, such as plotting historical values or checking status.
\end{itemize}

\begin{figure}
\centering
\includegraphics[width=.7\textwidth]{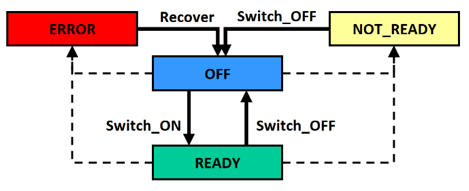}
\caption{FSM flow diagram for turning ON/OFF a detector component.}
\label{fig:dcs_fsm}
\end{figure}

For basic control operations, the DCS uses the Finite State Machine (FSM) component from JCOP~\cite{Gaspar2006ToolsSystems}. The FSM acts as an abstraction layer that simplifies control of the detector. Based on a well defined set of states and transitions, each part of the detector can be controlled hierarchically. Once the topology of the detector is defined, the FSM component is responsible for propagating the actions sent to the different parts of the detector and obtaining, as a result of these actions or asynchronous incidents, their corresponding states. An example of the FSM used for the detector power control is shown in Figure~\ref{fig:dcs_fsm}. Access Control, implemented in the FSM, ensures the safe operation of the detector.

For advanced detector operations, specific and more detailed panels have been designed. Rather than using an FSM for putting the detector – or any of its components – in a preset state, the advanced panels allow users with appropriate authorisation full control over all parts of the detector. The advanced panels control the lowest-level architecture of the detector, allowing the authorised experts to modify operational parameters, load pre-set configurations from files, set limits for alerts, and/or directly control critical devices beyond the scope of the FSM.

\subsubsection{DCS Architecture}

\begin{figure}
\centering
\includegraphics[width=1.\textwidth]{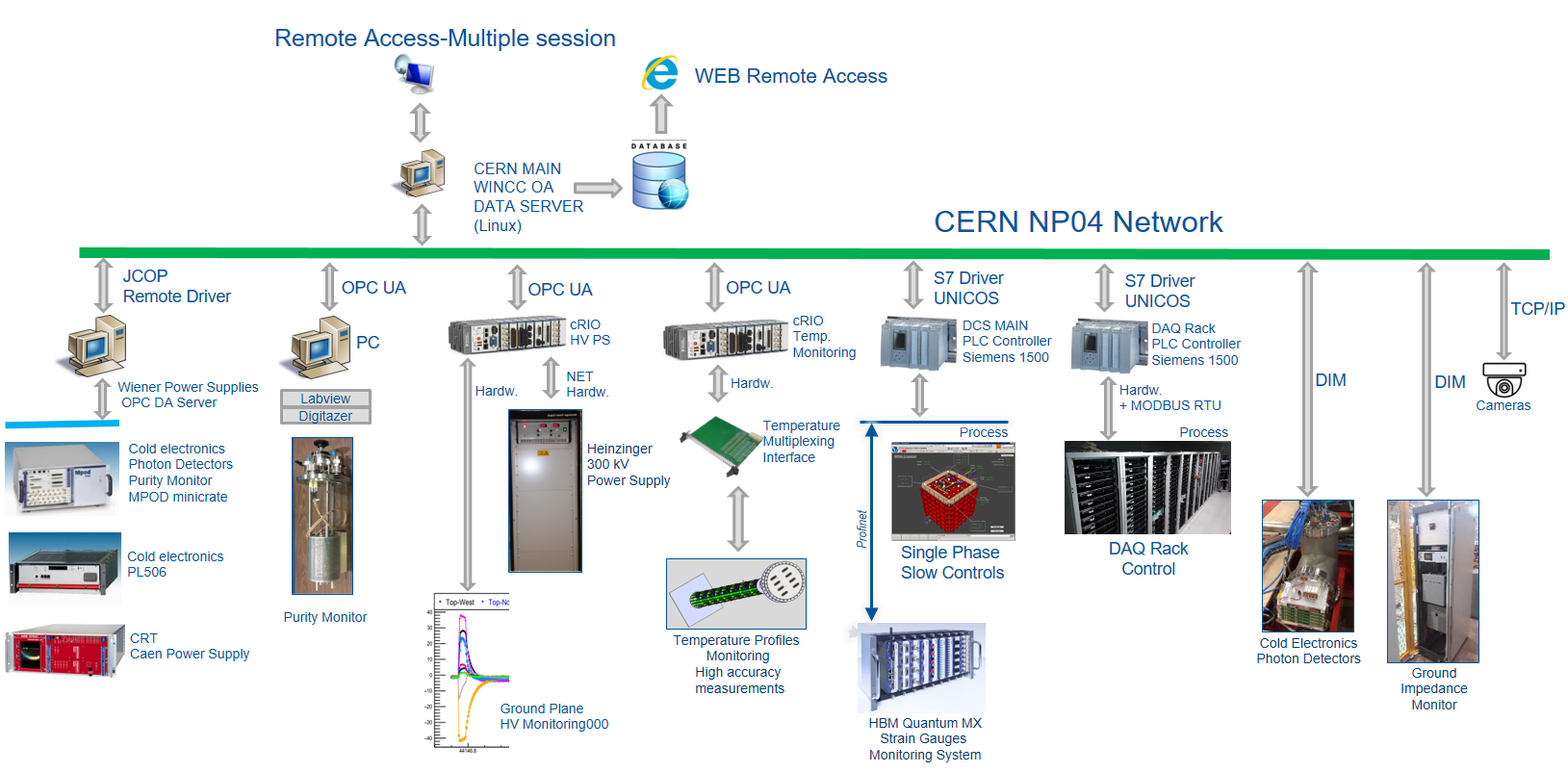}
\caption{DCS architecture and communication protocols.}
\label{fig:dcs_structure} 
\end{figure}

The DCS subsystems are connected to a dedicated \pdsp{} detector network, 
known as the NP04 network. The DCS can communicate with the individual subsystems using a variety of communication protocols, indicated by the white arrows in Figure~\ref{fig:dcs_structure}. Furthermore, the DCS is accessible over the Internet using a remote connection protocol, and the most relevant data are published to a web server for real-time off-site monitoring. A brief description of the different subsystems, and how the DCS manages them, is given below.

\begin{itemize}

\label{subsec_Power}
        \item{\emph{The Detector Power Control (DPC):}} The DPC is a set of power supplies that provide power to the different parts of the detector, including the cold electronics, the PDS, the HV for the field cage, and the CRT. These power supplies are connected to the main DCS application by means of the OPC DA driver~\cite{OPCFoundationOPCSpecifications}.
        The JCOP mass configuration tool is used to set all the power supply parameters, e.g., operating values, addresses, and archiving parameters, and to populate the default control panels.

\label{subsec_WIB}  
        \item{\emph{TPC cold electronics (CE):}} To assess their status, the DCS continuously monitors the main parameters of the CE crates, WIBs, and FEMBs through the detector network by means of the DIM protocol. 
        A hardware interlock between the CE cooling fans and the CE low-voltage power is in place to prevent powering on of the CE in case of fan failure.

\label{subsec_SSP}
        \item{\emph{Photon Detector system (PDS):}} The DCS provides an interface with the PDS for monitoring its working conditions and tuning its configuration for accurate operation. These configurations can be stored in a file and imported by the DCS. The interface is based on the DIM protocol. Critical interlocks have been implemented in the DCS to prevent any potential damage to the PDS when the purity monitors' xenon lamp flashes or the cryostat LEDs are turned on.

\label{subsec_HV}
         \item{\emph{HV system:}} 
        The DCS for the HV system is composed of two tightly correlated components: the controls of the Heinzinger power supply~\cite{HeinzingerelectronicGmbHHeinzinger} that provides HV to the TPC cathode, and the current monitors on the ground planes and beam plug.  A dedicated National Instruments cRIO FPGA~\cite{NationalInstrumentsCompactRIOSystems} performs the HV control process through a high-resolution analogue interface. A novel feature, called Recovery Mode, was developed to resolve streamer events (extended discharges, see Section~\ref{sec:detcomp:inner:hv:hvgp}), which are characterised by high current draw. Recovery mode automatically lowers the HV setpoint while continuously monitoring the current draw at the power supply until the current draw is below the current limit. It then restores the HV setpoint to its nominal value.
        
         An example of an automatic streamer recovery is given in Figure~\ref{fig:HV_streamer}. The current signals from the 12 ground planes and the beam plug are connected to the same HV control system. In the case of a `stream' or discharge event, the DCS records the main HV parameters at 20 kHz for 5 seconds with a pre-trigger of one second into a fast-acquisition file that it sends to the DAQ system for off-line analysis.
         

\label{subsec_PrM}
       \item{\emph{Purity monitors:}} The purity monitor system is controlled locally by a PCIe Alazartech ATS310 digitiser installed in a computer dedicated to this task. The local user interface  
       was developed in LabVIEW software, which  includes an OPC UA server, communicating with the main DCS as a client. The DCS provides a dedicated user panel for monitoring its status and setting the configuration for operation. The DCS can select different internal digitiser working modes and choose the appropriate HV configuration. 
        Once configured, the DCS supports automatically running the purity monitor DAQ with one click, after which it displays all the relevant information. As previously mentioned, to protect the PDS, the DCS implements an interlock system to prevent the purity monitors from operating during data-taking runs.

\label{subsec_Temp}
        \item{\emph{Gradient Temperature sensors:}} Temperature sensors (PT100) are installed in the \pdsp{} cryostat to monitor the temperature gradient of the liquid with high precision ($\pm$ 2mK) (see Section~\ref{sec:ci:t-pro}). A customised electronic interface developed by the collaboration is connected to a National Instruments cRIO FPGA that publishes all temperature values to the main DCS database through an OPC UA server. The same interface controls the position of the dynamic gradient sensor.  The data stored in the database is used for off-line calculations and simulations. The DCS also publishes the maximum and minimum temperatures of the static gradient via DIP; these are used by the cryogenics system for alarm reporting.
        Figure~\ref{fig:dcs_temp_dcs} shows an example of a liquid argon temperature measurement taken during standard operations. A comparison with the predicted values from the Computational Fluid Dynamics (CFD) simulation is also shown.  
       
\begin{figure}
\centering
\includegraphics[width=.7\textwidth]{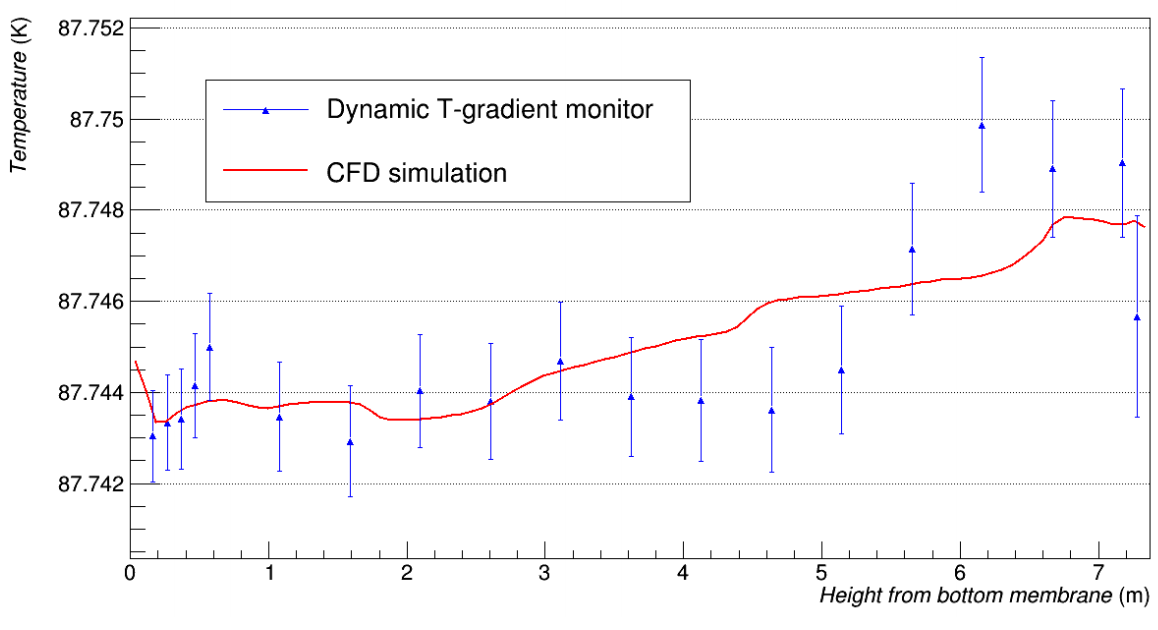}
\caption{Temperature measurement and predicted value as a function of the height of cryostat.  }
\label{fig:dcs_temp_dcs} 
\end{figure}        
\label{subsec_Membrane}
    
    \item{\emph{Ground Impedance Monitor: }} As described in Section~\ref{sec:assy:gr} the detector ground is isolated from the building ground, and the DCS monitors the ground decoupling through a Ground Impedance Monitor (Zmon). This device is connected to the main DCS by means of a DIM protocol. If a grounding problem is detected, the ground impedance monitor sounds an alarm. The DCS alerts operators and records incoming data.

\label{DCS_PLC}    

    \item{\emph{Multi-purpose PLC}:} A S7-1500 Siemens PLC connected to the main DCS application by means of the S7 driver\cite{SiemensAGSIMATICSystems} performs several tasks using the CERN UNICOS framework:

    \begin{itemize}
        \item{} controls and monitors the electrical network and ambient parameters (temperature, atmospheric pressure, humidity); 
        \item{} performs the interlock logic of the CE LV power supply according to the cooling crate status;
        \item{} reads the relative and absolute pressures inside the cryostat;
        \item{} performs the control and interlock logic of the LED light installed inside the cryostat;
        \item{} operates the cameras and the associated heaters;
        \item{} controls heaters on the CE and PDS flanges to dry out water condensation; and
        \item{} ensures the connection to the HBM system that reads the cryostat instrumentation gauges (see Section~\ref{sec:mechtests}) and publishes the values for each sensor to the DCS. 
\end{itemize}

\label{subsec_DAQRack}
        \item{\emph{DAQ environment control:}} Monitoring the racks where the DAQ system is installed is also critical for ensuring the correct recording of the \pdsp{} data. One Siemens S7-1500 PLC controls and supervises the environmental conditions and the main electrical and cooling parameters. In case of a fault, this PLC turns off all DAQ rack power to avoid overheating. A second Siemens S7-1500 PLC is connected to all internal DAQ rack door controls by means of a MODBUS network that acquires the individual rack door statuses and configuration parameters. Both PLCs are integrated in the main DCS through a Siemens S7 driver. They were developed using the CERN UNICOS framework.

    
    \item{\emph{Cryogenics system:}} The DCS and cryogenics systems continuously exchange information; any incorrect cryogenic condition may affect the detector's operation. As the cryogenics control system is installed in a different network, the data exchange is performed by means of a DIP protocol that only allows data subscription by an external system.
    
\end{itemize}


\section{Conclusions} 
\label{sec:conclusions}

The \pdsp{} detector with a total LAr mass of 0.77\,kt is the largest LArTPC operated to date. In the span of five years, from the time of approval in June 2015 to the end of data taking in July 2020, DUNE collaborators designed, constructed, commissioned, and operated \pdsp{}. In successfully validating the technologies planned for DUNE-SP, developing the logistics and installation procedures that will be needed, demonstrating operational stability, developing calibration strategies and techniques, and establishing operational parameters, DUNE has reached every goal it set forth for this prototype detector, an experiment in its own right.

The \pdsp{} cryostat implements for the first time in a 1 kt scale detector, a technology previously used only in liquid cryogen industrial settings and transport, known as membrane technology.  
The membrane cryostat was assembled with no problems and it was validated 
through several test campaigns. A sophisticated cryogenics and purification infrastructure 
ensured successful 
operation at the desired temperature and pressure. 
The flexibility of the system 
allowed for the implementation of alternative solutions during detector operations when needed while still ensuring the same high-quality LAr and control of the various detector operation stages. 
Each of the full-scale detector components underwent a rigorous process of design, manufacture, quality assurance, handling, packaging, transport, assembly, and installation. 
This experience 
has provided valuable input 
for optimising these procedures 
for the first DUNE single phase far detector module. In addition to the detector components immersed in the LAr, 
the detector readout and control system demonstrated excellent 
performance.

The DUNE collaboration gained immeasurable experience with each of the detector components during the period of \pdsp{} operations. The same is true for the CERN and LBNF team that provided the cryogenics infrastructure.
Beyond its role in validating designs, procedures, and strategies planned for DUNE-SP, \pdsp{} provided an excellent test stand for additional 
large-scale detector R\&D, e.g., doping the LAr with xenon, 
which will allow for further improvements in LArTPC technology. 

In summary, \pdsp{}  has successfully implemented this 
mature LArTPC technology 
in a scalable design, demonstrating its suitability 
for an APA-based single-phase DUNE far detector module. 
Its completion in the extremely tight schedule 
attests to the efficiency and reliability of the design and construction techniques. 
\pdsp{} data exhibits 
superb detector performance and signal-to-noise ratio, 
as described in the \pdsp{} performance paper~\cite{Abi:2020mwi}. 
The achievement of \pdsp{} represents a major milestone for DUNE. Further, the knowledge obtained and processes developed during its design, construction, and operation  
are sure to carry forward well beyond this experiment, benefiting other future experiments as well.
 
\section*{Acknowledgements} 
 %
%
The ProtoDUNE-SP detector was constructed and operated on the CERN Neutrino Platform.
We gratefully acknowledge the support of the CERN management, and the
CERN EP, BE, TE, EN and IT Departments for NP04/Proto\-DUNE-SP.
%
%
This document was prepared by the DUNE collaboration using the
resources of the Fermi National Accelerator Laboratory 
(Fermilab), a U.S. Department of Energy, Office of Science, 
HEP User Facility. Fermilab is managed by Fermi Research Alliance, 
LLC (FRA), acting under Contract No. DE-AC02-07CH11359.
%
%
This work was supported by
CNPq,
FAPERJ,
FAPEG and 
FAPESP,                         Brazil;
CFI, 
IPP and 
NSERC,                          Canada;
CERN;
M\v{S}MT,                       Czech Republic;
ERDF, 
H2020-EU and 
MSCA,                           European Union;
CNRS/IN2P3 and
CEA,                            France;
INFN,                           Italy;
FCT,                            Portugal;
NRF,                            South Korea;
CAM, 
Fundaci\'{o}n ``La Caixa'',
Junta de Andaluc\'ia-FEDER, and 
MICINN,                         Spain;
SERI and 
SNSF,                           Switzerland;
T\"UB\.ITAK,                    Turkey;
The Royal Society and 
UKRI/STFC,                      United Kingdom;
DOE and 
NSF,                            United States of America.
%
%
This research used resources of the 
National Energy Research Scientific Computing Center (NERSC), 
a U.S. Department of Energy Office of Science User Facility 
operated under Contract No. DE-AC02-05CH11231.
%

\cleardoublepage

\bibliographystyle{JHEP}
\bibliography{citedb}

\end{document}